\documentclass[a4paper,11pt]{article}
\usepackage{amsfonts}
\usepackage{latexsym}
\usepackage{amsmath}
\usepackage{amssymb}
\usepackage{amssymb}
\usepackage{slashed}
\usepackage{upgreek}
\usepackage{mathrsfs}
\usepackage{bbm}

\usepackage{graphicx}

\usepackage{tikz}

\usepackage[toc,page]{appendix}
\usepackage{amsthm}

\theoremstyle{remark}
\newtheorem{remark}{\bf Remark}[section]
\theoremstyle{definition}
\newtheorem{definition}{Definition}[section]

\newtheorem{theorem}{Theorem}[section]
\newtheorem{corollary}{Corollary}[section]
\newtheorem{lemma}[theorem]{Lemma}
\newtheorem{example}{Example}[section]
\newtheorem{prop}{{\bf Proposition}}[section]

\usepackage[utf8]{inputenc}
\usepackage[english]{babel}

\allowdisplaybreaks

\usepackage{relsize}
\usepackage{graphicx}
\usepackage{comment}
\usepackage{enumerate}

\renewcommand{\thefootnote}{\fnsymbol{footnote}}

\def\appendix#1{\addtocounter{section}{1}\setcounter{equation}{0}
\renewcommand{\thesection}{\Alph{section}}
\section*{Appendix \thesection\protect\indent \parbox[t]{11.15cm}{#1}}
\addcontentsline{toc}{section}{Appendix \thesection\ \ \ #1}}

\font\mybb=msbm10 at 11pt

\def\bb#1{\hbox{\mybb#1}}

\def\bQ {\bb{Q}}
\def\bR {\bb{R}}

\def\bN {\bb{N}}

\def\M {{M}}
\def\S {{S}}
\def\N {{N}}
\def\K {{K}}
\def\B {{B}}
\def\T {{T}}
\def\X {{X}}
\def\Y {{Y}}
\def\A {{A}}
\def\W {{W}}
\def\O{{O}}

\def\vt {\vec{t}}
\def\vv{\vec{v}}
\def\vu {\vec{u}}
\def\va {\vec{a}}
\def\vz {\vec{z}}
\def\vw {\vec{w}}
\def\vk {\vec{k}}
\def\vn {\vec{n}}
\def\vy{\vec{y}}
\def\vl{\vec{\ell}}
\def\vj{\vec{z}}

\def\mC{\mathcal {C}}

\def\bv {{\bf{v}}}

\def\rsq {]\kern -2.0pt]}
\def\lsq {[\kern -2.0pt[}

\def\be{\begin{equation}}
\def\ee{\end{equation}}

\def\rsq {]\kern -2.0pt]}
\def\lsq {[\kern -2.0pt[}

\newcommand{\bea}{\begin{eqnarray}}
\newcommand{\eea}{\end{eqnarray}}

\usepackage{chngcntr}
\counterwithout{equation}{section}

\begin{document}

\begin{center}
%\today
\vspace*{-1.0cm}
\begin{flushright}
%\normalsize{\texttt{ZMP-HH/17-24}}\\
\end{flushright}
%\hfill hep-th/yymmnnn \\
%\hfill UB-ECM-PF-06-43 \\
%\hfill DMUS--MP--13/06 \\

\vspace{2.0cm} {\Large \bf Notes on the singularity theorems of general relativity } \\[.2cm]

\vskip 2cm
 Georgios  Papadopoulos
\\
\vskip .6cm

\begin{small}
\textit{Department of Mathematics
\\
King's College London
\\
Strand
\\
 London WC2R 2LS, UK}\\
\texttt{george.papadopoulos@kcl.ac.uk}
\end{small}
\\*[1.5cm]

\emph{To the memory of Kellogg S. Stelle}
\\
\end{center}

\vskip 2.0 cm

\begin{abstract}
\noindent

    The  notes  begin with an  exploration of the local geometry of spacetime and  proceed with an investigation of its global causal structure.  The results obtained are used to present a proof of the Hawking cosmological and the Penrose  black hole  singularity theorems.  The aim is to give a pedagogical, but still technical,  proof of the statements. The notes include  a large number of drawings to aid the understanding of the material and many examples to explain key concepts.
Some of the mathematical tools utilised in the proof of the singularity theorems are also illustrated. These include an extensive description of the topological and analytic properties  of curves and their spaces,  of the  Arzel\'a-Ascoli theorem and of the properties of semi-continuous functions. The Hopf-Rinow theorem, which describes the conditions  for geodesic completeness in the context of Riemannian manifolds,  is also partially demonstrated.

 %.  A comparison of the behaviour  of geodesics in  Lorentzian and Riemannian manifolds, in relation to the singularity theorems, is given.

\end{abstract}

%\vskip 1cm

%{\small Keywords: spacetime geometry; black holes; special Lorentzian structures; G-structures}

\newpage

\renewcommand{\thefootnote}{\arabic{footnote}}

%%%%%%%%%%%%%%%%%%%%%%%%%%%%%%%%%%%%%%%%%%%%%%%%%%%%%%%%%%%%%%%%%%%%%%%%%%
%\setcounter{section}{0}\setcounter{equation}{0}
%%%%%%%%%%%%%%%%%%%%%%%%%%%%%%%%%%%%%%%%%%%%%%%%%%%%%%%%%%%%%%%%%%%%%%%%%%

%\section{Introduction}
%%\counterwithout{equation}{chapter}
%\maketitle %% A custom one
%\frontmatter
\tableofcontents
%\listoftables
%\listoffigures

 \setcounter{section}{0}

\numberwithin{equation}{section}

\section{Introduction}\label{intro}

General relativity is one of the pillars of modern theoretical and mathematical physics. It underpins all our efforts  to understand  the universe and its internal structure. The main assertion is that the force of gravity has  geometric origin.  Because of this, techniques from differential geometry have been widely  used to explore the theory. The apparent strength of this synergy between general relativity and differential geometry  is the ability to concisely formulate and provide complete proofs to statements.

During the more than 100 years history of the theory, many astonishing results have been discovered. Perhaps one of the most surprising results   of the theory  is
the singularity theorems of Penrose and Hawking \cite{penrose}-\cite{hawkingpenrose}.  These state  that under some reasonable  and physically justifiable assumptions\footnote{These assumptions can be recast via the use of Einstein equations as conditions on the energy momentum tensor of a theory.} on their geometry, spacetimes  exhibit unavoidable  singularities.

The generality of these results is far reaching\footnote{They also have philosophical implications. If the spacetime exhibits unavoidable singularities, nature may not be deterministic even at the macroscopic scale.}.  It settles once and for all in (classical) general relativity the question on  whether or not the singularities of solutions to the theory,  which  had been observed before,  are  generic or are artifacts of the usually made symmetry assumptions on the fields to solve the field equations. The singularity theorems demonstrate that under certain conditions  singularities  are generic. However,  this does not mean that there do not exist  smooth solutions to the Einstein equations with non-trivial matter. Many smooth solutions are known, e.g. the de-Sitter and anti-de-Sitter solutions.

The proof of these theorems is an engaging  journey in topology, differential geometry and analysis. But the results go far beyond this pursue. There is a substantial body of observational evidence that our universe exhibits such singularities in the form of black holes. These are prolific as the remnants of collapsed stars and many of them are located at the centre of galaxies.  The universe itself is thought to have emerged from an initial singularity.

The singularity theorems are existence theorems and give very few clues about the nature of the singularities themselves. Their proof typically relies on demonstrating incompleteness for one or more time-like or null geodesics.  It is known that these singularities form at what appears to be a perfectly normal spacetime enjoying sufficient (at least $C^2$)  smoothness.  This is in direct contrast with the situation in Riemannian geometry where as a consequence of Hopf-Rinow theorem Cauchy completeness, a rather weak and plausible assumption, is sufficient for geodesic completeness.

These notes grew from an informal set of lectures that I gave to PhD students in 2022-23 and 2023-24 at King's College London.  The main sources used are the book by Hawking and Ellis \cite{HE}, the book by Wald  \cite{Wald}, the book by Beem et. al. \cite{beem},  the article by Geroch \cite{Geroch}, the review article by Senovilla \cite{Senovilla} and the more recent review by Minguzzi \cite{Minguzzi}. The singularity theorems had already been well understood by the early 1970's. Since then, there has been a lot of work done by many authors exploring several aspects and extensions of the proof from different perspectives. A very partial list  of  papers, reviews, some with historical summaries, and  books is \cite{Geroch1972} - \cite{Natario2021}; for more recent review articles see \cite{Witten} and \cite{Minguzzi}. Singularity theorems can also be proved under energy conditions which are more suitable for quantum theory \cite{CJFGJG} and \cite{CJFEAK}.

In these notes, a pedagogical approach is adopted for the proof of singularity theorems maintaining at the same time their technical aspects. The objective is to describe the proof of the Hawking cosmological singularity theorem and the Penrose black hole singularity theorem. These are not the most general of singularity theorems but they form a firm basis to understand the area. For this,  we have expanded on the proof of the various statements in an effort to shorten the gap between the logical steps in the arguments and to make them as complete as possible. Inevitably, some results are simply stated.  But the proof of key statements has been given in detail and often from different points of view.
We have elaborated on the various topologies that can be put on curves and their spaces, and provided some illustrative examples to point out the differences. Traditionally, in general relativity compactness arguments  on the space of future directed causal curves rely on the limit curve theorem. We mostly adopt this approach in the proof of the vast majority of statements.

However, we have also commented on the role Arzel\'a-Ascoli theorem which  more recently is used to prove similar key statements.  This is especially the case when we investigate the analogous  result in Riemannian geometry. We have also expanded on the proof of the upper semi-continuity of the length functional, which is essential for the proof of existence of  geodesics with maximal proper time\footnote{We shall also refer to those as \emph{maximal geodesics}.} between two points of a spacetime. The analogous result in Riemannian geometry is more straightforward and we use for its proof  the Arzel\'a-Ascoli theorem and the lower semi-continuity of the energy functional. Also, the non-maximality of geodesics with a conjugate point has been reworked both via a distribution type of argument and a smooth calculation.

To illustrate the various results, we have made use of many drawings. These are useful from a pedagogical point of view. But, they should come with a warning that neither they constitute a proof of a statement  nor they depict accurately the complexity of the underlying mathematical structure. The technical proof of a statement is irreplaceable.

To contrast the properties of geodesics in general relativity  with those of Riemannian geometry, we have included in the appendices a partial proof  of  the Hopf-Rinow theorem. This states, amongst other results,  that Cauchy completeness of a Riemannian manifold implies geodesic completeness.  In addition, we also prove the existence of a minimal geodesics between two points in a (complete) Riemannian manifold.  This is done using a compactness argument on the space of curves and the lower semi-continuity of the energy functional.

  This document is a set of notes and  it is not aimed to provide a comprehensive description of  the progress that has been made in the field the last 60 years after the singularity theorems  were first established. So, many significant recent contributions made by researchers in the field are not included and the list of references is very limited.

 In section two, the normal coordinates  on a spacetime are introduced and the definition of normal and convex normal neighbourhoods is given. The properties of the exponential map are also explained.  Then, the properties of geodesic congruences are described for both time-like and null geodesics. The evolution equations for shear, rotation and expansion are derived -- the latter is known as the Raychaudhuri equation. For time-like geodesics these are given in proposition \ref{prop:sre} and for null geodesics in subsection \ref{ssec:nullcongruences}.  Then, the geodesic deviation equation is introduced and the definition of conjugate and focus points is given. The relation between conjugate and focus points with the Raychaudhuri equation is described in propositions \ref{prop:expansion}, \ref{prop:genericnull} and \ref{prop:focuspoint}.
 The section concludes with a detailed description of the local geometry and  local causal structure of a spacetime.  The results are summarised in proposition \ref{prop:lcs} and a detailed proof is given. The approach taken in \cite{Senovilla} is mostly followed.

 In section three, the properties of causal curves are explored. After some definitions that establish the basic concepts, the limit curve theorem is demonstrated in theorem \ref{prop:limitcurve}.  This is a key result and it is widely used to prove  many of global properties of a spacetime. Its significance cannot be overstated.   The theorem establishes the existence  of a causal curve that passes through the accumulation point of a sequence of other causal curves.  Two different proofs are given. One is that described in \cite{HE} that represents the traditional approach.  The other uses that Arzel\'a-Ascoli theorem that it is considered to be a more recent development. The latter gives the opportunity to discuss some subtle issues regarding the behaviour of end points of causal curves as limits are taken. These issues have been known for sometime and have been examined in detail in \cite{Minguzzi2008}.  Another significant result included in this section is presented in proposition \ref{prop:causaltime}. This states that any causal curve that it is not null geodesic can be approximated  by time-like curve included in a neighbourhood of the original causal curve.  Again, this result is indispensable in the investigation of the global causal structure of a spacetime.

 In section four, the global causal structure of a spacetime is investigated. The definitions of chronological and causal future and past of a set in a spacetime are given.  After that the properties of all these sets and their boundaries are described in proposition \ref{prop:chronos}. Because of the importance of this proposition in what follows, an effort has been made to give a detailed proof of the described statements followed by many diagrams that hopefully elucidate the underlying structures.  Then, the concept of achronicity of a set is presented followed by the proof that achronal boundaries admit an appropriate manifold structure.  The latter is given in proposition \ref{prop:boundary}. This statement is used later in the proof of the Penrose black hole singularity theorem.  The section concludes with the proof of proposition \ref{prop:Iboundary} that describes the nature of the boundary of the chronological future of a set. It states that points on such a boundary lie on null geodesics that either are past inextendible or end on the set, i.e. it states that such boundaries are null hyper-surfaces.

 In section five, we begin with the definition of causal, strongly causal and stably causal spacetimes. The differences between the three different notions of causality are elucidated including the use of figures. Most of the results explained are focused on the properties of strongly causal spacetimes. Then, the definitions of  future and past domain of dependence of an achronal closed set of a spacetime are given.  There is an extensive investigation of the properties of these domains following mostly the results of \cite{Geroch} as well as \cite{HE} and \cite{Wald}. These properties include a description of the closure of the domain of dependence explained in proposition \ref{prop:dopen} and their boundaries  -- these results mirror those established in proposition \ref{prop:chronos} for the chronological future of sets though there are significant differences.  Then, the definition of the Cauchy horizon of an achronal closed set is given, which is part of the boundary of its domain of dependence.  It is  demonstrated in theorem \ref{th:cauchyhorizons}  that future Cauchy horizons are achronal sets and their points lie on either past inextendible null geodesics  or on null geodesics with past end point on the edge of the closed set.  All these results are accompanied with a plethora of figures that aim to provide a pictorial explanation of the properties of all these sets.

 In section six, we begin with the definition of globally hyperbolic manifolds including the notion of a Cauchy surface and explore some of their properties. The definition for globally hyperbolic manifolds we use is that of \cite{Wald}.  Then, the behaviour of inextendible causal curves in globally hyperbolic manifolds is explored in proposition \ref{prop:cauchy1}.  We also demonstrate that globally hyperbolic manifolds are strongly causal -- in fact they are stably causal which is a stronger notion of causality. We continue with establishing the first compactness results associated to globally hyperbolic spacetimes.  First in theorem \ref{th:compact}, we demonstrate that the space of future directed causal curves between two causally related points in a globally hyperbolic spacetime is compact. This result is very significant and the proof follows that the steps as described in  \cite{Wald}, see also \cite{HE}.
 Second in theorem \ref{th:jjcompact}, we give the proof that the intersection of the causal past and the causal chronological future of two points in a globally hyperbolic manifold is also compact. Third in theorem \ref{th:Kqcompact}, the latter result is generalised to include the compactness of the  intersection of the chronological future and past of points and compact subsets in a globally  hyperbolic  spacetime.  These results are essential to demonstrate that there are geodesics with maximal proper time between two points in a globally hyperbolic spacetime, i.e. maximal geodesics.  The existence of such curves is essential in the proof of singularity theorems.

 In section seven, we begin with the proof of the statement that the proper time  of time-like geodesics that exhibit a conjugate point between two points or a focus point between a spacelike hypersurface and a point in a spacetime  is less than that between the two points or the hypersurface and the point. These results are established in theorems \ref{th:conjugategeodesic1} and \ref{th:focusgeodesic1}, respectively.  These are the so called ``round the corner'' theorems. Part of the proof of the former is included in theorem \ref{th:lengthmin}  in the appendices.  In fact two proofs for this are given. One is distributional and the other smooth. Then, the attention is turned to globally hyperbolic manifolds. The compactness properties of globally hyperbolic manifolds established in section six are used to prove the existence of geodesics with maximal proper time either between two points or a Cauchy surface and a point. These results are stated in the theorems \ref{th:maxgeodesic1} and \ref{th:maxgeodesic2}. An alternative proof of the former result using the Arzel\'a-Ascoli theorem is also described in appendix \ref{ap:dx}.  Finally, the properties of time-like geodesics in globally hyperbolic spacetimes are summarised in theorem \ref{th:sumhyperbolic}.

 In section eight, the Hawking's cosmological singularity theorem \ref{th:singularity1} is demonstrated by comparing the results of lemma \ref{lemma:irrotational} with those of the theorems \ref{th:conjugategeodesic1} and \ref{th:sumhyperbolic}. The resolution of the contradiction in the comparison of all these results is that the spacetime must be geodesically incomplete. In theorem \ref{th:singularity2}, an extension of the previous singularity theorem is described where the assumption that the spacetime is globally hyperbolic has been removed. Instead, it is assumed the it contains a compact hyper-surface. After an appropriate adaptation of the round the corner procedure from time-like geodesics with conjugate points to  null geodesics  described in detail in  theorem \ref{th:nulldeformation}, the black hole singularity theorem of Penrose is demonstrated in theorem \ref{th:singularity3}.  It should be stressed that these are not the most general singularity theorems known.  There is a more general singularity theorem explained in \cite{HE} and more recent such results are described in e.g. \cite{Minguzzi}.  However, these theorems are  milestones towards the development of more general results.

 In section nine, we give an outline of the main results demonstrated in these notes. In addition, we use the machinery developed to sketch a proof of Hawking's \emph{black hole horizon area theorem}.  This is a natural step to take as all the necessary ingredients for the proof have already been described on the way to demonstrate the singularity theorems.  As it is well known, this theorem is key in the development of black hole thermodynamics.

 In appendix \ref{ap:topspaces}, we have gather results from general topology that have been utilised  throughout these notes. Their use is essential to established the key results that have lead to the proof of the singularity theorems.

 In appendix \ref{ap:aa}, we have included a detailed discussion on the topologies that can be put on the space of curves including explanations of their various continuity properties. We also give several examples to describe the differences between the various definitions. In addition, we present,  without a proof,  the Arzel\'a-Ascoli theorem.  This is a key compactness argument on the space of functions widely used in minimalisation or extremisation problems (variational analysis).  We also include the proof of proposition \ref{prop:gepropertime} that provides a key estimate that bounds the proper time of a curve with its length with respect to any auxiliary Riemannian metric on the spacetime. This estimate is useful to show, under certain conditions, that two points in a spacetime time are not at infinite proper time apart.

 In appendix \ref{ap:minimal}, we state the Hopf-Renow theorem that addresses the question of geodesic completeness in the context of Riemannian manifolds and demonstrate some of the  statements relevant to the singularity theorems. In the process, we prove the existence of geodesics that  have length the distance between two points in a complete Riemannian manifold using the minimisation of the energy functional in combination with the Arzel\'a-Ascoli theorem.  We also present the ``round the corner'' theorem in the context of Riemannian geometry.
 Finally in appendix \ref{ap:dx}, we give an alternative proof of the existence of time-like geodesics  with proper time the maximal proper time between two points in a globally hyperbolic manifold.

\section{Geodesic congruences and the local geometry of spacetime}

\subsection{Preliminaries}\label{sec:pre}

The spacetime\footnote{Sometimes, we shall denote the spacetime with just $\M$ and neglect the reference to the metric $g$. Also, unless it is explicitly stated,  $\M$ does not have a boundary.} $(\M,g)$ is a Hausdorff, second countable,  connected,  4-dimensional manifold\footnote{Since manifolds are locally compact, $\M$ is paracompact, see appendix \ref{ap:topspaces}. } $\M$, at least $C^2$,  equipped with a  metric $g$ of Lorentzian signature $(-+++)$.  Therefore, the tangent space $T_p\M$ at every $p\in \M$ can be identified with Minkowski space and vectors at $p$ can be characterised as time-like, null or spacelike. We shall also assume that $\M$ is Cauchy complete. Although in many examples below, especially those depicted in the drawings, we shall use Cauchy incompleteness to illustrate some of the properties of the structures that we shall put on $(\M, g)$.

\begin{definition}
 $(\M, g)$ is \emph{ oriented}, iff it is oriented as a manifold, i.e. it admits a nowhere vanishing 4-form. Similarly, $(\M,g)$ is \emph{ time-oriented},  iff it admits a nowhere vanishing time-like vector field $\vt$. \hfill{$\triangle$}
 \end{definition}

 For a time-oriented spacetime with respect to the vector field $\vt$,  the length of $\vt$ with respect to $g$ is nowhere vanishing on $\M$,  $g(\vt, \vt)\not=0$.  As a result,  $\vt$ can always be normalised such that
 \be
 g(\vt,\vt)=-1~.
 \ee
 On time-oriented spacetimes, it is possible to distinguish between future and past directed light-cones. The notions of orientation and time-orientation of the spacetime are distinct and one does not imply the other.

 \begin{definition}
 Let $(\M,g)$ be a time-oriented spacetime and $p\in \M$.
The time-like and null vectors in $\vu_p\in T_p\M$ can be characterised as future or past directed,  iff $g(\vu_p, \vt_p)<0$ or $g(\vu_p, \vt_p)>0$, respectively.  Similarly,  time-like and null vector fields on $\M$ are future or past directed provided  that they have the corresponding property at every point  $p\in\M$.\hfill{$\triangle$}
 \end{definition}

A curve $\gamma$ is a map from an interval $I\subset \bR$ into the manifold $\M$, where $I$ can be closed, semi-closed or open.  Typically, no other restriction is put on $\gamma$ apart from the usual regularity conditions, like demanding for example to be continuous or $C^k$. General $C^1$ curves are not necessarily emersions, i.e the tangent vector $\vec{\dot\gamma}$ is not required to be everywhere non-vanishing.  In local coordinates with respect to a patch (or chart) $(U, x)$, the tangent vector of a curve is given by
\be
\vec{\dot\gamma}=\frac{d\gamma^\mu(r)}{d r} \frac{\partial}{\partial x^\mu}\Big|_\gamma~,~~~\gamma^\mu=x^\mu\circ \gamma~,
\ee
where $r\in I$.

\begin{definition}
A $C^k$, $k\geq1$,  \emph{ arc} is a curve $\gamma$ with a \emph{ nowhere vanishing} tangent vector field. If, in addition, the closure $\bar I$ is compact, then $\gamma$ is called a \emph{ segment}. A \emph{ continuous piecewise curve} $C^k$, $k\geq1$,  curve is a finite collection of $C^k$ arcs $\{\gamma_i; i=1, \dots, n\}$ defined on the intervals $\{I_i; i=1,\dots, n\}$ such that the last point of $\gamma_i$ coincides with the first point on $\gamma_{i+1}$. \hfill{$\triangle$}
\end{definition}

It is clear that a continuous piecewise curve $C^k$ curve is $C^k$ smooth everywhere apart from the joining points of $\gamma_i$ and $\gamma_{i+1}$, where there may exist two possible tangent vectors and so, although continuous, it is not $C^1$. These points are called \emph{ corners}.

\begin{definition}
  A continuous piecewise $C^k$,  $k\geq1$, curve $\gamma$ is a \emph{  time-like}, \emph{ null} or \emph{ spacelike} curve provided that all  its tangent vectors are time-like, null or spacelike, respectively. A curve is \emph{ causal}, iff its tangent vectors are either time-like or null. If $\M$ is time-oriented, time-like and casual curves are future or past directed provided that their tangent vectors are future or past pointing, respectively. \hfill{$\triangle$}
\end{definition}

\begin{remark}
It should be stressed that according to the definitions, the tangent vectors of ($C^1$-differentiable) time-like and casual curves vanish \emph{nowhere} on the curve, i.e. they are immersions of intervals in $\M$. The above definition of time-like and causal curves has to be generalised  later because taking the limits of curves leads to more general causal curves\footnote{One can show that the most general definition of causal curves used in the literature, see \cite{HE} page 184 and definition \ref{def:ccausal},  satisfy   locally the Lipschitz condition \cite{beem}, page 75 and 76. This means that the curve is \emph{continuous} and   a  tangent vector field can be \emph{defined} almost everywhere on the curve, see appendix \ref{ap:b12}. So we may as well assume that causal curves are locally Lipschitz, or at least absolutely continuous, from the beginning. Nevertheless, for smooth enough manifolds, like $C^2$, the above definition of causal curves suffices on many occasions as all the curves mentioned here can be approximated by smooth curves.} than continuous piecewise $C^k$, $k>0$.  From now on when we refer to causal curves in general, we shall mean that they are either continuous piecewise $C^k$ or locally Lipschitz -- the latter property will be more precisely  specified when it is required. \hfill{$\Diamond$}
\end{remark}

\subsection{Normal coordinates}

Let $(\M, g)$ be a spacetime with metric $g$. We denote with $\nabla$ the Levi-Civita connection\footnote{If we consider $\M$ with   more than than one metric $g$, then the Levi-Civita  of $g$ will be denoted with $\nabla^g$ to avoid confusion.}.

\begin{definition}
A geodesic is a $C^2$ arc $\gamma$ whose tangent vector field $\vu=\vec{\dot\gamma}$ satisfies $\nabla_{\vu} \vu=f(t) \vu$ for some function $f$ of the parameter $t$ of the curve. Equivalently, in local coordinates
\be
\frac{d^2\gamma^\mu}{dt^2}+\Gamma^\mu_{\nu\rho}(\gamma) \frac{d\gamma^\nu}{dt}\frac{d\gamma^\rho}{dt}=f(t) \frac{d\gamma^\mu}{dt}~,
\ee
where $\gamma^\mu=x^\mu\circ \gamma$ and $(U, x)$ is a chart of $\M$. \hfill{$\triangle$}
\end{definition}
It is well known that after a change of parameterisation $t=t(s)$, the geodesic equation becomes
\be
\frac{d^2\gamma^\mu}{ds^2}+\Gamma^\mu_{\nu\rho}(\gamma) \frac{d\gamma^\nu}{ds}\frac{d\gamma^\rho}{ds}=0~,
\ee
where $s$ is the \emph{ affine} parameter of the geodesic.  This parameterisation is unique up to scaling and shifting $s$ by a constants, i.e.
$s\rightarrow a s+b $, with $a, b\in \bR$  and $a\not=0$.

The geodesic equation is a system of second order ordinary differential equations.  We have assumed sufficient regularity such that such that there is a unique solution
 \be
 \gamma(s)= F(s; p, \vu)~,
 \ee
 subject to boundary conditions
 \be
 \gamma(0)=F(0; p, \vu)=p~,~~~\vec{\dot\gamma}_p=\frac{d}{ds} F(s; p, \vu)\big|_{s=0}=\vu~,
 \ee
 for $s\in [0, c)$ with $c$ sufficiently small.

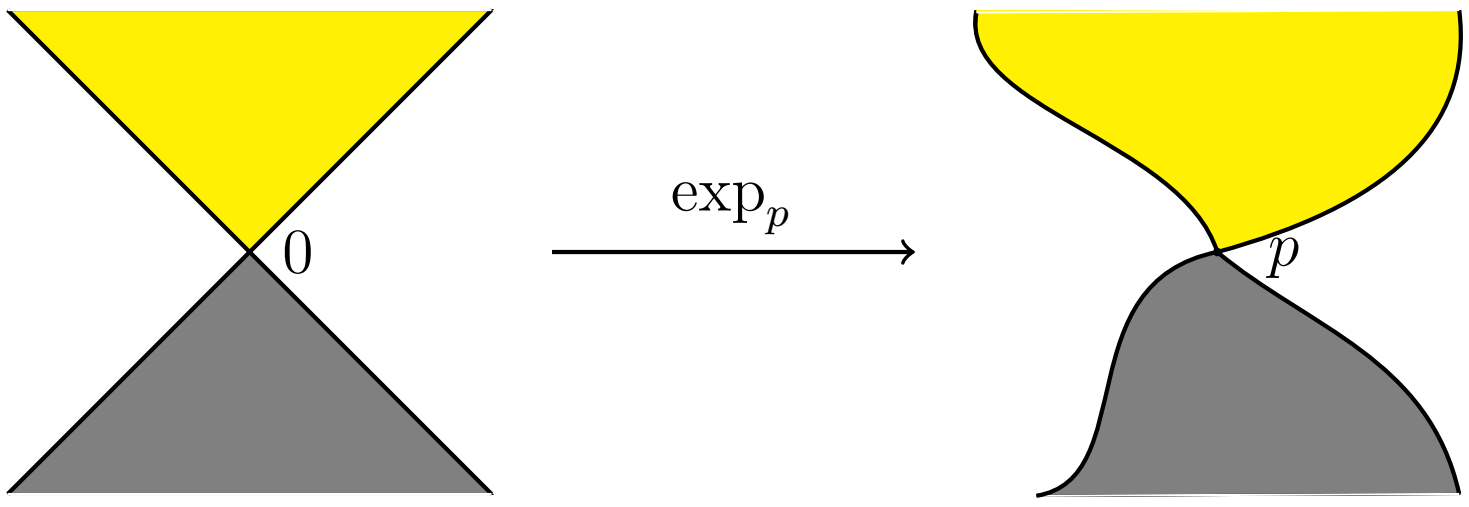
\begin{figure}[htbp]%\label{fig:exp}
\begin{center}
\begin{eqnarray*}
\begin{tikzpicture}[scale=0.4]
  % Coordinates
  \coordinate (A) at (4.00, -4.00);
  \coordinate (B) at (-4.00, 4.00);
  \coordinate (C) at (-4.00, -4.00);
  \coordinate (D) at (4.00, 4.00);
  \coordinate (E) at (5.00, 0.00);
  \coordinate (F) at (11.00, 0.00);
  \coordinate (G) at (16.00, 0.00);
  \coordinate (H) at (12.02, 3.96);
  \coordinate (I) at (11.70, 2.38);
  \coordinate (J) at (15.41, 1.86);
  \coordinate (K) at (13.01, -4.03);
  \coordinate (N) at (20.00, 4.00);
  \coordinate (O) at (18.75, 0.75);
  \coordinate (P) at (20.25, 2.00);
  \coordinate (Q) at (20.00, -4.00);
  \coordinate (R) at (17.50, -1.25);
  \coordinate (S) at (19.50, -1.75);
  \coordinate (L) at (14.75, -3.75);
  \coordinate (M) at (13.50, -0.50);
   \coordinate (M1) at (8,00, 0.10);

  % Drawing
   \draw [fill = yellow] (-4.00, 4.00)-- (4.00, 4.00) -- (0.00, 0.00);
   \draw [fill = gray] (-4.00, -4.00)-- (4.00, -4.00) -- (0.00, 0.00);
   \draw [fill = yellow] (12.00, 4.00) .. controls (11.70, 2.38) and (15.41, 1.86) .. (16.00, 0.00) .. controls (18.75, 0.75) and (20.25, 2.00) .. (N);
   \draw [fill = gray](K) .. controls (14.75, -3.75) and (13.50, -0.50) .. (G) .. controls (17.50, -1.25) and (19.50, -1.75) .. (Q) --(K)--(Q);
  \draw[thick] (A) -- (B);
  \draw[thick] (C) -- (D);
  \draw[->, thick] (E) -- (F);
  \draw[white] (B) -- (D);
  \draw[white] (C) -- (A);
  \draw[thick] (H) .. controls (11.70, 2.38) and (15.41, 1.86) .. (G);
  \draw[thick] (G) .. controls (18.75, 0.75) and (20.25, 2.00) .. (N);
  \draw[thick] (G) .. controls (17.50, -1.25) and (19.50, -1.75) .. (Q);
  \draw[white] (H) -- (N);
  \draw[white] (K) -- (Q);
  \draw[thick] (K) .. controls (14.75, -3.75) and (13.50, -0.50) .. (G);
  \fill[black] (G) circle (2pt);
  \node[above, font=\large] at (M1) {$\mathrm{exp}_p$};
  \node[right, font=\large] at (0.20, 0.00) {$0$};
   \node[right, font=\large] at (16.50, -0.10) {$p$};
\end{tikzpicture}
\end{eqnarray*}
\end{center}
%\begin{figure}[htbp]
\caption{\small{The figure depicts the action of the exponential map, $\mathrm{exp}$, that acts on $T_p\M$ with values in $\M$. The shaded regions indicate what is the domain and image of the interior of the past lightcone $D^-_p$ (in gray) and the interior of the future lightcone $D^+_p$ (in yellow).  The boundary of the lightcone  is also indicated within a normal neighbourhood of the point $p$.  }}
\centering
\label{fig:exp}
\end{figure}

It  is straightforward to observe that $F(as; p, \vu)=F(s; p, a\vu)$, $a\in \bR_{>0}$ -- both functions satisfy the geodesic equation with the same boundary conditions and the equality follows from the uniqueness of the solutions for small enough $s$.  This can be used to define a map from a  open set $U$ in $T_p\M$  to $\M$ as follows:
\begin{definition}
The \emph{exponential map}, $\exp_p$, from a  open set $U$  in $T_p\M$ that includes $\vec 0$, into $\M$    is given by
\be
\exp_p\equiv F_p :  U\subset T_p\M\rightarrow M~,~\text{where}~~\exp_p(u)=F_p(u)\equiv F(1; p, \vu)~,
\ee
where $\vu$ is the position vector of the point $u\in U$. \hfill{$\triangle$}
\end{definition}

The \emph{exponential map}, which is a $C^2$ diffeomorphism from $U$ to an open set $W_p\subset M$ with $p\in W_p$.  It turns outs that $\{(W_p, F_p^{-1}); p\in M\}$ is an atlas on $\M$ with charts or patches $(W_p, F_p^{-1})$.  In this atlas, the local coordinates of $M$ are the coordinates $u$ of $T_pM$.  These are also known as \emph{normal coordinates} of $\M$.

\begin{remark}
The maximal neighbourhood of $p$ that the exponential map is a diffeomorphism is called the \emph{maximal normal neighbourhood} of $p$ denoted by $N_p$. There are also refinements of this, like the notion of \emph{convex neighbourhood} $C_p$ of $p$ for which any two points $x, y\in C_p$   are joined by a geodesic $\gamma$ that is contained entirely in $C_p$. Moreover, a \emph{convex  normal  neighbourhood} of $p$ is convex neighbourhood for which the geodesic that joins any two points $x$ and $y$ is unique.  Such neighbourhoods always exist for sufficiently smooth manifolds.\hfill{$\Diamond$}
\end{remark}

Clearly as $F(s; p, \vu)=F(1; p, s\vu)$, in these new coordinates, the geodesics that pass through $p$ are the straight lines that pass through $\vec 0\in T_p\M$. Thus, the equation for the geodesics that pass through $p$ in normal coordinates is
\be
u(s)= s \vu~.
\ee
As the length of the tangent vector of geodesics remain constant along the geodesic, we have
\be
g_{\mu\nu}(s\vu) \vu^\mu \vu^\nu=g_{\mu\nu}(0) \vu^\mu \vu^\nu~.
\ee
Let $u$ be the coordinates of the point with position vector $s\vu$. After multiplying the above equation with $s^2$, we deduce that
\be
g_{\mu\nu}(u) u^\mu u^\nu=g_{\mu\nu}(0) u^\mu u^\nu~.
\label{ncon1}
\ee
Furthermore, the geodesic equation implies
\be
\Gamma^\rho_{\mu\nu}(s\vu) \vu^\mu\vu^\nu=0~.
\ee
This gives that
\be
\Gamma^\rho_{\mu\nu}(0) =0~, ~~~ \Gamma^\rho_{\mu\nu}(u) u^\mu u^\nu=0~.
\label{ncon2}
\ee
Taking the derivative of (\ref{ncon1}) with respect to the coordinate $u$, we find that
\be
\partial_\rho g_{\mu\nu}(u)  u^\mu u^\nu+ 2 g_{\rho\mu}(u) u^\mu=2 g_{\rho\mu}(0) u^\mu~.
\ee
Using the second condition in (\ref{ncon2}), we find that the equation above can be rewritten as
\be
2 \partial_\mu g_{\rho\nu}(u)  u^\mu u^\nu+ 2 g_{\rho\mu}(u) u^\mu-2 g_{\rho\mu}(0) u^\mu=0~.
\ee
Upon setting $u=\vu s$, this can be rewritten as
\be
 \partial_\mu g_{\rho\nu}(s\vu)  s \vu^\mu \vu^\nu+  g_{\rho\mu}(s\vu) \vu^\mu- g_{\rho\mu}(0) \vu^\mu=\frac{d}{ds} \big( g_{\rho\mu}(s\vu) s \vu^\mu-g_{\rho\mu}(0) s \vu^\mu\big)= 0~.
 \ee
 Therefore $g_{\rho\mu}(s\vu) s \vu^\mu-g_{\rho\mu}(0) s \vu^\mu$ must be constant along the geodesic generated by $\vu$ for any $\vu$. This implies that
 \be
 g_{\mu\nu}(u) u^\nu- g_{\mu\nu}(0)u^\nu=0~,
 \label{thomasf}
 \ee
 for every $u$ -- the constant must vanish as it must be true for $\vu=0$ as well.  This is known as the \emph{Thomas formula}.

 \begin{prop}\label{pro:ortho}
 Within a normal neighbourhood $U$ centred at $p$, the time-like geodesics emanating from $p$ with unit tangent vectors are orthogonal to the hypersurface $g_{\mu\nu}(0) u^\mu u^\nu=\text{const}$.
 \end{prop}
 \begin{proof}
 The tangent vectors of geodesics emanating from $p$ are $\partial/\partial s=\partial_s$. Thus
 \be
 \partial_s=\frac{\partial u^\mu}{\partial s} \frac{\partial}{\partial u^\mu}=\vu^\mu \frac{\partial}{\partial u^\mu}~.
 \ee
 This is not a unit tangent because
 \be
 g(\partial_s, \partial_s)|_u= g(\vu, \vu)|_u=g(0)(\vu, \vu)\equiv g_{\mu\nu}(0) \vu^\mu\vu^\nu~,
 \ee
 where (\ref{ncon1}) has been used to establish the last equality.
 Thus the unit tangent vector is
 \be
 \vv=\frac{1}{ \sqrt{-g(0)(\vu, \vu)}} \vu^\mu \frac{\partial}{\partial u^\mu}=\frac{1}{\sqrt {- g(0)(u, u)}} u^\mu \frac{\partial}{\partial u^\mu}~.
 \ee
 Thus,  the associated 1-form is
 \be
 \bv=\frac{1}{\sqrt {- g(0)(u, u)}} g_{\mu\nu}(u) u^\mu du^\nu=\frac{1}{\sqrt {- g(0)(u, u)}} g_{\mu\nu}(0) u^\mu du^\nu=- d \sqrt {- g(0)(u, u)}~,
 \ee
 where the Thomas formula (\ref{thomasf}) has been used to prove the second equality.
 This is equivalent to stating that the congruence of time-like geodesics,  see definition \ref{def:congruence} below, emanating from $p$ are orthogonal to the hyper-surface $s=\text{const.}$.  This proves the result, see also lemma 4.5.2 in \cite{HE} and \cite{Senovilla}.
 \end{proof}

\subsection{Geodesic congruences}

The proof of the singularity theorems is based on the behaviour of families of curves in regions of spacetime with positive curvature. A particular family of curves are the congruences whose properties will be explored below.

\begin{definition}\label{def:congruence}
A \emph{congruence of curves} $\gamma$ is a three parameter  family of $C^k$, $k\geq 1$, curves in an open submanifold $U$ of $\M$ such that at every point $p\in U$, there is only one curve from the congruence that passes through $p$.  The \emph{tangent vector field of the congruence} is generated by the tangent vectors of all the curves in the congruence. The congruence is time-like or null depending on whether the tangent vector field is time-like or null, respectively.
\hfill{$\triangle$}
\end{definition}
It is implicitly assumed that the tangent vector field of a congruence is nowhere vanishing on $U$. For the singularity theorems, the main focus will be in geodesic congruences.

\subsubsection{Time-like geodesic congruences}

For time-like congruences of curves, the tangent vector field $\vu$ of the congruence  can always be normalised as
\be
g(\vu, \vu)=-1~,
\label{vnorm}
\ee
because it is nowhere vanishing on $U\subset \M$  -- we shall refer to such time-like vector fields as \emph{normalised} time-like vector fields.
A coordinate $\tau$ can be adapted to the normalised tangent vector field $\vu$ by solving the first order partial differential equation
\be
\vu(\tau)=1~.
\ee
$\tau$ is also  called the \emph{ affine} parameter of the congruence.  Notice that if  $\tau$ is a solution, the shift of $\tau$ with a constant is also a solution as well.

 Conversely, given a normalised time-like vector field $\vu$, and so nowhere vanishing, one can solve the differential equation
\be
\frac{dx^\mu}{d\tau}= \vu^\mu(x)~,
\ee
to yield a congruence  constructed from integral curves $\gamma(x^\mu_p, \tau)$ of $\vu$ with  $\gamma(x^\mu_p, 0)=p$. It appears that the integral curves  of $\vu$ depend on four parameters,  the coordinates $x^\mu_p$  of the point $p$ at $\tau=0$, but one of the parameters can be absorbed in the definition of $\tau$.

Given a normalised time-like vector field as $\vu$ above, i.e. $g(\vu, \vu)=-1$, we can define\footnote{In index free notation, the associated 1-form to a vector field $\vu$ with respect to $g$ will be denoted with $\vu^\flat$ and similarly the associated vector field to a 1-form $\alpha$ will be denoted with $\alpha^\flat$.  But if indices appear explicitly, the additional labelling indicated will be neglected.}
\be
h_{\mu\nu}\equiv \vu_\mu \vu_\nu+ g_{\mu\nu}~,~~~\vu_\mu=g_{\mu\nu} \vu^\nu~,
\ee
which can be thought of as the projection of $g$ in directions orthogonal to $\vu$. Clearly $h$ is a symmetric tensor and it is immediate to demonstrate that
\be
\vu^\mu h_{\mu\nu}=0~,~~~h^\mu{}_\rho h^\rho{}_\nu= h^\mu{}_\nu~,~~~h^\mu{}_\mu=g^{\mu\nu} h_{\mu\nu}=3~,
\ee
with $h^\mu{}_\nu=g^{\mu\lambda} h_{\lambda \nu}$.
The \emph{ acceleration} of $\vu$ is defined as
\be
\va^\mu\equiv\nabla_{\vu} \vu^\mu=\vu^\nu \nabla_\nu \vu^\mu~.
\ee
Notice that $\va$ is orthogonal to $\vu$, $\va^\mu \vu_\mu=0$.  This can be easily verified upon using that $\vu$ is normalised, i.e. by taking the $\nabla_{\vu}$ derivative of equation (\ref{vnorm}). If the acceleration $\va$ of $\vu$ vanishes, $\va=0$,  then the congruence is a \emph{geodesic congruence}.

The covariant derivative of $\nabla \vu$ can be decomposed as
\be
\nabla_\nu \vu_\mu= -\vu_\nu \va_\mu+ \sigma_{\mu\nu}+\frac{1}{3} \theta h_{\mu\nu}+\omega_{\mu\nu}~,
\label{nvudec}
\ee
in terms of the acceleration,  shear $\sigma$, rotation $\omega$ and expansion $\theta$, where
\begin{align}
\sigma_{\mu\nu}&=\frac{1}{2} h^\rho{}_\mu h^\lambda{}_\nu \Big(\nabla_\rho \vu_\lambda+ \nabla_\lambda \vu_\rho\Big)-\frac{1}{3} \theta h_{\mu\nu}
\cr
 \omega_{\mu\nu}&=\frac{1}{2} h^\rho{}_\mu h^\lambda{}_\nu \Big(\nabla_\rho \vu_\lambda- \nabla_\lambda \vu_\rho\Big)~,~~~\theta=\nabla^\mu \vu_\mu~,
 \end{align}
 Notice that the acceleration $\va$, shear $\sigma$ and rotation $\omega$ are all spacelike.

The congruence is \emph{ irrotational} (or integrable),  iff $\vu^\flat\wedge d\vu^\flat=0$.  This in turn implies that locally there are functions $t$ and $f$ such that $\vu^\flat=f d t$.  Then, the congruence is orthogonal to the hypersurfaces defined by $t=\text{const}$.  If $\vu^\flat$ is a gradient, i.e. $\vu^\flat=-dt$,  then the acceleration vanishes as well $\va=0$, i.e. the congruence  is a \emph{geodesic} congruence.

\begin{lemma}\label{lemma:fdrot}
A forward directed time-like geodesic congruence emanating orthogonally from a hypersurface $\Sigma$ is irrotational.
\end{lemma}
\begin{proof}
Suppose that the hypersurface $\Sigma$ is (locally) given by the equation $t(x^\mu)=0$. The normal 1-form to hypersurface is $ dt\vert_{\Sigma}$.   Each geodesics passes through the hypersurface $\Sigma$ at a point labeled by the coordinates $y^i$ in a local patch of $\Sigma$.  Extend this coordinate system to a coordinate system of the congruence  such that the coordinates that label each geodesic crossing $\Sigma$ at $y^i$ remain the same along the geodesic in a neighbourhood of $\Sigma$, i.e the points on a geodesic passing through $\Sigma$ at the point $p$ with coordinates $y^i$ are labelled by $(t, y^i)$.  It is clear from this that in this coordinate system the normalised  tangent vector field $\vu$ of the congruence is proportional to $\frac{\partial}{\partial t}$
\be
\vu= p \frac{\partial}{\partial t}~,
\ee
where $p$ is a nowhere vanishing function.
Moreover
\be
\vu_i dy^i\vert_{\Sigma}= p g(\frac{\partial}{\partial y^i}, \frac{\partial}{\partial t}) dy^i\vert_{\Sigma}=0~,
\label{vucomp}
\ee
as the geodesics are taken orthogonal to the hypersurface.

We have to show that\footnote{In  index free notation, we denote with $\vu^\flat$ the associated 1-form to the vector field $\vu$, i.e. $\vu^\flat_\mu\equiv \vu_\mu=g_{\mu\nu} \vu^\nu$. Similarly, we denote with $\alpha^\flat$ the associated vector field to the 1-form $\alpha$.} $\omega\equiv d\vu^\flat=0$. First, using that the congruence is a geodesic congruence and $\vu$ is normalised, we have that
\be
\vu^\mu (\nabla_\mu \vu_\nu-\nabla_\nu \vu_\mu)=0~,
\label{oti}
\ee
i.e. in the coordinate basis introduced above $\omega_{ti}=0$.  As $d^2\vu=d\omega=0$,  we have that
\be
\partial_t \omega_{ij}+\partial_j \omega_{ti}+\partial_i \omega_{jt}=\partial_t\omega_{ij}=0~.
\ee
It follows that the components  $\omega_{ij}$ do not depend on $t$.  So if they vanish somewhere, they will vanish everywhere along the geodesics.
 But from (\ref{vucomp}),  $d\vu\vert_{\Sigma}=\omega\vert_{\Sigma}=0$ as $\vu\vert_{\Sigma}=0$.  Therefore,  the components    $\omega_{ij}=0$ and  as $\omega_{ti}=0$ from (\ref{oti})  as well,    $\omega$ vanishes everywhere.  Thus, the geodesic congruence orthogonal to a hypersurface  is irrotational.
\end{proof}

Making  the same assumptions as in the Lemma, note that   in the coordinates $(\tau, y^i)$, where $\vu(\tau)=1$, i.e. the coordinate $\tau$ is adapted to the vector field $\vu$, the metric takes the form
\be
g=- d\tau^2+ g_{ij} (\tau, y) dy^i dy^j~.
\label{gaussiancor}
\ee
These are the \emph{Gaussian} coordinates of the metric $g$.

\begin{prop}\label{prop:sre}
The shear, rotation and expansion of a future directed time-like geodesic congruence with normalised tangent vector field $\vu$ satisfy the equations
\begin{align}
\vu^\mu \nabla_\mu\sigma_{\nu\lambda}&=-\frac{2}{3}\theta \sigma_{\mu\nu}- \sigma_\nu{}^\mu \sigma_{\mu\lambda}+\omega_\nu{}^\mu \omega_{\lambda\mu}+ \frac{1}{3} h_{\nu\lambda} (\sigma_{\mu\rho} \sigma^{\mu\rho}-\omega_{\mu\rho} \omega^{\mu\rho})+ W_{\mu\nu\lambda\rho} \vu^\mu \vu^\rho+\frac{1}{2} \tilde R_{\nu\lambda}~,
\cr
\vu^\mu \nabla_\mu\omega_{\nu\lambda}&=-\frac{2}{3} \theta \omega_{\nu\lambda}-\omega_{\nu \mu} \sigma^\mu{}_\lambda+\omega_{\lambda \mu} \sigma^\mu{}_\nu~,
\cr
\vu^\mu \nabla_\mu\theta&=-\frac{1}{3} \theta^2-\sigma_{\mu\nu} \sigma^{\mu\nu}+ \omega_{\mu\nu} \omega^{\mu\nu}- R_{\mu\nu} \vu^\mu \vu^\nu~,
\label{sot}
\end{align}
where $W$ is the Weyl tensor and
\be
\tilde R_{\mu\nu}=h_{\mu\lambda} h_{\nu\rho} R^{\lambda\rho}-\frac{1}{3} h_{\mu\nu} h_{\rho\lambda} R^{\rho\lambda}~.
\ee
\end{prop}
\begin{proof}
The above equations can be derived by taking the covariant derivative of $\nabla\vu$ along $\vu$ and rearranging the terms. In particular, we have
\be
\vu^\mu \nabla_\mu \nabla_{\nu} \vu^\lambda=\vu^\mu \nabla_\nu \nabla_\mu \vu^\lambda+ R_{\mu\nu}{}^\lambda{}_\rho \vu^\mu  \vu^\rho=-\nabla_\nu \vu^\mu  \nabla_\mu \vu^\lambda+ R_{\mu\nu}{}^\lambda{}_\rho  \vu^\mu \vu^\rho~,
\ee
where to establish the last equality we have used the equation for geodesics $\nabla_{\vu} \vu=0$ and $R$ is the curvature of $g$.  Next, using the decomposition of $\nabla\vu$  as in (\ref{nvudec}) and noting that $\va=0$, as this is a geodesic congruence, we have that
\be
\vu^\mu \nabla_\mu \nabla_{\nu} \vu_\lambda=-(\sigma_\nu{}^\mu +\omega_\nu{}^\mu +\frac{1}{3} h_{\nu}{}^\mu \theta) (\sigma_{\mu\lambda}+\omega_{\mu\lambda}+\frac{1}{3} h_{\mu\lambda} \theta)+ R_{\mu\nu\lambda\rho}  \vu^\mu \vu^\rho~.
\ee
Contracting this equation with either $g^{\nu\lambda}$ or $h^{\nu\lambda}$, one derives the last equation in (\ref{sot}) for the expansion $\theta$ -- notice that $\nabla_{\vu} g=\nabla_{\vu} h=0$.  The equation for $\theta$ is known as the \emph{Raychaudhuri} equation.

Skew-symmetrising the indices $\nu$ and $\lambda$ and using the definition of $\omega$, one finds the second equation in (\ref{sot}). Finally, the equation for the shear is found by symmetrising the indices $\nu$ and $\lambda$ and then subtracting the trace using the relation
\be
R_{\mu\nu\lambda\rho}= W_{\mu\nu\lambda\rho}+\frac{1}{2} (g_{\mu\lambda} \mathring R_{\nu\rho}+g_{\nu\rho} \mathring R_{\mu\lambda}-g_{\nu\lambda} \mathring R_{\mu\rho}-g_{\mu\rho} \mathring R_{\nu\lambda}+\frac{1}{12} R (g_{\mu\lambda} g_{\nu\rho}-g_{\nu\lambda} g_{\mu\rho})~,
\ee
between the Riemann curvature $R$ and Weyl tensor $W$ in four dimensions, where $\mathring R_{\mu\nu}=R_{\mu\nu}-1/4 g_{\mu\nu} R$ is the trace free part of the Ricci tensor.
\end{proof}

\begin{remark}
Our conventions for the curvature are
\be
\nabla_\mu\nabla_\nu\vv^\rho-\nabla_\nu\nabla_\mu \vv^\rho=R_{\mu\nu}{}^\rho{}_\sigma \vv^\sigma~,
\ee
where $\vv$ is a vector field on $\M$.
Furthermore,  $R(\vv,\vu)\vw$ denotes the vector field
\be
R(\vv,\vu)\vw= R_{\mu\nu}{}^\rho{}_\sigma \vv^\mu\vu^\nu \vw^\sigma \partial_\rho~.
\ee
This notation will be used later.
\hfill{$\Diamond$}
\end{remark}

Let us investigate the behaviour of the equation for the expansion. Adapting a coordinate $\tau$ to the geodesic congruence vector field $\vu$, $\vu(\tau)=1$,  the last equation in (\ref{sot}) can be rearranged as
\be
\frac{d}{d\tau}\theta+\frac{1}{3} \theta^2=-\sigma_{\mu\nu} \sigma^{\mu\nu}+ \omega_{\mu\nu} \omega^{\mu\nu}- R_{\mu\nu} \vu^\mu \vu^\nu~.
\label{exptwo}
\ee
Note that $\tau$ is an affine parameter that can be identified with the proper time of a geodesic.
If the right hand side is negative, one derives the differential inequality
\be
\frac{d}{d\tau}\theta+\frac{1}{3} \theta^2\leq 0~,
\ee
which can be solved to yield
\be
\theta^{-1}-\theta_0^{-1}\geq \frac{1}{3} \tau~.
\label{thetatau}
\ee
Therefore if the initial value of the expansion $\theta_0$ is negative $\theta_0<0$ at some point, say $\tau=0$, on a geodesic $\gamma$ in the congruence, then $\theta(\tau)\rightarrow -\infty$ as $\tau\rightarrow 3|\theta_0|^{-1}$. In other words, if the geodesics of the congruence start focusing, i.e. $\theta$ negative at some point $p$, then the congruence will develop a singularity within proper time $\tau\leq 3|\theta_0|^{-1}$ of the geodesics $\gamma$ that passes through  $p$.  This is not a singularity of the spacetime. It is just a singularity of the congruence.

\begin{lemma}\label{lemma:irrotational}
Consider a future directed time-like geodesics congruence emanating orthogonally from a hypersurface $\Sigma$ of the spacetime with normalised tangent vector field $\vu$. If $R_{\mu\nu} \vu^\mu\vu^\nu\geq 0$ and at some point $p$ in a geodesics $\gamma$ the expansion  $\theta$ becomes negative $\theta_p=\theta_0<0$, then $\theta\rightarrow -\infty$ within at most proper time $\tau\leq 3|\theta_0|^{-1}$ from $p$.
\end{lemma}
\begin{proof}
We have demonstrated in Lemma \ref{lemma:fdrot} that such geodesic congruences are irrotational, i.e. $\omega=0$. As from the hypothesis $R_{\mu\nu} \vu^\mu\vu^\nu\geq 0$, the right hand side of the equation for the expansion (\ref{exptwo}) is negative and the argument above based on (\ref{thetatau}) can be repeated to demonstrate the lemma.
\end{proof}

\subsubsection{Null geodesic congruences}\label{ssec:nullcongruences}

Before we proceed with the description of null geodesic congruences, it is instructive to mention some properties of null vectors. Given a null tangent vector
$\vu_p\in T_pM$, $\vu_p\not=0$, one can define the orthogonal vector space to $\vu_p$ in $T_pM$ by taking
\be
T^\perp_pM=\{\vw_p\in T_pM~~\vert~~ g(\vu_p, \vw_p)=0\}~.
\ee
This is a 3-dimensional vector subspace of $T_pM$.  Of course because $\vu$ is null, $\vu_p\in T^\perp_pM$.  Using this, one can define an equivalence relation in $T^\perp_pM$ be declaring that $\vw_p\sim \vw_p'$, iff there is a $c\in \bR$ such that $\vw'_p=\vw_p+ c \vu_p$.  Given this equivalence relation, denote the vector space of  equivalence classes of this relation with $\widehat {T_p\M}$, i.e. $\widehat { T_p\M}=T_p\M/\sim$ or equivalently
\be
0\rightarrow \bR\langle \vu_p\rangle\rightarrow T^\perp_p\M \rightarrow \widehat {T_p\M}\rightarrow 0
\label{nullseq}
\ee
It is clear from this that $\widehat {T_p\M}$ is not a natural subspace of either $T_p\M$ or $T^\perp_p\M$. This means that a vector $\vw_p\in T_p\M$ does \emph{not} have a natural decomposition with components to belong in these spaces.

Similarly, if $\alpha_p\in T^*_p\M$ is a 1-form at $p$, one can define an element $\alpha^\perp$ in $(T_p^\perp \M)^*$ by just restricting $\alpha$ on $T_p^\perp \M$. Clearly,  $\alpha_p^\perp(\vu_p)=0$, where $\vu$ is the null vector we have previously considered.  Again, one can define an equivalence relation on $(T_p^\perp \M)^*$ by declaring that $\alpha'^\perp_p\sim \alpha^\perp_p$, iff there a $c\in \bR$, such that $\alpha'^\perp_p=\alpha^\perp_p+\,c\, \vu_p^\flat$, where $\vu_p^\flat\in T^*_p\M$.  Then, define the space of equivalence classes of this relation as $(\hat T_p M)^*=(T_p^\perp M)^*/\sim$. As for the tangent space, $(\hat T_p M)^*$ should not be thought as a subspace of either  $(T_p^\perp \M)^*$ or $T_p^*\M$.

\begin{remark}
Although, it is not strictly necessary in what follows, it is worth pointing out that it is possible to decompose $T_p\M$ in the presence of a null vector $\vu_p$ provided some additional structure is introduced.
Suppose we choose another null vector $\vv_p$, $g(\vv_p, \vv_p)=0$,  such that $g(\vu_p, \vv_p)=-1$.  Such a vector is not unique as for every constant $c\in \bR$, $\vv_p'=\vv_p+c \vu$ satisfies the same conditions above as $\vv_p$. With a choice of $\vv_p$, every vector $\vw_p\in T_pM$ admits a unique decomposition as
\be
\vw_p= a \vu_p+ b\vv_p+\widehat{\vw}_p~,~~~a, b\in \bR~,
\label{nulldec}
\ee
where $\widehat {\vw}_p$ is orthogonal to both $\vu_p$ and $\vv_p$. The choice of $\vv_p$ defines a so call splitting of the sequence (\ref{nullseq}), i.e. it gives an inclusion $\iota$ of $\hat T_pM$ in $T^\perp_pM$ and so in $T_pM$. Then $\hat \vw_p$ is identified as the component of $\vw_p$ in $\iota (\hat T_pM)\subset T_pM$. Notice though that the decomposition of $\vw_p$ in (\ref{nulldec}) is not natural as it depends on the choice of $\vv_p$.

Moreover, a similar (non natural) decomposition can be made for $T^*_pM$ by choosing $\vv_p^\flat$ to split the sequence and identify an inclusion of $(\widehat {T_p \M})^*$ in $(T_p^\perp \M)^*$ and so in $T^*_p\M$. The above constructions can be generalised to tensor fields. In particular, the spacetime metric can be decomposed as
\be
g=-\vu^\flat\otimes \vv^\flat-\vv^\flat\otimes \vu^\flat+ k~,
\ee
with $k$ orthogonal both to $\vu$ and $\vv$, i.e. $k$ is the fibre metric of $\iota (\widehat {T_p\M})$. Furthermore, $k$ satisfies
\be
k^\mu{}_\mu\equiv k_{\mu\nu} g^{\mu\nu}=2~,~~~k^\mu{}_\lambda k^\lambda{}_\nu=k^\mu{}_\nu~,~~~k_{\mu\nu}=k_{\nu\mu}~,
\label{trhn}
\ee
where $k^\mu{}_\nu=g^{\mu\lambda} k_{\lambda\nu}$ and $k^{\mu\nu} =g^{\mu\rho} g^{\nu\sigma} k_{\rho\sigma}$. In many computations a splitting, as above, is chosen to perform the calculations. At the end it is required to demonstrate that the end result does not depend on the choice of the splitting.
\hfill{$\Diamond$  }
\end{remark}

\begin{definition}
A null geodesic congruence is a congruence $\gamma$ with tangent vector field $\vu$ that  is null everywhere on $U\subset \M$, i.e.
\be
\nabla_{\vu} \vu=0~,~~~g(\vu, \vu)=0~.
\ee
\hfill{$\triangle$}
\end{definition}
As $\vu$ is null, one can perform the analysis above at every point $p\in U\subseteq \M$ in the congruence to decompose $TU$ and $T^*U$ as in (\ref{nullseq}).  The decomposition can also be extended to the  associated tensor bundles.

\begin{remark}
 The equivalence relation on the sections of $(T^\perp U)^*$ is preserved by $\nabla_{\vu}$.  In particular, if two 1-forms $\alpha$ and $\alpha'$ are in the same equivalence class, $\alpha \sim \alpha'$, $\alpha(\vu)=\alpha'(\vu)=0$, and so are related by $\alpha'=\alpha+ f \vu^\flat$, where $f$ is a function of $U$, then $\nabla_{\vu}\alpha'=\nabla_{\vu} \alpha+ \nabla_{\vu} f \vu^\flat$ and so $\nabla_{\vu}\alpha'\sim \nabla_{\vu}\alpha$.  This also applies to vector fields and to tensors as well. This means that the
directional derivative $\nabla_{\vu}$ descents as a directional derivative on the sections of the $\widehat {T^{p,q} U}$ bundle. \hfill{$\Diamond$}
\end{remark}

As in the time-like case, consider the second fundamental form of the null geodesic congruence
\be
B_{\mu\nu}= \nabla_\nu \vu_\mu~,
\label{sff}
\ee
and observe that $\vu^\mu B_{\mu\nu}=\vu^\nu B_{\mu\nu}=0$.  Thus, denoting the equivalence class represented by $B$ with $\widehat B$, $\nabla_{\vu} B$ also represents another equivalence class $\widehat{\nabla_{\vu} B}$.  A similar calculation as in the time-like case reveals that
\be
\nabla_{\vu} B_{\mu\nu}+ B_{\mu\lambda} B^\lambda{}_\nu= R_{\lambda \mu\nu\rho} \vu^\lambda \vu^\rho~,
\ee
which upon taking the equivalence class, descents to
\be
\nabla_{\vu} \widehat {B}_{\mu\nu}+ \widehat {B}_{\mu\lambda} \widehat {B}^\lambda{}_\nu= \widehat{R_{\lambda \mu\nu\rho} \vu^\lambda \vu^\rho}~.
\ee
This in turn gives
\begin{align}
\nabla_{\vu} \theta&=-\frac{1}{2} \theta^2- \widehat{\sigma}_{\mu\nu} \widehat{\sigma}^{\mu\nu}+\widehat{\omega}_{\mu\nu}\widehat{\omega}^{\mu\nu}- R_{\mu\nu} \vu^\mu\vu^\nu~,
\cr
\nabla_{\vu} \widehat{\sigma}&=-\theta \widehat{\sigma}_{\mu\nu}+ \widehat{W_{{\lambda \mu\nu\rho}} \vu^\lambda \vu^\rho}~,
\cr
\nabla_{\vu} \widehat{\omega}_{\mu\nu}&=-\theta \widehat{\omega}_{\mu\nu}~,
\label{nullexp}
\end{align}
where the second fundamental form has been decomposed as
\be
\widehat {B}_{\mu\nu}=\frac{1}{2}\theta  k_{\mu\nu}+\widehat {\sigma}_{\mu\nu}+\widehat{\omega}_{\mu\nu}~,
\ee
in terms of the expansion $\theta$, shear $\sigma$ and rotation $\omega$
with
\begin{align}
\theta\equiv k^{\mu\nu} \widehat {B}_{\mu\nu}~,~~~\widehat {\sigma}_{\mu\nu}\equiv \widehat {B}_{(\mu\nu)}-\frac{1}{2} k_{\mu\nu} \theta~,~~~\widehat{\omega}_{\mu\nu}\equiv \widehat{ B}_{[\mu\nu]}~,
\end{align}
and $W$ is the Weyl tensor.
The above equations are similar to those we have derived for the congruences with a time-like tangent vector. One noticeable difference is the factor of $1/2$ that occurs in the $\theta^2$ term in the first equation in (\ref{nullexp}), instead of factor of $1/3$ that occurs in (\ref{exptwo}) for the time-like case. The reason for this is that the trace of $k$ here is $2$ instead of that of $h$ which is $3$, see (\ref{trhn}).

Adapting a coordinate along $\vu$, $\vu(\tau)=1$, where $\tau$ is the affine time of the null geodesics,  the first equation in (\ref{nullexp}) can be re-arranged as
\be
\frac{d}{d\tau}\theta+\frac{1}{2} \theta^2=- \widehat{\sigma}_{\mu\nu} \widehat{\sigma}^{\mu\nu}+\widehat{\omega}_{\mu\nu}\widehat{\omega}^{\mu\nu}- R_{\mu\nu} \vu^\mu\vu^\nu~.
\label{nullexpansion}
\ee
If the right hand side is negative semi-definite, the differential inequality can be solved, as in the time-like case, to yield
\be
\theta^{-1}-\theta_0^{-1}\geq \frac{1}{2} \tau~.
\label{nullexpansionx}
\ee
Therefore, if the initial value of the expansion $\theta_0$ is negative $\theta_0<0$ at some point on a geodesic $\gamma$, say at $\tau=0$,  in the congruence, then $\theta(\tau)\rightarrow -\infty$ as $\tau\rightarrow 2|\theta_0|^{-1}$. In other words, if the geodesics of the congruence start focusing, i.e. $\theta$ negative at some point $p$, then the congruence will develop a singularity within affine time $\tau\leq 2|\theta_0|^{-1}$ of the geodesics $\gamma$ that passes through  $p$.  Again, this is not a singularity of the spacetime but it is a singularity of the congruence.

There is a generalisation of Lemma \ref{lemma:fdrot} for hyper-surface orthogonal null geodesic congruences. In particular, such geodesics congruences are irrotational. The proof is identical to that given for  time-like congruences. The only difference is that the normalisation condition we have used for the tangent vector field of time-like congruences is now replaced by the null condition. As a result, there is an analogue of Lemma \ref{lemma:irrotational}. In particular, one can show the following:

\begin{lemma}\label{lemma:irrotationalnull}
Consider a future directed null geodesics congruence emanating orthogonally from a hypersurface $\Sigma$ of the spacetime with  tangent vector field $\vu$. If $R_{\mu\nu} \vu^\mu\vu^\nu\geq 0$ and at some point $p$ in a geodesics $\gamma$ the expansion $\theta$ becomes negative $\theta_p=\theta_0<0$, then $\theta\rightarrow -\infty$ within at most affine time $\tau\leq 2|\theta_0|^{-1}$ from $p$.\hfill{$\square$}
\end{lemma}

\begin{remark}
Coordinates can be adapted to a null geodesic congruence such that the metric takes the form
\be
g=d\lambda (d\tau+p_i\, dy^i+ q\, d\lambda)+\gamma_{ij} \,dy^i dy^j~,
\ee
where $\tau$ is the affine parameter of the null geodesics and $\lambda$ can be identified with the null Hamilton-Jacobi function, i.e. $g^{\mu\nu} \partial_\mu \lambda \partial_\nu \lambda=0$, with $y$ completing the rest of the spacetime coordinates. All non-specified components of the metric depend on all coordinates. \hfill{$\Diamond$}
\end{remark}

\subsection{Geodesic deviation equation}

\begin{figure}[htbp]%\label{fig:2}
\begin{center}
\begin{eqnarray*}
\begin{tikzpicture}[scale=0.6]
  % Coordinates
  \coordinate (A) at (0.00, 0.00);
  \coordinate (B) at (0.00, 6.00);
  \coordinate (C) at (0.00, 2.00);
  \coordinate (D) at (2.25, 3.75);
  \coordinate (E) at (1.00, 4.00);
  \coordinate (F) at (2.00, 4.00);
  \coordinate (G) at (1.00, 5.50);
  \coordinate (H) at (-0.50, 6.00);
  \coordinate (I) at (-0.50, 0.00);
  \coordinate (J) at (1.50, 2.50);
  \coordinate (K) at (-0.50, 2.00);
  \coordinate (L) at (0.46, 5.97);
  \coordinate (M) at (0.75, 0.00);
  \coordinate (N) at (2.50, 3.00);
  \coordinate (O) at (-0.75, 1.25);
  \coordinate (P) at (1.25, 6.00);
  \coordinate (Q) at (1.25, 0.00);
  \coordinate (R) at (3.00, 3.50);
  \coordinate (S) at (-0.50, 1.75);
  \coordinate (T) at (1.70, 6.07);
  \coordinate (U) at (2.00, 0.00);
  \coordinate (V) at (4.00, 4.25);
  \coordinate (W) at (0.25, 1.50);

  % Drawing
  \draw[blue, thick] (A) .. controls (0.00, 2.00) and (2.25, 3.75) .. (B);
  \draw[->, red] (E) -- (F);
  \draw[->, blue] (E) -- (G);
  \draw[thick] (H) .. controls (1.50, 2.50) and (-0.50, 2.00) .. (I);
  \draw[thick] (L) .. controls (2.50, 3.00) and (-0.75, 1.25) .. (M);
  \draw[thick] (P) .. controls (3.00, 3.50) and (-0.50, 1.75) .. (Q);
  \draw[thick] (T) .. controls (4.00, 4.25) and (0.25, 1.50) .. (U);
   \node[below, font=\large] at (A) {$\gamma$};
    \node[above, red, font=\large] at (F) {$\vz$};
     \node[above, blue, font=\large] at (G) {$\vu$};
\end{tikzpicture}
\end{eqnarray*}
\end{center}
%\begin{figure}[htbp]
\caption{\small{The figure depicts a family of time-like geodesics  $\gamma(\tau, t)$ parameterised by $t$.  The  time-like geodesic $\gamma(\tau)=\gamma(\tau, 0)$ has tangent vector field $\vu$.  The Jacobi vector field, or equivalently deviation vector field, $\vz$, is taken to be orthogonal to $\vu$ because $\vu$ is a trivial solution to the geodesic deviation equation.  As a result, $\vz$ is space-like.}}
\centering
\label{fig:2}
\end{figure}
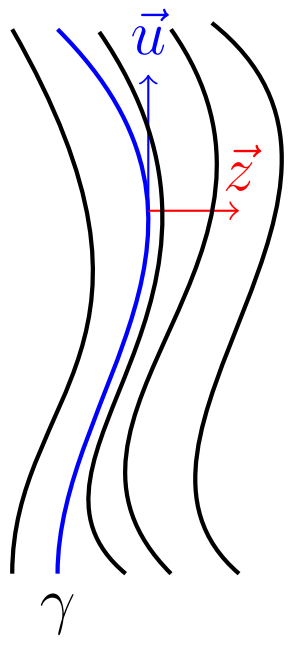

The geodesic deviation equation expresses the relative acceleration of a one-parameter family of geodesics in terms of the curvature of spacetime. Consider the family of geodesics $\gamma(\tau,t)$ parameterised by $t$, where $\tau$ is the affine parameter.  The tangent vector field of the geodesics is
\be
\vu\equiv \frac{\partial}{\partial \tau} \gamma(\tau,t)~,
\ee
and the displacement vector or Jacobi vector field is
\be
\vz\equiv \frac{\partial}{\partial t} \gamma(\tau,t)~.
\ee
It follows from the definitions that $[\vu, \vz]=0$.  The vector field  $\vz$ indicates the relative position of nearby geodesics.  Therefore, the relative acceleration of nearby geodesics is $\nabla_\tau^2 \vz$.  This can be expressed as follows:
\be
\nabla_\tau^2 \vz^\mu=\nabla_\tau \nabla_t \vu^\mu=R_{\rho\lambda}{}^\mu{}_\nu  \vu^\rho \vz^\lambda \vu^\nu~,
\label{gdeqn}
\ee
where to establish the first equality we have used that $\vu$ and $\vz$ commute and for the second equality we have used the standard curvature identity and the geodesic equation.

\begin{remark}
For a geodesic congruence, one expects to find six linearly independent non-trivial  Jacobi vector fields. This is because (\ref{gdeqn}) is a second order equation and a solution is specified uniquely for small $\tau$ provided the position and velocity initial conditions are specified for $\tau=0$. However, one of the position boundary conditions can be absorbed in the definition of $\tau$. Therefore it is sufficient to specify only three position initial conditions.  Moreover from (\ref{gdeqn}), we have that $\vu$ solves the equation and so it can be thought of as one of the Jacobi vector fields. But, such a vector field will not give a relative displacement vector field. Instead it will correspond to a variation along the direction of the geodesic congruence. To exclude this possibility, the independent  Jacobi vector fields $\vz$ are taken to be orthogonal to $\vu$, $g(\vu, \vz)=0$.   Therefore, there are only six independent boundary conditions that can be imposed giving rise to \emph{six independent Jacobi fields}. Moreover, if the Jacobi fields vanish at a point, then there are only three linearly independent remaining.  However as we shall see below in some computations, it is convenient  not to impose the orthogonality condition $g(\vu, \vz)=0$ to  the Jacobi vector fields and include $\vu$ as one of the Jacobi fields. \hfill{$\Diamond$}
\end{remark}

It is apparent that if two geodesics that have a non-trivial relative displacement described by the Jacobi vector field $\vz$ on one of the geodesics, when they intersect $\vz$ vanishes at the intersection. In  a geodesic congruence with more than one Jacobi fields, such a behaviour of geodesics can be described by saying that a linear combination of the Jacobi fields vanishes.

\begin{definition}
Consider a geodesic congruence emanating for a hypersurface $\Sigma$. A \emph{focus point} of the congruence is a point that the  Jacobi vector fields of the congruence become linearly dependent.  Similarly, a \emph{conjugate point} $q$ of a geodesic congruence emanating from a point $p$ is a point  that a linear combination of the Jacobi vector fields vanishes.\hfill{$\triangle$}
\end{definition}

In the next two sections, we shall examine the relation of the conjugate and focus points to the expansion $\theta$ of geodesic congruences. This relation will yield a criterion when to expect for a conjugate or focus point to emerge in geodesic congruences.

\subsection{Conjugate  points}

\subsubsection{Time-like geodesic congruences}

There is a relation between the expansion $\theta$ of a geodesic congruence with the existence of conjugate points on a geodesic.  To see this, consider a geodesic congruence emanating from a point $p$ which is conjugate to a point $q$ on some geodesic $\gamma$ of the congruence, see figure \ref{fig:conj}(a). Set the affine parameter $\tau$ such that at $\tau=0$, $\gamma(0)=p$. As the geodesic congruence is emanating from $p$, we impose the boundary condition $\vz(0)=0$. This boundary conditions restricts the number of potentially linearly independent Jacobi vector fields to four provided that the tangent vector field $\vu$ of the congruence  is included.  The geodesic deviation equation is linear in the displacement vector $\vz$. So we can  write the solutions as
\be
\vz(\tau)^\mu=A^\mu{}_\nu(\tau) \dot{\vz}^\nu(0)~,
\ee
where $\dot{\vz}(0)\equiv \nabla_\tau\vz\vert_{\tau=0}$ is the boundary condition for the derivative of the Jacobi vector fields  at $p$. Assuming that in addition to the three independent boundary conditions for $\dot{\vz}$ at $\tau=0$, we add also $\dot{\vz}(0)=\vu(0)$, the boundary condition for $A$ can be chosen as $A^\mu{}_\nu(0)=\delta^\mu{}_\nu$.

Substituting this into the geodesic deviation equation, we find that
\be
\nabla_\tau^2 A^\mu{}_\kappa =R_{\rho\lambda}{}^\mu{}_\nu  \vu^\rho A^\lambda{}_\kappa  \vu^\nu~,
\ee
where we have used that $(\dot\vz(0), \vu(0))$ are four linearly independent vector fields.

Next,  we have that
\be
\vu^\mu \nabla_\mu \vz^\nu=\vz^\mu \nabla_\mu \vu^\nu\Longrightarrow \nabla_\tau A^\nu{}_\rho \dot\vz^\rho(0)=A^\mu{}_\rho \dot\vz^\rho(0) \nabla_\mu \vu^\nu\Longrightarrow \nabla_\tau A^\nu{}_\rho=A^\mu{}_\rho \nabla_\mu \vu^\nu~.
\ee
This expresses the shear, rotation and expansion of the geodesic congruence in terms of geodesic deviation data.  In particular for the expansion $\theta$, we have that
\be
\theta=\frac{d}{d\tau}\ln |\det A|~.
\label{Atheta}
\ee
This formula relates, the conjugate points along a geodesic with the singularities of the expansion $\theta$.  In particular, if $\det A=0$, then the expansion $\theta\rightarrow -\infty$.

\begin{prop}\label{prop:expansion}
Let $\M$ be a spacetime with a future directed time-like geodesic congruence emanating from a point $p\in \M$. If $R_{\mu\nu}\vu^\mu\vu^\nu\geq 0$ for all time-like vector fields $\vu$ and there is a point $r$ along a geodesic $\gamma$ of the congruence such that the expansion $\theta_0\equiv \theta\vert_r<0$, then within an affine distance $\tau\leq 3|\theta_0|^{-1}$ from $r$ there is a conjugate point $q$ to $p$ provided that $\gamma$ extends that far.
\end{prop}
\begin{proof}
First note that as a consequence of proposition \ref{pro:ortho}, the geodesic congruence emanating from $p$ can be viewed as a geodesic congruence orthogonally emanating from a hypersurface within a normal neighbourhood of $p$. Then,
under the hypotheses of the proposition and as a consequence of Lemma \ref{lemma:irrotational}, there is a point $q$ along the geodesic $\gamma$ such that $\theta\rightarrow -\infty$.    On the other hand,
for a conjugate point $q$ to $p$ to exist along $\gamma$,  a linear combination of the three independent Jacobi fields emanating from $p$ vanishes at $q$.  This is indeed the case as a consequence of (\ref{Atheta}).  This is because as $\theta\rightarrow -\infty$, $\det A=0$ and so  the Jacobi fields because linearly independent.  The converse statement is also true.
\end{proof}

\begin{figure}[htbp]%\label{fig:2xxx}
\begin{center}
\begin{eqnarray*}
\begin{tikzpicture}[scale=0.6]
  % Coordinates
  \coordinate (A) at (0.00, 0.00);
  \coordinate (B) at (-0.14, -0.09);
  \coordinate (C) at (-6.00, 0.00);
  \coordinate (D) at (6.00, 0.00);
  \coordinate (E) at (-2.35, -4.12);
  \coordinate (F) at (2.92, -3.86);
  \coordinate (G) at (-2.00, 4.00);
  \coordinate (H) at (2.00, 4.00);
  \coordinate (I) at (0.00, 6.00);
  \coordinate (J) at (0.00, -6.00);
  \coordinate (K) at (4.00, 3.00);
  \coordinate (L) at (4.00, -3.00);
  \coordinate (M) at (-1.62, 4.63);
  \coordinate (N) at (-0.45, 5.69);
  \coordinate (P) at (-3.50, -3.00);
  \coordinate (Q) at (-3.50, 2.75);
  \coordinate (O) at (-0.75, 5.75);
  \coordinate (R) at (-1.50, 5.00);
  \coordinate (S) at (2.50, -2.50);
  \coordinate (T) at (-2.75, 2.25);
  \coordinate (U) at (7.00, 2.00);
  \coordinate (V) at (7.00, -2.00);
  \coordinate (W) at (2.45, 3.00);
  \coordinate (X) at (3.00, 3.25);
  \coordinate (Y) at (3.00, 0.00);
  \coordinate (Z) at (4.50, 0.00);
  \coordinate (AA) at (2.25, -3.00);
  \coordinate (AB) at (3.00, -3.50);

  % Drawing
  \draw[thick, rotate=-147.3] (-0.07, -0.04) ellipse (6.00 and 6.00);
  \draw[green, thick] (C) .. controls (-2.35, -4.12) and (2.92, -3.86) .. (D);
  \draw[green, thick, dashed] (C) .. controls (-2.00, 4.00) and (2.00, 4.00) .. (D);
  \draw[red, thick] (I) .. controls (4.00, 3.00) and (4.00, -3.00) .. (J);
  \draw (I) .. controls (-0.45, 5.69) and (-0.75, 5.75) .. (M);
  \draw[blue, thick, loosely dashdotted] (J) .. controls (-3.50, -3.00) and (-3.50, 2.75) .. (M);
  \draw[red, thick, loosely dashed] (I) .. controls (-0.75, 5.75) and (-1.50, 5.00) .. (M);
  \draw[orange, thick] (I) .. controls (7.00, 2.00) and (7.00, -2.00) .. (J);
  \draw[->, thick] (W) -- (X);
  \draw[->, thick] (Y) -- (Z);
  \draw[->, thick] (AA) -- (AB);
  \fill[red] (2.55, -2.55) circle (2.3pt);
  \fill[blue] (-2.70, 2.39) circle (2.3pt);
   \node[above, font=\large] at (I) {$n$};
    \node[ below right, red] at (2.55, -2.55) {$x$};
    \node[below right , blue] at (-2.70, 2.39) {$y$};
    \node[below, red, font=\large] at (J) {$s$};
    \node[below] at (Z) {$\vz$};
     \fill[red] (-1.62, 4.63) circle (2.5pt);
  \fill[red] (J) circle (2.5pt);
   \node[below, red, font=\large] at (-1.62, 4.63) {$q$};
    \node[left, red, font=\large] at (2.60, 1.00) {$\gamma$};
  \end{tikzpicture}
\end{eqnarray*}
\end{center}
%\begin{figure}[htbp]
\caption{\small{The figure depicts a geodesic  $\gamma$ that passes through the south pole, the point $x$ at the equator, the north pole $n$ and extends to the point $q$ beyond the north pole of a 2-sphere. The vector field $\vz$ is the Jacobi field of the congruence of great circles emanating either from the south pole $s$ or from the equator of the sphere and vanishes at the north pole -- the equator is depicted by the green line. In either case, the north pole $n$ is a conjugate point of the congruence emanating from $s$ on $\gamma$ or the focus point of the congruence emanating from the equator on $\gamma$. For later use, notice that the geodesic $\gamma$ is not the curve with minimal length from $s$ to $q$ or from the equator to $q$. The portion of the great circle, coloured in blue, from $s$ or the equator that passes through $y$ to $q$ has less length that $\gamma$. In any case, the theorem says that there must be other curves in the neighbourhood of $\gamma$, not necessarily geodesics, that have length less than $\gamma$ from $s$ to $q$ or from equator to $q$. }}
\centering
\label{fig:2xxx}
\end{figure}
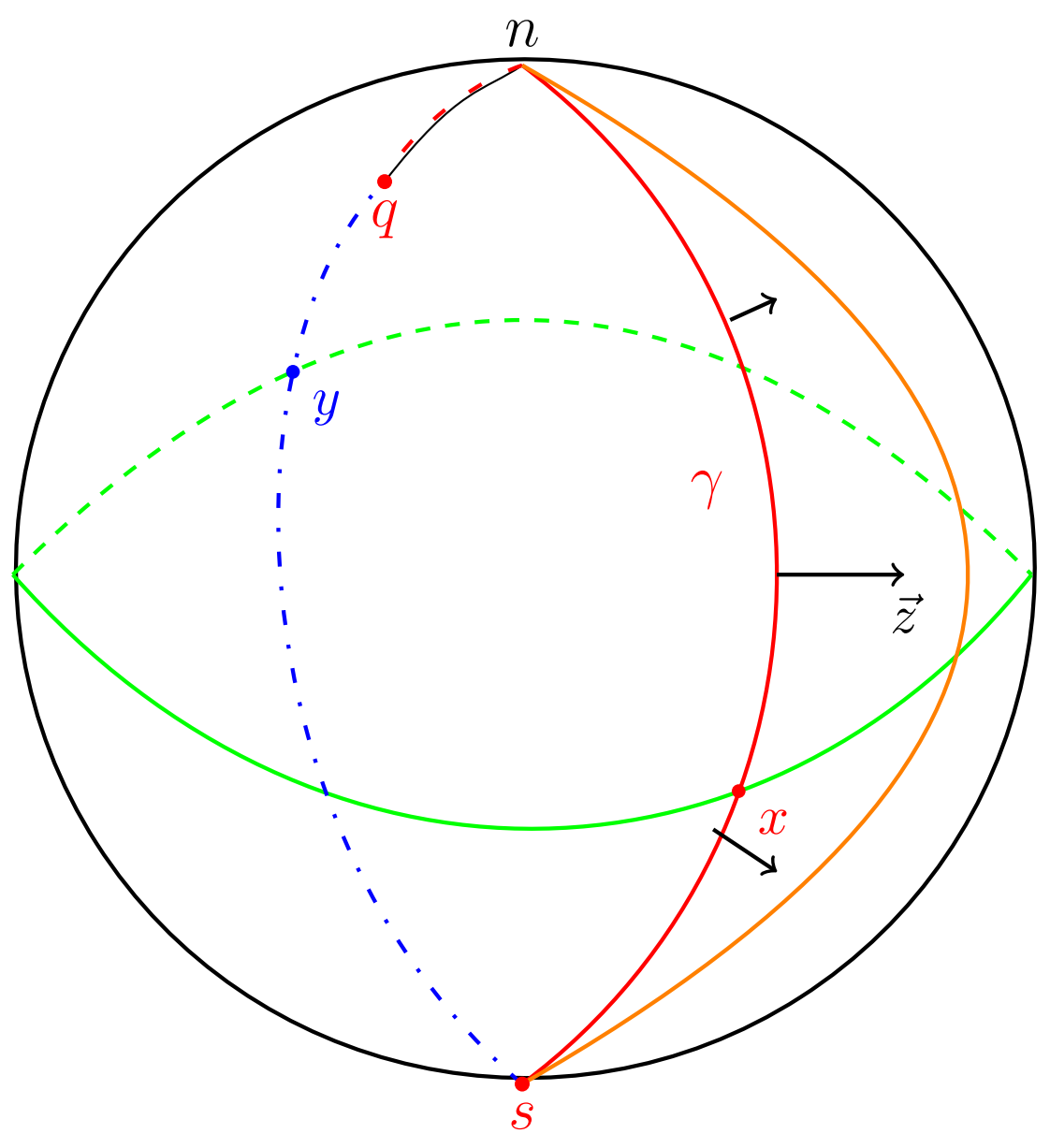

\begin{example}  The most famous example of conjugate points of geodesic congruences  comes from Riemannian geometry for $\M=S^2$ equipped with the round metric, see figure \ref{fig:2xxx}. Consider the geodesic congruence of great circles emanating from the south pole. These have a conjugate point at the north pole of the sphere.  This geodesic congruence can also be seen as emanating from the equator. Then, the north pole is a focus point.   The geodesics are orthogonal to the equator and so the geodesic congruence is irrotational.  Notice that despite the presence  of conjugate and focus points in the geodesics congruences considered, the underlying space is smooth. The singularities that occur are properties of the congruence and they do not affect the smoothness of the sphere. \hfill{$\triangledown$}
\end{example}

%\begin{minipage}
%\end{minipage}

The assumption of the proposition \ref{prop:expansion} that the expansion $\theta$ must have a negative value at some point $r$ of the geodesic $\gamma$ can be weaken.  This can be replaced by the \emph{time-like generic condition} which states that every time-like geodesic of the spacetime $M$ has at least one point such that $R_{\mu\nu\lambda\rho} \vu^\nu \vu^\rho \not=0$ for all time-like vectors $\vu$.  The proof of this statement is given in \cite{HE}. Putting this result together with those of proposition \ref{prop:expansion}, one concludes the following.

\begin{prop}\label{prop:generic}
Suppose that the spacetime $M$ satisfies the time-like generic condition and $R_{\mu\nu} \vu^\mu\vu^\nu\geq0$ for all time-like vector fields $\vu$. Then every time-like geodesic of $M$ that can be sufficiently extended has a pair of conjugate points.
\end{prop}

\begin{remark}
The time-like generic condition is equivalent to the statement  that the ``electric'' part of the gravitational field strength does not identically vanish.  While the condition $R_{\mu\nu} \vu^\mu\vu^\nu\geq0$ asserts that gravity is attractive. Thus, the statement above states that if the electric part of the gravitational field strength does not vanish everywhere and gravity is attractive, then the time-like geodesics exhibit conjugate points.
\hfill{$\Diamond$}
\end{remark}

\subsubsection{Null geodesic congruences}

So far, we have investigated the conjugate points of time-like geodesic congruences. Next we shall turn to the investigation of conjugate points of null geodesic congruences. First, let us clarify when a null geodesic has conjugate points.  Let $\vu$ be the tangent vector of a null geodesic congruence and $\vz$ be the displacement vector. Using both the geodesic deviation equation, the null geodesic equation and the Bianchi identities of the Riemann tensor, one can establish that
\be
\nabla_{\vu} \nabla_{\vu}(g(\vu, \vz))=\nabla^2_\lambda g(\vu, \vz)=0~.
\ee
where $\vu(\lambda)=1$ with $\lambda$ the affine parameter.  If the null geodesic $\gamma$ has a pair of conjugate points $p$ and $q$, then $g(\vu, \vz)|_p=g(\vu, \vz)|_q=0$.  But $g(\vu, \vz)$ satisfies a second order differential equation above, which implies that $g(\vu, \vz)=0$ along the whole of $\gamma$, i.e. the displacement vector must be orthogonal to the tangent vector of the geodesic. However, if $\vz$ is a displacement vector, so is $\vz'=\vz+ (a+b\lambda) \vu$. As $\vu$ is tangent along the geodesics, the \emph{independent} displacement vector fields should be taken to be both orthogonal to $\vu$ and linearly independent from $\vu$, i.e. they should be sections $\hat\vz$ of $\widehat {TM}$. This motivates the following definition.

\begin{definition}
Let $\vu$ be the tangent vector field of a null geodesic congruence. Then, the points $p$ and $q$ of a null geodesic $\gamma$ are conjugate, iff the displacement vector $\vz$ of the congruence, which is orthogonal to $\vu$, has the property that $\hat\vz$ vanishes at both $p$ and $q$, i.e. $\vz$ vanishes up to a possible addition of $\vu$ at those points.\hfill{$\triangle$}
\end{definition}

As for time-like congruences at conjugate points the expansion $\theta$ of the null geodesic congruence diverges to $-\infty$. In view of this, there is a extension of the proposition \ref{prop:expansion} to null geodesic congruencies. Furthermore, there is also an extension  of the proposition \ref{prop:generic}.  In particular for the latter, the spacetime $M$ satisfies the \emph{null generic condition}, iff every null geodesic has a point such that $\vu_{[\mu} R_{\nu]\rho\sigma[\lambda} \vu_{\delta]} \vu^\rho \vu^\sigma\not=0$ for every null vector $\vu$.

\begin{prop} \label{prop:genericnull}
Suppose that the spacetime $M$ satisfies the null generic condition and $R_{\mu\nu} \vu^\mu\vu^\nu\geq0$ for all null vector fields $\vu$. Then,  every null geodesic of $M$ that can be sufficiently extended  has a pair of conjugate points.
\end{prop}

\begin{figure}[htbp]%\label{fig:conj}
\begin{center}
\begin{eqnarray*}
\begin{tikzpicture}[scale=0.85]
  % Coordinates
  \coordinate (A) at (0.00, 0.00);
  \coordinate (B) at (2.00, 6.00);
  \coordinate (C) at (-0.25, 3.25);
  \coordinate (D) at (2.50, 3.75);
  \coordinate (E) at (2.50, 6.96);
  \coordinate (F) at (2.50, 6.50);
  \coordinate (G) at (2.25, 7.00);
  \coordinate (J) at (0.75, 2.50);
  \coordinate (K) at (-0.25, 4.75);
  \coordinate (L) at (0.50, 6.00);
  \coordinate (M) at (-1.00, 2.00);
  \coordinate (H) at (1.16, 2.98);
  \coordinate (I) at (2.00, 3.00);
  \coordinate (N) at (0.75, 3.75);
  \coordinate (O) at (2.75, 1.25);
  \coordinate (P) at (2.50, 4.00);
  \coordinate (Q) at (2.50, 1.00);
  \coordinate (R) at (1.19, 4.00);
   % Coordinates
  \coordinate (A1) at (7.00, 0.00);
  \coordinate (B1) at (16.00, 0.00);
  \coordinate (C1) at (10.00, -1.00);
  \coordinate (D1) at (13.25, 1.00);
  \coordinate (E1) at (11.00, 6.00);
  \coordinate (F1) at (12.00, 0.00);
  \coordinate (G1) at (11.50, 2.25);
  \coordinate (H1) at (12.75, 4.25);
  \coordinate (I1) at (10.25, -0.25);
  \coordinate (J1) at (9.75, 1.50);
  \coordinate (K1) at (12.25, 4.00);
  \coordinate (L1) at (12.75, 0.20);
  \coordinate (M1) at (11.25, 1.25);
  \coordinate (N1) at (14.75, 4.75);
  \coordinate (O1) at (13.50, 0.25);
  \coordinate (P1) at (12.00, 2.25);
  \coordinate (Q1) at (16.50, 4.50);
  \coordinate (R1) at (11.98, 3.00);
  \coordinate (S1) at (13.00, 2.90);
  \coordinate (T1) at (12.15, 4.60);
  \coordinate (U1) at (10.00, 7.00);
  \coordinate (V1) at (10.50, 6.50);
  \coordinate (W1) at (10.75, 7.00);
% Drawing
  \draw[blue, thick] (B) .. controls (-0.25, 3.25) and (2.50, 3.75) .. (A);
  \draw[blue, thick] (B) .. controls (2.50, 6.50) and (2.25, 7.00) .. (E);
  \draw[thick] (A) .. controls (0.75, 2.50) and (-0.25, 4.75) .. (B);
  \draw[thick] (B) .. controls (0.50, 6.00) and (-1.00, 2.00) .. (A);
  \draw[->, red] (H) -- (I);
  \draw[thick] (B) .. controls (0.75, 3.75) and (2.75, 1.25) .. (A);
  \draw[thick] (B) .. controls (2.50, 4.00) and (2.50, 1.00) .. (A);
  \draw[->, blue] (H) -- (R);
  \fill[black] (A) circle (2pt);
  \fill[black] (B) circle (2pt);
  \node[above,blue, font=\large] at (2.10, 6.60) {$\gamma$};
   \node[above,blue, font=\large] at (R) {$\vu$};
    \node[below, font=\large] at (A) {$p$};
     \node[above, font=\large] at (B) {$q$};
      \node[above, red, font=\large] at (I) {$\vz$};
       \node[below, font=\large] at (0.00, -1.00) {Fig. (a)};
        % Drawing
  \draw[ultra thick] (A1) .. controls (10.00, -1.00) and (13.25, 1.00) .. (B1);
  \draw[blue, thick] (F1) .. controls (11.50, 2.25) and (12.75, 4.25) .. (E1);
  \draw[thick] (I1) .. controls (9.75, 1.50) and (12.25, 4.00) .. (E1);
  \draw[thick] (L1) .. controls (11.25, 1.25) and (14.75, 4.75) .. (E1);
  \draw[thick] (O1) .. controls (12.00, 2.25) and (16.50, 4.50) .. (E1);
  \draw[->, red] (R1) -- (S1);
  \draw[->, blue] (R1) -- (T1);
  \draw[blue, thick] (E1) .. controls (10.50, 6.50) and (10.75, 7.00) .. (U1);
  \fill[black] (E1) circle (2pt);
  \node[right, font=\large] at (B1) {$\S$};
   \node[right, font=\large] at (10.90, 6.30) {$q$};
  \node[right, blue, font=\large] at (12.15, 4.60) {$\vu$};
  \node[right, red, font=\large] at (13.00, 2.90) {$\vz$};
    \node[below, blue, font=\large] at (10.35, 6.70) {$\gamma$};
      \node[below, font=\large] at (F1) {$p$};
      \node[below, font=\large] at (12.00, -1.00) {Fig. (b)};
    \end{tikzpicture}
\end{eqnarray*}
\end{center}
%\begin{figure}[htbp]
\caption{\small{Figure (a): the point $q$ is conjugate to the point $p$ on the future directed geodesic $\gamma$ with tangent vector field $\vu$ and vice versa $p$ is the conjugate point of $q$ in the past direction. At the conjugate points $p$ and $q$ a linear combination of the Jacobi fields $\vz$ vanishes.
Figure (b): the point $q$ is the focus point of a geodesic congruence emanating from the spacelike (hyper)surface $\S$ lying on the geodesic $\gamma$. Again a linear combination of Jacobi fields vanishes at $q$. }}
\centering
\label{fig:conj}
\end{figure}

\subsection{Focus points}

Consider a  smooth space-like hypersurface $\S$ in the spacetime $\M$. Let $\vt$ be the normalised  tangent vector field of a time-like geodesic congruence emanating orthogonally from $\S$, see figure \ref{fig:conj}(b). The \emph{extrinsic curvature} or equivalently \emph{second fundamental form} of $\S$ is defined as
\be
K_{\mu\nu}=\nabla_\mu \vt_\nu|_{\S}~.
\label{sffh}
\ee
It is clear that $\vt^\mu K_{\mu\nu}=\vt^\nu K_{\mu\nu}=0$ and in addition $K_{\mu\nu}=K_{\nu\mu}$ as the congruence is taken to be hypersurface orthogonal.  It is clear from this that the expansion $\theta$ of the congruence restricted on $\S$ is given by the extrinsic curvature of $\S$, i.e.
\be
\theta|_{\S}=g^{\mu\nu} K_{\mu\nu}\equiv K=h^{\mu\nu} K_{\mu\nu}~.
\ee
Then, a consequence of this and Lemma \ref{lemma:irrotational} is the  following:

\begin{prop}\label{prop:focuspoint}
Let $\M$ be a spacetime with a hypersurface $\S$ with $K<0$ at some point $p\in \S$. If in addition $R_{\mu\nu} \vw^\mu\vw^\nu\geq 0$ for all time-like vector fields $\vw$, then within proper time $\tau\leq 3|K|^{-1}$ there will be a focus point $q$ of $\Sigma$ along the the geodesic $\gamma$ of the congruence  that passes through $p$ provided that $\gamma$ extends that far.\hfill{$\square$}
\end{prop}

\subsection{The local structure of spacetime}\label{sec:localstructure}

Locally the causal properties of a sufficiently smooth spacetime, at least $C^2$,  are identical to those of Minkowski space. To demonstrate this choose a point $p\in \M$ and consider a  normal neighbourhood (patch) of $p$, $N_p\subset T_p\M$ with coordinates $u$.  Then, the time-like interior $C_p$ and the lightcone $\partial C_p$ of $p$ are defined as
\be
C_p=\{u\in N_p\vert\, g_{\mu\nu}(0) u^\mu u^\nu<0\}~,~~~\partial C_p=\{u\in N_p\vert\, g_{\mu\nu}(0) u^\mu u^\nu=0\}~,
\ee
where we have used the same symbol for the points of $N_p$ and their coordinates. Moreover, we shall denote with $C_p^+$, $\partial C_p^+$ and $C_p^-$, $\partial C_p^-$ the future and the past of these sets, respectively. Note that $N_p$ is a subspace of a vector space.

\begin{figure}[htbp]%\label{fig:normal1}
\begin{center}
\begin{eqnarray*}
\begin{tikzpicture}[scale=0.6]
  % Coordinates
  \coordinate (A) at (-1.00, 7.00);
  \coordinate (B) at (5.00, 1.00);
  \coordinate (C) at (-1.00, 1.00);
  \coordinate (D) at (5.00, 7.00);
  \coordinate (E) at (2.00, 4.00);
  \coordinate (F) at (2.00, 7.00);
  \coordinate (G) at (1.25, 5.50);
  \coordinate (H) at (2.50, 6.00);
  \coordinate (I) at (1.50, 5.00);
   \coordinate (J) at (3.00, 0.00);
   % Coordinates
    \coordinate (A1) at (7.00, 7.00);
  \coordinate (B1) at (13.00, 1.00);
  \coordinate (C1) at (7.00, 1.00);
  \coordinate (D1) at (13.00, 7.00);
  \coordinate (E1) at (10.00, 4.00);
  \coordinate (F1) at (8.97, 7.01);
  \coordinate (G1) at (8.25, 5.50);
  \coordinate (H1) at (9.50, 6.00);
   \coordinate (J1) at (11.00, 0.00);

  % Drawing
  \draw [fill = yellow] (-1.00, 7.00) -- (5.00, 7.00) --  (2.00, 4.00);
  \draw [fill = yellow] (-1.00, 1.00) -- (5.00, 1.00) --  (2.00, 4.00);
  \draw[thick] (A) -- (B);
  \draw[thick] (C) -- (D);
  \draw[red, thick] (E) .. controls (1.25, 5.50) and (2.50, 6.00) .. (F);
  \draw[white] (A) -- (D);
  \draw[white] (C) -- (B);
  \draw[->, blue] (E) -- (I);
   \node[left] at (E) {$p$};
   \node[left, font=\large] at (J) {Fig.(a)};
    % Drawing
  \draw [fill = yellow] (7.00, 7.00) -- (13.00, 7.00) --  (10.00, 4.00);
  \draw [fill = yellow] (7.00, 1.00) -- (13.00, 1.00) --  (10.00, 4.00);
  \draw[thick] (A1) -- (B1);
  \draw[thick] (C1) -- (D1);
  \draw (E1) -- (E1);
  \draw[red, thick] (E1) .. controls (8.25, 5.50) and (9.50, 6.00) .. (F1);
  \draw[white] (A1) -- (D1);
  \draw[white] (C1) -- (B1);
  \node[left] at (E1) {$p$};
  \node[left] at (9.50, 4.50) {$q$};
  \node[left, font=\large] at (J1) {Fig.(b)};
  \end{tikzpicture}
\end{eqnarray*}
\end{center}
%\begin{figure}[htbp]\label{fig:normal1}
\caption{\small{In figure (a), if a causal curve is future directed and time-like at the centre $p$ of normal coordinates, it remains inside the interior $C^+_p$ of the future lightcone  neighbourhood of $p$. The interiors of past $C^-_p$ and future $C^+_p$ lightcone of $p$ within a normal neightbourhood of $p$ are depicted with the shaded regions. In figure (b), if a causal curve enters at the point $q$ the interior $C^+_p$ of future lightcone in a normal neighbourhood of a point $p$, it remains in the interior $C^+_p$ from then on.}}
\centering
\label{fig:normal1}
\end{figure}

\begin{prop}\label{prop:lcs}
Consider any continuous piecewise $C^1$ future directed causal curve $\gamma$ starting at $p$ and  entirely contained in a normal neighbourhood  $N_p\subset T_p\M$.

\begin{enumerate}[(i)]
\item If $\gamma$ is time-like at $p$, then it entirely lies in $C^+_p$.

\item If a causal curve enters $C_p^+$, it stays in $C_p^+$ from then on. Moreover, if at a point $q$  $q\in \gamma\cap \partial C^+_p$, the tangent vector $\vec{\dot\gamma}_q$ is time-like, then it enters $C^+_p$ and stays within $C_p^+$ thereafter.

    \item All forward directed causal curves in $N_p$ that pass through $p$  lie in the closure $\overline{C_p^+}$ of $C_p^+$.

    \item All forward directed causal curves that completely lie in $\partial C^+_p$ are null geodesics.
    \end{enumerate}

Similar statements apply to past directed causal curves.
\end{prop}
\begin{proof}

(i): Let $t$ be a parameter of $\gamma$ with $\gamma(0)=p$ and denote with $\vec\gamma(t)$ the position vector of the point $\gamma(t)\in N_p$ in normal coordinates, $\vec\gamma(0)=0$. Suppose that $\gamma$ is a future directed causal curve that it is time-like at $p$, see figure \ref{fig:normal1}(a).  Then, we shall show that it lies entirely in $C_p^+$. Indeed, let $\gamma$ be a curve that satisfies the hypothesis. This implies that
\be
g_{\mu\nu}(\gamma) \vec{\dot\gamma}^\mu \vec{\dot\gamma}^\mu\leq 0~,~~~g_{\mu\nu}(0) \vec{\dot\gamma}^\mu(0) \vec{\dot\gamma}^\mu(0)< 0~,
\label{ncon3}
\ee
where the first condition implies that $\gamma$ is causal and the second implies that $\gamma$ is time-like at $p$. Moreover, $\vec{\dot\gamma}$ is the tangent vector field of $\gamma$, where the ``dot'' on $\gamma$ denotes the differentiation of $\gamma$ with respect to the parameter $t$.

We must demonstrate in normal coordinates that
\be
g_{\mu\nu}(0) \vec\gamma^\mu(t) \vec\gamma^\nu(t)<0~,~~~\forall t ~\text{such that}~ \gamma(t)\in N_p~,
\label{ncon3a}
\ee
which is the condition for $\gamma$ to lie in $C_p^+$.
 As $\vec\gamma(0)=0$, one has that
\be
\vec\gamma^\mu(t)= t( \vec{\dot\gamma}(0)+ \vec v(t))~,
\ee
for some vector $\vec v$ with $\vec v(0)=0$.
As the tangent of $\gamma$ at the origin is time-like and future directed, by continuity it will remain time-like (and future directed) for a small interval $[0,\epsilon)$, i.e. the second condition in (\ref{ncon3}) will also be valid for small $t$. Therefore, for small enough $\epsilon$,
\be
\vec{\dot\gamma}(t)= \vec{\dot\gamma}(0)+ \vec v(t)+\mathcal{O}(\epsilon)~,~~~g_{\mu\nu}(0)\big(\vec{\dot\gamma}(0)+ \vec v(t)\big)^\mu \big(\vec{\dot\gamma}(0)+ \vec v(t)\big)^\nu +\mathcal{O}(\epsilon)< 0~,
\ee
where we have identified $\dot{\vec \gamma}$ and $\vec{\dot\gamma}$.
Multiplying the last condition with $t^2$, we deduce (\ref{ncon3a}) for $t\in [0, \epsilon)$.

Next using this, we have that
\be
\frac{d}{dt} \big(g_{\mu\nu}(0) \vec\gamma^\mu (t) \vec\gamma^\nu(t)\big)=2 g_{\mu\nu}(0) \vec\gamma^\mu(t) \vec{\dot \gamma}^\nu=2 g_{\mu\nu}(\gamma) \vec\gamma^\mu(t) \vec{\dot \gamma}^\nu<0~,~~~t\in [0,\epsilon)~,
\label{dtder}
\ee
where in the last equality we have used the Thomas formula  and the last inequality holds as both the tangent and the position vector of the curve are future directed time-like vectors. As a result the function $g_{\mu\nu}(0) \vec\gamma^\mu (t) \vec\gamma^\nu(t)$ is decreasing along the curve in the interval $[0, \epsilon)$. It will continue to do the same   for all points of $\gamma$ in $N_p$.  This will be the case even if a corner is encountered.  There will be a discontinuity in the derivative in (\ref{dtder}) but it will remain negative.  Therefore $\gamma$ lies in $C_p^+$.

(ii): To prove that a causal curve that enters $C_p^+$, it stays in $C_p^+$ from then on,  suppose that for some value of the parameter $t=t_0$, $\gamma(t_0)=q \in C_p^+$. Joining $p$ and $q$ with the forward directed time-like geodesic from $p$ to $q$ and extend it beyond $q$ with $\gamma$, we construct a piecewise $C^1$ causal future directed curve that it is time-like at $p$.  From the previous statement, this new curve will stay with $C^+_p$ and so the extension of $\gamma$ beyond $q$ will do the same.

Another consequence of this is that if at a point $\gamma(t_0)=q\in \gamma\cap \partial C^+_p$, $\vec{\dot\gamma}(t_0)$ is time-like, then it enters $C^+_p$ and stays within $C_p^+$ thereafter, see figure \ref{fig:normal1}(b). To demonstrate this, from the hypothesis, we have $g_{\mu\nu}(0) \vec\gamma^\mu(t_0)\vec\gamma^\nu(t_0)=g_{\mu\nu}(\gamma(t_0)) \vec\gamma^\mu(t_0)\vec\gamma^\nu(t_0)=0$ and $g_{\mu\nu}(\gamma(t_0)) \vec{\dot\gamma}^\mu(t_0)\vec{\dot\gamma}^\nu(t_0)<0$. Then, we have
\be
\frac{d}{dt}\Big(g_{\mu\nu}(0) \vec\gamma^\mu(t) \vec\gamma^\nu(t)\Big)\vert_{t=t_0}=2 g_{\mu\nu}(0) \vec{\dot\gamma}^\mu(t_0) \vec\gamma^\nu(t_0)=2 g_{\mu\nu}(\gamma(t_0)) \vec{\dot\gamma}^\mu(t_0) \vec\gamma^\nu(t_0)<0~,
\ee
as both $\vec\gamma^\nu(t_0)$ and $\vec{\dot\gamma}^\mu(t_0)$ are future directed with the former vector null and the latter time-like. Thus $\vec\gamma(t)$ becomes time-like for $t>t_0$ and so the curve enters $C^+_p$ and stays in $C_p^+$ beyond that.

(iii): To prove that future directed causal curves in $N_p$ lie in the closure of $C^+_p$, $\overline{C^+_p}=C^+_p\cup \partial C^+_p$,  it suffices to demonstrate that every future directed causal curve in $N_p$ is the limit of time-like curves.  Let $\gamma$ be the causal curve. Define the functions $\psi$ as the unique solution of the equation
\be
\frac{d\psi^\mu}{dt}+\frac{1}{2} g^{\mu\nu}(\gamma(t))\partial_\rho g_{\nu\sigma}(\gamma(t))\, \psi^\rho(t)\, \vec{\dot\gamma}(t)=\vec w^\mu(t)~,~~~\vec\psi(0)=0~,
\label{ncon5}
\ee
where $\vec w$ is a time-like vector on $\gamma$.

Next, consider the family of curves
\be
\vec\gamma_s(t)=\vec\gamma(t)+s\, \vec\psi(t)~,
\ee
parameterised by $s$.

Then upon using (\ref{ncon5})
\be
\frac{d}{ds}\Big( g_{\mu\nu}(\gamma_s) \vec{\dot\gamma}^\mu_s \,\vec{\dot\gamma}^\nu_s\Big)\Big|_{s=0}=
\partial_\rho g_{\mu\nu}\, \psi^\rho\, \vec{\dot\gamma}^\mu\, \vec{\dot\gamma}^\nu+ 2 g_{\mu\nu}(\gamma)\, \vec{\dot\psi}\, \vec{\dot\gamma}^\nu= 2g_{\mu\nu}(\gamma) \,\vec{w}^\mu \vec{\dot\gamma}^\nu<0~,
\ee
where the inequality arises because  $\gamma$ is a future directed causal curve while $\vec w$ is a future directed vector field on $\gamma$. As the tangent vectors of causal curves are either time-like or null, the above equation implies that there is an $\epsilon$ such that the curves $\gamma_s$ in the interval $s\in(0, \epsilon)$ are time-like with $\gamma_0=\gamma$, i.e. $\gamma$ is the limit of the time-like curves $\gamma_s$  as $s$ goes to zero. Therefore, the every neighbourhood of a point  $q=\gamma(t_0)$  includes  points $\gamma_s(t_0)$ of the time-like curves $\gamma_s$, which entirely lie in $C^+_p$. Therefore, $q$ must lie in  $\overline{C^+_p}$.  So the entire curve $\gamma$ must lie in  $\overline{C^+_p}$.

(iv):  It remains to show that a forward directed causal curve $\gamma$ that lies entirely in the light-cone $\partial C_p^+$ must be a null geodesic.
By the hypothesis
\be
g_{\mu\nu}(0) \vec\gamma^\mu(t)\, \vec\gamma^\nu(t)=0~,~~~g_{\mu\nu}(\gamma(t))\, \vec{\dot\gamma}^\mu(t)\, \vec{\dot\gamma}^\nu(t)\leq 0~.
\label{ncon6}
\ee
If at some point in the curve, the tangent vector becomes time-like, it follows from the statement (ii) of the proposition that it must enter $C_p^+$ and so it cannot be entirely in the lightcone. This contradict the hypothesis and so the tangent of the curve must be null everywhere.

From the first equation in (\ref{ncon6}) and the Thomas formula, $g_{\mu\nu}(\gamma) \vec\gamma^\mu\, \vec\gamma^\nu=0$. Taking the derivative of the first formula in (\ref{ncon6}), we find that
\be
g_{\mu\nu}(0) \vec{\dot\gamma}^\mu\, \vec\gamma^\nu=g_{\mu\nu}(\gamma) \vec{\dot\gamma}^\mu\, \vec\gamma^\nu=0~.
\label{ncon7}
\ee
Therefore, the position vector  of the curve  at each point $q\in \gamma$, $\vec \gamma_q$ and  $\vec{\dot\gamma}_q$ are both null and from the last condition in (\ref{ncon7}) orthogonal.  This implies that there must be co-linear, i.e. there is  a function $a$ such that
\be
\vec\gamma(t)=  a(t) \vec{\dot\gamma}(t)~,
\ee
with $a(0)=0$ as $\vec\gamma(0)=0$. Such curves are null geodesics.
\end{proof}

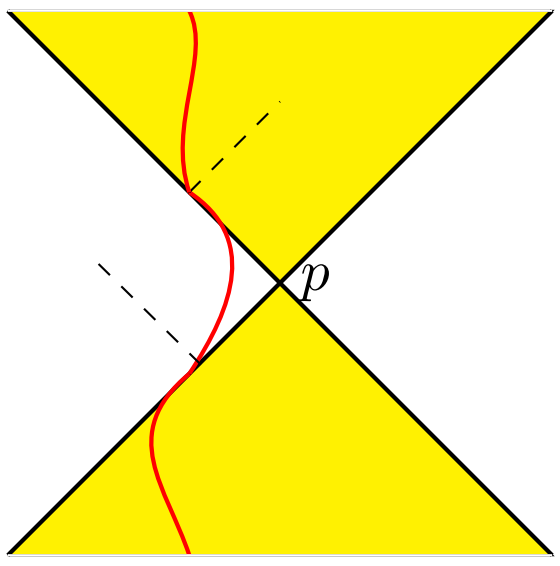
\begin{figure}[htbp]%\label{fig:2aa}
\begin{center}
\begin{eqnarray*}
\begin{tikzpicture}[scale=0.6]
  % Coordinates
  \coordinate (J) at (8.00, 7.00);
  \coordinate (K) at (14.00, 1.00);
  \coordinate (L) at (8.00, 1.00);
  \coordinate (M) at (14.00, 7.00);
  \coordinate (N) at (10.00, 1.00);
  \coordinate (O) at (10.00, 3.00);
  \coordinate (P) at (9.75, 1.75);
  \coordinate (Q) at (9.20, 2.33);
  \coordinate (R) at (10.00, 5.00);
  \coordinate (S) at (10.50, 3.75);
  \coordinate (T) at (10.75, 4.50);
  \coordinate (U) at (11.00, 6.00);
  \coordinate (V) at (10.00, 7.00);
  \coordinate (W) at (9.75, 5.75);
  \coordinate (X) at (10.25, 6.50);
  \coordinate (Y) at (9.00, 4.21);
  \coordinate (O_1) at (10.25, 2.98);
  \coordinate (J1) at (11.00, 4.00);
  \coordinate (J2) at (11.00, 0.00);

  % Drawing
  \draw [fill = yellow] (8.00, 7.00)-- (14.00, 7.00) -- (11.00, 4.00);
   \draw [fill = yellow] (8.00, 1.00)-- (14.00, 1.00) -- (11.00, 4.00);
  \draw[thick] (J) -- (K);
  \draw[thick] (L) -- (M);
  \draw[red, thick] (N) .. controls (9.75, 1.75) and (9.20, 2.33) .. (O);
  \draw[red, thick] (O) .. controls (10.50, 3.75) and (10.75, 4.50) .. (R);
  \draw[dashed] (R) -- (U);
  \draw[red, thick] (R) .. controls (9.75, 5.75) and (10.25, 6.50) .. (V);
  \draw[white] (J) -- (M);
  \draw[white] (L) -- (K);
  \draw[dashed] (Y) -- (O_1);
  \node[right] at (J1) {$p$};
  % \node[below] at (J2) {Fig. (a)};
\end{tikzpicture}
\end{eqnarray*}
\end{center}
%\begin{figure}[htbp]
\caption{\small{The figure depicts the behaviour of a causal curve in a normal neighbourhood of $p$ as enters as a time-like curve from the past, enter the boundary $\partial D^-_p$ of the past lighcone, continuous as a time like curve, enters the boundary $\partial D^+_p$ of the future fightcone and final enters the interior $D^+_p$ of the future lightcone.  After that, it will remain in $C_p^+$.  }}
\centering
\label{fig:2aa}
\end{figure}

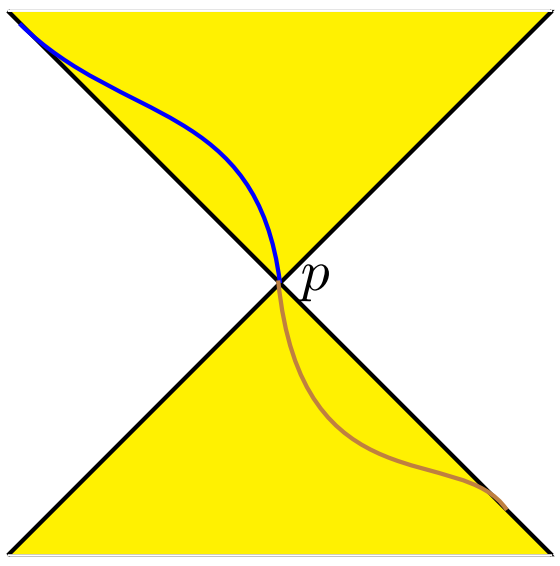
\begin{figure}[htbp]%\label{fig:2a}
\begin{center}
\begin{eqnarray*}
\begin{tikzpicture}[scale=0.6]
  % Coordinates
  \coordinate (A1) at (-1.00, 7.00);
  \coordinate (B1) at (5.00, 1.00);
  \coordinate (C1) at (-1.00, 1.00);
  \coordinate (D1) at (5.00, 7.00);
  \coordinate (E1) at (2.00, 4.00);
  \coordinate (F1) at (-0.87, 6.86);
  \coordinate (G1) at (1.75, 6.00);
  \coordinate (H1) at (0.25, 5.75);
  \coordinate (I1) at (4.50, 1.50);
  \coordinate (J1) at (1.98, 4.03);
  \coordinate (K1) at (4.00, 2.25);
  \coordinate (L1) at (2.23, 1.50);
  \coordinate (J11) at (2.00, 4.00);
  %\coordinate (J21) at (2.00, 0.00);

  % Drawing
  \draw [fill = yellow] (-1.00, 7.00)-- (5.00, 7.00) -- (2.00, 4.00);
   \draw [fill = yellow] (-1.00, 1.00)-- (5.00, 1.00) -- (2.00, 4.00);
  \draw[thick] (A1) -- (B1);
  \draw[thick] (C1) -- (D1);
  \draw[blue, thick] (E1) .. controls (1.75, 6.00) and (0.25, 5.75) .. (F1);
  \draw[brown, thick] (I1) .. controls (4.00, 2.25) and (2.23, 1.50) .. (J1);
  \draw[white] (A1) -- (D1);
  \draw[white] (C1) -- (B1);
  \node[right] at (J11) {$p$};
   %\node[below] at (J21) {Fig. (b)};
\end{tikzpicture}
\end{eqnarray*}
\end{center}
%\begin{figure}[htbp]\label{fig:2a}
\caption{\small{In the figure, the curves depicted in the normal neighbourhood of $p$ are not causal. The first one enters the interior $C^-_p$ of the past lightcone from the boundary $\partial C^-_p$, while the second one enters the boundary $\partial D^+_p$  from the interior $D^+_p$ of the normal neighbourhood of $p$.}}
\centering
\label{fig:2a}
\end{figure}

\begin{remark}
Causal curves that consist of null geodesics with corners are not considered to be null geodesics. This is a consequence of statement (ii) of the above proposition.  After a minor modification of the proof,   if a null geodesic starting at $p$ develops a corner later on and continuous after the corner as a curve with a null tangent vector, it will enter $C^+_p$. \hfill{$\Diamond$}
\end{remark}

 \begin{remark}
 It should be emphasised that the global causal structure of a spacetime can be very different from the local causal structure presented in the proposition above. One simple example is the cylinder $\M=S^1\times \bR$ with the flat metric $g=-dx^2+dy^2$, where the coordinate $0\leq x< 2\pi$ is periodically identified.  Such a spacetime is locally isometric to Minkowski space, so it enjoys the same local causal structure. However the global picture is very different. At every normal neighbourhood of  $p\in \M$, $\partial C^+_p$ is intersected by forward directed time-like  curves emanating from $p$, see also figure \ref{fig:noncausal}. This appears to be a direct contradiction to the results of the proposition above. However, these time-like curves go around $\M$ and so they lie outside the normal neighbourhood (patch) of $p$.\hfill{$\Diamond$}
\end{remark}

\section{Causal curves}

\subsection{Properties and the limit curve theorem}

Curves in a spacetime $\M$ can ``end'' in a variety of ways. They end because we just demand that they finish at some point in $\M$ or wind around forever in a spiral towards a point or they ``end'' because   a point  is missing from  $\M$. A definition that encapsulates all these possibilities is the following.

\begin{definition}
A point $p\in \M$ is the \emph{right end point} of a curve $\gamma: I\rightarrow \M$, iff for every neighbourhood of $p$, $U_p$, there is a $t_0\in I$ such that $\gamma(t)\in U_p$ for $t>t_0$. The \emph{left  end point} can be defined in a similar way. For future directed causal curves, the right (left) end points are the future (past) end points. If a curve \emph{does not have end points}, it is called \emph{inextendible} -- this notion can be extended to past and future inextendability for future and past directed causal curves. \hfill{$\triangle$}
\end{definition}

It should be emphasized that for a point $p$ to \emph{qualify as the end point} of a curve $\gamma$, it \emph{must lie} in $\M$. Curves have unique end points (provided that $\M$ is Hausdorff).  Also, the end point of a curve may not be a point in the curve. But, it will always be a limit of  points in the curve.

Before, we proceed further it is worth clarifying some points that arise in the investigation of curves in the context of general relativity. Traditionally, two curves that are related by a reparameterisation in the context of general relativity are identified. To be precise, let us give a definition.

\begin{definition}\label{def:reparameterisation}
Two causal  curves $\gamma: I\rightarrow \M$ and $\tilde \gamma: J\rightarrow \M$ are related by a reparameterisation $\phi$, iff $\phi: J\rightarrow I$ is a diffeomorphism that preserves the orientation of $I$ and $J$ and $\tilde\gamma=\gamma\circ \phi$, where $I, J\subset \bR$, e.g. $I=(a,b)$ and $J=(a', b')$ or $I=[a,b]$ and $J=[a', b']$ if end points are included.
\end{definition}

\begin{remark}
It should be stressed that although two causal curves related by a reparameterisation, as described in the previous definition, have the same image in $\M$. \emph{The converse is not true}. Different curves up to a reparameterisation can have the same image in $\M$, i.e. such curves \emph{cannot be distinguished} by their images alone. To illustrate this point,
 a simple example is to consider the curve $\gamma: (0,1)\rightarrow \M$ and $\phi: (0,2)\rightarrow (0,1)$, $t=\phi(s)=1/2 s$. Then, $\tilde \gamma(s)= \gamma\circ \phi(s)=\gamma(\frac{1}{2} s)$. Clearly $\gamma$ and $\tilde \gamma$ have the same image in $\M$ viewed as a subset of $\M$.

However, the converse is not true.  Indeed,  consider a periodic curve $\gamma: [0,1]\rightarrow \M$, i.e. $\gamma(0)=\gamma(1)$ and another curve $\lambda(t)= \gamma(2t)$ with $t\in[0,1]$.  Clearly, $\mathrm{Im}\,\gamma=\mathrm{Im}\,\lambda$ as subsets of $\M$ but $\gamma$ and $\lambda$ are \emph{distinct} as curves because the map $t\rightarrow 2t$ is not a diffeomorphism of $I$.
 \hfill{$\Diamond$}
\end{remark}

Another issue that arises in the investigation of curves, more specifically the convergence properties of sequences of curves, is how we take limits, i.e. what is the topology we use to define the limit of sequences. First, let us answer this question for the convergence of sequences of points in the spacetime $\M$ and then we shall deal with this issue for the curves.

\begin{remark}
By construction, the manifold $\M$ is a topological space, so the apparent answer for convergence of sequences of points in $\M$ is that we can use the topology we have originally put on $\M$. However, it is also convenient to view $\M$ as a topological metric space.  For this,    we use the distance function
  \be
  d_{g_e}(p, q)\equiv\inf_{\gamma\in \mC(p,q)}\, L_{g_e}(\gamma; p,q)~,
  \ee
  defined between two points $p,q$ in the spacetime  $\M$ with respect to a Riemannian metric $g_e$, where $\mC(p,q)$ is the set of all the continuous curves between $p$ and $q$ and $L_{g_e}(\gamma; p,q)$ is their length with respect to $g_e$, see definition \ref{def:c0length}.  All Riemannian metrics $g_e$ give rise to the same topology on $\M$. Under some reasonable topological assumptions, there always exist  Riemannian metrics on manifolds and so such a convergence can always be defined.  Furthermore, the topology defined on $\M$ by a Riemannian metric induces  on a chart a topology that it is equivalent to the usual Euclidean topology on $\bR^n$. Thus, on many occasions limits can be taken in the same way as if they were taken in $\bR^n$.\hfill {${\Diamond}$}
%{\bf This needs more explanation}
\end{remark}

Traditionally in general relativity, the convergence of sequences of curves $(\gamma_n)$ is described in terms of the convergence of sequences of points in the spacetime $\M$. This is in contrast to the alternative  that is to define a topology on the space of curves and then use it to define the convergence of sequences of curves  -- this point of view is clearly explained  in remark \ref{rem:curvtop} in relation to the limit curve theorem below.  To begin, we shall explore the traditional approach first.

\begin{definition}
Let $(\gamma_n)$ be a sequence of curves, then $p$ is an \emph{accumulation point} of the sequence, iff every neighbourhood of $p$, $U_p$, intersects an infinite number of curves of the sequence. On the other hand $p$ is a \emph{convergence point} of $(\gamma_n)$,  iff for every $U_p$, there is a $n_0$ such that all the curves with $n>n_0$ intersect $U_p$. \hfill {${\triangle}$}
\end{definition}

\begin{definition}\label{def:limitcurve}
 A sequence $(\gamma_n)$ \emph{converges to a curve} $\gamma$, iff every $p\in \gamma$ is a convergence point of the sequence. Furthermore, a curve $\gamma$ is a \emph{limit curve} of the sequence $(\gamma_n)$, iff there is a subsequence $(\gamma_{n_k})$ that converges to $\gamma$.
 \hfill {${\triangle}$}
\end{definition}

\begin{remark}
There are some subtle differences in the definition of accumulation and convergence points in the literature. In \cite{HE} and \cite{Wald}, the `accumulation point'' defined above  coincides with the definition of the  ``limit point'' of the sequence $(\gamma_n)$. Moreover, the definition of the ``convergence'' point of a sequence $(\gamma_n)$ above coincides with that of \cite{Wald}. The definition of the limit curve of a sequence $(\gamma_n)$ above coincides with those of \cite{HE} and \cite{Wald}.  Although, these definitions of the notion of convergence of a sequence of curves are sufficient to prove the limit curve theorem below, additional restrictions on $\M$ are needed to prove the upper semi-continuity of the proper time functional, see remark \ref{rem:curvtop}.  \hfill{$\Diamond$}
\end{remark}

\begin{figure}[htbp]%\label{fig:limitcurve}%\label{fig:2}
\begin{center}
\begin{eqnarray*}
\begin{tikzpicture}[scale=0.5]
  % Coordinates
  \coordinate (A) at (0.00, 0.00);
  \coordinate (B) at (0.00, 0.00);
  \coordinate (C) at (0.00, 0.00);
  \coordinate (D) at (0.01, -0.11);
  \coordinate (E) at (1.00, 6.50);
  \coordinate (F) at (-0.50, -6.00);
  \coordinate (G) at (1.75, 3.00);
  \coordinate (H) at (-2.00, -3.25);
  \coordinate (I) at (1.50, 6.50);
  \coordinate (J) at (0.00, -6.00);
  \coordinate (K) at (2.25, 0.50);
  \coordinate (L) at (-1.25, 1.75);
  \coordinate (M) at (2.00, 6.50);
  \coordinate (N) at (0.50, -6.00);
  \coordinate (O) at (3.25, 1.25);
  \coordinate (P) at (-1.25, 2.25);
  \coordinate (Q) at (2.50, 6.50);
  \coordinate (R) at (1.00, -6.00);
  \coordinate (S) at (4.25, 0.50);
  \coordinate (T) at (-1.50, 3.00);
  \coordinate (U) at (3.00, 6.50);
  \coordinate (V) at (1.50, -6.00);
  \coordinate (W) at (5.75, 0.25);
  \coordinate (X) at (-1.75, 4.00);

  % Drawing
  \draw[thick] (A) circle (4.95);
  \draw[thick] (A) circle (3.50);
  \draw[thick, rotate=-84.8] (0.01, -0.06) ellipse (2.00 and 2.00);
  \draw[red, thick] (E) .. controls (1.75, 3.00) and (-2.00, -3.25) .. (F);
  \draw[thick] (I) .. controls (2.25, 0.50) and (-1.25, 1.75) .. (J);
  \draw[thick] (M) .. controls (3.25, 1.25) and (-1.25, 2.25) .. (N);
  \draw[thick] (Q) .. controls (4.25, 0.50) and (-1.50, 3.00) .. (R);
  \draw[thick] (U) .. controls (5.75, 0.25) and (-1.75, 4.00) .. (V);
   \node[below, font=\large] at (-3.00, 3.50) {$r_1$};
    \node[below, font=\large] at ( -2.70, 1.95) {$r_2$};
     \node[below, font=\large] at (-1.50, 1.00) {$r_k$};
     \node[left, red, font=\large] at (1.10, 5.40) {$p_{r_1}$};
     \node[left, red, font=\large] at ( 0.95, 3.90) {$p_{r_2}$};
      \node[left, red, font=\large] at (0.50, 2.40) {$p_{r_k}$};
       \node[below,  font=\large] at (1.00, -6.00) {$(\gamma_n)$};
       \node[left, red, font=\large] at (-0.5, -6.00) {$\gamma$};
       \node[left, red, font=\large] at (0.00, 0.00) {$p$};
        \fill[red] (1.10, 4.83) circle (2pt);
        \fill[red] ( 0.89, 3.44) circle (2pt);
        \fill[red] (0.52, 1.97) circle (2pt);
        \fill[red] (0.00, 0.00) circle (2pt);
\end{tikzpicture}
\end{eqnarray*}
\end{center}
%\begin{figure}[htbp]\label{fig:2}
\caption{\small{In the figure, $p$ is the accumulation point of the sequence $(\gamma_n)$. Subsequences  $(\gamma_{n_k})$  of $(\gamma_n)$,  with $(\gamma_{n_k})\subseteq (\gamma_{n_{k'}})$ for $k<k'$, intersect the boundary of balls that have radii $r_k$, which are rational numbers.   These  intersection points accumulate at the points $p_{r_k}$. The closure of all these points yields the limit curve $\gamma$ that passes through  $p$ depicted in red.  }}
\centering
\label{fig:limitcurve}
\end{figure}
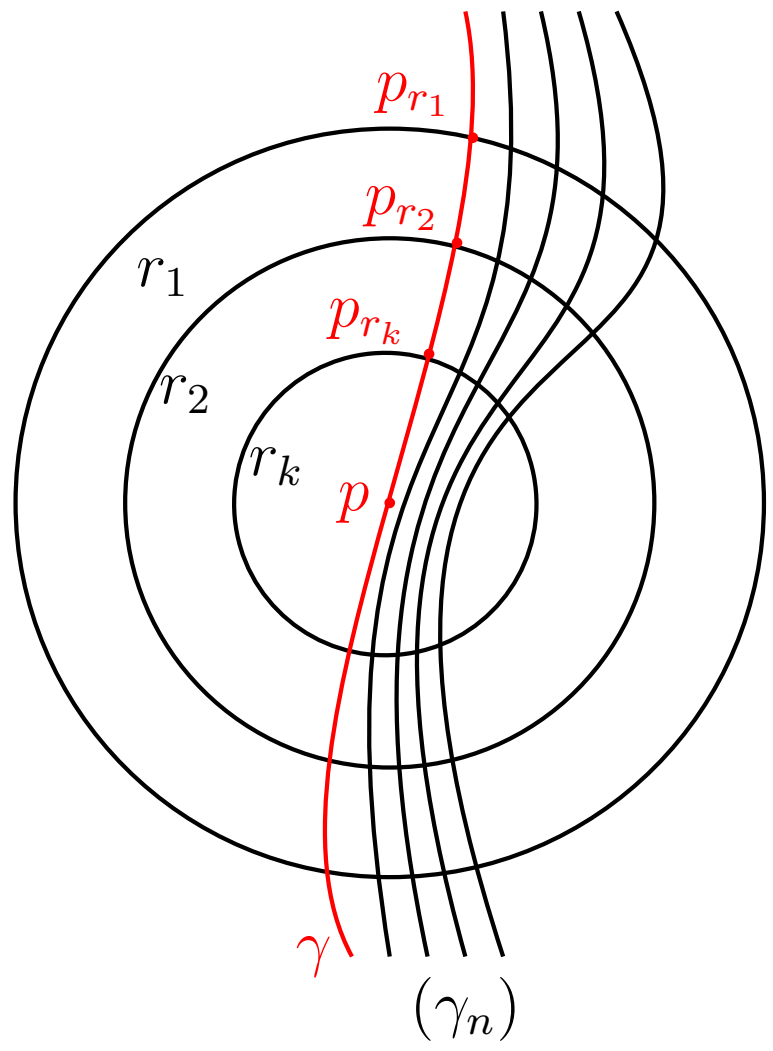

\begin{theorem}\label{prop:limitcurve}
Let $p\in \M$ be an accumulation point of a sequence $(\gamma_n)$ of future directed inextendible causal curves. Then, there is a future directed inextendible causal limit curve $\gamma$ of  $(\gamma_n)$ that passes through $p$.
\end{theorem}
\begin{proof}
Choose a normal covex neighbourhood of $p$, $N_p$ with coordinates $u$ and the Euclidean balls,
\be
B(p,r)=\{ u\in N_p | (u^0)^2+(u^1)^2+(u^2)^2+(u^3)^2\leq r\}
\ee
with radii  rational numbers, $r\in \bQ$,  and $0\leq r\leq r_1$. As $p$ is an accumulation point of $(\gamma_n)$, there is a subsequence $(\gamma_{n_k})$ that converges to $p$. Moreover, $(\gamma_{n_k})$ intersect $\partial B_{r_1}$ at infinite points. As $\partial B(p, r_1)$ is compact, these intersection points have an accumulation point $p_{r_1}$.  Because of the local causal structure of spacetime described in proposition \ref{prop:lcs}, $p_{r_1}\in \overline{C_p^+}\cap \partial B(p,r_1)$. Choosing a subsequence in $(\gamma_{n_k})$ with convergence point $p_{r_1}$, one proceed to find a point $p_r$, $r<r_1$,  $p_r\in \overline{C_p^+}\cap \partial B(p,r)$ and so on for every $r\in \bQ$, $0\leq r\leq r_1$, see also figure \ref{fig:limitcurve}.  Taking the closure of the union of all these points, one constructs a curve $\gamma_{r_1} \subset \overline {C_p^+}\cap N_p$. Therefore, this curve is causal and forward directed with future end point $p_{r_1}$.

As $p_{r_1}$ is an accumulation point of $(\gamma_n)$, the process can be repeated but now for $p_{r_1}$ to construct a causal forward directed curve $\gamma_{r_2}\subset \overline{C_{p_{r_1}}^+}\cap N_{p_{r_1}}$ and so on. Joining these curves together as $\gamma=\dots \cup \gamma_{r_2}\cup \gamma_{r_1}$,  $\gamma$  is a forward directed causal curve  that has accumulation point $p$. Moreover, if the curves $(\gamma_n)$ do not have an end point, the process described can be repeated indefinitely and $\gamma$ will not have a future end point as well, i.e $\gamma$ will be inextendible.
\end{proof}

\begin{remark}
The above theorem is often referred to as the \emph{limit curve  theorem}. The proof described above is that given in  \cite{HE}. There are many aspects to it beyond   the traditional treatment of the result   given above. An extensive discussion can be found in \cite{Minguzzi2008}. Below, we also explore the limit curve theorem from another perspective using the Arzel\'a-Ascoli theorem.
\hfill{$\Diamond$}
\end{remark}

Another aspect of taking limits of curves is that even if one begins with a smooth sequence of causal curves, the curve at the limit may not even be $C^1$ differentiable. However, it is expected that the curve at the limit to be continuous at least continuous up to a reparameterisation.   However, if this is the case, there is the need to generalise the notion of causality to more general than $C^1$ curves.  One way to do this, looking at the details of the proof of the theorem, is as follows (see also \cite{HE} page 184):

\begin{definition}\label{def:ccausal}
A continuous curve $\gamma: J\rightarrow \M$, not necessarily $C^1$, is causal at the point $\gamma(t_0)$, $t_0\in \mathrm{int} J$,  if there is an open interval $I\subset J$ with $t_0\in I$ and a normal neighbourhood $U$ of $\gamma(t_0)$ with $\gamma\vert_I\subset U$ such that for every $\gamma(t_1), \gamma(t_2)\in U$ with $t_1<t_2$, $t_1, t_2\in I$, $\gamma(t_1)\not=\gamma(t_2)$, there is another $C^k$ causal curve $\lambda$ that begins at $\gamma(t_1)$ and ends at $\gamma(t_2)$ and entirely lies in $U$.\hfill{$\triangle$}
\end{definition}

It can be shown that continuous causal curves, as defined above,  are locally Lipschitz \cite{beem} in an appropriate parameterisation, see next section and also appendix \ref{ap:b12}. This means that they are continuous and their velocity can be defined almost everywhere on their domain. Moreover, their (Euclidean) length can be computed by the usual integration of their speed as in (\ref{gelength}).  Furthermore, their proper time is bounded as in  proposition \ref{prop:gepropertime}.

\begin{example}\label{ex:c0limit}
  To illustrate these points, let us demonstrate with an example that the (usual) limit of a sequence of $C^\infty$ functions may not be continuous even  on compact domains. Imitating, the properties of a time-like future directed curves, consider the sequence $(\gamma_n)$ with $\gamma: [0,1]\rightarrow \bR$, where
\be
\gamma_n(t)=t^n+t~,
\ee
 and $\bR$ is thought as the one-dimensional Minkowski spacetime. The tangent vector field of $\gamma_n$ vanishes nowhere on $\gamma$ and the curve is clearly future directed and time-like. The pointwise limit as $n\rightarrow +\infty$ gives the limit curve $\gamma$
 \be
 \gamma(t)=t~,~~~t\in [0,1)~~~~\gamma(1)=2~.
 \ee
 Clearly, $\gamma$ is not continuous.  The convergence of this sequence is not absolutely continuous.

 However, consider these curves up to a parameterisation.  Clearly, using the diffeomeorphism $\phi$ of the domain $[0,1]$, where
 \be
 s=\phi(t)=\frac{1}{2} (t^n+t)~,
 \ee
 we conclude that
 \be
 \tilde\gamma_n(s)=\gamma_n\circ \phi (t)=2s~.
 \ee
 Clearly, the limit of the sequence $(\tilde\gamma_n)$ is $\tilde\gamma(s)=2s$,  which is continuous on $[0,1]$. \hfill{$\triangledown$}
\end{example}

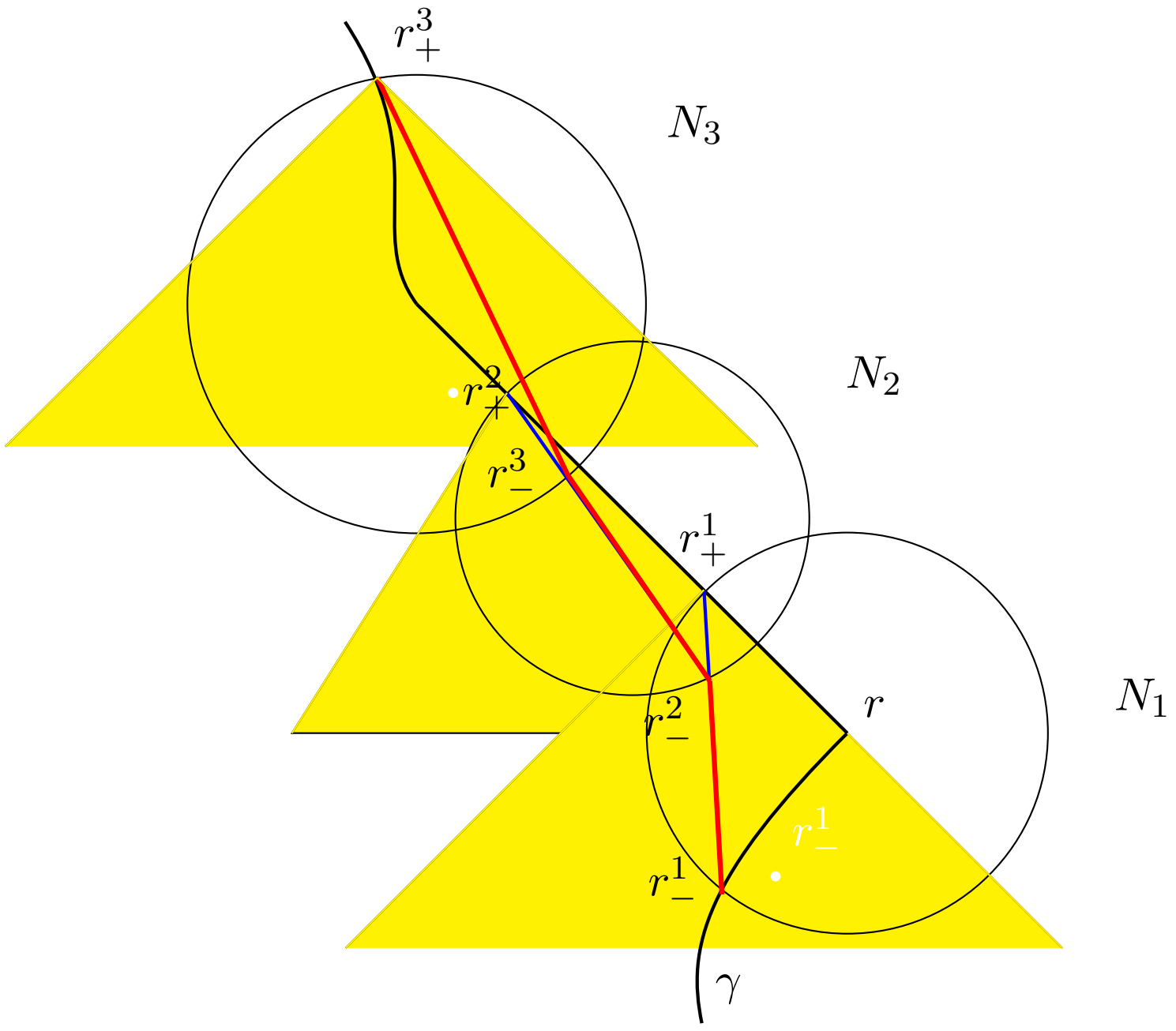
\begin{figure}[htbp]
\begin{center}
\begin{eqnarray*}
\begin{tikzpicture}[scale=0.6]
  % Coordinates
  \coordinate (A) at (-2.03, -4.05);
  \coordinate (B) at (0.00, 0.00);
  \coordinate (C) at (-2.25, -3.00);
  \coordinate (D) at (-2.00, -2.00);
  \coordinate (E) at (-6.01, 5.99);
  \coordinate (F) at (0.00, 0.00);
  \coordinate (J) at (-3.00, 3.00);
  \coordinate (K) at (-3.00, 3.00);
  \coordinate (G) at (-7.01, 9.93);
  \coordinate (H) at (-6.75, 7.00);
  \coordinate (I) at (-5.75, 8.00);
  \coordinate (L) at (-6.01, 5.99);
  \coordinate (M) at (-1.75, -2.25);
  \coordinate (N) at (-2.00, 2.00);
  \coordinate (O) at (-1.92, 0.73);
  \coordinate (P) at (-4.75, 4.75);
  \coordinate (Q) at (-7.00, -3.00);
  \coordinate (R) at (3.00, -3.00);
  \coordinate (S) at (-3.90, 3.60);
  \coordinate (T) at (-6.56, 9.15);
  \coordinate (U) at (-11.75, 4.00);
  \coordinate (V) at (-1.25, 4.00);
  \coordinate (W) at (-7.75, 0.00);
  \coordinate (X) at (-1.25, -4.00);
  \coordinate (Y) at (3.50, 0.00);
  \coordinate (Z) at (-0.25, 4.50);
  \coordinate (AA) at (-2.75, 8.00);
  \coordinate (AB) at (-1.00, -2.00);
  \coordinate (AC) at (-2.00, 2.75);
  \coordinate (AD) at (-1.50, 0.50);
  \coordinate (AE) at (-4.00, 2.75);
  \coordinate (AF) at (-5.50, 4.75);
  \coordinate (AG) at (-6.25, 9.75);

  % Drawing
  \draw [fill = yellow] (U)-- (T) -- (V);
   \draw [fill = yellow] (B)-- (W)--(P);
  \draw [fill = yellow] (R)-- (N)--(Q);
  \draw[thick] (A) .. controls (-2.25, -3.00) and (-2.00, -2.00) .. (B);
  \draw[thick] (E) -- (B);
  \draw (B) circle (2.80);
  \draw (J) circle (2.47);
  \draw[thick] (E) .. controls (-6.75, 7.00) and (-5.75, 8.00) .. (G);
  \draw (E) circle (3.20);
  \draw[blue, thick] (M) -- (N);
  \draw[blue, thick] (O) -- (P);
  \draw[solid, yellow] (N) -- (Q);
  \draw[solid, yellow] (B) -- (R);
  \draw[red, very thick] (S) -- (T);
  \draw[solid, yellow] (T) -- (U);
  \draw[solid, yellow] (T) -- (V);
  \draw[solid, yellow] (W) -- (P);
  \draw[red, very thick] (O) -- (S);
  \draw[red, very thick] (M) -- (O);
  \fill[white] (X) circle (2pt);
  \node[above left, black] at (X) {$\gamma$};
  \fill[white] (Y) circle (2pt);
  %\node[above right, black] at (Y) {$\gamma$};
  \fill[white] (Z) circle (2pt);
  \node[above right, white] at (Z) {$N_2$};
  \fill[white] (AA) circle (2pt);
  \node[above right, black] at (AA) {$N_3$};
  %\fill[white] (AB) circle (2pt);
  %\node[above right, black] at (AB) {$r_-^1$};
  \node[above left, black] at (-1.85, -2.70) {$r_-^1$};
  \fill[white] (AC) circle (2pt);
  \node[above right, white] at (AC) {$r_-^2$};
 % \fill[white] (AD) circle (2pt);
  %\fill[white] (AE) circle (2pt);
  \node[above, black] at (-2.00, 2.10) {$r_+^1$};
  \fill[white] (AF) circle (2pt);
  %\fill[white] (AG) circle (2pt);
  \node[right, black] at (-5.60, 4.75) {$r_+^2$};
  \fill[white] (Y) circle (2pt);
  \node[above right, black] at (Y) {$N_1$};
  \fill[white] (Z) circle (2pt);
  \node[above right, black] at (Z) {$N_2$};
  \fill[white] (AB) circle (2pt);
  \node[above right, white] at (AB) {$r_-^1$};
  %\fill[white] (AC) circle (2pt);
  \node[below left, black] at (-1.92, 0.73)  {$r_-^2$};
  %\fill[white] (AE) circle (2pt);
  \node[left, black] at (-4.10, 3.60) {$r_-^3$};
  %\fill[white] (AG) circle (2pt);
  \node[above right, black] at (-6.56, 9.15) {$r_+^3$};
  \node[above right, black] at (B) {$r$};
\end{tikzpicture}
\end{eqnarray*}
\end{center}
%\begin{figure}[htbp]
\caption{\small{The  figure depicts the construction of proposition \ref{prop:causaltime}  a time-like curve $\tau$ consisting of the segment of $\gamma$ in the past $r_-^1$ and the time-like curves $\lambda$ coloured in red passing through the points $r_-^1, r_-^2, r_-^3$ and $ r_+^3]$,  and continuing with $\gamma$ after $r_+^3$. The point $r$ is the first point that $\gamma$ is not time-like and exhibits a null corner.  The curve $\gamma$ is depicted in black.}}
\centering
\label{fig:causaltime}
\end{figure}

\begin{prop}\label{prop:causaltime}
If there is a future directed causal curve $\gamma$ from $p$ to $q$, with $p,q\in \M$, and $\gamma$ is not a null geodesic, then there is a time-like curve $\tau$ from $p$ to $q$.
\end{prop}
\begin{proof}
If $\gamma$ passes through $q$ several times, consider only the first portion of $\gamma$ that reaches $q$ which is not a null geodesics -- denote this portion again with $\gamma$. Consider a cover of $\gamma$ with convex normal neighbourhoods $N_p$ centred at the points $p\in \gamma$ such that they have compact closure in the corresponding normal neighbourhood. As the curve $\gamma$ is compact, this cover will have a finite refinement $\{N_i; i=1, \dots, k\}$.

If $\gamma$ is not a null geodesic, there must be a point $r\in \gamma$ such that with $\vec{\dot\gamma}_r$ is time-like or $\gamma$ has a corner at $r$.  By continuity, in the former case, $r$ must lie in an open sub-segment of $\gamma$ that it is not a null geodesic. In either case, $r$ is contained in one of the sets of the cover, say $N_1$, $r\in N_1$. Consider the two end points $r^1_+, r^1_-$ of $ \gamma\cap \overline N_1$. As $r^1_+$ and $r^1_-$ are jointed by either a time-like curve or two null curves with a corner or a combination of the two, it is a consequence of proposition \ref{prop:lcs} that $r^1_+\in C^+_{r^1_-}$ and so there is a time-like geodesic segment $\lambda_1$ from $r^1_-$ to $r^1_+$. If $r^1_+\not=q$, suppose that $r^1_+\in N_1\cap N_2\not=\emptyset$ and consider the curve constructed from $\lambda$ prolonged by the original curve $\gamma$ denoted by $\gamma\cup \lambda$. This curve is not a null geodesics and the construction can be repeated. Indeed,  $\gamma\cup \lambda$ has two end points $r^2_-, r^2_+\in \overline{N_2}$ and there exists a time-like geodesics $\lambda_2$ with end points $r^2_-$ and $r^2_+$. This construction can clearly be repeated. The time-like curve $\tau$ that joins $p$ and $q$, which  is constructed by this process, is
\be
\tau=\dots\cup \lambda_1\vert_{[r^1_-, r^2_-]}\cup \lambda_2\vert_{[r^2_-, r_-^3]}\cup \dots~,
\ee
see also figure \ref{fig:causaltime}.
The construction  is made in a finite number of steps as there are only finite number of normal neighbourhoods $N_i$.
\end{proof}

\subsection{Limit curve theorem revisited}

 Although, we have used the traditional approach to prove the limit curve theorem \ref{prop:limitcurve}, it is instructive to restate a proof using the Arzel\'a-Ascoli theorem. One difference is that the curve at the limit is explicitly constructed in theorem  \ref{prop:limitcurve} while in the Arzel\'a-Ascoli theorem only its existence is  demonstrated.
It has been shown in \cite{beem} page 75 and 76, see also \cite{Minguzzi},  that continuous causal curves satisfy the local Lipschitz property in some parameterisation, see appendix \ref{ap:b123} and \ref{ap:dx}.  This means  that in a suitable parameterisation of the continuous causal curve $\gamma$, one has that
\be
d_{g_e}(\gamma(t), \gamma(s))\leq C |t-s|~,
\label{lipseq}
 \ee
where $\gamma$ is  defined on an interval $I=(a,b)$ with compact closure, $t,s\in I$, $C>0$  is a constant that may depend on $I$  and $g_e$  is an auxiliary Riemannian metric\footnote{In many computations, an auxiliary Euclidean signature metric $g_e$ is used. However, it can be shown that the results do not depend on the choice of $g_e$.} $g_e$ on the spacetime $\M$   see appendices \ref{ap:b12} and \ref{ap:dx}.

 To outline the main idea behind a proof of this statement,  suppose that $\gamma(s)$ is in the future of $\gamma(t)$ lying is an normal convex open neighbourhood of $\gamma(t)$. Then, there is a constant $K>0$ such that
 \be
 d_{g_e}(\gamma(t), \gamma(s))\leq K \sqrt{(u^0)^2+\sum_i (u^i)^2}\leq \sqrt{2} K |u^0|~,
 \ee
 where $(u^0, u^i)$ is the position of $\gamma(s)$ in normal coordinates with centre $\gamma(t)$. The second inequality follows because $\gamma(s)$ lies $C^+_{\gamma(t)}$ that in normal coordinates is expressed as $(u^0)^2\geq \sum_i (u^i)^2$, see section \ref{sec:localstructure}. Assuming that the parameterisation of the curve is with respect to $u^0$, one derives the Lipschitz inequality.
This proof can be extended to include the possibility that $\gamma(t)$ and $\gamma(s)$ do not lie in the same normal convex neighbourhood by covering $\gamma$ with such neighbourhoods and then extracting a finite subcover as the domain, and so the image of $\gamma$, is compact.

 Lipschitz curves  is a more general class than piece-wise $C^1$ curves and more special than absolutely continuous ones. The property of causal curves to exhibit the local Lipschitz property  has some advantages. First, in addition to be continuous,  their velocity of such curves is defined almost everywhere on the curve.
  Moreover, their speed $|\vec{\dot\gamma}|_{g_e}$ is integrable and their (Riemannian) length can be found by computing the integral
  \be
  \ell=L_{g_e}(\gamma, [0, s]\equiv \int_0^s dt\,~|\vec{\dot\gamma}|_{g_e}(t)~.
  \label{gelength}
  \ee
  As a result, they can be reparameterised using their arc length $\ell$.  In this new parameterisation, the Lipschitz property is evident because their speed
  \be
|\vec{\dot \gamma}|_{g_e}=1~.
\ee
Thus,
\be
d_{g_e}(\gamma(s), \gamma(t))\leq \int^s_t |\vec{\dot \gamma}|_{g_e}= |s-t|~,
\label{loclipschitz}
\ee
where now $s,t$ are values of the length parameter $\ell$ of $\gamma$ with respect to $g_e$.

The convergence of sequences of such curves can be uniform over compact subsets of their parameter space. This happens whenever the constant $C$ in (\ref{lipseq}) does not depend on the members of the sequence -- such sequences are also known as equicontinuous and the curves are called equi-Lipschitz, see theorem \ref{th:aath}. Moreover, the curve at the limit is also Lipschitz. This means that the curve  at the limit  is continuous and admits a tangent vector almost every where on the curve -- sequences of velocities  converge weakly, see section \ref{ap:ac} .
So from now on in this section, we shall assume that the causal curves we are dealing with have the above mentioned convergence properties of Lipschitz curves.

\begin{prop}
Let $(\gamma_n): [a_n, b_n]\rightarrow \M$ be a sequence of future directed causal curves parameterised with their length with respect to an auxiliary Riemannian metric $g_e$ on the spacetime $\M$.  Suppose that the sequence $(\gamma_n(0))$ has an accumulation point $p\in \M$, where $0\in [a_n,b_n]$, and there is a compact interval $J$, $0\in \mathrm{Int} J$,  such that $J\subset [a_n, b_n]$. Then, there is a subsequence $(\gamma_{n_k})$
that converges to a limit curve $\gamma: (a, b)\rightarrow \M$, $J\subset (a, b)$,  that passes through $p$.
%If either $b=+\infty$ or $a=-\infty$, the curve $\gamma$ is future or past inextendible.
\end{prop}
\begin{proof}
Possibly choosing  a subsequence, let us take $p$ to be a limit point of the sequence $(\gamma_n)$.
As the curves $\gamma_n$ are parameterised with their Riemannian length $s$, their speed with respect to auxiliary Riemannian metric $g_e$ is one,
\be
|\vec{\dot \gamma}_n|_{g_e}=1~.
\ee
As a result the curves in the sequence are Lipschitz and equicontinuous as
\be
d_{g_e}(\gamma_n(s), \gamma_n(t))\leq \int^s_t |\vec{\dot \gamma}_n|_{g_e}= |s-t|~,
\ee
where $d_{g_e}$ is the length function associated to $g_e$ for $t,s\in J$ with $J\subset [a_n, b_n]$ with non-empty interior and $0\in J$.  This fulfills the first criterion of the Arzel\'a-Ascoli theorem \ref{th:aath}.
Moreover, for any compact $J\subset [a_n, b_n]$, the length of the curves $\gamma_n$ is bounded by the length $L$ of $J$ which is finite.  Therefore, all these curves with domain $J$ lie in a ball $B(p, L+1)$ with centre $p$ and radius $L+1$. As the closure of balls in complete manifolds are compact, $(\gamma_n)$ fulfills
the second condition of the Arzel\'a-Ascoli theorem and so, upon passing to a subsequence $(\gamma_{n_k})$,  there is a limit curve $\gamma$.

It turns out that  $\gamma$ passes through $p$. Indeed choose $k>k_0$ such that $d_{g_e}(\gamma_{n_k}(0), p)\leq \epsilon/2$ and  $d_{g_e}(\gamma_{n_k}, \gamma)\equiv \sup_{t\in J} d_{g_e}(\gamma_{n_k}(t), \gamma(t))\leq \epsilon/2$. Then by the triangular inequality, one finds that
\be
d_{g_e}(\gamma(0), p)\leq d_{g_e}(\gamma_{n_k}(0), p)+d_{g_e}(\gamma_{n_k}(0), \gamma(0))\leq d_{g_e}(\gamma_{n_k}(0), p)+d_{g_e}(\gamma_{n_k}, \gamma)\leq \epsilon~.
\ee
Therefore, at the limit $\gamma(0)=p$ and $\gamma$ passes through $p$.

It remains to show that at the limit curve $\gamma$ is causal.  To show this consider a compact interval $J$.  Since $\gamma$ is continuous, $\gamma(J)$ is compact.  Therefore, for every  cover of $\gamma(J)$ with open convex normal neighbourhoods centred at  the points in $\gamma(J)$, there is a finite subcover $(V_i)$. In every $V_i$, $\gamma$ will be the limit of a sequence of causal curves $(\gamma_{n_k}|_{V_i})$.  Consider now $s, t\in J$ with $\gamma_n(t), \gamma_n(s) \in V_i$.
Inside $V_i$,  $p_n=\gamma_n(t)$ and $q_n=\gamma_n(s)$ are points on  future directed causal curves $\gamma_n$, so there is a unique causal future directed geodesic from $p_n$ to $q_n$.  Using the exponential map $q_n=\exp_{p_n} \vv_n$, where $\vv_n$ is unique forward directed causal vector, see proposition \ref{prop:lcs}.
Moreover as $p, q\in V_i$, where $p=\gamma(t)$ and $q=\gamma(s)$, there is a unique $\vv$ such that $q=\exp_p(\vv)$ with $\vv$ the limit of the sequence $(\vv_n)$ because  the exponential map, $\exp$,  is a local diffeomorphism.
 Next, $g_{p_n} (\vv_n, \vv_n)\leq 0$ and so by continuity $g_p(\vv, \vv)\leq 0$.  Therefore,   $\vv$ is a causal future directed vector. Thus, the geodesic from $p$ to $q$ is causal and future directed.  As a result all the small segments of $\gamma$ are causal and future directed that in turn imply that $\gamma$ is future directed.
\end{proof}

\begin{remark}
Controlling the end points of the curve $\gamma$ at the limit of the sequence $(\gamma_n)$, $\gamma_n: [a_n, b_n]\rightarrow \M$,  is significant in many arguments that follow. Suppose that $a_n\rightarrow 0$ and $p$ is an accumulation point of the sequence with limit curve $\gamma$ with $\gamma(0)=p$. It is possible to parameterise the curves $\gamma_n$ in such a way  that in the new parameterisation $\gamma_n(0)=p_n$, where $p_n=\gamma_n(a_n)$ in the original parameterisation. So let us declare that $\gamma_n$ is defined in the interval $[0, \ell_n)$ and ask the question what happens at the limit of $\ell_n$, where $\ell_n$ is  the length of the curve $\gamma_n$ from $0$ with respect to $g_e$.  If $\lim_n\ell_n=L$ and $L$ is finite, then all curves $\gamma_n$ can be extended to the interval $[0,L]$.  Indeed, assuming  that $\ell_n\leq L$, each curve $\gamma_n$ can be extended to a new curve $\tilde \gamma_n$ on $[0,L]$ by assigning to it the value $\gamma_n(\ell_n)$ in the interval $[\ell_n, L]$. The new curve $\tilde \gamma_n$ is Lipschitz on $[0,L]$ and the sequence $(\tilde \gamma_n)$ remains equicontinuous. In this case, the limit curve $\gamma=\tilde \gamma$ is defined in the interval $[0,L]$ and the end point will be $q=\lim_n \gamma(\ell_n)=\lim_n\tilde \gamma(\ell_n)$. However, if $L=+\infty$ the curve will be future inextendible and the limit $\ell_n\rightarrow +\infty$ cannot be controlled, for more details see \cite{Minguzzi2008, Minguzzi}
 \hfill{$\Diamond$}
 \end{remark}

\begin{example}  To give an example how the end points behave in the limits, consider the sequence of curves $\gamma_n: [0,n]\rightarrow \bR^2$ with
\be
\gamma(t)=(t, \frac{t}{n})~.
\ee
One can view $\bR^2$ as the Minkowski space. For any fix $T$, the sequence $(\gamma_n)$ converges uniformly to $\gamma: [0, +\infty)\rightarrow \bR^2$ with $\gamma(t)=(t,0)$.  Indeed
\be
d_{g_e}(\gamma_n, \gamma)=\sup_{t\in [0, T]}|\gamma_n(t)-\gamma(t)|=\frac{T}{n}\rightarrow 0
\ee
as $n$ goes to infinity.  Thus at every compact subset $[0, T]$ of the domain the end points of the uniform limit curve of the sequence are $(0,0)$ and  $(T, 0)$.  However, if the domain is allowed to be non-compact and in particular $[0, n)$, as $n\rightarrow +\infty$, $\gamma_n(n)=(n, 1)$ and so one of the end points escapes to infinity. \hfill{$\triangledown$}
\end{example}

\begin{remark}
A related topic to explore is the behaviour of sequences of future (or past) directed causal curves with fixed end points that also exhibit an accumulation point $w$. To explain this, suppose we consider a sequence of curves $(\gamma_n)$ each parameterised with parameter $s_n$ with $s_n\in [a_n, b_n]$ with $\gamma_n(a_n)=p$ and $\gamma_n(b_n)=q$, where $p$ and $q$ are fixed points of the spacetime. Next suppose that for the  value $t_n\in [a_n, b_n]$ of the parameter $s_n$  the sequence $(\gamma_n(t_n))$ has accumulation point $w$. If $q$ is in the future of $p$, assume that $w$ is in the future of $p$ and the past of $q$.

Now consider a new parameterisation of the curves $\gamma_n$ with parameter $r_n=s_n-t_n$. Denoting the reparameterised curves  with $\tilde\gamma_n$, they are  defined on the new intervals $[\alpha_n, \beta_n]=[a_n-t_n, b_n-t_n]$ and the sequence that accumulates to $w$ is $(\tilde \gamma_n(0))$.  Going to a subsequence $\tilde\gamma_{n_k}$ that $(\tilde \gamma_{n_k}(0))$ converges to $w$, there are several possibilities. Both $(\alpha_{n_k})$ and $(\beta_{n_k})$ converge to $\alpha$ and $\beta$ with both $\alpha$ and $\beta$ to have  finite values. Then, the curve $\tilde\gamma$ at the limit has end points $p$ and $q$. If either $(\alpha_{n_k})$ or $(\beta_{n_k})$ converges to a finite value that the other goes to infinity, then $\tilde \gamma$ will be inextendible in the past or the future with end point either $p$ ($\alpha$ finite) or ($q$ $\beta$ finite).
In  such a case, the limit curve $\gamma$ does not have to pass from $q$ or $p$, respectively. If both $\alpha$ and $\beta$ are infinite, then the curve $\tilde\gamma$ is not expected to  pass pass through $p$ and $q$.  Though in all cases, $\tilde\gamma$ passes through the accumulation point $w$. A very detailed exploration of all these issues regarding the limit curve theorem can be found in \cite{Minguzzi2008}.

Clearly such a behaviour is applicable in the investigation of sequences of causal curves $(\gamma_n)$ with fixed end points $\gamma_n\in \mC(p,q)$, where $\mC(p,q)$ is the set of future directed  causal curves from $p$ to $q$ up the reparameterisations. The points $p$ and $q$ can be seen as accumulation points  of any such sequence $(\gamma_n)$.  However, the  limit curve $\gamma$ of a sequence with accumulation point $p$ may not pass from $q$ and vice versa even though $p$ and $q$ may be points (in the interior) of spacetime and the proper time of the curves $\gamma_n$ between $p$ and $q$ is finite.
\hfill{$\Diamond$}
\end{remark}

\begin{remark}
It is clear that the Lipschitz property of the curves induces a topology on the space of maps from intervals to $\M$ instead of curves that it is finer than that we have described and used in the traditional proof limit curve theorem. This can be seen in the example \ref{ex:c0limit} where in the original parameterisation the curves are not Lipschitz as they do not converge uniformly on compact sets. But they converge in the sense of \cite{HE} and \cite{Wald} as used in the proof of the limit curve theorem.  It also clear that the Lipschitz property depends on the parameterisation of the curves. Despite this, it has an advantage in relation to proving the existence of maximal geodesics without assuming that $\M$ is strongly causal.\hfill{$\Diamond$}
\end{remark}

\section{Global causal structure of spacetime}

\subsection{Chronological and causal future}

The global causal structure of a spacetime $\M$ may differ in a significant way from the local causal structure of $\M$ as described in proposition \ref{prop:lcs}. Consequently, the techniques required for the investigation of the global casual structure are different. For this, we begin with some definitions.

\begin{definition}
Let $p$ be a point in the spacetime $\M$, $p\in \M$. The \emph{chronological future} of $p$, $I^+(p)$, is the subset of $\M$
\be
I^+(p)\equiv\{x\in \M\vert \exists \,\, \text{future directed time-like curve}\, \gamma\, \text{from}\, p \, \text{to}\,  x\}~.
\nonumber
\ee
The \emph{causal future} of $p$, $J^+(p)$, is the subset of $\M$
\be
J^+(p)\equiv\{x\in \M\vert \exists \,\, \text{future directed causal curve}\, \gamma\, \text{from}\, p \, \text{to}\,  x\}~,
\nonumber
\ee
and the \emph{future horismos} of $p$ is the
\be
E^+(p)=J^+(p)-I^+(p)~,
\nonumber
\ee
subset of $\M$. \hfill{$\triangle$}
\end{definition}

\begin{remark}
In the definition of $I^+(p)$, $p$ is not included in $I^+(p)$, in the sense that the trivial curve, where the whole interval $I$ is mapped to $p$, is not considered as a time-like curve.  Though, there are spacetimes that admit non-trivial future directed  time-like curves that start and end at $p$ and so in such a case, $p\in I^+(p)$. On the other hand, $p\in J^+(p)$ as the trivial curve above is considered as a null curve.
\hfill{$\Diamond$}
\end{remark}

\begin{remark}
The above definitions can be easily extended to the past chronological, causal and horismos subsets associated to $p\in \M$ and denoted by
$I^-(p)$, $J^-(p)$ and $E^-(p)$, respectively.  Moreover, given a subset $\S\subseteq \M$, one defines the sets
\be
I^+(\S)=\cup_{p\in\S} I^+(p)~,~~~J^+(\S)=\cup_{p\in\S} J^+(p)~,~~~E^+(\S)=J^+(\S)-I^+(\S)~,
\nonumber
\ee
which are the chronological future, causal future and future horismos of the set $\S$, respectively. Similarly, one defines $I^-(\S)$, $J^-(\S)$ and $E^-(\S)$.
\hfill{$\Diamond$}
\end{remark}

There are some apparent properties that these sets have. In particular, if $q\in I^+(p)$, then $p\in I^-(q)$ and similarly if $q\in J^+(p)$, then $p\in J^-(q)$.
The following propositions describe some additional properties of these sets.

\begin{figure}[htbp]
\begin{center}
\begin{eqnarray*}
\begin{tikzpicture}[scale=0.3]
  % Coordinates
  \coordinate (A) at (6.00, 1.00);
  \coordinate (B) at (12.00, 7.00);
  \coordinate (C) at (-1.00, 1.00);
  \coordinate (D) at (-7.00, 7.00);
  \coordinate (G) at (-6.75, 5.25);
  \coordinate (H) at (-4.00, 2.00);
  \coordinate (I) at (10.00, 3.75);
  \coordinate (J) at (2.00, 0.00);
  \coordinate (JF) at (2.00, -2.00);
  \coordinate (E) at (-4.00, 4.00);
  \coordinate (F) at (-5.00, 3.50);
   \coordinate (A1) at (31.00, 1.00);
  \coordinate (B1) at (37.00, 7.00);
  \coordinate (C1) at (24.00, 1.00);
  \coordinate (D1) at (18.00, 7.00);
  \coordinate (G1) at (17.50, 6.00);
  \coordinate (H1) at (21.00, 2.00);
  \coordinate (I1) at (35.00, 3.75);
   \coordinate (JF1) at (27.00, -2.00);
  \coordinate (J1) at (27.00, 0.00);
  \coordinate (E1) at (21.00, 4.00);
  \coordinate (F1) at (20.00, 3.50);
   \draw [fill = yellow] (12.00, 7.00) -- (6.00, 1.00) --  (-1.00, 1.00) --(-7.00, 7.00);
 \draw [fill = yellow] (37.00, 7.00) -- (31.00, 1.00) --  (24.00, 1.00) --(18.00, 7.00);
% Drawing
%\draw[white] (A)--(D);
  \draw[red, thick] (A) -- (B);
  \draw[very thick] (A) -- (C);
  \draw[white,  thick] (D) -- (E);
 % \draw[rotate=-135.0] (-4.50, 3.50) ellipse (0.25 and 0.10);
  \draw[red, thick] (C) -- (E);
 % \node[below left] at (G) {$A$};
  \node[below left] at (H) {$B$};
  \node[below right] at (I) {$C$};
  \node[below] at (J) {$\S$};
  \node[below] at (JF) {$\mathrm{Fig.}~(a)$};
  \node[below left] at (F) {$p$};
  \fill[black] (E) circle (4pt);
   \draw[red, thick] (A1) -- (B1);
  \draw[very thick] (A1) -- (C1);
  \draw[red,  thick] (D1) -- (E1);
 % \draw[rotate=-135.0] (-4.50, 3.50) ellipse (0.25 and 0.10);
  \draw[red, thick] (C1) -- (E1);
  \node[below left] at (G1) {$A$};
  \node[below left] at (H1) {$B$};
  \node[below right] at (I1) {$C$};
  \node[below] at (J1) {$\S$};
  \node[below] at (JF1) {$\mathrm{Fig.}~(b)$};
  \node[below left] at (F1) {$p$};
  \fill[black] (E1) circle (4pt);
\end{tikzpicture}
\end{eqnarray*}
\end{center}
%\begin{figure}[htbp]
\caption{\small{In figure (a), $J^+(\S)$ is depicted and includes the shaded region as well as the boundaries $B$, $C$ and $\S$, where the point $p$ is not included in the spacetime. Clearly $J^+(\S)$ is not a closed subset of spacetime. While in figure (b), $\overline{I^+(\S)}= \overline{J^+(\S)}$ is depicted and includes the shaded region as well as the boudaries $A$, $B$, $C$ and $\S$. $I^+(\S)$ is depicted by only the shaded region and it is an open subset of spacetime.  The subspaces $A$, $B$ and $C$ of the boundary are disjoint because the spacetime described is two-dimensional. This will not be the case in higher dimensions.}}
\centering
\label{fig:onex}
\end{figure}
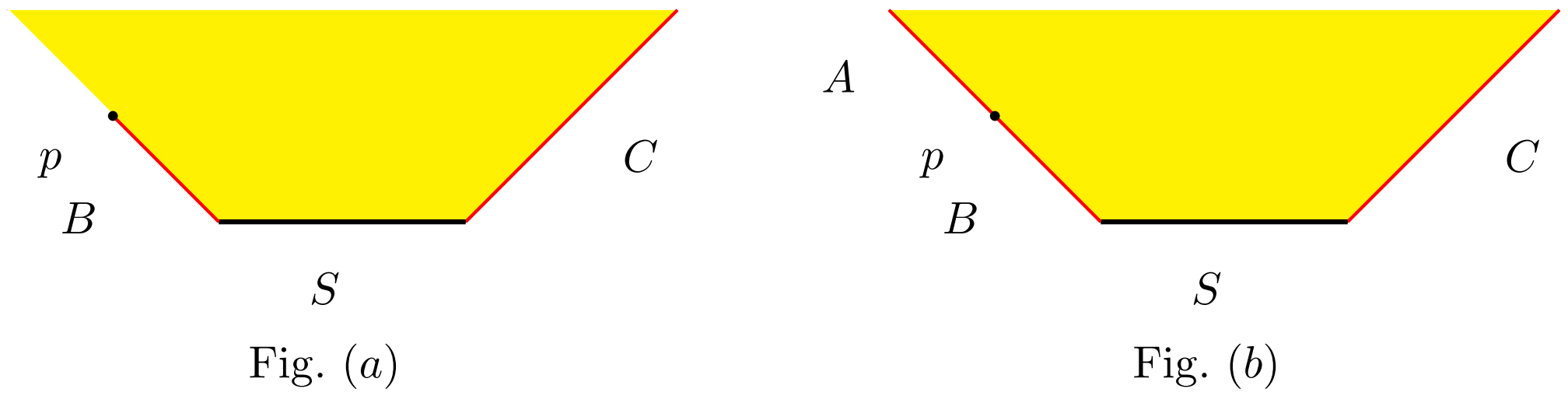

\begin{prop}\label{pro:nullgeo}
As a consequence of proposition  \ref{prop:causaltime}, the chronological and causal future sets have  the following properties:
\begin{enumerate} [(i)]
\item If $q\in E^+(\S)$, then $q$ lies in a null geodesic segment from $\S$. In particular, if $q\in J^+(p)-I^+(p)$ with $p,q\in \M$, then any causal curve that connects  $p$ and $q$ must be a null geodesic segment.

    \item If $q\in J^+(p)$ and $r\in I^+(q)$, then $r\in I^+(p)$.

\item if $q\in I^+(p)$ and $r\in J^+(q)$, then $r\in I^+(p)$.

\item if $q\in J^+(p)$ and $r\in J^+(q)$, then $r\in J^+(p)$.
\end{enumerate}
\end{prop}
\begin{proof}
(i) If $q\in E^+(\S)$, there is a causal curve for $\S$ to $q$. If the curve is not a null geodesics, then from proposition \ref{prop:causaltime} there must exist a time-like curve from $\S$ to $q$, i.e. $q\in I^+(\S)$, which is a contradiction. The proof for $\S =\{p\}$ is similar.

(ii) Joining the causal curve from $p$ to $q$ with the time-like curve from $q$ to $r$, this curve is a causal curve, which is not a null geodesic segment. Therefore, there must be a time-like curve from $p$ to $r$ and so $r\in I^+(p)$.  The proof for (iii) is similar.

(iv) This statement follows from the definitions.
\end{proof}

\begin{figure}[htbp]
\begin{center}
\begin{eqnarray*}
\begin{tikzpicture}[scale=0.3]
  % Coordinates
  \coordinate (C) at (10.00, 4.00);
  \coordinate (D) at (5.44, -5.08);
  \coordinate (E) at (-3.11, 5.08);
  \coordinate (F) at (-2.12, 1.91);
  \coordinate (G) at (-3.00, 3.00);
  \coordinate (A) at (6.00, -6.00);
  \coordinate (B) at (-6.00, 6.00);
  \coordinate (H) at (11.00, 6.00);
  \coordinate (I) at (11.00, -6.00);
  \coordinate (J) at (-1.00, -2.00);
  \coordinate (K) at (6.00, 2.00);
  \coordinate (L) at (-4.00, 3.00);
   \coordinate (FL) at (8.50, -8.00);
   % Coordinates
  \coordinate (C1) at (35.00, 4.00);
  \coordinate (D1) at (30.44, -5.08);
  \coordinate (E1) at (22.11, 5.08);
  \coordinate (F1) at (23.12, 1.91);
  \coordinate (G1) at (22.00, 3.00);
  \coordinate (A1) at (31.00, -6.00);
  \coordinate (B1) at (19.00, 6.00);
  \coordinate (H1) at (36.00, 6.00);
  \coordinate (I1) at (36.00, -6.00);
  \coordinate (J1) at (24.00, -2.00);
  \coordinate (K1) at (31.00, 2.00);
  \coordinate (L1) at (21.00, 3.00);
  \coordinate (LL1) at (18.00, 5.50);
   \coordinate (FL1) at (33.50, -8.00);
% Drawing
  \draw [fill = yellow] (-6.00, 6.00)-- (11.00, 6.00) -- (11.00, -6.00) -- (6.00, -6.00);
  \draw[very thick] (C) .. controls (-3.11, 5.08) and (-2.12, 1.91) .. (D);
  \draw[red, very thick] (G) -- (A);
  \draw[white, very thick] (B) -- (G);
  \draw[white] (B) -- (H);
  \draw[white] (H) -- (I);
  \draw[white] (A) -- (I);
  \node[above, font=\large] at (J) {$B$};
  \node[above, font=\large] at (K) {$\S$};
   \node[below] at (L) {$p$};
   \node[below] at (FL) {$\mathrm{Fig.}~(a)$};
  \fill[black] (G) circle (5pt);
  % Drawing
  \draw [fill = yellow] (19.00, 6.00)-- (36.00, 6.00) -- (36.00, -6.00) -- (31.00, -6.00);
  \draw[very thick] (C1) .. controls (22.11, 5.08) and (23.12, 1.91) .. (D1);
  \draw[red, very thick] (G1) -- (A1);
  \draw[red, very thick] (B1) -- (G1);
  \draw[white] (B1) -- (H1);
  \draw[white] (H1) -- (I1);
  \draw[white] (A1) -- (I1);
  \node[above, font=\large] at (J1) {$B$};
  \node[above, font=\large] at (K1) {$\S$};
   \node[below] at (L1) {$p$};
   \node[below] at (LL1) {$A$};
    \node[below] at (FL1) {$\mathrm{Fig.}~(b)$};
  \fill[black] (G1) circle (5pt);
\end{tikzpicture}
\end{eqnarray*}
\end{center}
%\begin{figure}[htbp]
\caption{\small{In figure (a), $J^+(\S)$ is depicted and includes the shaded region as well as the boundary $B$, where the point $p$ is excluded from the spacetime. All the points of boundary $B$ lie in past inextendible null geodesics.  Similarly in figure (b), $\overline{I^+(\S)}= \overline{J^+(\S)}$ is depicted. The points of the boundaries $A$ and $B$ again lie in past inextendible null geodesics. While the points of the boundaries $B$ and $C$ in the example of figure \ref{fig:onex} lie on null geodesics that have a past end point on $\S$. }}
\centering
\label{fig:2xy}
\end{figure}
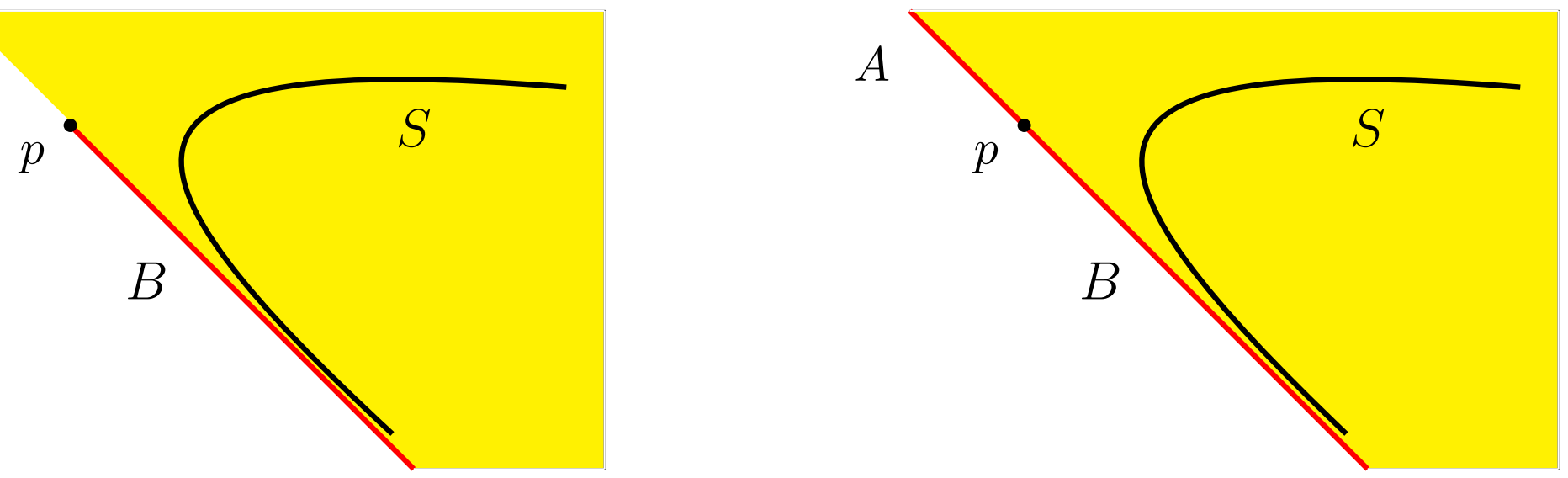

\begin{prop}\label{prop:chronos}
Let $\S\subseteq \M$, $\S\not=\emptyset$, the following statements hold:

\begin{enumerate} [(i)]

\item $I^+(\S)$ is open.

\item $I^+ (I^+(\S))=I^+(\S)$ and $J^+ (J^+(\S))=J^+(\S)$.

\item $I^+(\bar{\S})=I^+(\S)$~.

\item $\overline{I^+(\S)}=\{x\in \M\vert I^+(x) \in I^+(\S)\}$

\item $J^+(\S) \subseteq \overline{I^+(\S)}$

\item $\overline{I^+(\S)}=\overline{J^+(\S)}$, $\partial I^+(\S)=\partial J^+(\S)$ and $\text{int}\, J^+(\S)=I^+(\S)$
\end{enumerate}
where $\overline {A}$ denotes the closure of the subset  $A$ in $\M$ and $\text{int}\, A$ is the interior of the set $A$.
\end{prop}
\begin{proof}
(i) We shall demonstrate that $I^+(\S)$ is a union of open sets. Indeed, consider $q\in I^+(\S)$ and a future directed time-like curve $\gamma$ from $p\in \S$ to $q$. Let $N_q$ be an open normal neighbourhood of $q$ and a point $r\in N_q\cap \gamma$. Clearly $ C^+_r\cap N_q$ is open and $q\in C^+_r\cap N_q$. Moreover, there are future directed time-like curves from $r$ to any of the points of $ C^+_r\cap N_q$. Joining these time-like curves with $\gamma$ from $p$ to $r$, one can construct future directed time-like curves from $p$ to all the points of $ C^+_r\cap N_q$. Thus, $q\in C^+_r\cap N_q\subset I^+(p)\subseteq I^+(\S)$.  As this construction can be done for every $q\in I^+(\S)$, $I^+(\S)$ is the union of open sets and so it is open.

(ii) First, let us establish that $I^+(I^+(\S))\subseteq I^+(\S)$. Indeed, if $q\in I^+(I^+(\S))$, there is a future directed  time-like curve $\gamma$ from a point $r\in I^+(\S)$ to $q$. But $r\in I^+(\S)$, so there is a future directed  time-like curve $\lambda$ from a point $p\in \S$ to $r$. Joining $\lambda$ with $\gamma$, one can construct a future directed  time-like curve from $p$ to $q$ and so $q\in I^+(\S)$, which establishes the statement.

On the other hand if $q\in I^+(\S)$, there is a  future directed  time-like curve $\gamma$ from a point $p\in \S$ to $q$. Take any point $r\in \gamma$ between $p$ and $q$. Clearly $r\in I^+(\S)$. Moreover, $q\in I^+(r)$ because $q$ is joined to $r$ with $\gamma$. As $r\in I^+(\S)$, then $q\in I^+(I^+(\S))$ and so $I^+(\S)\subseteq I^+(I^+(\S))$.  Combining this with the statement above, we conclude that $I^+ (I^+(\S))=I^+(\S)$.  The proof for $J^+ (J^+(\S))=J^+(\S)$ is similar.

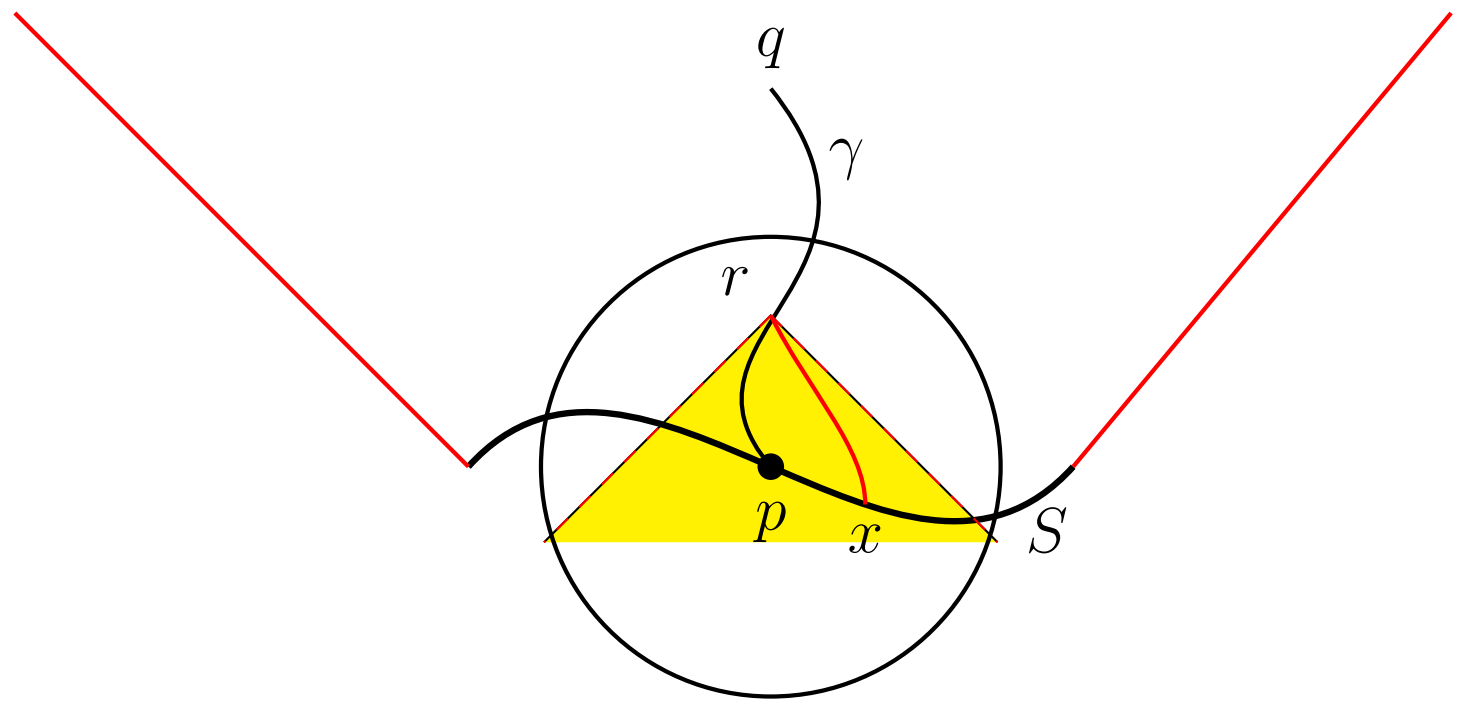
\begin{figure}[htbp]
\begin{center}
\begin{eqnarray*}
\begin{tikzpicture}[scale=0.5]
  % Coordinates
  \coordinate (A) at (0.00, 0.00);
  \coordinate (B) at (8.00, 0.00);
  \coordinate (C) at (2.25, 2.50);
  \coordinate (D) at (5.75, -2.50);
  \coordinate (E) at (-6.00, 6.00);
  \coordinate (F) at (13.00, 6.00);
  \coordinate (G) at (4.00, 0.00);
  \coordinate (H) at (4.00, 0.00);
  \coordinate (I) at (4.00, 5.00);
  \coordinate (J) at (2.50, 1.75);
  \coordinate (K) at (6.00, 2.50);
  \coordinate (N) at (4.00, 2.00);
  \coordinate (O) at (7.00, -1.00);
  \coordinate (P) at (1.00, -1.00);
  \coordinate (Q) at (5.25, -0.50);
  \coordinate (R) at (5.25, 0.25);
  \coordinate (S) at (4.50, 1.00);

  % Drawing
  \draw [fill = yellow] (P)-- (N) -- (O);
  \draw[very thick] (A) .. controls (2.25, 2.50) and (5.75, -2.50) .. (B);
  \draw[red, thick] (A) -- (E);
  \draw[red, thick] (B) -- (F);
  \draw[thick] (G) circle (3.04);
  \draw[thick] (G) .. controls (2.50, 1.75) and (6.00, 2.50) .. (I);
  \draw[red, dashed] (N) -- (O);
  \draw[red, dashed] (N) -- (P);
  \draw[red, thick] (Q) .. controls (5.25, 0.25) and (4.50, 1.00) .. (N);
  \fill[black] (G) circle (5pt);
  \node[above left, font=\large] at (N) {$r$};
  \node[above, font=\large] at (I) {$q$};
  \node[below, font=\large] at (4.00, -0.20) {$p$};
  \node[below, font=\large] at (Q) {$x$};
  \node[below, font=\large] at (5.00,4.60) {$\gamma$};
  \node[above left, font=\large] at (8.20,-1.40) {$\S$};
\end{tikzpicture}
\end{eqnarray*}
\end{center}
%\begin{figure}[htbp]
\caption{\small{The figure illustrates the proof of proposition \ref{prop:chronos}, item (iii). The point $p$ is a point of the spacetime $\M$ but not of $\S$. If $q\in I^+(\overline{\S})$, there is a time-like curve from $p$ to $q$. But there is another time-like curve that emanates from $\S$ and passes through $q$ that lies in the future of $p\in \overline{\S}$  constructed by joining the curve from $x\in\S$ to $r$ with $\gamma$ from $r$ to $q$. As a result $q\in I^+(\S)$ and $I^+(\S)=I^+(\overline{\S})$. }}
\centering
\label{fig:chronos1}
\end{figure}

(iii)  It is clear that as $\S\subseteq \bar{\S}$, one has that $I^+(\S)\subseteq I^+(\bar{\S})$.  Next consider $q\in I^+(\bar{\S})$. If $q\in I^+(\S)$, there is nothing to prove. Alternatively, let us consider $q\in I^+(\bar{\S})$ with $q\notin I^+(\S)$. In such a case, there is   $\gamma$  a future directed time-like curve that joins   $p\in \bar{\S}-\S$  with $q$. Next, choose an open normal neighbourhood $N_p$ of $p$ and a point $r\in N_p\cap\gamma$.  Clearly $p\in C^-_r$. As $C_r^-$ is open and $p$ is in the closure of $\S$, $C_r^-\cap\S\not=\emptyset$. Therefore, there is a future directed time-like curve $\lambda$ from a point $x\in C_r^-\cap\S$ to $r$. Joining $\lambda$ with $\gamma$ from $r$ to $q$, one constructs a future directed time-like curve from $x\in S$ to $q$ and so $I^+(\bar{\S}) \subseteq I^+(\S)$. Therefore, $I^+(\bar{\S})=I^+(\S)$ and the statement is established.

\begin{figure}[htbp]
\begin{center}
\begin{eqnarray*}
\begin{tikzpicture}[scale=0.5]
  % Coordinates
  \coordinate (A) at (0.00, 0.00);
  \coordinate (B) at (9.00, 0.00);
  \coordinate (C) at (3.00, 1.50);
  \coordinate (D) at (6.50, -1.75);
  \coordinate (E) at (-10.00, 9.96);
  \coordinate (F) at (18.94, 9.90);
  \coordinate (G) at (-4.00, 4.00);
  \coordinate (H) at (-6.00, 6.00);
  \coordinate (I) at (-2.02, 9.99);
  \coordinate (J) at (-6.00, 6.00);
  \coordinate (K) at (-5.00, 10.00);
  \coordinate (L) at (-6.25, 7.50);
  \coordinate (M) at (-5.25, 8.25);
  \coordinate (N) at (-9.00, 6.00);
  \coordinate (O) at (-1.00, 6.00);
  \coordinate (P) at (-4.45, 5.99);
  \coordinate (Q) at (-4.25, 8.50);
  \coordinate (R) at (-5.50, 6.75);
  \coordinate (S) at (1.00, 0.25);
  \coordinate (T) at (0.25, 3.50);
  \coordinate (U) at (-3.25, 4.00);
  \coordinate (V) at (-4.75, 6.25);
  \coordinate (W) at (4.00, 7.00);
  \coordinate (X) at (-2.00, 2.00);
  \coordinate (Y) at (1.00, 5.00);
  \coordinate (Z) at (-3.00, 3.00);
  \coordinate (AA) at (-1.00, 3.00);

  % Drawing
  \draw [fill = yellow] (E)-- (H) -- (I);
  \draw [fill = yellow] (K)-- (N) -- (O);
  \draw [fill = yellow] (X)-- (Z) -- (AA);
  \draw[very thick] (A) .. controls (3.00, 1.50) and (6.50, -1.75) .. (B);
  \draw[red, thick] (A) -- (E);
  \draw[red, thick] (B) -- (F);
  \draw[red, dashed] (H) -- (I);
  \draw[thick] (H) circle (1.90);
  \draw[thick] (H) .. controls (-6.25, 7.50) and (-5.25, 8.25) .. (K);
  \draw[red, dashed] (K) -- (N);
  \draw[red, dashed] (O) -- (K);
  \draw[red, thick] (K) .. controls (-4.25, 8.50) and (-5.50, 6.75) .. (P);
  \draw[red, thick] (P) .. controls (0.25, 3.50) and (-3.25, 4.00) .. (S);
  \draw[white] (N) -- (O);
  \draw[white] (E) -- (I);
  \draw[red, dashed] (X) -- (Y);
  \draw[white] (Z) -- (AA);
  \fill[black] (G) circle (3pt);
  \fill[black] (V) circle (2pt);
  \fill[white] (W) circle (2pt);
   \node[below left] at (X) {$y$};
    \node[below left] at (H) {$x$};
      \node[above] at (K) {$q$};
       \node[below] at (S) {$p$};
        \node[above] at (W) {$\overline{I^+(\S)}$};
         \node[below] at (6.00, -0.70) {$\S$};
           \node[below left] at  (-4.75, 6.25) {$r$};
    \node[left] at  (-5.40, 8.50) {$\gamma$};
        \node[below left] at  (G) {$z$};
\end{tikzpicture}
\end{eqnarray*}
\end{center}
%\begin{figure}[htbp]
\caption{\small{The figure illustrates the proof of proposition \ref{prop:chronos}, item (iv). First the point $z$ is not included in the spacetime $\M$. The point $q$ that lies in the future $x$, with $x$ in the boundary of $I^+(\S)$, is also in the future of $p\in \S$ and so in $I^+(\S)$. The time-like curve from $\S$ to $q$ is the red line constructed passing through $p$, $r$ and $q$. Notice also that $I^+(y)$ is also included in $I^+(\S)$. Thus $\overline{I^+(\S)}$ comprises of all the points of $x\in\M$  such that $I^+(x)$ are included in $I^+(\S)$.}}
\centering
\label{fig:chronos2}
\end{figure}
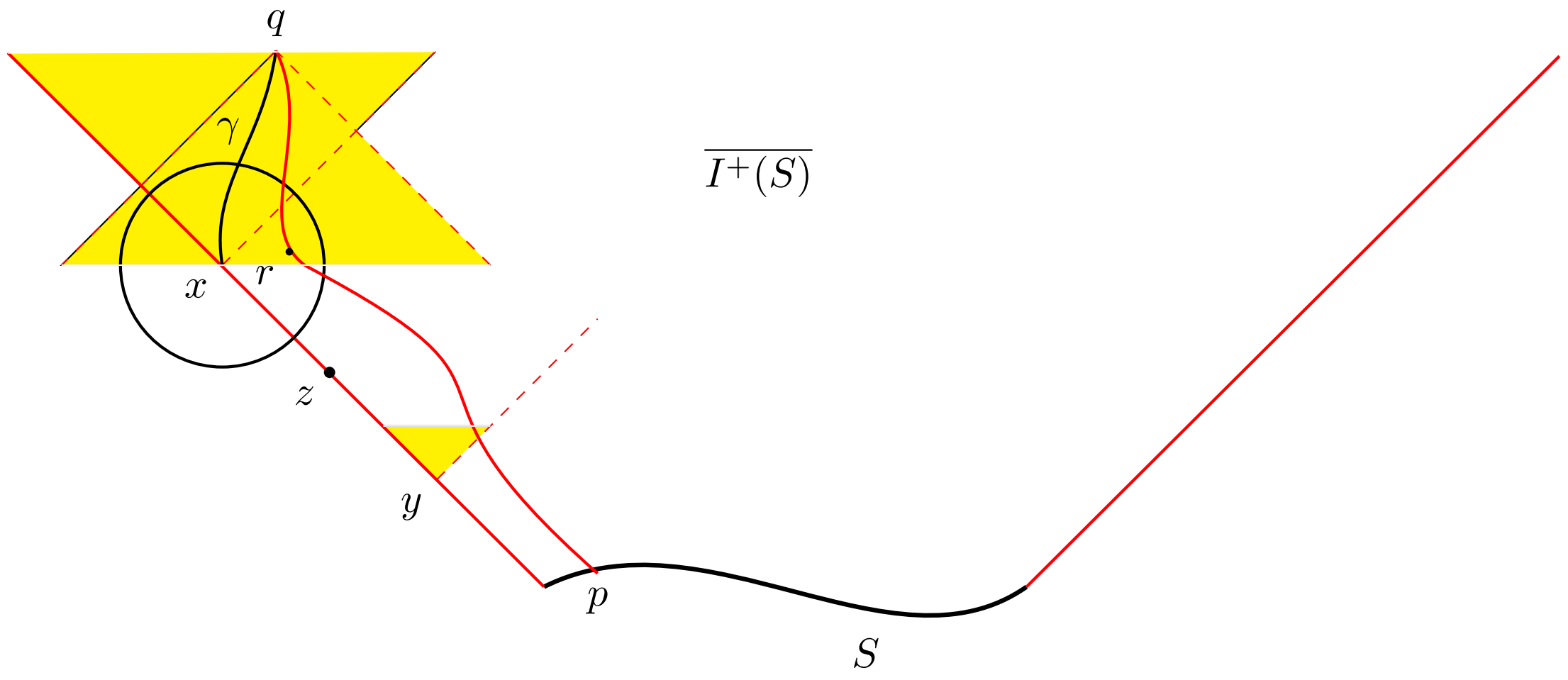

(iv) If $x\in I^+(\S)$, then clearly $I^+(x)\subset I^+(\S)\subset \overline{I^+(\S)}$.  So let us suppose that $x\in \overline{I^+(\S)}-I^+(\S)\subset \M$  -- we have to show that $I^+(x)\subset I^+(\S)$.  For this consider $q\in I^+(x)$. Therefore, there is a future directed time-like curve $\gamma$ from $x$ to $q$ and so $x\in I^-(q)$.  As $x$ is in the closure of $I^+(\S)$,  $x\in \overline{I^+(\S)}-I^+(\S)$, and $I^-(q)$ is open containing $x$, we have that $I^-(q)\cap I^+(\S)\not=\emptyset$. Consider a point $r\in I^-(q)\cap I^+(\S)$.  There is clearly a future directed time-like curve $\lambda$ from a point $p\in \S$ to $r$ and another future directed time-like curve $\lambda$ from $r$ to $q$. Joining these two curves, we conclude that $q\in I^+(\S)$ and so  $I^+(x)\subset I^+(\S)$, which establishes the statement in one direction. Conversely, if $I^+(x)\subseteq I^+(\S)$, then every neighbourhood of $x$ must intersect $I^+(\S)$. This is because every neighbourhood of $x$ must intersect $C_x^+$. But  $C_x^+\subset I^+(x)\subseteq I^+(\S)$. Thus every neighbourhood of $x$ intersects $I^+(\S)$ and so $x \in \overline{I^+(\S)}$.

(v) For this consider $x\in J^+(\S)$. Thus, there is a future directed causal curve $\gamma$ from a point $p\in \S$ to $x$. To establish the result, we shall use the property (iv). In particular, we shall show that $I^+(x)\subset I^+(\S)$, which will establish that $x\in \overline{I^+(\S)}$. Indeed, let $q\in I^+(x)$.  So, there is a future directed time-like curve $\lambda$ from $x$ to $q$. Joining the curves $\gamma$ and $\lambda$, one constructs a future directed causal curve for $p$ to $q$.  However, this curve is not a null geodesic.  As a result from proposition \ref{prop:causaltime}, there is a future directed time-like curve from $p$ to $q$.  This establishes that $I^+(x)\subset I^+(\S)$ and concludes the proof.

(vi) It follows from (v) that $\overline{J^+(\S)}\subseteq \overline{I^+(\S)}$.  But $I^+(\S)\subseteq J^+(\S)$, which in turn implies that $\overline{I^+(\S)}\subseteq \overline{J^+(\S)}$. Thus, $\overline{J^+(\S)}= \overline{I^+(\S)}$.  The rest follow from the properties of the closure of sets.
\end{proof}

\subsection{Achronal boundaries}

\begin{definition}
A set $\S$ is a \emph{ future set}, iff $I^+(\S)\subseteq \S$. A set $\S$ is an \emph{ open} future set, iff $I^+(\S)= \S$. \hfill{$\triangle$}
\end{definition}

There are some immediate consequences of the definition above are the following:

\begin{remark}\label{rem:aaa}
\begin{enumerate}
\item From the definition above and statement (iv) in the proposition \ref{prop:chronos}, it follows that if $\S$ is a future set, then $\S\subseteq \overline{I^+(\S)}$.

      \item If $\S$ is a future set, then $\text{Int}\, \S$ is a \emph{proper} future set. To see this, from the previous property
      $\text{Int}\,\S\subseteq \text{int}\,{\overline{I^+(\S)}}=I^+(\S)$. But as $I^+(\S)\subseteq \S$, $I^+(\S)$ open set and $\text{Int}\,\S$ is the maximal open set contained in $\S$, we conclude that $I^+(\S)\subseteq \text{Int}\,\S$. Thus, $\text{Int}\, \S=I^+(\text{Int}\, \S)$.

      \item If follows from the statement (ii) of the  proposition \ref{prop:chronos} that for any subset $\K$ of $\M$, $I^+(\K)$ is an open future set.\hfill{$\Diamond$}
      \end{enumerate}
      \end{remark}

\begin{definition}
A subset $\S$ of a spacetime $\M$,  $\S\subset \M$,  is \emph{ achronal}, iff $\S\cap I^+(\S)=\emptyset$.
\end{definition}

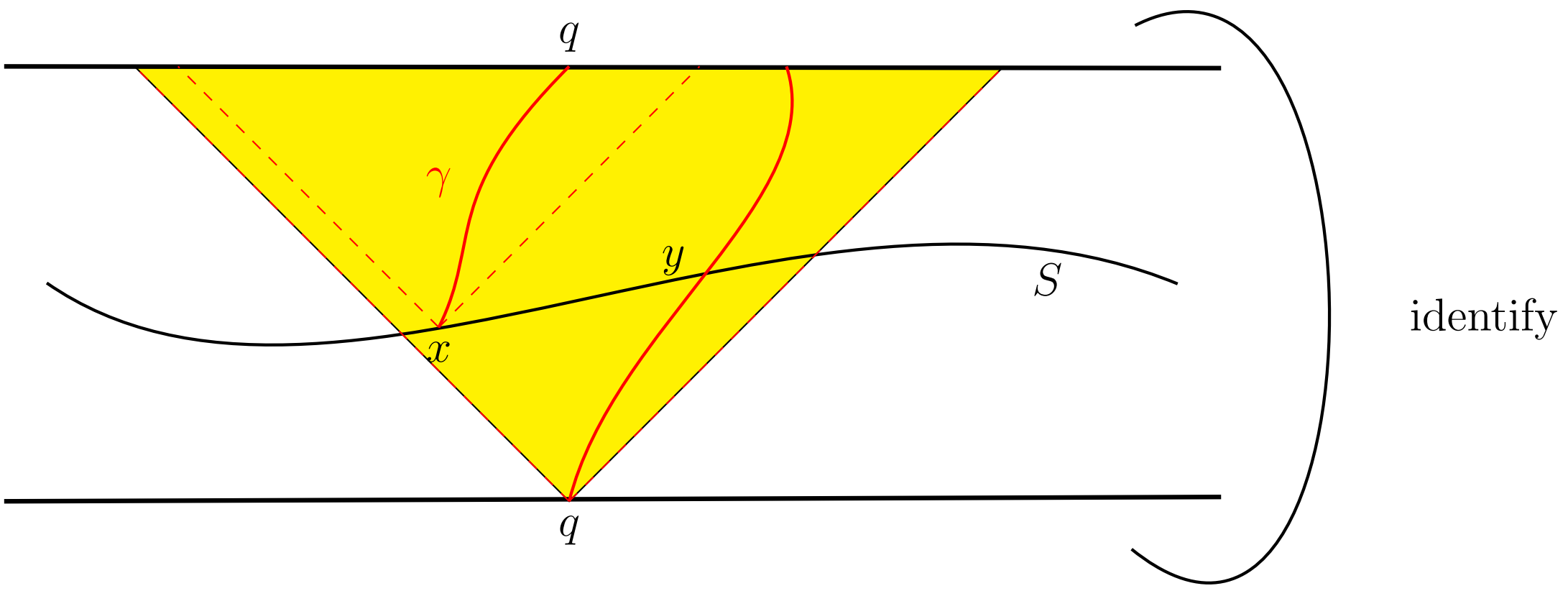
\begin{figure}[htbp]
\begin{center}
\begin{eqnarray*}
\begin{tikzpicture}[scale=0.8]
  % Coordinates
  \coordinate (A) at (0.00, 0.00);
  \coordinate (B) at (14.00, 0.05);
  \coordinate (C) at (0.00, 5.00);
  \coordinate (D) at (14.00, 4.98);
  \coordinate (E) at (13.01, 5.47);
  \coordinate (F) at (12.97, -0.55);
  \coordinate (G) at (16.00, 7.00);
  \coordinate (H) at (16.00, -3.00);
  \coordinate (I) at (0.49, 2.51);
  \coordinate (J) at (13.50, 2.50);
  \coordinate (K) at (3.75, 0.25);
  \coordinate (L) at (9.25, 4.25);
  \coordinate (M) at (5.00, 2.00);
  \coordinate (N) at (2.00, 5.00);
  \coordinate (O) at (8.00, 5.00);
  \coordinate (P) at (6.50, 5.00);
  \coordinate (Q) at (5.50, 3.00);
  \coordinate (R) at (5.00, 3.50);
  \coordinate (S) at (6.50, 0.00);
  \coordinate (T) at (11.50, 5.00);
  \coordinate (U) at (1.50, 5.00);
  \coordinate (V) at (9.00, 5.00);
  \coordinate (W) at (7.00, 2.00);
  \coordinate (X) at (9.50, 3.50);
  \coordinate (Y) at (16.00, 2.50);
% Drawing
  \draw [fill = yellow] (U)-- (S) -- (T);
  \draw[very thick] (A) -- (B);
  \draw[very thick] (C) -- (D);
  %\draw[Stealth-Stealth, thick] (E) .. controls (16.00, 7.00) and (16.00, -3.00) .. (F);
  \draw[black,  thick] (E) .. controls (16.00, 7.00) and (16.00, -3.00) .. (F);
  \draw[thick] (I) .. controls (3.75, 0.25) and (9.25, 4.25) .. (J);
  \draw[red, dashed] (M) -- (N);
  \draw[red, dashed] (M) -- (O);
  \draw[red, thick] (M) .. controls (5.50, 3.00) and (5.00, 3.50) .. (P);
  \draw[red, dashed] (S) -- (T);
  \draw[red, dashed] (S) -- (U);
  \draw[red, thick] (S) .. controls (7.00, 2.00) and (9.50, 3.50) .. (V);
  \fill[white] (Y) circle (2pt);
  \node[above, font=\large] at (P) {$q$};
  \node[below, font=\large] at (S) {$q$};
  \node[below, font=\large] at (M) {$x$};
  \node[below, font=\large] at (7.70,3.10) {$y$};
  \node[below right, font=\large] at (Y) {$\text{identify}$};
  \node[below, font=\large] at (12.00,2.90) {$\S$};
   \node[below, red, font=\large] at (5.00,4.00) {$\gamma$};

\end{tikzpicture}
\end{eqnarray*}
\end{center}
%\begin{figure}[htbp]
\caption{\small{The time-like curve $\gamma$ crosses the space-like surface $\S$ twice at $x$ and $y$. $\S$ is space-like but not achronal.}}
\centering
\label{fig:noncausal}
\end{figure}

\begin{remark}
The definition implies that a subset $\S$ of $\M$ is achronal,  if there \emph{does not exist} a future directed time-like curve that connects any two points of $\S$. This notion should not be confused with either that of space-like or that of null hyper-surfaces. It is possible for space-like and null hyper-surfaces not to be achronal.  Future directed time-like curves from these hyper-surfaces can go around the spacetime and intersect them again at some point violating the achronicity condition. An example of this has already been mentioned at the end of section \ref{sec:localstructure}, see also  figure \ref{fig:noncausal}.

It is also possible to define \emph{acausal} sets $\S$ by demanding that not two points of $\S$ can be connected by a \emph{ non-trivial} causal curve. But in what follows, we shall mostly investigate the properties of achronal sets.
\hfill{$\Diamond$}
\end{remark}

\begin{corollary}\label{cor:achronal}
The boundary of a future set $\S\subseteq \M$ is achronal.
\end{corollary}
\begin{proof}
Let $\S$ be a future set. We have to demonstrate that $\partial \S\cap I^+(\partial \S)=\emptyset$.  Indeed,
\be
\emptyset=\text{int}\, \S\cap \partial \S= I^+(\S)\cap \partial \S=I^+(\bar{\S})\cap \partial \S~,
\ee
where the equality follows from the second remark in \ref{rem:aaa}.
Therefore as $\partial\S\subset \bar{\S}$, it follows that $\partial \S\cap I^+(\partial \S)=\emptyset$.
\end{proof}

\begin{corollary}\label{cor:achronalbI}
Let $\S$ be a subset of $\M$. The boundary of $I^+(\S)$ is  achronal.
\end{corollary}
\begin{proof}
This is immediately follows because $I^+(\S)$ is an open future set, as a consequence of the statement (ii) of proposition \ref{prop:chronos}, and from the results  of the previous corollary.
\end{proof}

\begin{definition}
A \emph{ proper} achronal boundary is the boundary of a future set. \hfill{$\triangle$}
\end{definition}

\begin{figure}[htbp]
\begin{center}
\begin{eqnarray*}
\begin{tikzpicture}[scale=1]
  % Coordinates
  \coordinate (A) at (-2.00, 2.00);
  \coordinate (B) at (8.00, 4.00);
  \coordinate (C) at (1.50, 2.25);
  \coordinate (D) at (4.25, 6.25);
  \coordinate (E) at (3.00, 4.00);
  \coordinate (F) at (3.00, 4.00);
  \coordinate (G) at (0.00, 1.00);
  \coordinate (H) at (6.00, 1.00);
  \coordinate (I) at (4.75, 6.50);
  \coordinate (J) at (3.00, 1.00);
  \coordinate (K) at (9.00, 7.00);
  \coordinate (L) at (-3.00, 7.00);
  \coordinate (M) at (1.00, 4.00);
  \coordinate (N) at (5.00, 4.00);
  \coordinate (O) at (1.25, 5.75);
  \coordinate (P) at (5.00, 5.75);
  \coordinate (Q) at (1.00, 2.25);
  \coordinate (R) at (5.00, 2.50);
  \coordinate (S) at (1.50, 4.75);
  \coordinate (T) at (2.75, 2.75);
  \coordinate (U) at (1.75, 4.00);
  \coordinate (V) at (2.75, 3.25);
  \coordinate (W) at (1.75, 5.00);
  \coordinate (X) at (3.06, 2.64);
  \coordinate (Y) at (2.00, 3.75);
  \coordinate (Z) at (3.00, 3.50);
  \coordinate (AA) at (2.00, 5.25);
  \coordinate (AB) at (3.50, 2.75);
  \coordinate (AC) at (2.00, 4.50);
  \coordinate (AD) at (3.00, 3.50);
  \coordinate (AE) at (2.50, 5.25);
  \coordinate (AF) at (3.78, 2.80);
  \coordinate (AG) at (2.75, 4.50);
  \coordinate (AH) at (3.25, 3.50);
  \coordinate (AI) at (3.15, 5.41);
  \coordinate (AJ) at (3.98, 2.94);
  \coordinate (AK) at (3.50, 5.00);
  \coordinate (AL) at (4.25, 4.00);
  \coordinate (AM) at (3.75, 5.25);
  \coordinate (AN) at (4.48, 3.28);
  \coordinate (AO) at (4.50, 4.75);
  \coordinate (AP) at (4.75, 4.25);

  % Drawing
  \draw [fill = yellow] (L)-- (J) -- (K) ;
  \draw [fill = yellow] (G)-- (F) -- (H) ;
  \draw[very thick] (A) .. controls (1.50, 2.25) and (4.25, 6.25) .. (B);
  \draw (E) circle (4.00);
  \draw[red, dashed] (E) -- (G);
  \draw[red, dashed] (H) -- (E);
  \draw (I) -- (I);
  \draw[red, dashed] (J) -- (K);
  \draw[red, dashed] (J) -- (L);
  \draw[thick] (M) .. controls (1.25, 5.75) and (5.00, 5.75) .. (N);
  \draw[thick] (M) .. controls (1.00, 2.25) and (5.00, 2.50) .. (N);
  \draw[red, thick] (S) .. controls (1.75, 4.00) and (2.75, 3.25) .. (T);
  \draw[red, thick] (W) .. controls (2.00, 3.75) and (3.00, 3.50) .. (X);
  \draw[red, thick] (AA) .. controls (2.00, 4.50) and (3.00, 3.50) .. (AB);
  \draw[red, thick] (AE) .. controls (2.75, 4.50) and (3.25, 3.50) .. (AF);
  \draw[red, thick] (AI) .. controls (3.50, 5.00) and (4.25, 4.00) .. (AJ);
  \draw[red, thick] (AM) .. controls (4.50, 4.75) and (4.75, 4.25) .. (AN);
  \node[above right] at (F) {$p$};
  \node[below] at (J) {$q$};
  \node[above] at (3.00, 7.00) {$N_p$};
  \node[above] at (3.00, 5.80) {$C_q^+$};
  \node[above] at (3.00, 4.70) {$U_p$};
   \node[above] at (1.50, 1.40) {$I^-(p)$};
    \node[below] at (8.00, 3.90) {$\B$};
\end{tikzpicture}
\end{eqnarray*}
\end{center}
%\begin{figure}[htbp]
\caption{\small{In the figure, there is a depiction of the proof of proposition \ref{prop:boundary}. The red lines inside the neighbourhood $U_p$ of $p$ are the time-like geodesics with constant $y$. The cross $B$ once, since it is achronal.}}
\centering
\label{fig:boundary1}
\end{figure}
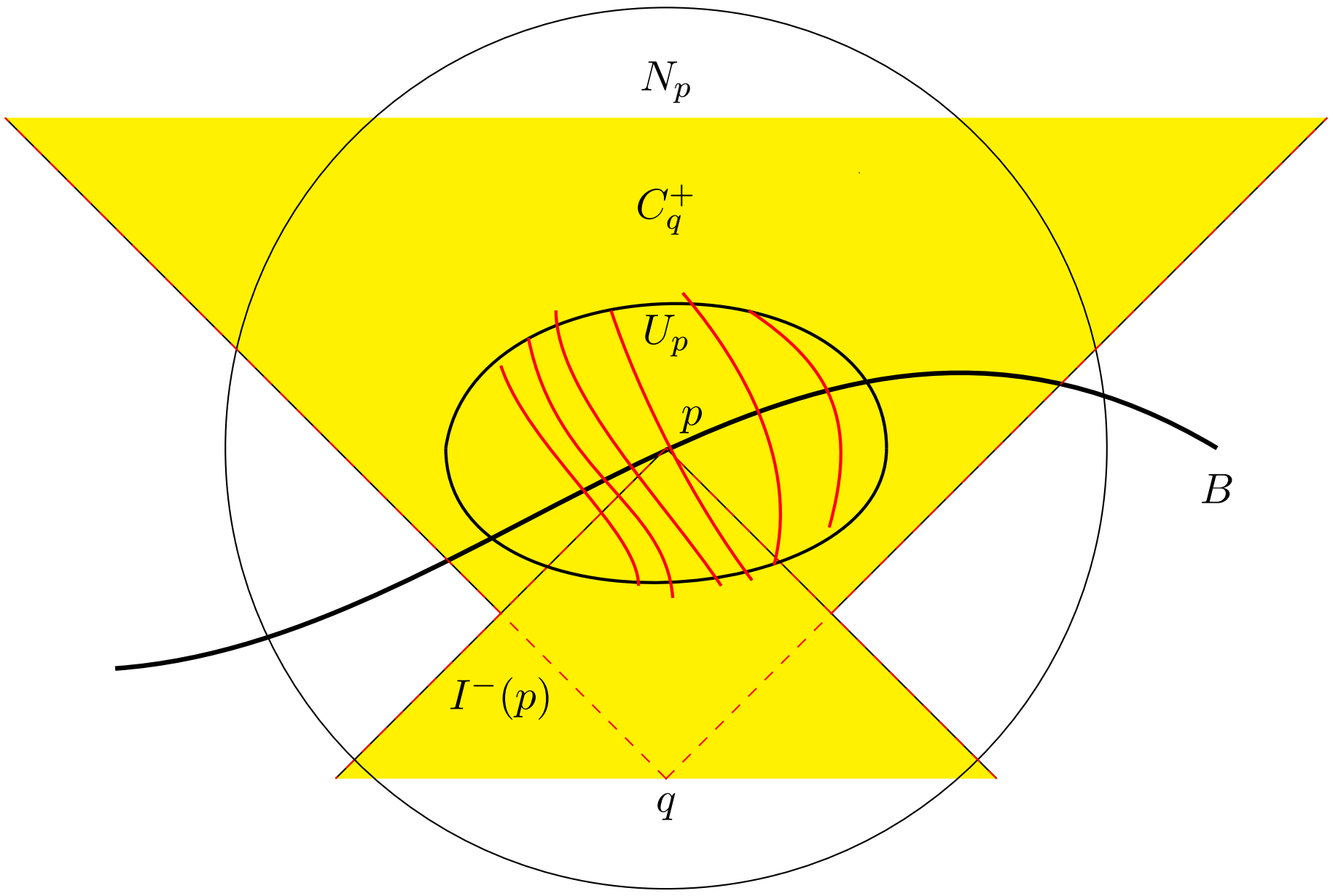

\begin{prop}\label{prop:boundary}
Any proper achronal boundary $\B$ is an embedded three dimensional $C^{1-}$ submanifold in $\M$, i.e. $\B$ is an embedded achronal hyper-surface without boundary in $\M$.
\end{prop}
\begin{proof}
Consider $p\in \B$ and the normal neighbourhood $\N_p$ of $p$. Then choose a $q\in I^-(p)\cap N_p$. Clearly $p\in C^+(q)$. Choose an open neighbourhood $U_p$ of $p$
such that $\bar U_p\subset C_q^+\cap N_p$. As $U_p$ is a proper subset of $C^+_q$, introduce Gaussian coordinates $(y^0, y^i)$, i.e.
\be
g=-(dy^0)^2+g_{ij}(y^0, y^i) dy^idy^j~,
 \ee
 such that the time-like geodesics from $I^-(p)\cap U_p$ to $I^+(p)\cap U_p$ are those with constant $y^i$. Every such geodesic crosses $B$ once as otherwise $B$ will not be achronal. For the geodesic with $y^i=u^i$ assign $u^i$ as the coordinates of the intersection point of $B$ with the geodesic. This gives a map from $B\cap U_p$ to $\bR^3$ which is continuous.  As this can be done continuously at every point $p$, $B$ becomes a topological manifold.

To prove that actually $B$ is a $C^{1-}$ manifold, let us prove that the embedding is a $C^{1-}$ map. The embedding can be described locally as
\be
y^0=\Phi^0(u^i)~,~~~y^i=\Phi^i(u)=u^i
\ee
for some function $\Phi$. Next take two points $q,r\in U_p\cap B$ and define the curve $\gamma$ with
\be
y^0=\gamma^0(t)~,~~~y^i(t)=t \Delta u^i+ u^i_q~,~~~\Delta u^i= u_r^i-u_q^i~.
\ee
This curve cannot be time-like as $B$ is achronal -- otherwise the points $q,r$ of $B$ would have been joined with a time-like curve.  Therefore,
\be
(\dot y^0)^2\leq g_{ij} \Delta u^i \Delta u^j\leq C^2 |\Delta u|^2
\ee
where $|\Delta u|$ is the usual Euclidean norm of $\Delta u$ in $\bR^3$ and $C$ some large enough positive constant $C>0$.  This inequality gives that $|\gamma^0(t)-\gamma^0(0)|\leq C  t |\Delta u|$, where $t\in [0,1]$. In turn, we have
\be
|\gamma^0(1)-\gamma^0(0)|=|\Phi^0(1)-\Phi^0(0)|\leq C |\Delta u|~,
\ee
which proves the statement.
\end{proof}

\begin{figure}[htbp]
\begin{center}
\begin{eqnarray*}
\begin{tikzpicture}[scale=0.5]
  % Coordinates
  \coordinate (A) at (0.00, 0.00);
  \coordinate (B) at (7.00, 0.00);
  \coordinate (C) at (2.25, 1.25);
  \coordinate (D) at (4.75, -1.25);
  \coordinate (E) at (-6.00, 6.00);
  \coordinate (F) at (13.00, 6.00);
  \coordinate (G) at (-5.00, 5.00);
  \coordinate (H) at (-3.00, 3.00);
  \coordinate (I) at (-2.00, 2.00);
  \coordinate (J) at (-4.47, 5.26);
  \coordinate (K) at (-3.98, 5.50);
  \coordinate (L) at (-2.93, 5.69);
  \coordinate (M) at (-2.26, 5.90);
  \coordinate (N) at (0.26, 0.15);
  \coordinate (O) at (-3.00, 4.50);
  \coordinate (P) at (-0.75, 1.75);
  \coordinate (Q) at (0.75, 0.25);
  \coordinate (R) at (-1.50, 4.75);
  \coordinate (S) at (-0.25, 1.75);
  \coordinate (T) at (1.25, 0.25);
  \coordinate (U) at (-1.25, 5.25);
  \coordinate (V) at (0.75, 3.00);
  \coordinate (W) at (2.00, 0.25);
  \coordinate (X) at (0.50, 4.50);
  \coordinate (Y) at (1.50, 3.00);

  % Drawing
  \draw[very thick] (A) .. controls (2.25, 1.25) and (4.75, -1.25) .. (B);
  \draw[red, thick] (A) -- (E);
  \draw[red, thick] (B) -- (F);
  \draw[thick] (J) .. controls (-3.00, 4.50) and (-0.75, 1.75) .. (N);
  \draw[thick] (K) .. controls (-1.50, 4.75) and (-0.25, 1.75) .. (Q);
  \draw[thick] (L) .. controls (-1.25, 5.25) and (0.75, 3.00) .. (T);
  \draw[thick] (M) .. controls (0.50, 4.50) and (1.50, 3.00) .. (W);
  \fill[black] (G) circle (3pt);
  \fill[black] (H) circle (4pt);
  \fill[black] (I) circle (2pt);
  \fill[black] (J) circle (2pt);
  \fill[black] (K) circle (2pt);
  \fill[black] (L) circle (2pt);
  \fill[black] (M) circle (2pt);
  \node[below left] at (G) {$p$};
  \node[below left] at (H) {$z$};
  \node[below left] at (I) {$w$};
   \node[below] at (-3.50,6.90) {$(q_n)$};
     \node[above] at (2.00, 3,00) {$(\lambda_n)$};
      \node[above] at (4.00, 4,00) {$\overline{I^+(\S)}$};
      \node[below] at (4.00, -1,00) {$\S$};
        \node[below left, red] at (-4.00, 4,00) {$\gamma$};
\end{tikzpicture}
\end{eqnarray*}
\end{center}
%\begin{figure}[htbp]
\caption{\small{This figure illustrates the proof of proposition \ref{prop:Iboundary}. The sequence of points $(q_n)$ converges to the point $p$ at the boundary of $I^+(\S)$. The associated sequence of time-like curves $(\lambda_n)$, which has $p$ as an accumulation point, converges by the limit curve theorem to a curve $\gamma$ on the boundary of $I^+(\S)$.  If the point $z$ is not in the spacetime $\M$, then $p$ lies on a past inextendible null geodesic, while the point $w$ lies on a null geodesic with end point on the closed set $\S$. If $z$ is a spacetime point, then both $p$ and $w$ lie on null geodesics that have past end points on $\S$.   }}
\centering
\label{fig:boundary2}
\end{figure}
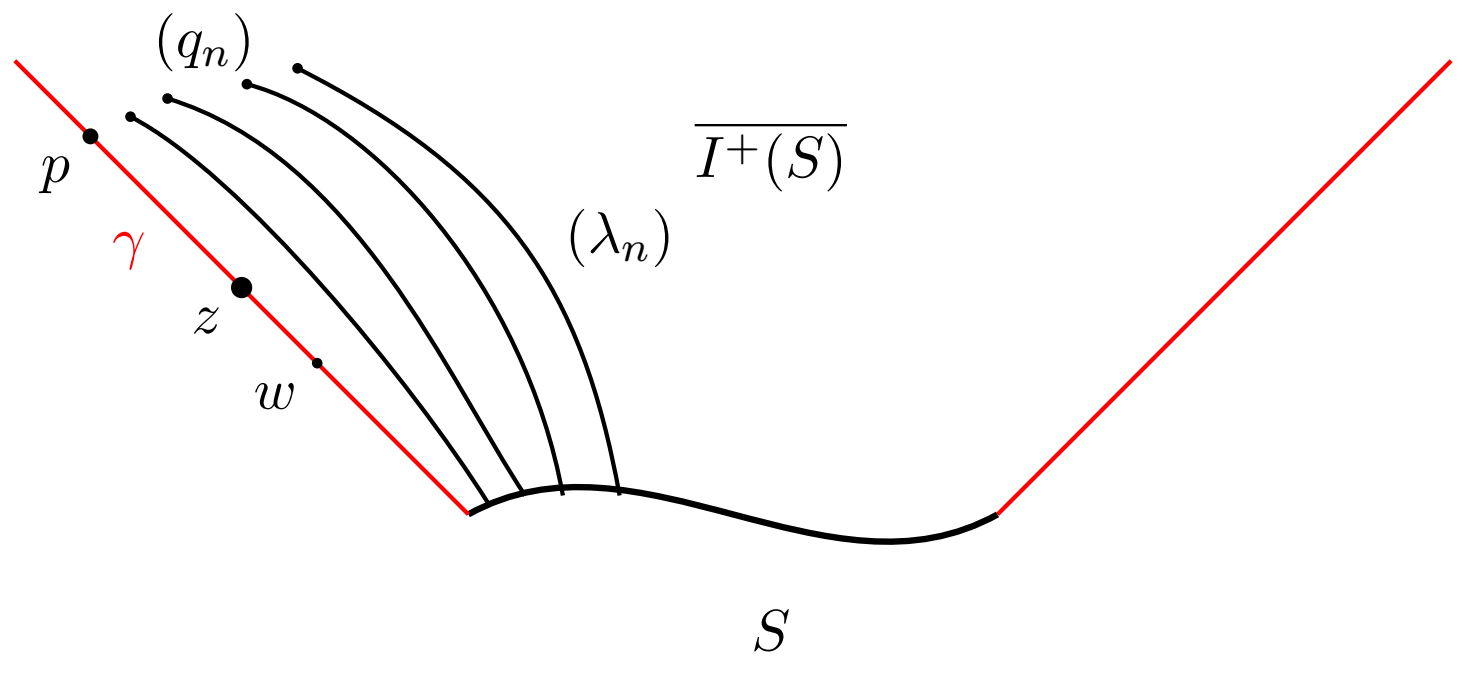

\begin{prop}\label{prop:Iboundary}
Let $\S$ be a closed subset of the spacetime $\M$. Then, any point of $p\in\partial I^+(\S)$ lies on a null geodesics which in turn is contained entirely in $\partial I^+(\S)$ and either is past inextendible or  has an end point in $\S$. ( See also figures \ref{fig:onex} and \ref{fig:2xy}.)
\end{prop}
\begin{proof}
Consider a point $p\in \partial I^+(\S)$.  As $p\in \partial I^+(\S)$, there is a sequence $(q_n)$ of points in $I^+(\S)$ with limit $p$. In addition, there are future directed time-like curves $(\lambda_n)$ that start from the points $q_n$ and end on $\S$. As $p$ is an accumulation point of $(q_n)$, it is also an accumulation point of $(\lambda_n)$ and from the limit curve  theorem \ref{prop:limitcurve},  there is a future directed causal curve $\gamma$ that passes through $p$ and $\gamma\in \overline{I^+(\S)}$.
Moreover,  since  $\S$ is closed, $\M-\S$ is an open set and so it is a submanifold of $\M$. On $\M-\S$, the time-like curves $\lambda_n$ are past inextendible  and as a result again of the theorem \ref{prop:limitcurve}, $\gamma$ must also be inextendible.  It follows then from proposition \ref{prop:causaltime} that either $\gamma$ is a null geodesic or it can be deformed to a time-like curve from $p$. However, the latter possibility is ruled out as otherwise $p\in I^+(\S)$, which is a contradiction.

Furthermore, $\gamma$ must lie entirely in $\partial I^+(\S)$. This is because $\gamma\in \overline{I^+(\S)}$  and if there is a point $q$ of $\gamma$  in $I^+(\S)$, then joining the time-like curve from $\S$ to $q$ with $\gamma$ will lead to a construction of a causal curve from $\S$ to $p$ that it is not a null geodesic.  Thus again from \ref{prop:causaltime}, there must be a time-like curve from $\S$ to $p$, which is again a contradiction.
Although $\gamma$ is past inextendible on $\M-\S$, on $\M$ will either have a past end point in $\S$ or it will remain past inextendible.
\end{proof}

\section{Causality}

\subsection{Strong causality}

\begin{definition}
A spacetime $\M$ is \emph{ causal}, if it does not admit closed causal curves.
\end{definition}

\begin{prop}
If the spacetime $\M$ is compact, then $\M$ contains closed time-like curves.  Therefore, compact spacetimes  are not causal.
\end{prop}
\begin{proof}
To begin, if $q\in I^+(p)$, then $I^+(q)\subseteq I^+(p)$ for any two points $p,q\in \M$. Thus, let us cover $\M$ with the open sets $\{I^+(p_i)\}$ such that $p_i\notin I^+(p_j)$ for $p_i\not=p_j$.  Since $\M$ is compact, this cover admits a finite refinement $\{I^+(p_n)\}$. Choose one of these points, say $p_1$. By construction of the cover, $p_1$ cannot be contained in any of the sets $I^+(p_n)$ with $n\not=1$.  But $p_1\in \M$. This implies that $p_1\in I^+(p_1)$ and so there must be a future directed non-trivial time-like curve from $p_1$ to $p_1$ violating the causality of spacetime.
\end{proof}

Even though a spacetime may be causal, it may come ``very close'' to violate causality. This leads to the definition of strong causality.

\begin{definition}
A spacetime $\M$ is \emph{strongly causal}, if for every $p\in \M$ and every neighbourhood $U$ of $p$, there is another neighbourhood $V\subset U$, such that there are no causal curves of $\M$ that intersect $V$ more than once.
\end{definition}

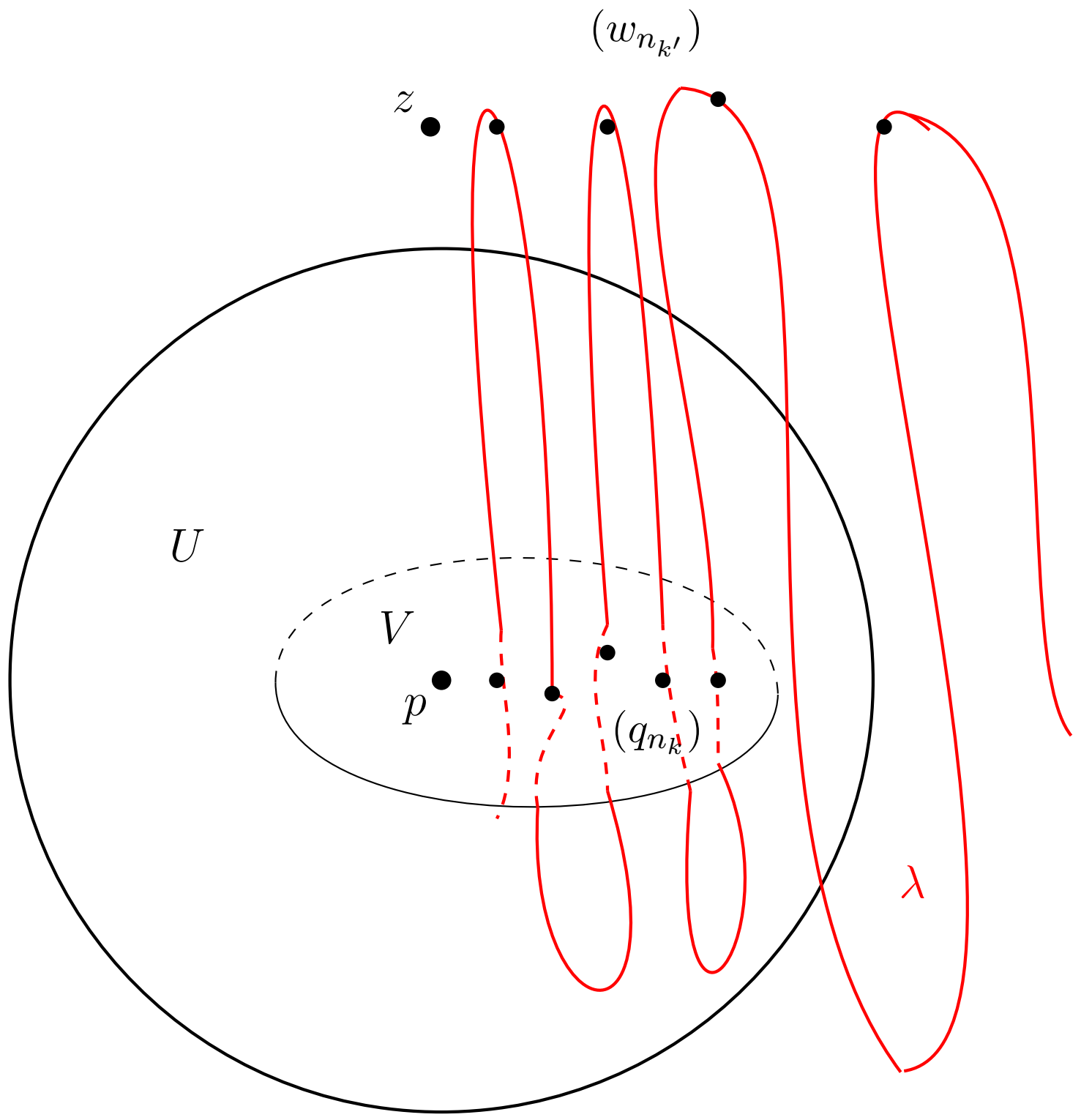
\begin{figure}[htbp]
\begin{center}
\begin{eqnarray*}
\begin{tikzpicture}[scale=1]
  % Coordinates
  \coordinate (A) at (5.90, 5.92);
  \coordinate (C) at (3.00, 4.98);
  \coordinate (D) at (7.54, 4.87);
  \coordinate (E) at (3.00, 6.50);
  \coordinate (F) at (7.50, 6.50);
  \coordinate (G) at (3.00, 3.50);
  \coordinate (H) at (7.50, 3.50);
  \coordinate (I) at (8.72, 10.11);
  \coordinate (J) at (8.65, 1.46);
  \coordinate (K) at (6.00, 5.50);
  \coordinate (L) at (10.50, 1.75);
  \coordinate (M) at (6.66, 10.35);
  \coordinate (N) at (6.75, 4.00);
  \coordinate (O) at (8.50, 10.25);
  \coordinate (P) at (6.95, 5.29);
  \coordinate (Q) at (6.00, 9.75);
  \coordinate (R) at (7.00, 7.00);
  \coordinate (S) at (4.50, 5.00);
  \coordinate (B_1) at (4.87, 4.92);
  \coordinate (U) at (7.00, 4.25);
  \coordinate (V) at (7.00, 5.00);
  \coordinate (W) at (7.00, 4.75);
  \coordinate (X) at (7.75, 2.75);
  \coordinate (Y) at (6.50, 1.00);
  \coordinate (Z) at (6.50, 5.50);
  \coordinate (AA) at (6.50, 5.00);
  \coordinate (AB) at (6.50, 5.50);
  \coordinate (AC) at (6.25, 11.75);
  \coordinate (AD) at (5.50, 11.75);
  \coordinate (AE) at (9.58, 5.26);
  \coordinate (AK) at (6.00, 4.00);
  \coordinate (AL) at (5.75, 5.00);
  \coordinate (AM) at (6.00, 4.50);
  \coordinate (AN) at (5.37, 3.85);
  \coordinate (AO) at (6.75, 1.50);
  \coordinate (AP) at (5.25, 1.75);
  \coordinate (AQ) at (5.50, 4.88);
  \coordinate (AR) at (5.25, 4.50);
  \coordinate (AS) at (5.87, 4.80);
  \coordinate (AT) at (5.04, 5.45);
  \coordinate (AU) at (5.50, 10.75);
  \coordinate (AV) at (4.25, 12.75);
  \coordinate (I_1) at (8.91, 9.97);
  \coordinate (K_1) at (7.25, 11.50);
  \coordinate (J_1) at (8.68, 1.47);
  \coordinate (T_1) at (10.33, 9.76);
  \coordinate (B_2) at (10.19, 4.50);
  \coordinate (B) at (5.00, 3.75);
  \coordinate (T) at (5.00, 5.00);
  \coordinate (AF) at (5.25, 4.25);
  \coordinate (AG) at (6.00, 5.25);
  \coordinate (AH) at (5.00, 10.00);
  \coordinate (AI) at (6.00, 10.00);
  \coordinate (AJ) at (7.00, 10.25);
  \coordinate (AW) at (8.50, 10.00);

  % Drawing
 % \draw[thick, rotate=-135.8] (5.38, 5.42) ellipse (3.91 and 3.84);
 \draw[thick] (S) circle (3.9);
  \draw[dashed] (C) .. controls (3.00, 6.50) and (7.50, 6.50) .. (D);
  \draw (C) .. controls (3.00, 3.50) and (7.50, 3.50) .. (D);
  \draw[red, thick] (I_1) .. controls (7.25, 11.50) and (10.50, 1.75) .. (J_1);
  \draw[red, thick] (J) .. controls (6.75, 4.00) and (8.50, 10.25) .. (M);
  \draw[red, thick] (M) .. controls (6.00, 9.75) and (7.00, 7.00) .. (P);
  \draw[red, thick, dashed] (P) .. controls (7.00, 5.00) and (7.00, 4.75) .. (U);
  \draw[red, thick] (U) .. controls (7.75, 2.75) and (6.50, 1.00) .. (N);
  \draw[red, thick, dashed] (N) .. controls (6.50, 5.00) and (6.50, 5.50) .. (Z);
  \draw[red, thick] (Z) .. controls (6.25, 11.75) and (5.50, 11.75) .. (K);
  \draw[red, thick] (I) .. controls (10.33, 9.76) and (9.58, 5.26) .. (B_2);
  \draw[red, thick, dashed] (K) .. controls (5.75, 5.00) and (6.00, 4.50) .. (AK);
  \draw[red, thick] (AK) .. controls (6.75, 1.50) and (5.25, 1.75) .. (AN);
  \draw[red, thick, dashed] (AN) .. controls (5.25, 4.50) and (5.87, 4.80) .. (AQ);
  \draw[red, thick] (AQ) .. controls (5.50, 10.75) and (4.25, 12.75) .. (AT);
  \draw[red, thick, dashed] (AT) .. controls (5.00, 5.00) and (5.25, 4.25) .. (B);
  \fill[black] (S) circle (2.5pt);
  \fill[black] (4.40, 10.00) circle (2.5pt);
  \fill[black] (V) circle (2pt);
  \fill[black] (AA) circle (2pt);
  \fill[black] (AQ) circle (2pt);
  \fill[black] (T) circle (2pt);
  \fill[black] (AG) circle (2pt);
  \fill[black] (AH) circle (2pt);
  \fill[black] (AI) circle (2pt);
  \fill[black] (AJ) circle (2pt);
  \fill[black] (AW) circle (2pt);
  \node[below left, font=\large] at (S) {$p$};
   \node[below left, font=\large] at (2.50, 6.50) {$U$};
    \node[above left, font=\large] at (4.40, 5.20) {$V$};
     \node[above left] at (7.00, 4.20) {$(q_{n_k})$};
       \node[above left] at (7.00, 10.50) {$(w_{n_{k'}})$};
       \node[above left, font=\large] at (4.40, 10.00) {$z$};
        \node[above right, red, font=\large] at (8.50, 2.90) {$\lambda$};
\end{tikzpicture}
\end{eqnarray*}
\end{center}
%\begin{figure}[htbp]
\caption{\small{The figure depicts the behaviour of the curve $\lambda$ explained in proposition \ref{th:compactendpoints}. If $p$ is an accumulation point, but not a convergence point,  of the sequence $(\lambda(t_n))$, there is a subsequence $(q_{n_k})=(\lambda(t_{n_k}))$ that converges to $q$. Moreover, another subsequence $(w_{n_{k'}})=(\lambda(t_{n_{k'}})$ of $(\lambda(t_n))$ will converge to another point $z$. The curve $\lambda$ has to intersect infinite many times any neighbourhood $V$ of $p$ while has to escape $U$ as the other accumulation point $z$ lies outside $U$. This behaviour of $\lambda$ violates strong causality.}}
\centering
\label{compactendpoints}
\end{figure}

\begin{prop}\label{th:compactendpoints}
Let $\M$ be a strongly causal spacetime and $\K$ be a compact subset of $\M$. Then, every causal curve entirely contained in $\K$ has a future and past end points  in $\K$.
\end{prop}
\begin{proof}
Let $\lambda$ be a causal curve in $\K$ with parameter $t\in (-\infty, +\infty)$.  Consider a sequence $(t_n)$ with $t_n\rightarrow +\infty$. As $\K$ is compact, the sequence $(\lambda(t_n))$ has an accumulation point $p\in \K$. If $p$ is a convergence of $(\lambda(t_n))$, then $p\in \K$ will be the  future end point that proves the statement. However, if $p$ is an accumulation point, and not a convergence point, then as $t\rightarrow +\infty$, the curve will oscillate between the different accumulation points.  Suppose that the subsequence $(q_{n_k})= (\lambda(t_{n_k}))$ of $(\lambda(t_n))$ converges at $p$ and there is another accumulation  point of the sequence  $(\lambda(t_n))$ at $z$. There is a neighbourhood $U$ of $p$ that does not contain $z$. The curve $\lambda$ cannot be entirely contained in $U$ for some $t>t_0$.  Then for every neighbourhood of $V$ of $p$, $V\subset U$, $\lambda$ will intersect $V$ infinite many times at $q_{n_k}\in V$ for large enough $k$ as $t\rightarrow +\infty$. Therefore, $\lambda$ intersects $V$ infinite many times and it is not contained in $U$.  This violates the strong causality condition for $\M$. Thus, $p$ must be a future end point. The proof for the existence of a past end point in $\K$ for $\lambda$ is similar.
\end{proof}

 \begin{example} To illustrate how curves with several accumulation points behave,  consider the curve $\gamma(t)=(\sin(2\pi t), t^{-1})$ in $\bR^2$. For $t=n$, $n\in \bN$,  the convergence point as $n\rightarrow +\infty$ is $(0,0)$,  while for $t=n+\frac{1}{4}$, the convergence  point as $n\rightarrow +\infty$  is $(1,0)$. This curve has accumulation points but no limit as $t\rightarrow +\infty$. Clearly such a curve will intersect infinite many times any neighbourhood of either $(0,0)$ or $(1,0)$ as $t\rightarrow +\infty$ and it will not be entirely contained in any  small enough neighbourhoods of these points.
\hfill{$\triangledown$ }
\end{example}

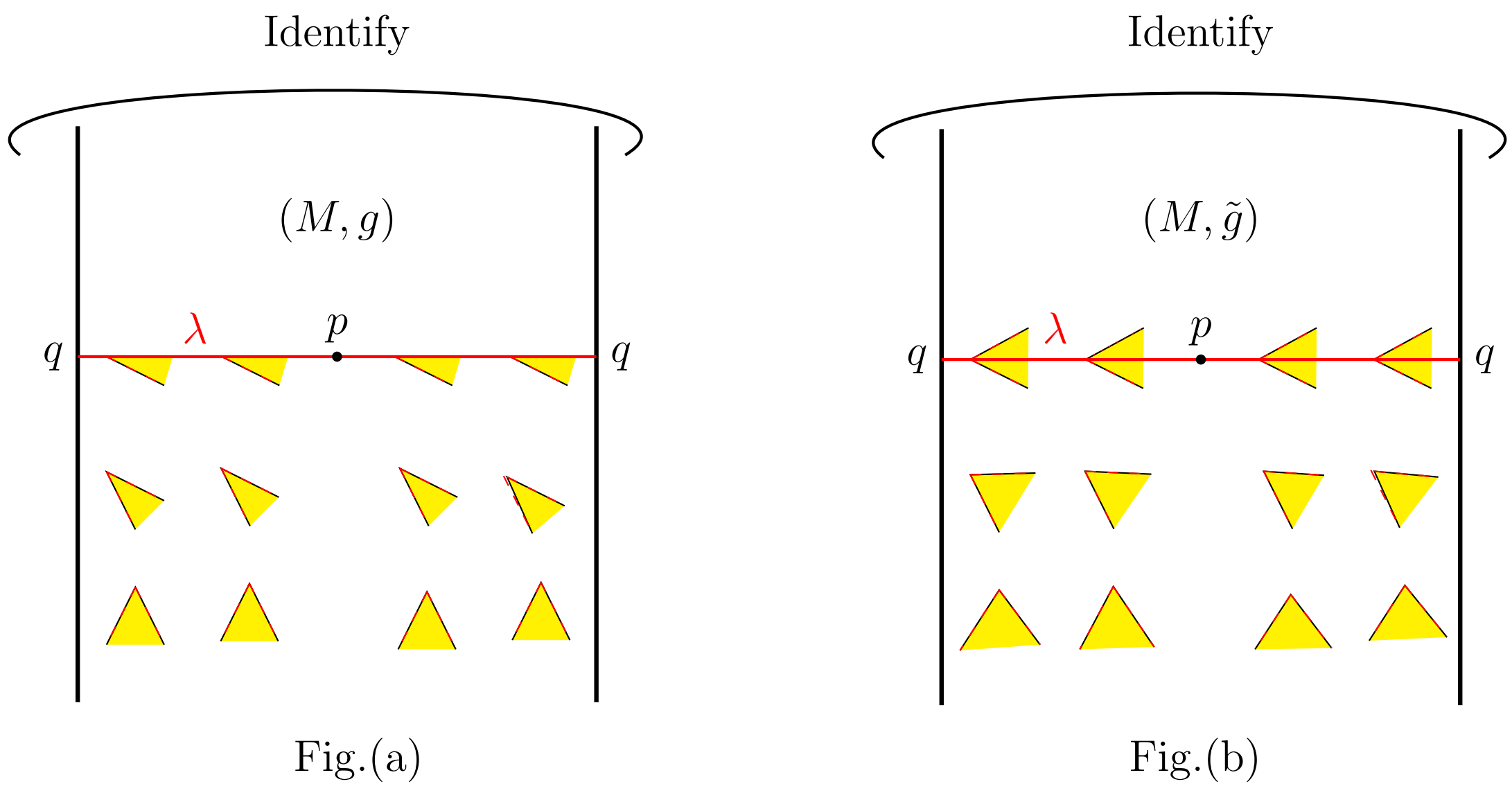
\begin{figure}[htbp]
\begin{center}
\begin{eqnarray*}
\begin{tikzpicture}[scale=0.55]
  % Coordinates
  \coordinate (A) at (0.00, 10.00);
  \coordinate (B) at (0.00, 0.00);
  \coordinate (C) at (9.00, 10.00);
  \coordinate (D) at (9.00, 0.00);
  \coordinate (E) at (-1.00, 9.50);
  \coordinate (F) at (9.50, 9.50);
  \coordinate (G) at (-3.00, 11.00);
  \coordinate (H) at (12.00, 11.00);
  \coordinate (I) at (0.00, 6.00);
  \coordinate (J) at (9.00, 6.00);
  \coordinate (K) at (4.50, 6.00);
  \coordinate (L) at (0.50, 6.00);
  \coordinate (M) at (1.50, 5.50);
  \coordinate (N) at (2.50, 6.00);
  \coordinate (O) at (3.50, 5.50);
  \coordinate (P) at (5.50, 6.00);
  \coordinate (Q) at (6.50, 5.50);
  \coordinate (R) at (7.50, 6.00);
  \coordinate (S) at (8.50, 5.50);
  \coordinate (T) at (0.50, 4.00);
  \coordinate (U) at (1.50, 3.50);
  \coordinate (V) at (1.00, 3.00);
  \coordinate (W) at (2.49, 4.06);
  \coordinate (X) at (3.49, 3.56);
  \coordinate (Y) at (2.99, 3.06);
  \coordinate (Z) at (5.59, 4.06);
  \coordinate (AA) at (6.59, 3.56);
  \coordinate (AB) at (6.09, 3.06);
  \coordinate (AC) at (7.39, 3.93);
  \coordinate (AD) at (8.45, 3.41);
  \coordinate (AE) at (7.89, 2.93);
  \coordinate (AC_1) at (7.45, 3.91);
  \coordinate (AF) at (1.00, 2.00);
  \coordinate (AG) at (0.50, 1.00);
  \coordinate (AH) at (1.50, 1.00);
  \coordinate (AI) at (2.98, 2.06);
  \coordinate (AJ) at (2.48, 1.06);
  \coordinate (AK) at (3.48, 1.06);
  \coordinate (AL) at (6.06, 1.92);
  \coordinate (AM) at (5.56, 0.92);
  \coordinate (AN) at (6.56, 0.92);
  \coordinate (AO) at (8.04, 2.08);
  \coordinate (AP) at (7.54, 1.08);
  \coordinate (AQ) at (8.54, 1.08);
  \coordinate (AR) at (14.99, 9.95);
  \coordinate (AS) at (14.99, -0.05);
  \coordinate (AT) at (23.99, 9.95);
  \coordinate (AU) at (23.99, -0.05);
  \coordinate (AV) at (13.99, 9.45);
  \coordinate (AW) at (24.49, 9.45);
  \coordinate (AX) at (11.99, 10.95);
  \coordinate (AY) at (26.99, 10.95);
  \coordinate (AZ) at (14.99, 5.95);
  \coordinate (BA) at (23.99, 5.95);
  \coordinate (BB) at (19.49, 5.95);
  \coordinate (BC) at (15.49, 5.95);
  \coordinate (BD) at (16.49, 5.45);
  \coordinate (BE) at (17.49, 5.95);
  \coordinate (BF) at (18.49, 5.45);
  \coordinate (BG) at (20.49, 5.95);
  \coordinate (BH) at (21.49, 5.45);
  \coordinate (BI) at (22.49, 5.95);
  \coordinate (BJ) at (23.49, 5.45);
  \coordinate (BK) at (15.49, 3.95);
  \coordinate (BL) at (16.62, 3.98);
  \coordinate (BM) at (15.99, 2.95);
  \coordinate (BN) at (17.48, 4.01);
  \coordinate (BO) at (18.63, 3.96);
  \coordinate (BP) at (17.98, 3.01);
  \coordinate (BQ) at (20.58, 4.01);
  \coordinate (BR) at (21.63, 3.94);
  \coordinate (BS) at (21.08, 3.01);
  \coordinate (BT) at (22.44, 4.03);
  \coordinate (BU) at (23.61, 3.91);
  \coordinate (BV) at (22.94, 3.03);
  \coordinate (BW) at (22.50, 4.01);
  \coordinate (BX) at (15.99, 1.95);
  \coordinate (BY) at (15.31, 0.90);
  \coordinate (BZ) at (16.70, 1.00);
  \coordinate (CA) at (17.97, 2.01);
  \coordinate (CB) at (17.39, 0.92);
  \coordinate (CC) at (18.68, 0.96);
  \coordinate (CD) at (21.05, 1.87);
  \coordinate (CE) at (20.43, 0.92);
  \coordinate (CF) at (21.76, 0.94);
  \coordinate (CG) at (23.03, 2.03);
  \coordinate (CH) at (22.41, 1.07);
  \coordinate (CI) at (23.76, 1.13);
  \coordinate (CJ) at (16.50, 6.50);
  \coordinate (CK) at (18.50, 6.50);
  \coordinate (CL) at (21.50, 6.50);
  \coordinate (CM) at (23.50, 6.50);

  % Drawing
  \draw [fill = yellow] (M)-- (L)--(1.65, 6.00);
   \draw [fill = yellow] (O)-- (N)--(3.65, 6.00);
    \draw [fill = yellow] (Q)-- (P)--(6.65, 6.00);
     \draw [fill = yellow] (S)-- (R)--(8.65, 6.00);
  \draw [fill = yellow] (V)-- (T)--(U);
  \draw [fill = yellow] (Y)-- (W)--(X);
  \draw [fill = yellow] (AB)-- (Z)--(AA);
  \draw [fill = yellow] (AE)-- (AC_1)--(AD);
  \draw [fill = yellow] (AG)-- (AF)--(AH);
  \draw [fill = yellow] (AJ)-- (AI)--(AK);
  \draw [fill = yellow] (AM)-- (AL)--(AN);
  \draw [fill = yellow] (AP)-- (AO)--(AQ);
  %figb
  \draw [fill = yellow] (BD)-- (BC)--(CJ);
  \draw [fill = yellow] (BF)-- (BE)--(CK);
  \draw [fill = yellow] (BH)-- (BG)--(CL);
  \draw [fill = yellow] (BJ)-- (BI)--(CM);
  \draw [fill = yellow] (BM)-- (BK)--(BL);
  \draw [fill = yellow] (BP)-- (BN)--(BO);
  \draw [fill = yellow] (BS)-- (BQ)--(BR);
  \draw [fill = yellow] (BV)-- (BW)--(BU);
  \draw [fill = yellow] (BY)-- (BX)--(BZ);
  \draw [fill = yellow] (CB)-- (CA)--(CC);
  \draw [fill = yellow] (CE)-- (CD)--(CF);
   \draw [fill = yellow] (CH)-- (CG)--(CI);
  \draw[very thick] (A) -- (B);
  \draw[very thick] (C) -- (D);
  \draw[thick] (E) .. controls (-3.00, 11.00) and (12.00, 11.00) .. (F);
  \draw[red, thick] (I) -- (J);
  \draw[red, dashed] (L) -- (M);
  \draw[red, dashed] (N) -- (O);
  \draw[red, dashed] (P) -- (Q);
  \draw[red, dashed] (R) -- (S);
  \draw[red, dashed] (T) -- (U);
  \draw[red, dashed] (T) -- (V);
  \draw[red, dashed] (W) -- (X);
  \draw[red, dashed] (W) -- (Y);
  \draw[red, dashed] (Z) -- (AA);
  \draw[red, dashed] (Z) -- (AB);
  \draw[red, dashed] (AC_1) -- (AD);
  \draw[red, dashed] (AC) -- (AE);
  \draw[red, dashed] (AF) -- (AG);
  \draw[red, dashed] (AF) -- (AH);
  \draw[red, dashed] (AI) -- (AJ);
  \draw[red, dashed] (AI) -- (AK);
  \draw[red, dashed] (AL) -- (AM);
  \draw[red, dashed] (AL) -- (AN);
  \draw[red, dashed] (AO) -- (AP);
  \draw[red, dashed] (AO) -- (AQ);
  \draw[very thick] (AR) -- (AS);
  \draw[very thick] (AT) -- (AU);
  \draw[thick] (AV) .. controls (11.99, 10.95) and (26.99, 10.95) .. (AW);
  \draw[red, thick] (AZ) -- (BA);
  \draw[red, dashed] (BC) -- (BD);
  \draw[red, dashed] (BE) -- (BF);
  \draw[red, dashed] (BG) -- (BH);
  \draw[red, dashed] (BI) -- (BJ);
  \draw[red, dashed] (BK) -- (BL);
  \draw[red, dashed] (BK) -- (BM);
  \draw[red, dashed] (BN) -- (BO);
  \draw[red, dashed] (BN) -- (BP);
  \draw[red, dashed] (BQ) -- (BR);
  \draw[red, dashed] (BQ) -- (BS);
  \draw[red, dashed] (BW) -- (BU);
  \draw[red, dashed] (BT) -- (BV);
  \draw[red, dashed] (BX) -- (BY);
  \draw[red, dashed] (BX) -- (BZ);
  \draw[red, dashed] (CA) -- (CB);
  \draw[red, dashed] (CA) -- (CC);
  \draw[red, dashed] (CD) -- (CE);
  \draw[red, dashed] (CD) -- (CF);
  \draw[red, dashed] (CG) -- (CH);
  \draw[red, dashed] (CG) -- (CI);
  \draw[red, dashed] (BC) -- (CJ);
  \draw[red, dotted] (BE) -- (CK);
  \draw[red, dashed] (BG) -- (CL);
  \draw[red, dashed] (BI) -- (CM);
  \fill[black] (K) circle (2.5pt);
  \fill[black] (BB) circle (2.5pt);
  \node[above, font=\large] at (K) {$p$};
  \node[above, font=\large] at (BB) {$p$};
   \node[above, font=\large] at (4.50,11.00) {$\mathrm{Identify}$};
   \node[above, font=\large] at (19.49, 11.00) {$\mathrm{Identify}$};
   \node[below, font=\large] at (4.50,9.00) {$(\M, g) $};
   \node[below, font=\large] at (19.49, 9.00) {$(\M, \tilde g)$};
   \node[left, font=\large] at (I) {$q$};
   \node[right, font=\large] at (J) {$q$};
   \node[left, font=\large] at (AZ) {$q$};
   \node[right, font=\large] at (BA) {$q$};
    \node[right, font=\large] at (3.50, -1.00) {$\mathrm{Fig. (a)}$};
     \node[right, font=\large] at (18.00, -1.00) {$\mathrm{Fig. (b)}$};
     \node[left, red, font=\large] at (2.50, 6.50) {$\lambda$};
     \node[right,red, font=\large] at (16.50, 6.50) {$\lambda$};
   \end{tikzpicture}
\end{eqnarray*}
\end{center}
%\begin{figure}[htbp]
\caption{\small{In figure (a), if the point $p$ is in the spacetime, $p\in \M$, the curve $\lambda$ is a closed null curve and $(\M, g)$ is not causal. If $p$ is removed from the spacetime, $(\M, g)$ becomes causal but not strongly causal as points in $\lambda-\{p\}$ have neighbourhoods that are intersected infinite many times by a time-like curve that goes around the spacetime. In figure (b) depicts a spacetime $(\M, \tilde g)$, where the lightcones of $\tilde g$ are opened up relative to those of $g$ in figure (a). The new spacetime $(\M, \tilde g)$ admits closed time-like curves and so $(\M, g)$ is not stably causal even if the point $p$ is not in the spacetime, $p\notin \M$.  }}
\centering
\label{fig:causalspace}
\end{figure}

\begin{definition}
The spacetime is \emph{ stably causal}, iff it admits a differentiable function $f$ such that the vector field associated to the 1-form $df$ is  time-like, past directed and nowhere vanishing; $f$ is referred to as either the \emph{time} or \emph{temporal} function.\hfill{$\triangle$ }
\end{definition}

\begin{lemma}
If $M$ is stably causal,  then it is strongly causal.
\end{lemma}
\begin{proof}
From the definition of stable causality, if $\vec{\dot\gamma}$ is the tangent vector of a causal future directed curve $\gamma$, then
\be
\vec{\dot\gamma}(f)>0~.
\ee
Thus, $f$ strictly increases along future directed causal curves.  Let $p\in\M$ and $U$   a neighbourhood of $p$, $p\in U$. If the strong causality condition is violated, then one can find a curve $\gamma_n$ leaving from point $p_n$ and, after going around the spacetime, re-entering at point $q_n$ of the open neighbourhood
 \be
 V_n=\{x\in U\vert\, |f(x)-f(p)|<1/n, ~~~n\in \bN\}~.
 \ee
 Clearly, $|f(p_n)-f(q_n)|<2/n$ and as $n\rightarrow +\infty$, $p_n, q_n\rightarrow p$. Therefore, $p$ is an accumulation point of the sequence $(\gamma_n)$ of future directed causal curves.  Therefore, after possibly choosing a subsequence, there is a future directed causal curve $\gamma$ that passes through $p$.

 There are several possibilities. (i) $\gamma$ starts and ends at $p$.  But this is not allowed as $f$ is strictly increasing along $\gamma$.  (ii) $\gamma$ is inextendible periodically passing through $p$. Again, this is not allowed because $f$ is strictly increasing along $\gamma$. (iii) $\gamma$ is future inextendible and at every loop around the spacetime approaches ever closer to $p$.  This means that $f$ evaluated on $\gamma$ will be approaching the value $f(p)$ after $\gamma$ has already passed through $p$. Again, this is not possible as $f$ is strictly increasing along $\gamma$.  A similar argument can be made if $\gamma$ is past inextendible.
 This proves the statement.
 \end{proof}

\begin{remark}
Stable causality can be understood geometrically by observing that lightcones of the metric
\be
\tilde g=-\alpha df\otimes df+ g~,~~~\alpha\geq 0
\ee
are more open than those of $g$.  This means that if $\vu$ is not space-like with respect to $g$, then $\vu$ is not space-like with respect to $\tilde g$. But the converse is not true.  As a result in the spacetime $(\M, \tilde g)$, there ``many more'' directions for causal curves that enhances the possibility of causality violations, see figure \ref{fig:causalspace}(b).
\end{remark}

\subsection{Domain of dependence}
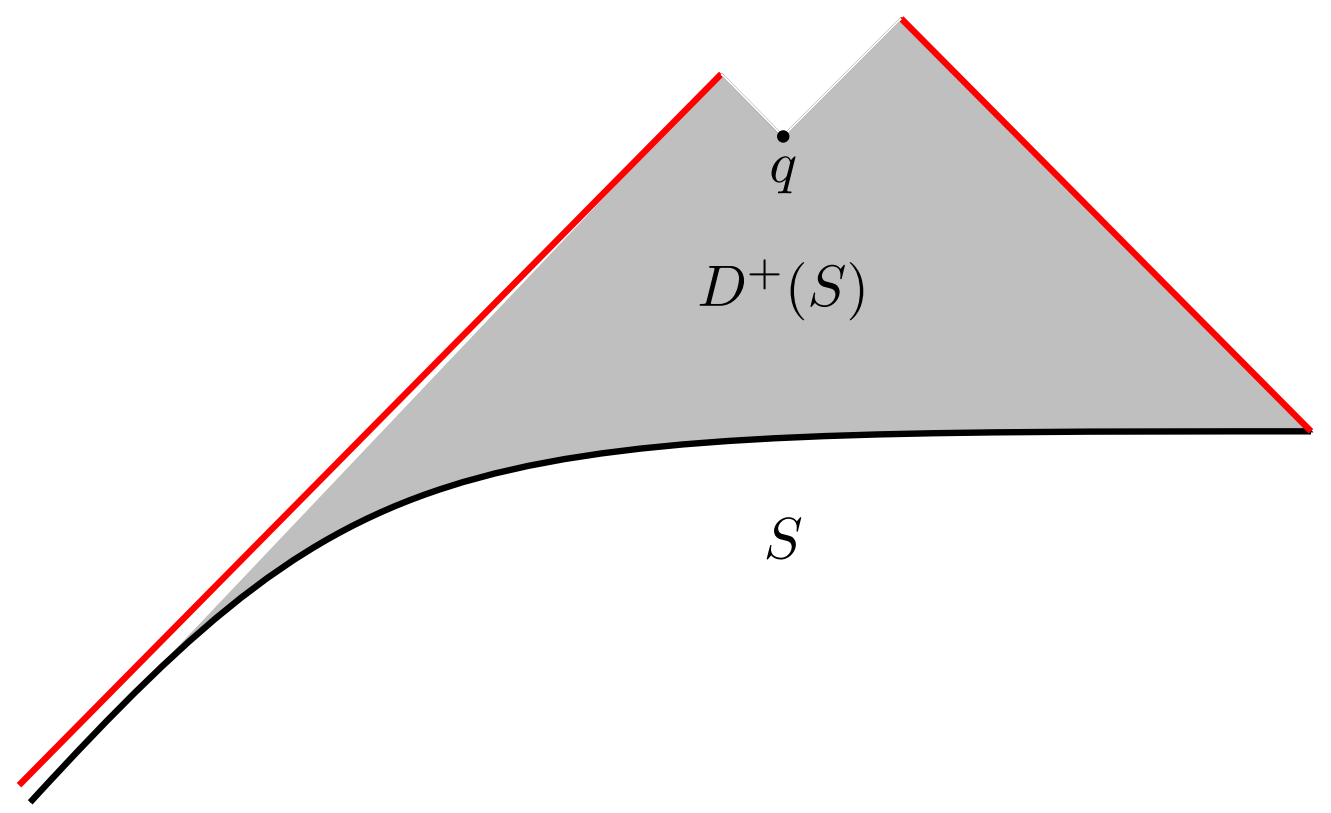
\begin{figure}[htbp]
\begin{center}
\begin{eqnarray*}
\begin{tikzpicture}[scale=0.39]
  % Coordinates
  \coordinate (Q) at (-5.94, -6.36);
  \coordinate (R) at (15.78, -0.07);
  \coordinate (S) at (-0.17, -0.07);
  \coordinate (T) at (1.33, -0.07);
  \coordinate (U) at (8.83, 6.93);
  \coordinate (V) at (6.83, 4.93);
  \coordinate (W) at (5.78, 5.99);
  \coordinate (X) at (-6.13, -6.07);

  % Drawing
  \draw [fill = lightgray] (Q) .. controls (-0.17, -0.07) and (1.33, -0.07) .. (R) -- (U) -- (V) -- (W);
  \draw[very thick] (Q) .. controls (-0.17, -0.07) and (1.33, -0.07) .. (R);
  \draw[red, very thick] (R) -- (U);
  \draw[white] (U) -- (V);
  \draw[red, very thick] (W) -- (X);
  \draw[white] (W) -- (V);
  \fill[black] (V) circle (3pt);
   \node[below] at (V) {$q$};
    \node[below] at (6.83, 3.20) {$D^+(\S)$};
     \node[below] at (6.83, -1.20) {$\S$};
\end{tikzpicture}
\end{eqnarray*}
\end{center}
%\begin{figure}[htbp]
\caption{\small{The figure depicts the domain of dependence $D^+(\S)$ of the (hyper)surface $\S$, where the point $q$ has been removed from the spacetime. $D^+(\S)$ is not a closed set.   }}
\centering
\label{dpluss}
\end{figure}

Another key subset related to the global causal structure of a spacetime $\M$  is the domain of communications. This is particularly valuable in the initial value problem of general relativity but this topic will not be explored in these notes.

\begin{definition}
Let $\S$ be a closed achronal set of $\M$ possibly with an edge. The \emph{ future domain of dependence} $D^+(\S)$ of $\S$ is the subset of $\M$ with the property
\be
D^+(\S)\equiv \{x\in \M\vert \text {every past inextendible  causal curve  through}~~x~~\text {intersects}~~\S\}~.
\ee
The definition of the \emph{past} domain of dependence  $D^-(\S)$ can be given in a similar way.  Furthermore, $D(\S)=D^+(\S)\cup D^-(\S)$.
\hfill{$\triangle$ }
\end{definition}

In what follows, it is also useful to define the edge of a set. In particular, we have the following:
\begin{definition}
Let $\S$ be a closed achronal  subset of $\M$. The \emph{ edge} of $\M$ is the collection of points $p\in \S$ such that for every open neighborhood $U$ of $p$, there is a point $x\in I^+(p)$,  a point $y\in I^-(p)$ and a time-like curve $\lambda$ from $y$ to $x$ contained in $U$ that  does not intersect $\S$.\hfill{$\triangle$ }
\end{definition}

\begin{remark}\hfill

\begin{enumerate}

\item Most of the results we shall describe below on the properties of the domain of dependence  are proven in the  paper \cite{Geroch} by Geroch.  The same applies for the description of globally hyperbolic manifolds presented in the next section.

\item From the definition follows that $\S\subseteq D^+(\S)\subseteq J^+(\S)$.

\item There are achronal subsets $\S$ of $\M$ that have a trivial domain of dependence, i.e. $D(\S)=\S$. To give an example consider the achronal space-like subspace $I=\{(0,x,0,0)\in \bR^4\,|\, x\in [0,1]\}$ of 4-dimensional Minkowski spacetime. Clearly $D(I)=I$.

\item The domain of dependence can be defined for subspaces of $\M$ that are either achronal or closed. But the current definition stated here will suffice.\hfill{$\Diamond$ }
\end{enumerate}

\end{remark}

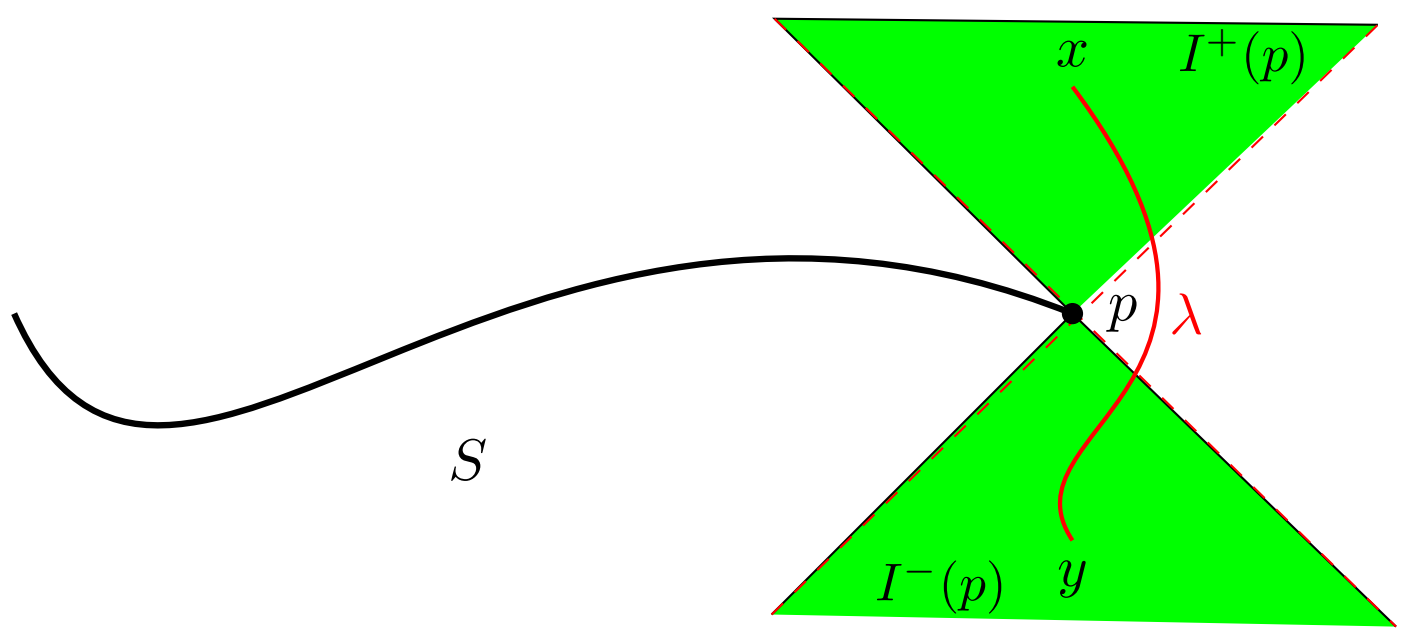
\begin{figure}[htbp]
\begin{center}
\begin{eqnarray*}
\begin{tikzpicture}[scale=1]
  % Coordinates
  \coordinate (A) at (0.00, 0.00);
  \coordinate (B) at (7.00, 0.00);
  \coordinate (C) at (1.00, -2.25);
  \coordinate (D) at (3.25, 1.50);
  \coordinate (E) at (5.03, 1.95);
  \coordinate (F) at (9.14, -2.07);
  \coordinate (G) at (5.01, -1.99);
  \coordinate (H) at (9.02, 1.91);
  \coordinate (I) at (7.00, 1.50);
  \coordinate (J) at (7.00, -1.50);
  \coordinate (K) at (8.50, -0.50);
  \coordinate (L) at (6.50, -0.75);

  % Drawing
  \draw [fill = green] (B)-- (E) -- (H);
   \draw [fill = green] (G)-- (B) -- (F);
  \draw[very thick] (A) .. controls (1.00, -2.25) and (3.25, 1.50) .. (B);
  \draw[red, dashed] (E) -- (F);
  \draw[red, dashed] (G) -- (H);
  \draw[red, thick] (I) .. controls (8.50, -0.50) and (6.50, -0.75) .. (J);
   \fill[black] (B) circle (2pt);
   \node[below] at (J) {$y$};
   \node[above] at (I) {$x$};
    \node[below] at (3, -0.70) {$\S$};
     \node[left, red] at (8.00, 0.00) {$\lambda$};
      \node[right] at (7.10, 0.00) {$p$};
      \node[left, font=\small] at (6.70, -1.80) {$I^-(p)$};
       \node[left, font=\small] at (8.70, 1.70) {$I^+(p)$};
\end{tikzpicture}
\end{eqnarray*}
\end{center}
%\begin{figure}[htbp]
\caption{\small{The figure illustrates the definition of edge of the set $\S$. The point $p$ is at the edge of $\S$ as there is a time-like curve $\lambda$ from a point $y\in I^-(p)$ to a point $x\in I^+(p)$ with $\lambda$ not intersecting $\S$. }}
\centering
\label{fig:edge}
\end{figure}

\begin{prop}
     Let $\S$ be an achronal subset of the spacetime $\M$.  Then $D^+(\S)\cap I^-(\S)=\emptyset$.
     \end{prop}
     \begin{proof}
     To see this suppose $D^+(\S)\cap I^-(\S)\not=\emptyset$ and consider $p \in D^+(\S)\cap I^-(\S)$. If in addition $p\in \S\subset D^+(S)$, it means that there is past directed time-like curve $\gamma$ that joins a point $p$ to another point $q\in \S$.  This violates the achronicity of $\S$. On the other hand if $p\in D^+(\S)\cap I^-(\S)-\S$, as $p\in I^-(\S)$  there is a past directed time-like curve $\gamma$ from a point $q\in \S$ to $p$. In addition as $p$ also a point in $D^+(\S)$, there is causal curve $\lambda$ from $p$ to a point $x\in \S$.  Joining $\gamma$ and $\lambda$, one can construct a causal curve from $q$ to $x$ that is not a null geodesics. It follows from proposition \ref{prop:causaltime} that there is a time-like curve that joins $q$ and $x$ violating the achronicity of $\S$.
\end{proof}

The domain of dependence of a closed achronal set $\S$ has  properties similar to some of those described in proposition \ref{prop:chronos} of chronological and causal future sets.

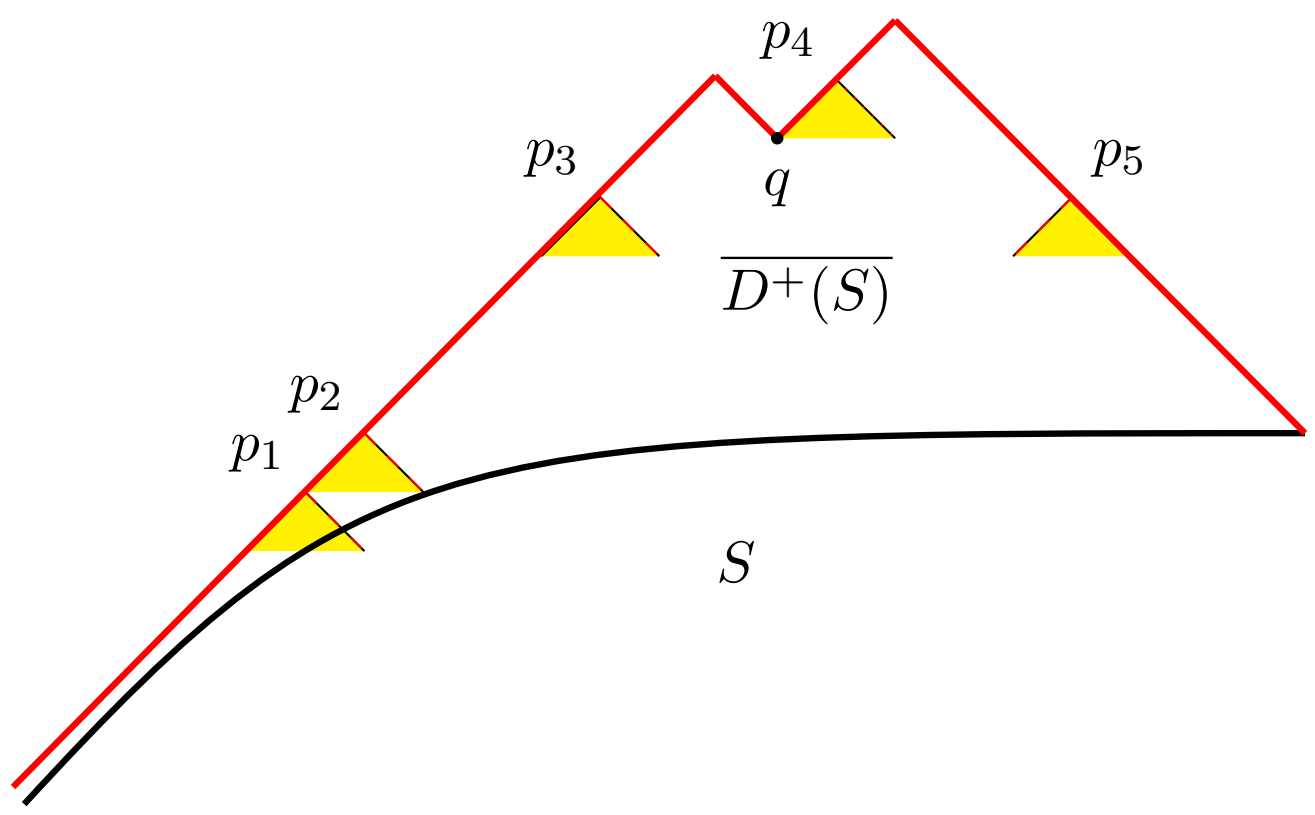
\begin{figure}[htbp]
\begin{center}
\begin{eqnarray*}
\begin{tikzpicture}[scale=0.39]
  % Coordinates
  \coordinate (A) at (-5.77, -6.29);
  \coordinate (B) at (15.95, -0.00);
  \coordinate (C) at (0.00, 0.00);
  \coordinate (D) at (1.50, 0.00);
  \coordinate (E) at (9.00, 7.00);
  \coordinate (F) at (7.00, 5.00);
  \coordinate (G) at (5.95, 6.06);
  \coordinate (A_1) at (-5.96, -6.00);
  \coordinate (H) at (1.00, -1.00);
  \coordinate (I) at (-1.00, -1.00);
  \coordinate (J) at (0.00, -2.00);
  \coordinate (K) at (8.00, 6.00);
  \coordinate (L) at (9.00, 5.00);
  \coordinate (M) at (4.00, 4.00);
  \coordinate (N) at (5.00, 3.00);
  \coordinate (O) at (12.00, 4.00);
  \coordinate (P) at (11.00, 3.00);

  % Drawing
  \draw [fill = yellow] (-2.00, -2.00)-- (I) -- (J);
  \draw [fill = yellow] (-1.00, -1.00)-- (C) -- (H);
  \draw [fill = yellow] (3.00, 3.00)-- (M) -- (N);
   \draw [fill = yellow] (7.00, 5.00)-- (K) -- (L);
    \draw [fill = yellow]  (P) -- (O) -- (13.00, 3.00);
  \draw[very thick] (A) .. controls (0.00, 0.00) and (1.50, 0.00) .. (B);
  \draw[red, very thick] (B) -- (E);
  \draw[red, very thick] (E) -- (F);
  \draw[red, very thick] (G) -- (A_1);
  \draw[red, very thick] (G) -- (F);
  \draw[red, dashed] (C) -- (H);
  \draw[red, dashed] (I) -- (J);
  \draw[red, dotted] (K) -- (L);
  \draw[red, dashed] (M) -- (N);
  \draw[red, dashed] (O) -- (P);
  \fill[black] (F) circle (3pt);
   \node[above left] at (I) {$p_1$};
     \node[above left] at (C) {$p_2$};
       \node[above left] at (M) {$p_3$};
         \node[above left] at (K) {$p_4$};
           \node[above right] at (O) {$p_5$};
    \node[above] at (7.50, 1.50) {$\overline{D^+(\S)}$};
     \node[above] at (7.00, 3.50) {$q$};
       \node[below] at (6.30, -1.50) {$\S$};
\end{tikzpicture}
\end{eqnarray*}
\end{center}
%\begin{figure}[htbp]
\caption{\small{The figure depicts the result of proposition \ref{prop:dopen} that any past inextendible time-like curve from a point in $\overline{D^+(\S)}$ intersects $\S$. Notice that all the past time-like directions of the points $p_1, \dots, p_5$ point towards the interior of $D^+(\S)$.  This statement is  valid even if  $q$  is not  a point in the spacetime. It also applies to points on the boundary of $\overline{D^+(\S)}$ that lie on past inextendible null geodesics, like $p_1$, i.e parts of the boundary that do not end on $\S$ and instead they are asymptotic to it.  }}
\centering
\label{boundarydomain}
\end{figure}

\begin{prop}\label{prop:dopen}
A point $p$ is in $\overline{D^+(\S)}$, i.e. $p\in \overline{D^+(\S)}$,  iff every past inextendible time-like curve from $p$ intersect $\S$.
\end{prop}
\begin{proof}
$\Longrightarrow$ Suppose that  $p\in \overline{D^+(\S)}$ and there is a past inextendible time-like curve $\gamma$ from $p$ that does not intersect $\S$. If this is the case, consider $q\in \gamma\cap I^-(p)$.  Clearly, $p$ is in the open set $I^+(q)$,  $p\in I^+(q)$.  From every point  $x\in I^+(q)$, there is a past inextendible time-like curve that does not intersect $\S$ -- just join the time-like curve from $x$ to $q$ with the rest of $\gamma$. As $I^+(q)$ is open and $p\in \overline{D^+(\S)}$, $I^+(q)\cap D^+(\S)\not=\emptyset$.  But this is a contradiction as all causal curves from points in $D^+(\S)$ intersect $\S$.

$\Longleftarrow$ Let $p\in \M$ such that all past directed time-like inextendible curves from $p$ intersect $\S$. In such a case, either $p\in \S\subset D^+(\S)$ or $p\in I^+(\S)$.  In the former case, there is nothing to prove as $\S\subset D^+(\S)\subset \overline{D^+(\S)}$. In the latter case, as $p\in \M-\overline{D^+(\S)}$ and $\M-\overline{D^+(\S)}$ open, because $\overline{D^+(\S)}$ is closed, there is an open neighborhood $U$ of $p$ with $U\cap \overline{D^+(\S)}=\emptyset$.  Consider a point $q\in I^-(p)\cap U\subset \M-\overline{D^+(\S)}$. As $q\notin \overline{D^+(\S)}$, there must exist a causal past directed curve $\lambda$ from $q$ that does not intersect $\S$.  There are two possibilities. Either (i) $\lambda$ remains in $I^+(\S)$ or (ii) $\lambda$ intersects $\partial I^+(\S)$.  In the former case, it follows from proposition \ref{prop:causaltime} that there is a past directed inextendible time-like curve $\gamma$ from $p$ with $\gamma\in I^+(\lambda)$.  This curve does not intersect $\S$ because $I^+(\lambda)\cap D^+(\S)=\emptyset$. Thus $\gamma$ is a past directed time-like inextendible curve that does not intersect $\S$, which is a contradiction to the hypothesis that all past directed time-like curves from $p$ must intersect $\S$. Thus, $p\in \overline{D^+(\S)}$.

Parenthetically to see that $I^+(\lambda)\cap D^+(\S)=\emptyset$, suppose that there is an $x\in I^+(\lambda)\cap D^+(\M)$.  This would imply that there is a past directed time-like curve $\xi$ from $x$ to a point $y\in \lambda$. Joining $\xi$ with $\lambda$ starting at $y$, one can construct a past directed inextendible causal curve from $x$ that does not intersect $\S$.  This is a contradiction  as $x\in D^+(\S)$.

Alternatively, the causal curve $\lambda$ from $q$ intersects $\partial I^+(\S)$ at a point $r$.  Repeating the same argument as above using proposition \ref{prop:causaltime}, there is a past directed time-like curve $\gamma$ from $p$ to $r$ that does not intersect $\S$ -- the argument is the same as for the curve $\gamma$ above. This can be extended beyond $r$ to a past directed inextendible time-like curve.  Furthermore, $\overline{D^+(\S)}\subset \overline{I^+(\S)}$ and so the whole curve $\gamma$ from $p$ including the extension from $r$ does not intersect $\S$. This is again a contradiction as all past directed time-like curves from all the points of $\overline{D^+(\S)}$  must intersect $\S$.
\end{proof}

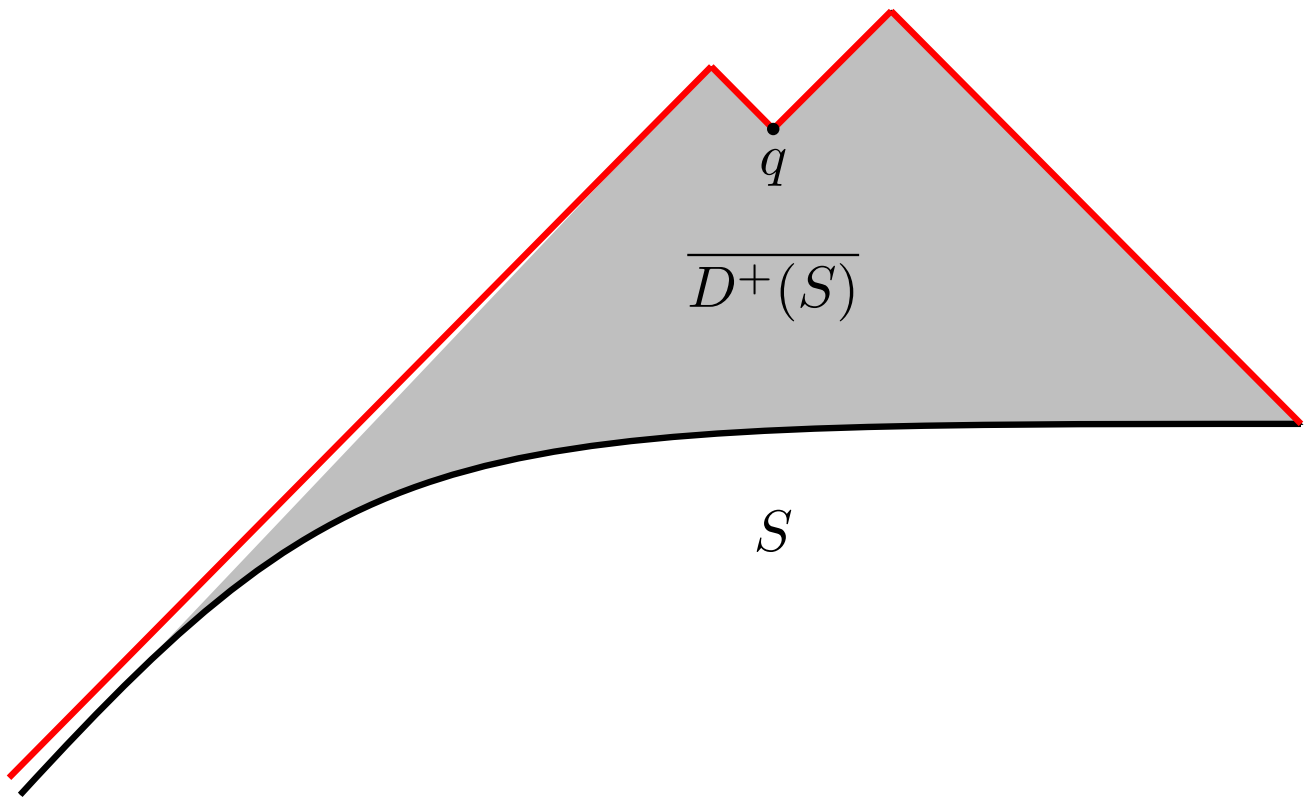
\begin{figure}[htbp]
\begin{center}
\begin{eqnarray*}
\begin{tikzpicture}[scale=0.39]
  % Coordinates
  \coordinate (Q) at (-5.94, -6.36);
  \coordinate (R) at (15.78, -0.07);
  \coordinate (S) at (-0.17, -0.07);
  \coordinate (T) at (1.33, -0.07);
  \coordinate (U) at (8.83, 6.93);
  \coordinate (V) at (6.83, 4.93);
  \coordinate (W) at (5.78, 5.99);
  \coordinate (X) at (-6.13, -6.07);

  % Drawing
  \draw [fill = lightgray] (Q) .. controls (-0.17, -0.07) and (1.33, -0.07) .. (R) -- (U) -- (V) -- (W);
  \draw[very thick] (Q) .. controls (-0.17, -0.07) and (1.33, -0.07) .. (R);
  \draw[red, very thick] (R) -- (U);
  \draw[red, very thick] (U) -- (V);
  \draw[red, very thick] (W) -- (X);
  \draw[red, very thick] (W) -- (V);
  \fill[black] (V) circle (3pt);
   \node[below] at (V) {$q$};
    \node[below] at (6.83, 3.20) {$\overline{D^+(\S)}$};
     \node[below] at (6.83, -1.20) {$\S$};
\end{tikzpicture}
\end{eqnarray*}
\end{center}
%\begin{figure}[htbp]
\caption{\small{The  figure  depicts the closure $\overline{D^+(\S)}$ of the domain of dependence $D^+(\S)$ of figure \ref{dpluss}. Notice that $\overline{D^+(\S)}$ is not compact.  The boundary of $\overline{D^+(\S)}$ consists of $\S$ and the lines coloured in red. The point $q$ is not in the spacetime and so it is not in the boundary either.}}
\centering
\label{dplussclosure}
\end{figure}

\begin{lemma}
The interior of $D^+(\S)$, $\text{int}\, D^+(\S)\equiv \overline{D^+(\S)}- \partial \overline{D^+(\S)}$, is $\text{int}\, D^+(\S)= I^-(D^+(\S))\cap I^+(\S)$. Similarly,
$\text{int}\, D(\S)=I^-(D^+(\S))\cap I^+(D^-(\S))$.
\end{lemma}
\begin{proof}
Let $p\in I^-(D^+(\S))\cap I^+(\S)$. This implies that there is a past directed time-like curve $\lambda$ from $p$ that intersects $\S$ because $p\in I^+(\S)$.  Moreover, there is a past directed time-like curve $\gamma$ from $p$ to a point $q\in D^+(\S)$. If there were a past directed inextendible time-like curve $\xi$ from $p$ that does not intersect $\S$, it would imply that the past directed time-like curve obtained by joining $\gamma$ and $\xi$ from $q$ will not intersect $\S$. But this is a contradiction as all past directed inextendible time-like curves from $q\in D^+(\S)$ intersect $\S$. Thus, all past directed time-like curves from $p$ must intersect $\S$ as well and so as a result of proposition \ref{prop:dopen}, $p\in \overline{D^+(\S)}$ .  Therefore, $I^-(D^+(\S))\cap I^+(\S)\subset \overline{D^+(\S)}$.

On the other hand if $p\in \text{int}\, D^+(\S)$, clearly $p\in I^+(\S)$.  Moreover as $\text{int}\, D^+(\S)$ is open, there is an open neighbourhood $U\subset \text{int}\, D^+(\S)$ of $p$ and, as a result of proposition \ref{prop:lcs}, a point $q\in U$ with $p\in I^-(q)$. This in turn implies that $p\in I^-(D^+(\S))$, i.e.
$p\in  I^-(D^+(\S))\cap I^+(\S)$. Thus, $\text{int}\, D^+(\S)\subseteq I^-(D^+(\S))\cap I^+(\S)$. As $\text{int}\, D^+(\S)$ is the largest open set contained in $\overline{D^+(\S)}$, we conclude that $\text{int}\, D^+(\S)= I^-(D^+(\S))\cap I^+(\S)$.
\end{proof}

\begin{definition}
Let $S$ be a closed achronal set. The future Cauchy horizon of $\S$, $H^+(\S)$, is
\be
H^+(\S)\equiv \overline{D^+(\S)}-I^-(D^+(\S))=\overline{D^+(\S)}\cap (\M-I^-(D^+(\S))).
\ee
The past Cauchy horizon of $\S$, $H^-(\S)$, is defined in a similar way. Also the total Cauchy horizon of $\S$ is $H(\S)=H^+(\S)\cup H^-(\S)$.\hfill{$\triangle$}
\end{definition}

\begin{remark}
The future Cauchy horizon of $\S$ is achronal. Indeed
\be
I^-(H^+(\S))\subset I^-(\overline{D^+(\S)})=I^-({D^+(\S)})\subset \M-H^+(\S)~.
\ee
Therefore, $I^-(H^+(\S))\cap H^+(\S)=\emptyset$ that establishes the result.\hfill{$\Diamond$}
\end{remark}

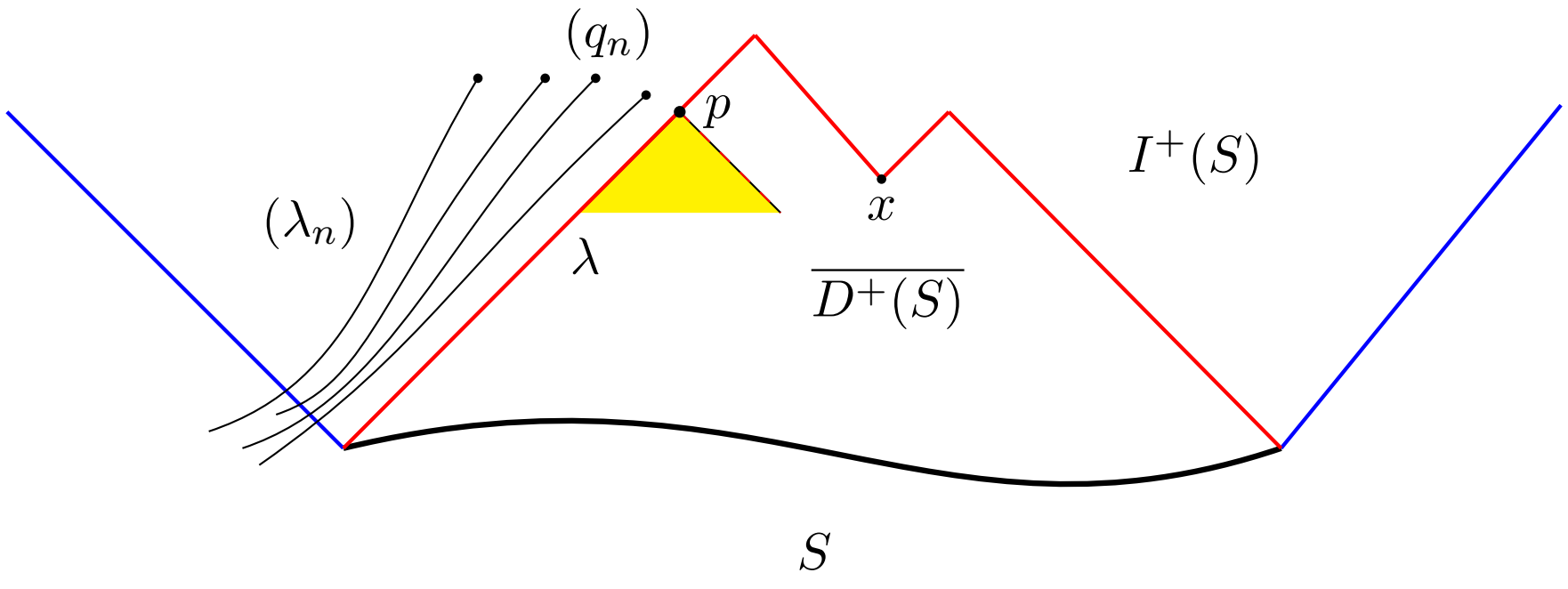
\begin{figure}[htbp]
\begin{center}
\begin{eqnarray*}
\begin{tikzpicture}[scale=0.5]
  % Coordinates
  \coordinate (A) at (0.00, 0.00);
  \coordinate (B) at (13.94, 0.00);
  \coordinate (C) at (6.25, 1.50);
  \coordinate (D) at (8.75, -1.75);
  \coordinate (E) at (-5.00, 5.00);
  \coordinate (F) at (18.10, 5.10);
  \coordinate (G) at (6.12, 6.14);
  \coordinate (H) at (9.00, 5.00);
  \coordinate (I) at (8.00, 4.00);
  \coordinate (J) at (5.00, 5.00);
  \coordinate (K) at (4.50, 5.25);
  \coordinate (L) at (3.75, 5.50);
  \coordinate (M) at (3.00, 5.50);
  \coordinate (N) at (2.00, 5.50);
  \coordinate (O) at (-1.25, -0.25);
  \coordinate (P) at (1.75, 2.75);
  \coordinate (Q) at (1.25, 1.50);
  \coordinate (R) at (-1.50, 0.00);
  \coordinate (S) at (1.50, 3.25);
  \coordinate (T) at (0.75, 0.75);
  \coordinate (U) at (-1.00, 0.50);
  \coordinate (V) at (0.50, 2.50);
  \coordinate (W) at (0.50, 1.00);
  %\coordinate (X) at (NaN, NaN);
 % \coordinate (Y) at (NaN, NaN);
  \coordinate (Z) at (-2.00, 0.25);
  \coordinate (AA) at (0.50, 3.00);
  \coordinate (AB) at (0.25, 1.00);
  \coordinate (AC) at (6.50, 3.50);

  % Drawing
    \draw [fill = yellow] (3.50, 3.50)-- (J) -- (AC);
  \draw[very thick] (A) .. controls (6.25, 1.50) and (8.75, -1.75) .. (B);
  \draw[blue, thick] (A) -- (E);
  \draw[blue, thick] (B) -- (F);
  \draw[red, thick] (A) -- (G);
  \draw[red, thick] (B) -- (H);
  \draw[red, thick] (G) -- (I);
  \draw[red, thick] (I) -- (H);
  \draw (K) .. controls (1.75, 2.75) and (1.25, 1.50) .. (O);
  \draw (L) .. controls (1.50, 3.25) and (0.75, 0.75) .. (R);
  \draw (M) .. controls (0.50, 2.50) and (0.50, 1.00) .. (U);
 % \draw (N) .. controls (0.50, 2.50) and (0.50, 1.00) .. (N);
  \draw (N) .. controls (0.50, 3.00) and (0.25, 1.00) .. (Z);
  \draw[dashed, red] (J) -- (AC);
  \fill[black] (I) circle (2pt);
  \fill[black] (J) circle (2.5pt);
  \fill[black] (K) circle (2pt);
  \fill[black] (L) circle (2pt);
  \fill[black] (M) circle (2pt);
  \fill[black] (N) circle (2pt);
  \node[right] at (5.10, 5.00) {$p$};
  \node[below] at (I) {$x$};
   \node[below] at (7,-1) {$\S$};
    \node[above right] at (M) {$(q_n)$};
     \node[below right] at (3.10, 3.40) {$\lambda$};
 \node[below left] at (0.50, 4.00) {$(\lambda_n)$};
 \node[below left] at (13.94, 5.00) {$I^+(\S)$};
 \node[above left] at (9.50, 1.50) {$\overline{D^+(\S)}$};
 (13.94, 0.00)
\end{tikzpicture}
\end{eqnarray*}
\end{center}
%\begin{figure}[htbp]
\caption{\small{The figure depicts the first part of the proof of theorem \ref{th:cauchyhorizons}. The point $x$ is not in the spacetime. The yellow shaded region that depicts $C^-_p$ is a reminder that all past directed time-like curves from $p$ enter the interior of $D^+(\S)$, which is critical to establish that the limit curve $\lambda$ lies in the boundary of $D^+(\S)$. }}
\centering
\label{xx}
\end{figure}

\begin{theorem}\label{th:cauchyhorizons}
Every point $p\in H^+(\S)$ lies on a null geodesic $\lambda$ entirely contained in $H^+(\S)$, which is either past inextendible or has a past end point on the edge of $\S$.
\end{theorem}
\begin{proof}
Let $p\in H^+(\S)$ with $p\notin \text{edge} \S$. Then either (i) $p\in I^+(\S)$ or (ii) $p\in \S$ with $p\not= \text{edge} \S$. The assertion (i) holds because all the past directed inextendible time-like curves from points of $\overline{D^+(\S)}$ intersect $\S$ and $H^+(\S)\subset \overline{D^+(\S)}$, thus $p\in I^+(\S)$.

Consider first (i), i.e. $p\in I^+(\S)$. As $p\notin I^-(D^+(\S))=I^-(\overline{D^+(\S)})$, from every $q\in I^+(p)$, there is a past directed inextendible causal curve $\lambda$ that does not intersect\footnote{If all past inextendible causal curves from $q$ intersected $\S$, it would have implied that $q\in D^+(\S)$. But $q\notin I^-(D^+(\S))$ as $q\in I^+(p)$ and   $p\notin I^-(D^+(\S))$. To see this, if there were an $x\in I^-(D^+(\S))$ with $q\in I^-(x)$, it would have implied that $p\in I^-(x)$ and so $p\in I^-(D^+(\S))$, which is a contradiction.   Thus the remaining option is that $q\in H^+(\S)$. But this is also a  contradiction as $H^+(\S)$ is achronal.}  $\S$. Using this, consider such a sequence of points $(q_n)$ that converge to $p$ and the sequence of past directed inextendible causal curves $(\lambda_n)$ with $\lambda_n$ passing through $q_n$ and does not intersect $\S$. As $p$ is an accumulation point of the sequence  of $(\lambda_n)$, there is a past directed inextendible causal curve $\lambda$ that passes through $p$.  $\lambda$ cannot enter $I^+(\S)\cap I^-(D(\S))$ because otherwise an infinite number of curves of the sequence $(\lambda_n)$ would have entered $I^+(\S)\cap I^-(D(\S))$. But this cannot be the case as if $\lambda_n$ enters $I^+(\S)\cap I^-(D(\S))$, it has to intersect $\S$, which is a contradiction. Thus, the limit curve $\lambda$ cannot enter $I^+(\S)\cap I^-(D(\S))$.

Moreover,  $I^-(p)\subset I^-(D^+(\S))$. Thus,  $\lambda$ is a causal curve passing through $p$ that cannot enter $I^-(p)$. It follows from proposition \ref{pro:nullgeo} that $\lambda$ must be a null geodesic.

Furthermore, $\lambda\cap I^+(\S)\subset H^+(\S)$. Indeed, suppose that there is a point $q\in \lambda\cap I^+(\S)$ such that a past directed inextendible time-like curve does not intersect $\S$.  If this is the case, we can construct a past directed inextendible time-like curve from $p$ with the same property. Proposition \ref{prop:dopen} would then imply that $p\notin \overline{D^+(\S)}$, which is a contradiction. Thus, $\lambda\cap I^+(\S)\subset  \overline{D^+(\S)}$. As we have already demonstrated that $\lambda\cap I^+(\S)\cap I^-(D^+(\S))=\emptyset$, it follows that $\lambda\cap I^+(\S)\subset H^+(\S)$.

(ii) Next let $p\in \S$ with $p\not= \text{edge} \S$. Since $p\notin \text{edge} \S$, there is an open neighbourhood $U$ of $p$ such that all the causal curves connecting points in $I^+(p)\cap U$ with points in $I^-(p)\cap U$ must intersect $\S$. The argument that proves the existence of a past directed null geodesic $\lambda$ that passes through $p$ entirely contained in $H^+(\S)$ proceed as in case (i).

Finally, suppose the null geodesic $\lambda$ has an end point $r$ in $H^+(\S)$. As $H^+(\S)$ is closed $r\in H^+(\S)$. If $r$ is not in $\text{edge}\S$, $\lambda$ could be extended beyond $r$ to a null geodesics $\lambda'$. There cannot be a corner at $r$ because otherwise there will be a time-like geodesic between points in  $\lambda$ and $\lambda'$. But this cannot be the case because $H^+(\S)$ is achronal. So $\lambda'$ must be a continuation of $\lambda$. Thus, $\lambda$ cannot have an end point unless it is in $\text{edge}\S$.
\end{proof}

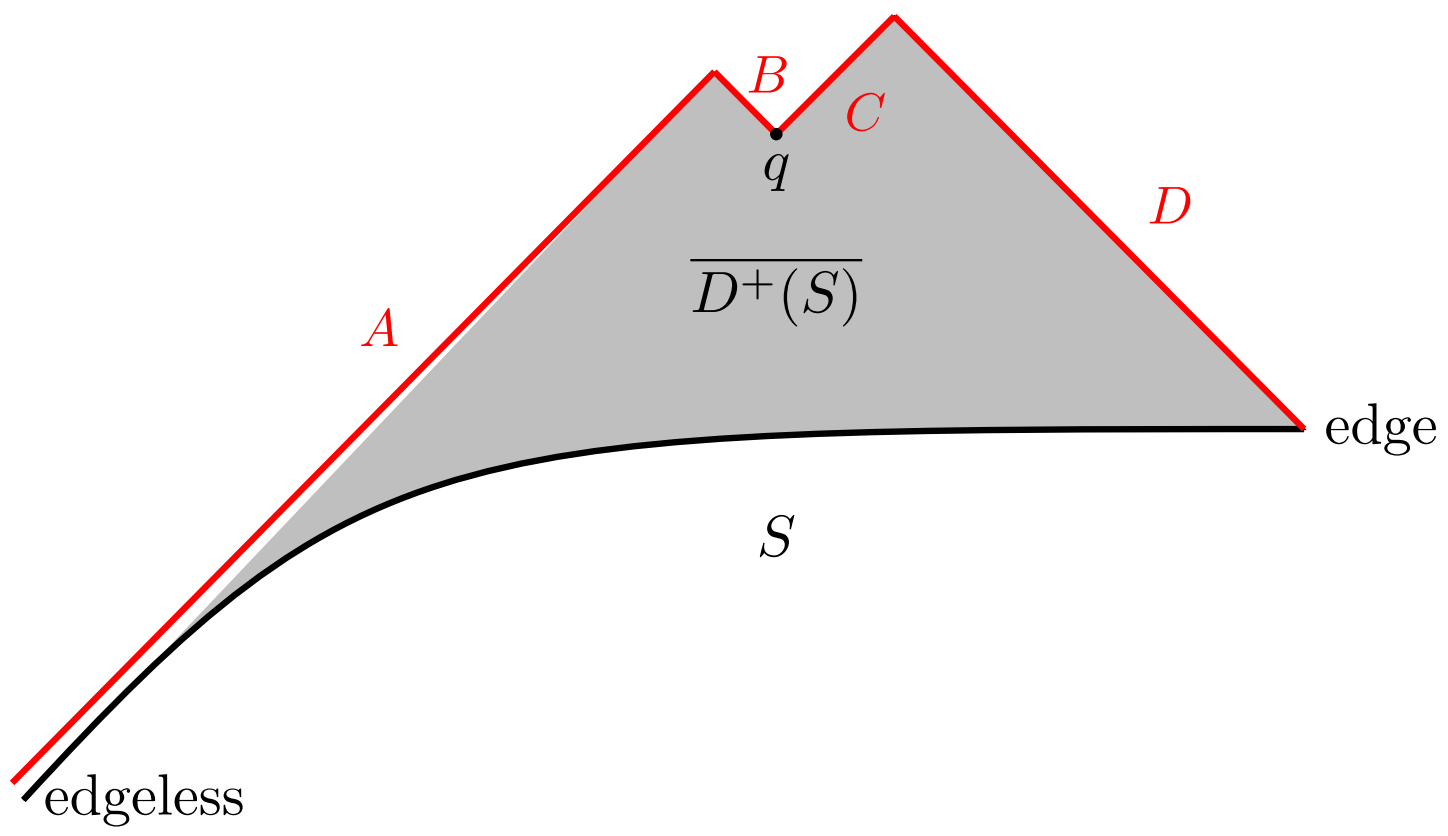
\begin{figure}[htbp]
\begin{center}
\begin{eqnarray*}
\begin{tikzpicture}[scale=0.39]
  % Coordinates
  \coordinate (Q) at (-5.94, -6.36);
  \coordinate (R) at (15.78, -0.07);
  \coordinate (S) at (-0.17, -0.07);
  \coordinate (T) at (1.33, -0.07);
  \coordinate (U) at (8.83, 6.93);
  \coordinate (V) at (6.83, 4.93);
  \coordinate (W) at (5.78, 5.99);
  \coordinate (X) at (-6.13, -6.07);

  % Drawing
  \draw [fill = lightgray] (Q) .. controls (-0.17, -0.07) and (1.33, -0.07) .. (R) -- (U) -- (V) -- (W);
  \draw[very thick] (Q) .. controls (-0.17, -0.07) and (1.33, -0.07) .. (R);
  \draw[red, very thick] (R) -- (U);
  \draw[red, very thick] (U) -- (V);
  \draw[red, very thick] (W) -- (X);
  \draw[red, very thick] (W) -- (V);
  \fill[black] (V) circle (3pt);
   \node[below] at (V) {$q$};
    \node[below] at (6.83, 3.20) {$\overline{D^+(\S)}$};
     \node[below] at (6.83, -1.20) {$\S$};
     \node[right] at (R) {$\mathrm{edge}$};
    \node[right] at (Q) {$\mathrm{edgeless}$};  %\coordinate (Q) at (-5.94, -6.36);
    \node[above left, red, font=\small] at (0.78, 0.99) {$A$};
     \node[above left, red, font=\small ] at (7.40 , 5.29) {$B$};
     \node[below right, red, font=\small ] at (7.64 , 5.96) {$C$};
      \node[above right, red, font=\small ] at (12.78, 3.07) {$D$};

\end{tikzpicture}
\end{eqnarray*}
\end{center}
%\begin{figure}[htbp]
\caption{\small{The  figure is a depiction of the results of theorem \ref{th:cauchyhorizons}.   The Cauchy horizon  $H^+(\S)$  of the domain of dependence $D^+(\S)$ in  figure \ref{dpluss},  with closure  as in figure \ref{dplussclosure},  is the union $H^+(\S)=A\cup B\cup C\cup D$. The points in $A$,$B$ and $C$ lie on  past inextendible null geodesics. While the points in $D$ lie on null geodesics with end point at the edge of $\S$.  }}
\centering
\label{dplussclosure}
\end{figure}

\begin{figure}[htbp]
\begin{center}
\begin{eqnarray*}
\begin{tikzpicture}[scale=0.39]
  % Coordinates
  \coordinate (Q) at (-5.94, -6.36);
  \coordinate (R) at (15.78, -0.07);
  \coordinate (S) at (-0.17, -0.07);
  \coordinate (T) at (1.33, -0.07);
  \coordinate (U) at (8.83, 6.93);
  \coordinate (V) at (6.83, 4.93);
  \coordinate (W) at (5.78, 5.99);
  \coordinate (X) at (-6.13, -6.07);

  % Drawing
  \draw [fill = lightgray] (Q) .. controls (-0.17, -0.07) and (1.33, -0.07) .. (R) -- (U) -- (V) -- (W);
  \draw[very thick] (Q) .. controls (-0.17, -0.07) and (1.33, -0.07) .. (R);
  \draw[red, very thick] (R) -- (U);
  \draw[red, very thick] (U) -- (V);
  \draw[red, very thick] (W) -- (Q);
  \draw[red, very thick] (W) -- (V);
  \fill[black] (V) circle (3pt);
   \node[below] at (V) {$q$};
    \node[below] at (6.83, 3.20) {$\overline{D^+(\S)}$};
     \node[below] at (6.83, -1.20) {$\S$};
     \node[right] at (R) {$\mathrm{edge}$};
    \node[right] at (Q) {$\mathrm{edgeless}$};  %\coordinate (Q) at (-5.94, -6.36);
    \node[above left, red, font=\small] at (0.78, 0.99) {$A$};
     \node[above left, red, font=\small ] at (7.40 , 5.29) {$B$};
     \node[below right, red, font=\small ] at (7.64 , 5.96) {$C$};
      \node[above right, red, font=\small ] at (12.78, 3.07) {$D$};

\end{tikzpicture}
\end{eqnarray*}
\end{center}
%\begin{figure}[htbp]
\caption{\small{The  figure is a depiction of the results of theorem \ref{th:cauchyhorizons}, where $H^+(\S)$ has a non-trivial intersection with $\S$ away from an edge. Observe that the subset  $A$ of the boundary of $H^+(\S)$ intersects with $\S$ -- it is not merely asymptotic to it as in the previous figure.  This case is dealt in the theorem \ref{th:cauchyhorizons}; it is  item (ii) of the proof. Again, the points in $A$, $B$ and $C$ part of the boundary lie in past inextendible null geodesics while those in $D$ lie on null geodesics with past end points at the edge of $\S$.  }}
\centering
\label{dplussclosure2}
\end{figure}
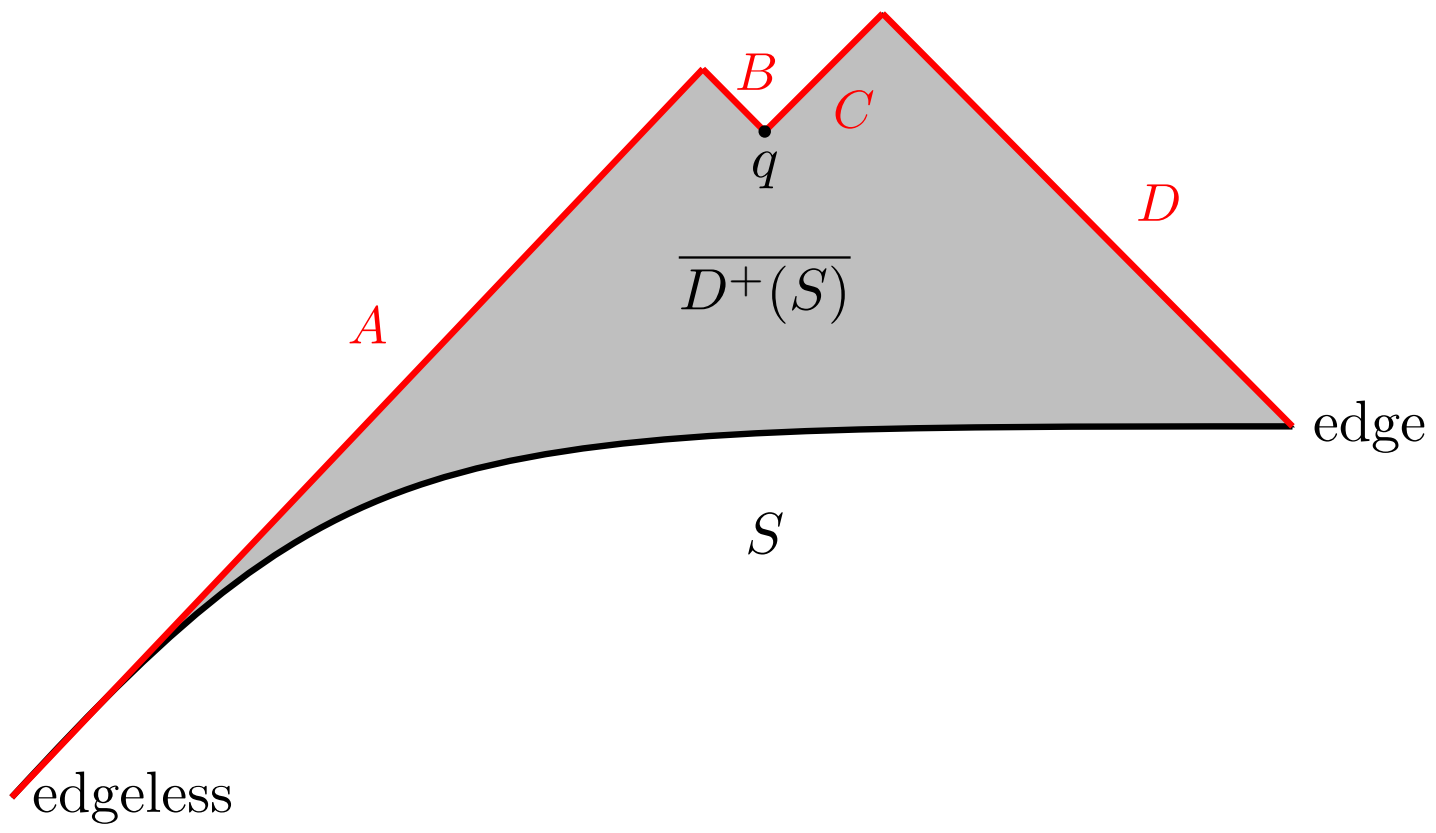

\section{Globally hyperbolic spacetimes}
\subsection{Definition and properties}
\begin{definition}
Let $\M$ be a spacetime. A \emph{ Cauchy surface} $\Sigma$  is a \emph{closed achronal} subset of $\M$ such that $\M=D(\Sigma)$.  A spacetime $\M$ with a Cauchy surface is called \emph{ globally hyperbolic}.\hfill{$\triangle$}
\end{definition}

\begin{remark}
Cauchy surfaces $\Sigma$ do not have edges.  If $\Sigma$ had an edge, there will be time-like curves from points $p$ of the spacetime that do not intersect $\Sigma$ contradicting the condition that $\M=D(\Sigma)$.  Moreover since $\Sigma$ is a closed achronal set without an edge, it is a $C^{1-}$ hyper-surface of $\M$.  The proof of this follows the same steps as those in the proof of proposition \ref{prop:boundary}.\hfill{$\Diamond$}
\end{remark}

\begin{corollary}
If the spacetime $\M$ is connected, then a non empty closed achronal set $\Sigma$ is a Cauchy surface, iff $H(\Sigma)=\emptyset$.
\end{corollary}
\begin{proof}
If $H(\Sigma)=\emptyset$, then $\overline{D(\Sigma)}=\text{int}\, (D(\Sigma))$. Thus $D(\Sigma)$ is both open and close and non-empty as $\Sigma\subset D(\Sigma)$. As $\M$ is connected, $\M=D(\Sigma)$ and $\Sigma$ is a Cauchy surface.
\end{proof}

\begin{prop}\label{prop:cauchy1}
Let $\Sigma$ be a Cauchy surface and let $\gamma$ be an inextendible causal curve. Then $\gamma$ intersects  $\Sigma$, $I^+(\Sigma)$ and $I^-(\Sigma)$.
\end{prop}
\begin{proof}
Recall that by the definition of the Cauchy surface,  $\Sigma$ is closed,  achronal and $\M=D(\Sigma)=D^+(\Sigma)\cup D^-(\Sigma)$. Suppose that $\gamma$ passes through a point $p\in D^+(\Sigma)\cap I^+(\Sigma)$ and assume that it does not enter $I^-(\Sigma)$  -- proof for the curve passing through $p\in D^-(\Sigma)\cap I^-(\Sigma)$ is similar. Then, as $\gamma$ is both future and, especially, past inextendible, it has to intersect $\Sigma$ as all such curves from $p$ must intersect $\Sigma$. From proposition \ref{prop:causaltime}, there is a past directed inextendible time-like curve $\lambda$ with $\lambda\subset I^+(\gamma)$ that passes through a point $q$ in an open neighbourhood of $p$ with $q\in I^+(p)\subset I^+(\Sigma)$. As $\gamma$ does not enter $I^-(\Sigma)$, but intersects both $\Sigma$ and $I^+(\Sigma)$, we have that $\lambda\subset I^+(\gamma)\subset I^+(\Sigma\cup I^+(\Sigma))=I^+(\Sigma)$, see proposition \ref{prop:chronos}.  Moreover, as $\Sigma$ is achronal $\Sigma\cap I^+(\Sigma)=\emptyset$, i.e. $\lambda$ does not intersect $\Sigma$ by extending it into the future. However, from the definition of $D^+(\Sigma)$ no such curve $\lambda$ can exist as all  past directed inextendible  causal curves from points $q\in D^+(\Sigma)$   intersect $\Sigma$. Thus, $\gamma$ must enter $I^-(\Sigma)$.
\end{proof}

\begin{theorem}
If $\Sigma$ is a closed achronal set without an edge and $\M$ connected, then $\Sigma$ is a Cauchy surface iff every inextendible null geodesic  intersects $\Sigma$ and enters both $I^-(\Sigma)$ and $I^+(\Sigma)$.
\end{theorem}
\begin{proof}
$\Longrightarrow$ If $\Sigma$ is a Cauchy surface and since null geodesics are causal curves, the result in this direction follows from that of proposition \ref{prop:cauchy1}.

$\Longleftarrow$ Suppose that $\Sigma$ is not a Cauchy surface with all the null geodesics intersecting $\Sigma$ and entering both $I^+(\Sigma)$ and $I^-(\Sigma)$.  As $\Sigma$ is not a Cauchy surface at least one of  $H^+(\Sigma)$ or $H^-(\Sigma)$ is non-empty, say $H^+(\Sigma)\not=\emptyset$. Since $\S$ has not an edge, theorem \ref{th:cauchyhorizons} implies that a null geodesic in $H^+(\S)$ is past directed and inextedible and never enters $I^-(\Sigma)$. This contradicts the hypothesis that all null geodesics enter $I^-(\Sigma)$. Therefore, $H(\Sigma)=\emptyset$ and $\Sigma$ is a Cauchy surface.
\end{proof}

\begin{lemma}\label{le:ghstrong}
Let $\M$ be a globally hyperbolic spacetime. Then, $\M$ is strongly causal.
\end{lemma}
\begin{proof}
As $\M$ is globally hyperbolic,  $\M=I^+(\Sigma)\cup I^-(\Sigma)\cup \Sigma$, where $\Sigma$ is  closed achronal and without an edge.  Suppose that strong causality is violated at $p\in I^+(\Sigma)$. If this is the case, one can find a convex normal neighbourhood $U$ of $p$ and a sequence $(V_n)$ of open sets $V_n\subset U$  that converge to $p$ such that there are future directed time-like curves $\lambda_n$ that start at $V_n$ leave $U$ and end at $V_n$. None of them enters $I^-(\Sigma)$ as this will violate the achronicity  of $\S$. Clearly $p$ is an accumulation point of the sequence $(\lambda_n)$ and  so there is a causal curve that passes through $p$. Although each $\lambda_n$ has end points, $\lambda$ can either be  inextendible or be a closed casual curve. In the latter case, it can become inextendible by going around infinite times. In either case, $\lambda$ does not enter $I^-(\Sigma)$ which directly contradicts proposition \ref{prop:cauchy1}.

A similar result can be established if $p\in I^-(\Sigma)$. Thus, it remains to consider the case that there is strong causality violation at  $p\in \Sigma$. In this case we choose $U$ a convex normal neighbourhood of $p$ and a sequence $(V_n)$ of open sets $V_n\subset U$  that converge to $p$. Next consider a sequence of time-like curves that start from points $q_n\in I^+(\S)\cap V_n$ leave $U$ and end at $V_n$. Again none of these curves will not enter $I^-(\Sigma)$ as this will violate the achronicity of $\Sigma$.  The limit curve $\lambda$ will not enter $I^-(\Sigma)$ and it be inextendible in $\M-\Sigma$. Thus it will be either periodic or it will have end points in $\Sigma$. This contradicts  either the proposition \ref{prop:cauchy1} or the achronicity of $\Sigma$.
\end{proof}

\begin{remark}
 It can be shown that globally hyperbolic spacetimes $\M$ are stably causal \cite{HE}. The temporal  function $f$ can be chosen such that each slice is a Cauchy surface.  Moreover,  $\M$ is homeomorphic, but not necessarily isometric,  to a product $\M=\bR\times \Sigma$.
\end{remark}

Traditionally in general relativity,  one works with the space of the causal curves  between the points $p$ and $q$, $\mC(p,q)$, that are identified \emph{  up to reparameterisations that preserve their causal structure} (unparameterised curves), see definition \ref{def:reparameterisation} and remark below.   As we shall explain below, there are several topologies that can be put on $\mC(p,q)$. For one such topology,
a sequence of such curves $(\gamma_n)$ in $\mC(p,q)$ converges to a curve $\gamma$, iff each point of $\gamma$ is a convergence point of the sequence $(\gamma_n)$ and the sequence $(\gamma_n)$ does not have other convergence points apart from those of $\gamma$. This topology can be defined more formally by declaring that a basis of  open sets $\mathcal{O}_U$ in $\mC(p,q)$  is
\be
\mathcal{O}_U=\{\gamma\in \mC(p,q)\vert \gamma\subset U, U \text{open set in} \M\}~,
\label{waldtop}
\ee
where the notation $\gamma\subset U$ is synonymous with $\mathrm{Im}\, \gamma\subset U$.
It can be shown that this defines a topology%\footnote{This topology is related to the Vietoris topology on the space of compact subsets of a topological space. For metric topological spaces, the Vietoris topology is identical to the topology induced by the Hausdorff distance on the subsets  of a topological space. This topology is  both Hausdorff and second countable. But is not the same as the topology defined  in (\ref{waldtop}), see \cite{Wald}, because unparameterised curves are not distinguished by their images in $\M$ alone.}
 on $\mC(p,q)$. This is the topology used to prove that under certain conditions, there are causal curves  that extremise the proper time between the points $p$ and $q$.

 \begin{remark}\label{rem:curvtop}
 If $\M$ is causal, then the topology on $\mC(p,q)$ with basis defined in (\ref{waldtop}) is both Hausdorff and second countable. To sketch an argument for this, if the causal curves $\lambda$ and $\tilde \lambda$ have different images in $\M$, there  exist open set $U$ and $\tilde U$, with $\lambda\subset U$ and $\tilde \lambda \subset \tilde U$ such that $\lambda$ does not lie entirely in $\tilde U$ and $\tilde \lambda$ does not lie entirely in $U$ -- this is the case even if $\lambda$ and $\tilde \lambda$ intersect at some points. In such a case, the neighbourhoods $O_U$ and $O_{\tilde U}$ constructed using $U$ and $\tilde U$ consist of curves  $\gamma \subset U$ and curves $\tilde \gamma \subset\tilde U$ with $\gamma$  not entirely lying  in $\tilde U$ and vice versa.  Therefore, one can find neighbourhoods\footnote{The open sets $O_U$ and $O_{\tilde U}$ have a non empty intersection, iff $O_U$ and $O_{\tilde U}$ have curves that entirely lie in both $U$ and $\tilde U$. If such curves exist, they must lie in $U\cap \tilde U$.}  $O_{U}$ and $O_{\tilde U}$ of $\lambda$ and $\tilde \lambda$
 with $O_{U}\cap O_{\tilde U}=\emptyset$.  However, if two curves have the same image, like for example one closed time-like curve and another by going twice around the first one, it is not possible to find neighbourhoods that separate two such curves. As a result if $\M$ is not causal, the topology on $\mC(p,q)$ is not Hausdorff.

 To illustrate that the  above topology on  $\mC(p,q)$ is second countable, one uses that $\M$ is second countable. If $\{V_1, V_2, \dots\}$ is a countable basis of the topology of  $\M$, then
 a countable basis of open sets in $\mC(p,q)$ is constructed by considering all the open sets
 \be
 O_{V_{j_1}, V_{j_2},\dots, V_{j_k}}=\{\gamma\in \mC(p,q)\vert \gamma\subset V_{j_1}\cup V_{j_2}\cup\cdots \cup V_{j_k}\}
 \ee
 where $j_1, j_2, \dots, j_k\in\bN$ are finite strings of natural numbers -- it is assumed that $\gamma$ enters each $V$ sequentially. In \cite{Geroch}, the countable basis in the topology was constructed using normal neighbourhoods of the curves.

 Convergence in the topology defined in (\ref{waldtop}) can be expressed as follows: the sequence $(\lambda_n)$ converges to a curve $\lambda$, iff for every $U$ with $\lambda\subset U$, there is an $n_0$ such that all the curves $\lambda_n$ with $n>n_0$ have the property that $\lambda_n\subset U$.  This definition of convergence coincides with that used in the proof of the limit curve theorem provided that $\M$ is \emph{strongly causal}. In such a case, the limit curve will be an accumulation point of a sequence $(\gamma_n)$ as the notion is defined in  the topology (\ref{waldtop})  on $\mC(p,q)$. For more details, see next remark. \hfill{$\Diamond$}
 \end{remark}

\begin{remark}\label{rem:curvtop2}
Let us explore more the relationship between  the convergence associated with the topology on $\mC(p,q)$ defined in  (\ref{waldtop}) with that  used in the proof of the limit curve theorem \ref{prop:limitcurve}, see figure \ref{fig:conv}. The topology used to prove the limit curve theorem is the so called lower Vietoris topology induced on the space of causal curves. Convergence in this topology can be stated as follows:  A sequence of curves $(\gamma_n)$ converges to a curve $\lambda$, iff for every open set $U\in \M$ with $U\cap\lambda\not=\emptyset$, there is an $N$ such that $\lambda_n\cap U\not=\emptyset$ for all $n>N$ -- $N$ can depend on $U$. On the other hand, convergence of the sequence $(\gamma_n)$ to a curve $\lambda$ in the topology used in the proof for the existence of curves with maximal proper time between two points on a spacetime, i.e. that with basis (\ref{waldtop}), also called upper Vietoris topology,  can be expressed as follows: For every open set $U\in \M$ such that $\lambda\in U$, there is an $N$ such that $\lambda_n\in U$ for $n>N$. If the causal curves have end points, convergence in either upper or lower Vietoris topology does not preserve them. If a sequence of casual curves converges both in both the upper and lower Vietoris topologies, then it converges in the Vietoris topology.  Convergence in the Vietoris topology preserves end points.

In general, upper and lower Vietoris topologies  are distinct. However, if $\M$ is strongly causal, they give rise to the same topology on $\mC(p,q)$. In fact, they coincide on J-convex spacetimes, i.e. spacetimes that every $p\in\M$ admits a neighbourhood $U$ such that $J^-(x)\cap J^+(y)\subseteq U$ for every causally ordered points $x, y\in U$, see \cite{beem} proposition 2.21 and appendix B of \cite{sorkin}.
\hfill{$\Diamond$}
 \end{remark}

\begin{figure}[htbp]
\begin{center}
\begin{eqnarray*}
\begin{tikzpicture}[scale=0.75]
  % Coordinates
  \coordinate (A) at (0.00, 4.00);
  \coordinate (B) at (0.00, -3.00);
  \coordinate (C) at (1.75, 2.50);
  \coordinate (D) at (-2.00, -1.00);
  \coordinate (E) at (-1.00, 4.00);
  \coordinate (F) at (-1.00, -4.00);
  \coordinate (G) at (1.50, 2.00);
  \coordinate (H) at (-2.50, -1.50);
  \coordinate (I) at (-2.00, 5.00);
  \coordinate (J) at (-2.00, -3.00);
  \coordinate (K) at (0.50, 2.25);
  \coordinate (L) at (-2.00, -0.50);
  \coordinate (M) at (0.50, 4.50);
  \coordinate (N) at (0.80, -5.45);
  \coordinate (O) at (1.25, 1.50);
  \coordinate (P) at (-0.75, -1.00);
  \coordinate (Q) at (1.00, 4.50);
  \coordinate (R) at (1.00, -4.00);
  \coordinate (S) at (2.75, 1.50);
  \coordinate (T) at (-0.75, -2.00);
  \coordinate (U) at (0.00, 3.00);
  \coordinate (V) at (0.00, 3.00);
  \coordinate (W) at (0.50, 0.50);
  \coordinate (X) at (0.50, 0.50);
  \coordinate (Y) at (-1.17, -1.28);
  \coordinate (Z) at (-1.17, -1.28);
  \coordinate (AA) at (0.00, -3.50);
  \coordinate (AB) at (0.00, -3.50);
  \coordinate (AC) at (9.00, 4.00);
  \coordinate (AD) at (9.00, -3.00);
  \coordinate (AE) at (7.50, 2.00);
  \coordinate (AF) at (10.50, -1.00);
  \coordinate (AG) at (8.00, 3.50);
  \coordinate (AH) at (8.00, -2.50);
  \coordinate (AI) at (7.75, 1.50);
  \coordinate (AJ) at (10.25, -0.50);
  \coordinate (AK) at (9.50, 4.00);
  \coordinate (AL) at (10.00, -3.00);
  \coordinate (AM) at (8.25, 1.50);
  \coordinate (AN) at (10.25, -1.00);
  \coordinate (AO) at (7.50, 4.00);
  \coordinate (AP) at (7.50, -2.50);
  \coordinate (AQ) at (6.75, 1.75);
  \coordinate (AR) at (9.50, 1.75);
  \coordinate (AS) at (11.55, 4.47);
  \coordinate (AT) at (11.47, -3.57);
  \coordinate (AU) at (13.25, 2.50);
  \coordinate (AV) at (12.25, -0.75);
  \coordinate (AW) at (6.50, 0.50);
  \coordinate (AX) at (10.50, 0.50);
  \coordinate (AY) at (6.50, 7.00);
  \coordinate (AZ) at (10.50, 7.00);
  \coordinate (BA) at (6.50, -8.00);
  \coordinate (BB) at (10.50, -8.00);

  % Drawing
  \draw[red, thick] (A) .. controls (1.75, 2.50) and (-2.00, -1.00) .. (B);
  \draw[blue, thick] (E) .. controls (1.50, 2.00) and (-2.50, -1.50) .. (F);
  \draw[blue, thick] (I) .. controls (0.50, 2.25) and (-2.00, -0.50) .. (J);
  \draw[blue, thick] (M) .. controls (1.25, 1.50) and (-0.75, -1.00) .. (N);
  \draw[blue, thick] (Q) .. controls (2.75, 1.50) and (-0.75, -2.00) .. (R);
  \draw[green] (U) circle (1.60);
  \draw[green] (W) circle (1.35);
  \draw[green] (Y) circle (1.50);
  \draw[green] (AA) circle (1.68);
  \draw[red, thick] (AC) .. controls (7.50, 2.00) and (10.50, -1.00) .. (AD);
  \draw[blue, thick] (AG) .. controls (7.75, 1.50) and (10.25, -0.50) .. (AH);
  \draw[blue, thick] (AK) .. controls (8.25, 1.50) and (10.25, -1.00) .. (AL);
  \draw[blue, thick] (AO) .. controls (6.75, 1.75) and (9.50, 1.75) .. (AP);
  \draw[blue, thick] (AS) .. controls (13.25, 2.50) and (12.25, -0.75) .. (AT);
  \draw[green] (AW) .. controls (6.50, 7.00) and (10.50, 7.00) .. (AX);
  \draw[green] (AW) .. controls (6.50, -8.00) and (10.50, -8.00) .. (AX);
   \node[below left] at (1.00, -7.00) {$\text{Fig. (a)}$};
   \node[below left, green] at (0.40, -4.00) {$U_1$};
   \node[ green] at (D) {$U_2$};
   \node[left, green] at (0.00, 4.00) {$U_k$};
    \node[below left] at (10.00, -7.00) {$\text{Fig. (b)}$};
    \node[below left, green] at (9.00, -4.50) {$U$};
     \node[below, red ] at (AD) {$\lambda$};
      \node[below left, red ] at (B) {$\lambda$};
       \node[below right, blue ] at (-3.50, -3.00) {$(\gamma_n)$};
        \node[below right, blue ] at (7.00, -2.70) {$(\gamma_n)$};
\end{tikzpicture}
\end{eqnarray*}
\end{center}
%\begin{figure}[htbp]
\caption{\small{ In figure (a), the convergence of a sequence of causal curves in the lower Vietoris topology is depicted. The curves in sequence $(\gamma_n)$ eventually intersect the open sets $U_k$ that the curve at the limit $\lambda$ also intersects. In figure (b), the convergence of a sequence of causal curves in the upper Vietoris topology is depicted. The curves in the sequence $(\gamma_n)$ are entirely included in the open set $U$ and also contains the curve $\lambda$ at the limit. }}
\centering
\label{fig:conv}
\end{figure}

\begin{remark}

Convergence in this topology (\ref{waldtop}) is different from that of the compact open topology (\ref{comopen}) on the space of continuous curves
\be
C^0(I, \M; p,q)=\{\gamma : I=[a,b]\rightarrow \M\,|\, \gamma(a)=p, \gamma(b)=q\}~,
\ee
 which   takes  the parameterisation of the curves into account. One can also extend the definition for open intervals allowing either $a$ or $b$ or both to appropriately  go to infinity.  If $I$ is compact,  convergence in the compact open topology  is the same as uniform convergence, i.e. convergence using  the distance function (\ref{uni}). Otherwise, if $I$ is not compact it becomes the same as uniform topology on compact subsets.  Thus the compact open topology is a topology on the space of maps instead of a topology on the space of images of maps.\hfill{$\Diamond$}
 \end{remark}

 \begin{example}A give a simple example that illustrates the difference between the spaces $\mC(p,q)$ and $C^0(I, \M; p,q)$, and their respective topologies, consider the maps $f_n(x)=x^n$ with $x\in [0,1]$. For any integer $n>0$, all these maps have the same image $[0,1]$ in $\bR$. Therefore, they represent a single point in $\mC(0,1)$.  However, there are different points as viewed as elements of $C^0(I, \M; p,q)$.  In fact, one can calculate the $||\cdot||_{\infty}$ norm between the maps $f_1$ and $f_n$, $n>1$, to find
\be
||f_n-f_1||_{\infty}\equiv \sup_{x\in[0,1]} |x^n-x|=(n-1) n^{-\frac{n}{n-1}}~,
\label{normxx}
\ee
which is not zero even though the two curves have the same image.

Furthermore, the sequence $(f_n)$ does not converge in the $||\cdot ||_{\infty}$ topology  -- the point-wise limit of the sequence is not a continuous function. Alternatively, the norm in (\ref{normxx}) does not converge to zero as $n\rightarrow +\infty$. Clearly, the topology defined by $||\cdot||_{\infty}$ is finer than that defined by (\ref{waldtop}) on $\mC(p,q)$. \hfill{$\triangledown$}
\end{example}

 \begin{example}To illustrate the differences between the set of curves and the set of their images, let us investigate  the set of geodesics in $\bR^3$. As geodesics in $\bR^3$ are straight lines, they are given by $\gamma(s)=u s+ v$, $u, v\in \bR^3$ and $s$ is the affine parameter.  Thus, the set of oriented geodesics are parameterised by $(u,v)\in \bR^3\times \bR^3$ and so the set of all oriented geodesics is $ \bR^3\times \bR^3$.  On the other hand,  if the tangent vector $\vu=\vec{\dot\gamma}=u$ is chosen to have length one, then the space of oriented geodesics is $S^2\times \bR^3$ -- geodesics with co-linear tangent vectors have the same image. Still, this is not the space of images of geodesics because there is a remaining freedom to choose their origin $v$.  This can be fixed by declaring that $v$ is orthogonal to $u$, $v\perp u$. The end result is that the space of images of oriented geodesics in $\bR^3$, or equivalently the geodesics as subsets  in $\bR^3$,  is the total space of the tangent bundle of $S^2$, $TS^2$.\hfill{$\triangledown$}
 \end{example}

\begin{remark}
Even though,  the  topology on $\mC(p,q)$ defined by the basis (\ref{waldtop}) is traditionally used  to prove key statements in general relativity,  the compact open topology on the space of curves can also be useful.  This is  especially the case whenever  the Arzel\'a-Ascoli theorem is used to prove a statement, like for example in the proof of existence of geodesics with maximum proper time. The problem that one faces though is that the curves must be parameterised with the same parameter.   The typical choice made is the length of the curves with respect to an auxiliary Riemannian metric $g_e$ on the spacetime $\M$. However, this introduces another set of subtleties, see e.g. \cite{Minguzzi}.  Nevertheless, it also has  some advantages including some control over the convergence of the  velocities of the curves.  This   is essential to establish the existence of minimal length (Riemannian manifolds)  or maximal proper time (Lorentzian manifolds) geodesics --  the velocities   converge weakly,  see  e.g. appendix \ref{ap:aa}  for the Riemannian manfolds and appendix \ref{ap:dx} for Lorentzian manifolds. \hfill{$\Diamond$}
\end{remark}

\subsection{Compact causal diamonds and globally hyperbolic manifolds}

\begin{figure}[htbp]
\begin{center}
\begin{eqnarray*}
\begin{tikzpicture}[scale=0.8]
  % Coordinates
  \coordinate (A) at (-2.00, 5.00);
  \coordinate (B) at (12.11, 4.96);
  \coordinate (C) at (1.75, 7.00);
  \coordinate (D) at (7.00, 3.00);
  \coordinate (E) at (5.00, 10.00);
  \coordinate (F) at (-0.94, 3.97);
  \coordinate (G) at (11.00, 4.00);
  \coordinate (H) at (6.00, 1.00);
  \coordinate (I) at (0.00, 6.00);
  \coordinate (J) at (11.25, 5.00);
  \coordinate (K) at (16.25, 5.00);
  \coordinate (L) at (7.75, 6.50);
  \coordinate (M) at (8.25, 3.75);
  \coordinate (N) at (6.25, 6.50);
  \coordinate (O) at (7.00, 3.75);
  \coordinate (P) at (3.00, 6.50);
  \coordinate (Q) at (2.50, 5.00);
  \coordinate (R) at (3.25, 4.25);
  \coordinate (S) at (5.25, 4.50);
  \coordinate (T) at (4.75, 6.00);
  \coordinate (U) at (6.00, 4.00);
  \coordinate (V) at (5.25, 6.00);
  \coordinate (W) at (7.25, 5.75);
  \coordinate (X) at (7.25, 3.75);
  \coordinate (Y) at (6.00, 5.25);
  \coordinate (Z) at (6.25, 3.50);
  \coordinate (AA) at (3.00, 6.25);
  \coordinate (AB) at (4.00, 4.25);

  % Drawing
  \draw[very thick] (A) .. controls (1.75, 7.00) and (7.00, 3.00) .. (B);
  %\draw[blue, thick] (E) -- (F);
  %\draw[blue, thick] (E) -- (G);
 % \draw[blue, thick] (H) -- (I);
  %\draw[blue, thick] (H) -- (J);
  \draw[thick] (E) .. controls (7.75, 6.50) and (8.25, 3.75) .. (H);
  \draw[thick] (E) .. controls (6.25, 6.50) and (7.00, 3.75) .. (H);
  \draw[thick] (E) .. controls (3.00, 6.50) and (2.50, 5.00) .. (H);
  \draw[thick] (E) .. controls (3.25, 4.25) and (5.25, 4.50) .. (H);
  \draw[red, thick] (E) .. controls (4.75, 6.00) and (6.00, 4.00) .. (H);
  \draw[green, thick] (E) .. controls (7.25, 5.75) and (7.25, 3.75) .. (H);
  \draw[green, thick] (E) .. controls (6.00, 5.25) and (6.25, 3.50) .. (H);
  \draw[green, thick] (E) .. controls (3.00, 6.25) and (4.00, 4.25) .. (H);
  \fill[black] (E) circle (2pt);
 % \fill[black] (K) circle (2.5pt);
  \fill[black] (V) circle (2pt);
  \node[above, font=\large] at (E) {$q$};
  \node[below, font=\large] at (H) {$p$};
   \node[right, font=\large] at (V) {$r$};
   \node[right, font=\large] at (11.11, 4.40) {$\Sigma$};
    \node[right, font=\large] at (7.95, 3.70) {$I^-(\Sigma)$};
     \node[right, font=\large] at (7.95, 5.40) {$I^+(\Sigma)$};
      \node[right, font=\large] at (4.50, 6.80) {$\lambda'$};
 \node[right, font=\large] at (5.25, 3.60) {$\lambda$};
  \node[right, green, font=\large] at (7.35, 6.80) {$(\lambda_{n_k})$};
   \node[right,  font=\large] at (3.00, 2.50) {$(\lambda_{n})$};

\end{tikzpicture}
\end{eqnarray*}
\end{center}
%\begin{figure}[htbp]
\caption{\small{The figure gives an illustration of the proof of the third statement of theorem \ref{th:compact}, where the points $p$ and $q$ are below and above the Cauchy surface $\Sigma$. The subsequence $(\lambda_{n_k})$ of $(\lambda_n)$ has $r$ as a limit point. Therefore, any limit curve, like $\lambda'$, constructed  by possibly considering a further subsequence of $(\lambda_{n_k})$,  has to pass from $r$. }}
\centering
\label{fig:compact}
\end{figure}
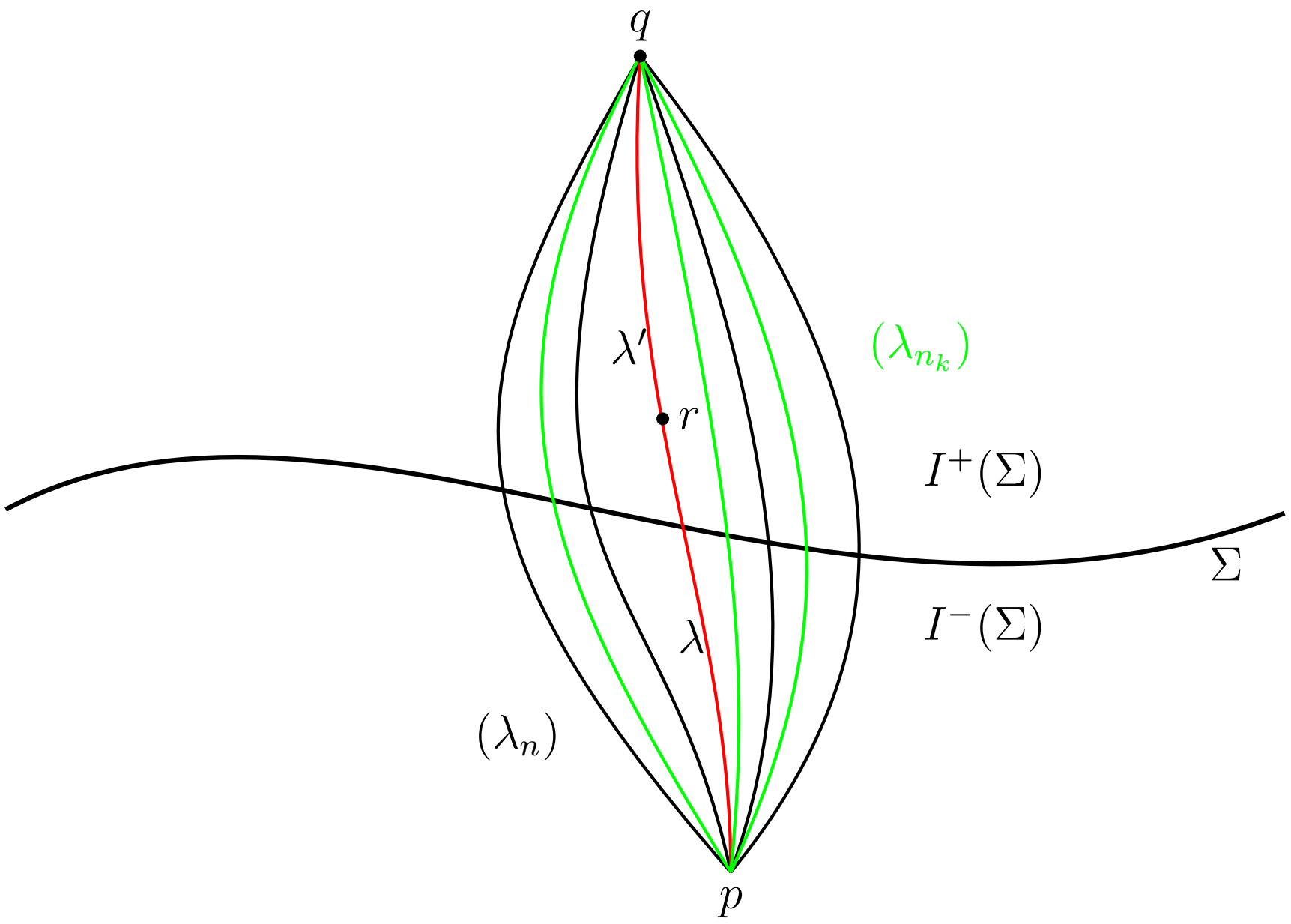

\begin{theorem}\label{th:compact}
Let $\M$ be a globally hyperbolic spacetime. The space of future directed causal curves from $p$ to $q$, $\mC(p,q)$, is compact.
\end{theorem}
\begin{proof}
The topology of $\mC(p,q)$ is second countable and so it suffices to show for compactness that every sequence of curves in $\mC(p,q)$ has a convergent subsequence, see Theorem \ref{th:bw}. Let $(\lambda_n)$ be a sequence of future directed causal curves from $p$ to $q$. First suppose that both $p,q\in I^-(\Sigma)$, where $\Sigma$ is the Cauchy surface.  We shall demonstrate that there is always a converging subsequence to a future directed causal curve $\lambda$ from $p$ to $q$.  Consider a sequence $(\lambda_n)$ with $\lambda_n\in \mC(p,q)$. Each such curve is inextendible in the space $\M-\{q\}$, i.e. the spacetime $\M$ with the point $q$ removed. Thinking of $p$ (trivially) as an accumulation point of $(\lambda_n)$, it follows from the limit curve theorem, i.e.  Lemma \ref{prop:limitcurve}, that there is a future directed inextendible causal curve $\lambda\subset M-\{q\}$ that passes through $p$. Restoring $q$, $\lambda$ will either (i) remain inextendible in $M$ or (ii) will have an end point at $q$. However (i) cannot occur. As all curves in the sequence remain in $I^-(\Sigma)$, $\lambda\subset I^-(\Sigma)$. In particular, it will not enter $I^+(\Sigma)$. However, this contradicts proposition  \ref{prop:cauchy1}. Thus $\lambda$ must have an end point at $q$, i.e. $\lambda\in \mC(p,q)$.

The proof of the statement for $p,q\in I^+(\Sigma)$ is similar.  The same applies whenever either $p$ or $q$ are in $\Sigma$.  In particular for $q\in \Sigma$, notice that the intersection point $w_n$ of the $\lambda_n$ curve of the sequence with $\partial B(q, r)$  are in $\overline{I^-(q)\cap \partial B(q, r)}$.  This is again compact and none of the curves $(\lambda_n)$ enters $I^+(\Sigma)$. Therefore the limit curve does not enter $I^+(\Sigma)$ either and the argument can be repeated.

It remains to demonstrate the theorem for the case that $p\in I^-(\Sigma)$ and $q\in I^+(\Sigma)$. Remove $q$ from $\M$ and from a sequence of curves $(\lambda_n)$ passing through both $p$ and $q$ construct a future directed inextendible causal curve $\lambda$ in $\M-\{q\}$  from $p$ that enters $I^+(\Sigma)$ -- if $\lambda$ ends at the point $q$, there is nothing to prove but in general it will not. Let $r\in \lambda\cap I^+(\Sigma)$ and  $(\lambda_{n_k})$ be the subsequence of $(\lambda_n)$ that has as a convergence point $r$, e.g. a subsequence of the sequence that has convergence curve $\lambda$. Then, remove the point $p\in M$ and use $(\lambda_{n_k})$ to construct a past directed causal  limit curve $\lambda'$ that starts from $q$. However, $r$ is a convergence point of $(\lambda_{n_k})$ and so $\lambda'$ must pass through $r$. Joining $\lambda$ from $p$ to $r$ with $\lambda'$ from $r$ to $q$, one constructs  a future directed causal limit curve from $p$ to $q$ and so $\mC(p,q)$ is compact.  Note that this argument would not have worked, if future inextendible curves from $p$ could not  enter $I^+(\Sigma)$.
\end{proof}

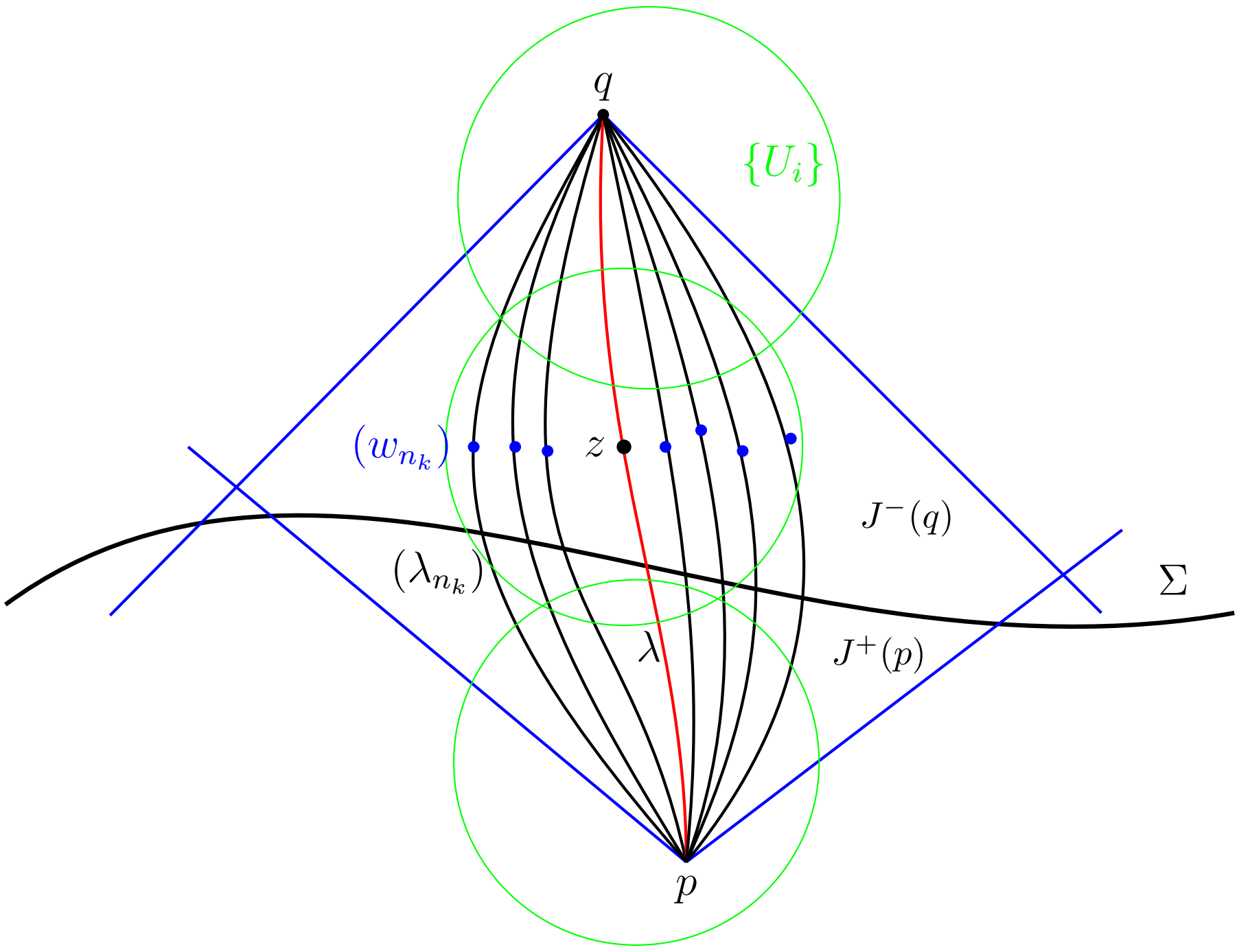
\begin{figure}[htbp]
\begin{center}
\begin{eqnarray*}
\begin{tikzpicture}[scale=0.8]
  % Coordinates
  %\coordinate (A) at (-2.00, 5.00);
  %\coordinate (B) at (12.11, 4.96);
   \coordinate (A) at (-2.20, 4.10);
  \coordinate (B) at (12.60, 4.00);
  \coordinate (C) at (1.75, 7.00);
  \coordinate (D) at (7.00, 3.00);
  \coordinate (E) at (5.00, 10.00);
  \coordinate (F) at (-0.94, 3.97);
  \coordinate (G) at (11.00, 4.00);
  \coordinate (H) at (6.00, 1.00);
  \coordinate (I) at (0.00, 6.00);
  \coordinate (J) at (11.25, 5.00);
  \coordinate (K) at (16.25, 5.00);
  \coordinate (L) at (7.75, 6.50);
  \coordinate (M) at (8.25, 3.75);
  \coordinate (N) at (6.25, 6.50);
  \coordinate (O) at (7.00, 3.75);
  \coordinate (P) at (3.00, 6.50);
  \coordinate (Q) at (2.50, 5.00);
  \coordinate (R) at (3.25, 4.25);
  \coordinate (S) at (5.25, 4.50);
  \coordinate (T) at (4.75, 6.00);
  \coordinate (U) at (6.00, 4.00);
  \coordinate (V) at (5.25, 6.00);
  \coordinate (W) at (7.25, 5.75);
  \coordinate (X) at (7.25, 3.75);
  \coordinate (Y) at (6.00, 5.25);
  \coordinate (Z) at (6.25, 3.50);
  \coordinate (AA) at (3.00, 6.25);
  \coordinate (AB) at (4.00, 4.25);

  % Drawing
  \draw[very thick] (A) .. controls (1.75, 7.00) and (7.00, 3.00) .. (B);
  \draw[blue, thick] (E) -- (F);
  \draw[blue, thick] (E) -- (G);
  \draw[blue, thick] (H) -- (I);
  \draw[blue, thick] (H) -- (J);
  \draw[thick] (E) .. controls (7.75, 6.50) and (8.25, 3.75) .. (H);
  \draw[thick] (E) .. controls (6.25, 6.50) and (7.00, 3.75) .. (H);
  \draw[thick] (E) .. controls (3.00, 6.50) and (2.50, 5.00) .. (H);
  \draw[thick] (E) .. controls (3.25, 4.25) and (5.25, 4.50) .. (H);
  \draw[red, thick] (E) .. controls (4.75, 6.00) and (6.00, 4.00) .. (H);
  \draw[ thick] (E) .. controls (7.25, 5.75) and (7.25, 3.75) .. (H);
  \draw[ thick] (E) .. controls (6.00, 5.25) and (6.25, 3.50) .. (H);
  \draw[ thick] (E) .. controls (3.00, 6.25) and (4.00, 4.25) .. (H);
  \draw[green] (V) circle (2.15);
  \draw[green]  (5.55, 9.00) circle (2.30);
  \draw[green] (5.40, 2.20) circle (2.20);
  \fill[black] (E) circle (2pt);
 % \fill[black] (K) circle (2.5pt);
  \fill[black] (V) circle (2.5pt);
   \fill[blue] (3.44, 6.00) circle (2pt);
  \fill[blue] (3.94, 6.00) circle (2pt);
   \fill[blue] (4.33, 5.95) circle (2pt);
     \fill[blue] (6.68, 5.95) circle (2pt);
       \fill[blue] (7.26, 6.10) circle (2pt);
         \fill[blue] (6.18, 6.20) circle (2pt);
           \fill[blue] (5.75, 6.00) circle (2pt);

  \node[above, font=\large] at (E) {$q$};
  \node[below, font=\large] at (H) {$p$};
  % \node[right, font=\large] at (V) {$r$};
   \node[right, font=\large] at (11.51, 4.40) {$\Sigma$};
    \node[right, font=\small] at (7.58, 3.50) {$J^+(p)$};
     \node[right, font=\small] at (7.92, 5.15) {$J^-(q)$};
      %\node[right, font=\large] at (4.50, 6.80) {$\lambda'$};
 \node[right, font=\large] at (5.23, 3.60) {$\lambda$};
 \node[left, font=\large] at (5.20, 6.00) {$z$};
 % \node[right, green, font=\large] at (7.35, 6.80) {$(\lambda_{n_k})$};
   \node[right,  font=\large] at (2.27, 4.50) {$(\lambda_{{n_k}})$};
 \node[left, blue, font=\large] at (3.35, 6.00) {$(w_{{n_k}})$};
 \node[right, green, font=\large] at (6.49, 9.40) {$\{U_i\}$};
\end{tikzpicture}
\end{eqnarray*}
\end{center}
%\begin{figure}[htbp]
\caption{\small{The figure gives an illustration of the proof of the  theorem \ref{th:jjcompact}. The subsequence $(\lambda_{n_k})$ that pass through $(w_{n_k})$ has a limit curve $\lambda$ and in turn the subsequence  $(w_{n_k})$ has an accumulation point $z\in \lambda$.  }}
\centering
\label{fig:jjcompact}
\end{figure}

\begin{theorem}\label{th:jjcompact}
Let $\M$ be a globally hyperbolic manifold. Then, $J^+(p)\cap J^-(q)$ is compact.
\end{theorem}
\begin{proof}
We shall demonstrate that for every sequence of points $(w_n)$ in $J^+(p)\cap J^-(q)$ has a accumulation point. Indeed consider the sequence of causal curves $(\lambda_n)$ in $\mC(p, q)$, where $\lambda_n$  passes through $w_n$. As $\mC(p,q)$ is compact, there is a subsequence $(\lambda_{n_k})$ that converges to a curve $\lambda\in \mC(p,q)$. $\lambda$ as subspace of $J^+(p)\cap J^-(q)$ is compact. Choose a cover $\{U_i\}$ of  $\lambda$ with compact closure and use the compactness of $\lambda$ to extract a finite subcover. Taking the union of the open sets of the finite subcover, one constructs an open set $U$ with $\lambda\subset U$ and $\bar U$ compact.  As $(\lambda_{n_k})$  converges to $\lambda$, there is an $N$ such that for $n_k>N$, $\lambda_{n_k}\subset U\subset \bar U$. Thus $w_{n_k}\in \bar U$ for $n_k>N$. But $\bar U$ is compact, and so the sequence $(w_{n_k>N})$ has an accumulation point $z$. This is also an accumulation point of $(w_n)$.
\end{proof}

\begin{remark}\label{rem:compclosed}
From general topology results, as $J^+(p)\cap J^-(q)$ is compact, $J^+(p)\cap J^-(q)$ is also \emph{closed}. It can also be shown that for $K$ compact,  $J^+(K)\cap J^-(q)$ is also compact, see theorem \ref{th:Kqcompact} below. These spaces are also known as \emph{causal diamonds}. \hfill{$\Diamond$}
\end{remark}

\begin{corollary}\label{cor:closejp}
Let $\M$ be globally hyperbolic. Then $J^+(p)$ is closed.
\end{corollary}
\begin{proof}
Suppose that $J^+(p)$ is not closed and there is a $w\in {J^+(p)}$ with $w\notin \overline{J^+(p)}$. Consider a point $q\in I^+(w)$. Clearly,
$w\in J^+(p)\cap J^-(q)=\overline {J^+(p)\cap J^-(q)}$, where the closure of $J^+(p)\cap J^-(q)$ has been used, see remark \ref{rem:compclosed}. This is  contradicts  the hypothesis that $w\notin \overline{J^+(p)}$. Therefore, $J^+(p)$ is closed.
\end{proof}

\begin{remark}
A basis in the topology of $\mC(K,q)$ is given by the requirement that
\be
\mathcal{O}_{U, V}=\{\gamma\in \mC(K,q)\vert \gamma\subset U;  \gamma\, \text{has past end point in}\, V;  U, V \text{open sets in}\, \M\}~.
\label{waldtop2}
\ee
This is an adaptation of the basis of open sets in $\mC(p,q)$ given in (\ref{waldtop}).\hfill{$\Diamond$}
\end{remark}

The above results can be generalised as follows:
\begin{theorem}\label{th:Kqcompact}
Let $\M$ be a globally hyperbolic spacetime, $\Sigma$ be a Cauchy surface of $\M$ and $\K$ be a compact subset of $\M$. Then, the following statements hold
\begin{enumerate}
\item $J^-(q)\cap J^+(\K)$ is a compact subset of $\M$.

\item $J^-(q)\cap J^+(\Sigma)$ is a compact subset of $\M$, where $q\in D^+(\Sigma)$.

\item $J^+(\K)$ is a closed subset of $\M$.
\end{enumerate}
\end{theorem}
\begin{proof}
1)\,To prove that $J^-(q)\cap J^+(\K)$ is compact, we first demonstrate that the space of future directed causal curves from $K$ ro $q$, $\mC(K,q)$, is compact. The proof is the same as for $\mC(p,q)$. Indeed first suppose that $\{q\}, K\subset I^-(\Sigma)$.  Then one considers a sequence of curves $(\lambda_n)$  with $\lambda_n\in \mC(K,q)$. The sequence of curves is future inextendible in the manifolds $\M-\K$ with (trivial) accumulation point $p$.  By the limit curve theorem, Lemma \ref{prop:limitcurve}, there is a limit curve $\lambda$ that either ends on $\K$ or it is future inextendible. The latter possibility does not occur as all curves $\lambda_n$ are contained in $I^-(\Sigma)$ while future inextendible curves must cross to $I^+(\Sigma)$ as $\M$ is globally hyperbolic. Thus $\lambda$ must have an end point at $\K$, i.e. $\lambda\in \mC(\K,q)$. The case with $K\subset J^-(\Sigma)$ and $q\in J^+(\Sigma)$ can be proved as in the $\mC(p,q)$. Similarly, the compactness of $J^-(q)\cap J^+(\K)$ can be proved.

2)\,Next to prove that $J^-(q)\cap J^+(\Sigma)$ is compact, let us first prove that $\mC(\Sigma, q)$ is compact.  Note that all causal curves from $q$ to $\Sigma$ can always be extended to past inextendible causal curves that cross $\Sigma$ once and enter $J^-(\Sigma)$. To continue,  consider a sequence of curves $(\lambda_n)$ with $\lambda_n\in C(\Sigma, q)$.  All of these have a trivial accumulation point $q$ can be extended to past inextendible curves as mentioned above. Therefore  by the limit curve theorem, there is   curve $\gamma$ that passes through $q$. Moreover, $\gamma$ can be constructed to be past inextendible using the extension of the sequence $(\lambda_n)$ to $J^-(\Sigma)$.  Thus, $\gamma$ must be past inextendible causal curve and so it has to intersect $\Sigma$  as a consequence of Proposition \ref{prop:cauchy1}. Restricting $\gamma$ in $J^-(q)\cap J^+(\Sigma)$ gives a limit curve   for a subsequence of $(\lambda_n)$ to converge and so $C(\Sigma, q)$ is compact.

Then,  the compactness of $J^-(q)\cap J^+(\Sigma)$ can be proved using the finite cover argument on the limit curve constructed from the sequence of causal curves $(\lambda_n)$ with $\lambda_n\in C(\Sigma, q)$ that pass through the sequence of points $(x_n)$, $x_n\in J^-(q)\cap J^+(\Sigma)$. This is identical to the proof given for the compactness of $J^-(q)\cap J^+(q)$ in Theorem \ref{th:jjcompact}.

Finally, we shall not elaborate on the proof of the final statement of the theorem. It is similar to the proof given above in corollary \ref{cor:closejp} that $J^+(p)$ is closed.
\end{proof}

\begin{remark}
Notice that for globally hyperbolic manifolds  with a Cauchy surface $\Sigma$, $J^-(q)\cap \Sigma= J^-(q)\cap J^+(\Sigma)\cap \Sigma$  is compact. This is because $\Sigma$ is closed and so $J^-(q)\cap J^+(\Sigma)\cap \Sigma$ is a closed subset of the compact set $J^-(q)\cap J^+(\Sigma)$.\hfill{$\Diamond$}
\end{remark}

\begin{remark}
 We have already mentioned that globally hyperbolic spacetimes are stably causal. So, there is a temporal function $t$, strictly increasing such that every level set of $t$ is a Cauchy surface. We shall make use of this later in appendix \ref{ap:dx}  to prove the existence of geodesics with maximal proper time in $\mC(p,q)$ for globally hyperbolic manifolds using the Arzel\'a-Ascoli theorem. \hfill{$\Diamond$}
\end{remark}

\section{Curves of maximum proper time}

\subsection{Proper time}

Let $\gamma$ be a smooth\footnote{One could also have assumed that $\gamma$ is absolutely continuous or Lipschitz.} (at least $C^1$) casual curve between two points $p,q\in \M$.  The \emph{proper time} or \emph{proper length} calculated by an observer to travel from point $p$ to point $q$ along the curve $\gamma$ is
\be
L_g(\gamma; p, q, [a,b]) \equiv\int^b_a \,dt\, \sqrt{-g(\vec{\dot\gamma}, \vec{\dot\gamma})}~,
\ee
where $t$ is a parameterisation of the curve with $\gamma(b)=q$ and $\gamma(a)=p$.  Observe that $L_g(\gamma)$ is independent from the parameterisation of the curve $\gamma$.

\begin{remark}
If there is no source for confusion, we shall also denote the proper time functional $L_g(\gamma; p, q, [a,b])$ with  either $L_g(\gamma;  [a,b])$ or $L_g(\gamma; p, q)$. We shall even use just $L_g(\gamma)$ to denote the proper time functional. \hfill{$\Diamond$}
\end{remark}

Clearly this definition can be extended to all smooth curves $\widetilde {\mC(p,q)}\subset \mC(p,q)$ from $p$ to $q$. It turns out that $\widetilde {\mC(p,q)}$ is dense in $\mC(p,q)$. Using this, we can define the following:

\begin{definition}\label{def:properdist}
The proper distance, $L_g(p,q)$,  between the points $p$ and $q$ with $q\in J^+(p)$ is defined as
\be
L_g(p,q)\equiv \sup_{\gamma}\{L_g(\gamma)\vert \gamma\in \widetilde {\mC(p,q)}\}
\ee
\hfill{$\triangle$}
\end{definition}

\begin{remark}
It is not apparent from this definition \ref{def:properdist} that there is a $\lambda\in \mC(p,q)$ such that $L_g(\lambda)=L_g(p,q)$. A proof of this will be an analogue of the Hopf-Rinow
theorem of Riemannian geometry described in  appendix \ref{ap:bb}. Observe also that the definition maximises the proper distance between the two points instead of minimising it as in the case of Riemannian geometry.  Using as a rough guide the proof of the Hopf-Rinow theorem in appendix \ref{ap:bb}, the proof of existence of a curve $\lambda$ with proper time equal to the proper distance $L_g(p,q)$ between the points $p$ and $q$ will require three ingredients.

\begin{itemize}
\item First, we must prove a compactness theorem on the space of curves, so that if we prepare a sequence of curves $(\gamma_n)$ with proper time $L_g(\gamma_n)$ that approaches $L_g(p,q)$, it has a convergence subsequence $(\gamma_{n_k})$ with limit curve $\lambda$.

\item Second, we must make sure that the sequence of velocities $(\vec{\dot\gamma}_{n+k})$ has also a subsequence that converges so that the proper time of the curve $\lambda$ is defined as well -- typically the sequence of velocities converges weakly.

\item Finally, one has to demonstrate that $L_g(\lambda)=L_g(p,q)$.  This is where the semi-continuity of the  functional $L_g$ is required.
\end{itemize}

There is a description of all three points in appendices \ref{ap:ab}, \ref{ap:ac} and \ref{ap:bb}.  The latter appendix gives an extensive explanation of these results in the context of  Riemannian case.\hfill{$\Diamond$}
\end{remark}

\subsection{Time-like geodesics with conjugate or focus points}\label{sec:timeconjfocus}

%\subsubsection{Time-like geodesics}

\begin{theorem}\label{th:conjugategeodesic1}
Let $\gamma$ be a time-like smooth curve  passing through the points $p$ and $q$ in a spacetime $\M$. Then, $\gamma$ locally maximizes the proper time between $p$ and $q$, via smooth variations of the curve, provided that $\gamma$ is a geodesic without a conjugate point between $p$ and $q$.
\end{theorem}
\begin{proof}
Clearly geodesics $\gamma$ that pass through $p$ and $q$ are critical points of the proper time functional. We shall demonstrate that if there is a conjugate point between $p$ and $q$, then $\gamma$ is a saddle point instead of maximum of the proper time functional. The proof involves the calculation of the Hessian of the proper time functional. For a maximum, one expects all directions of the Hessian to be negative. However,  we shall demonstrate that if there is a conjugate point between $p$ and $q$, the Hessian of the proper time functional has positive directions.  This implies the existence of nearby time-like curves going through $p$ and $q$ that have bigger proper time than that of the geodesic $\gamma$.

The proper time functional is
\be
L_g(\gamma)\equiv \int dt \sqrt{-g(\vec{\dot\gamma}, \vec{\dot\gamma})}=\int dt \sqrt{-g_{\mu\nu}(\gamma) \dot \gamma^\mu \dot\gamma^\nu}~.
\ee
Consider a 1-parameter family of curves $\gamma(s)$ parameterised by $s$ with $\gamma(0)=\gamma_0$.  The first derivative of the proper time functional is
\begin{align}
\frac{d}{ds}L_g(\gamma(s))&=-\int^b_a\, dt  f^{-1} g(\nabla_s \vec{\dot \gamma}, \vec{\dot \gamma})
\cr
&= -\int^b_a\, dt \partial_t g(\partial_s \gamma, f^{-1} \vec{\dot \gamma})+\int^b_a\, dt  g(\partial_s \gamma, \nabla_t( f^{-1}\vec{\dot \gamma}))
\cr
&=\int^b_a\, dt  g(\partial_s \gamma, \nabla_t( f^{-1}\vec{\dot \gamma}))~,
\label{fvpt}
\end{align}
where $\gamma(s, a)=p$ and $\gamma(s, b)=q$, $t$ parameterises the curves  and $f= \sqrt{-g(\vec{\dot\gamma},\vec{\dot\gamma}) }$. Clearly, these boundary conditions imply that
$ \partial_s\gamma\vert_{t=a}=\partial_s\gamma\vert_{t=b}=0$.  This establishes the last equality as the surface term vanishes.

Taking the second derivative of the proper time functional, we get
\begin{align}
\frac{d^2}{ds^2}L_g(\gamma(s))&=\int^b_a\, dt  [g(\nabla_s\partial_s \gamma, \nabla_t( f^{-1}\vec{\dot \gamma}))+\int^b_a\, dt  g(\partial_s \gamma, \nabla_s\nabla_t( f^{-1}\vec{\dot \gamma}))
\cr
&=\int^b_a\, dt  g(\nabla_s\partial_s \gamma, \nabla_t( f^{-1}\vec{\dot \gamma}))
\cr
&\qquad\qquad+\int^b_a\, dt\,  g(\partial_s \gamma, \nabla_t\nabla_s( f^{-1}\vec{\dot \gamma})+R(\partial_s \gamma, \vec{\dot \gamma}) f^{-1}\vec{\dot \gamma} )]~.
\label{svpt}
\end{align}
Evaluating the above expression at $s=0$ and choosing the affine parameterisation for the geodesic $\gamma_0$, i.e. $f=1$, we find that
\begin{align}
\frac{d^2}{ds^2}L_g(\gamma(s))\Big|_{s=0}&=\int^{b'}_{a'}\, d\tau   g(\vz, \nabla^2_\tau \vz+R(\vz, \vec{\dot \gamma}) \vec{\dot \gamma} )
\cr
&=-\int^{b'}_{a'}\, d\tau  [ g(\nabla_\tau\vz, \nabla_\tau \vz)-g(\nabla_\tau\vz, R(\vz, \vec{\dot \gamma}) \vec{\dot \gamma} )]~,
\end{align}
where $\vz=\partial_s\gamma\vert_{s=0}$,  we have used the geodesic equation for $\gamma_0$ and adjusted the boundaries of the integral to the affine parameterisation with $\tau$, i.e. $\gamma_0(\tau)\vert_{\tau=a'}=p$ and $\gamma_0(\tau)\vert_{\tau=b'}=q$. Also to establish the last equation, we have performed an integration by parts  -- the boundary term vanishes as $\vz$ vanishes at both $p$ and $q$.

Observe that the expression for the Hessian of the proper time functional is exactly the same as that of the energy functional in  appendix  \ref{ap:bb2} up to an overall sign.  As a result a similar argument to that described for  Riemannian manifolds in the  appendix \ref{ap:bb2} can be used to prove that the Hessian has some positive directions.  This  demonstrates that the presence of a conjugate point between $p$ and $q$ leads to the conclusion that although $\gamma_0$ is a geodesic, so a critical point of the proper time functional, it is a saddle point instead of a local maximiser.
\end{proof}

\begin{figure}[htbp]
\begin{center}
\begin{eqnarray*}
\begin{tikzpicture}[scale=0.6]
  % Coordinates
  \coordinate (A) at (5.00, 10.00);
  \coordinate (B) at (6.90, -2.05);
  \coordinate (C) at (9.00, 6.67);
  \coordinate (D) at (0.00, 4.25);
  \coordinate (E) at (5.00, 0.00);
 \coordinate (E1) at (4.25, 2.00);
  \coordinate (F) at (5.75, 7.00);
  \coordinate (G) at (6.75, 2.50);
  \coordinate (H) at (7.25, 5.25);
  \coordinate (I) at (5.50, 9.50);
  \coordinate (J) at (2.00, 6.25);
  \coordinate (K) at (7.25, 6.50);

  % Drawing
  \draw[thick] (A) .. controls (9.00, 6.67) and (0.00, 4.25) .. (B);
  \draw[blue, thick] (E) .. controls (6.75, 2.50) and (7.25, 5.25) .. (F);
  \draw[red, thick] (E1) .. controls (2.00, 6.25) and (7.25, 6.50) .. (I);
   \node[right] at (5.50,0.00) {$p$};
   \node[right] at (5.30, 10.00) {$q$};
   \node[left] at (4.00, 2.00) {$w$};
   \node[right] at (6.25, 7.00) {$r$};
    \node[right] at (6.00,-1.00) {$\gamma$};
     \node[right, blue] at (7.00,3.50) {$\tau$};
      \node[right, red] at (3.30,5.70) {$\lambda$};
\end{tikzpicture}
\end{eqnarray*}
\end{center}
%\begin{figure}[htbp]
\caption{\small{The curve $\lambda$ rounds up the corner at $r$, the conjugate point of $p$ on time-like geodesic $\gamma$. It begins at $w$ before the conjugate point  $r$ on $\gamma$ and ends again on $\gamma$ before the point $q$.  }}
\centering
\label{timeconj}
\end{figure}
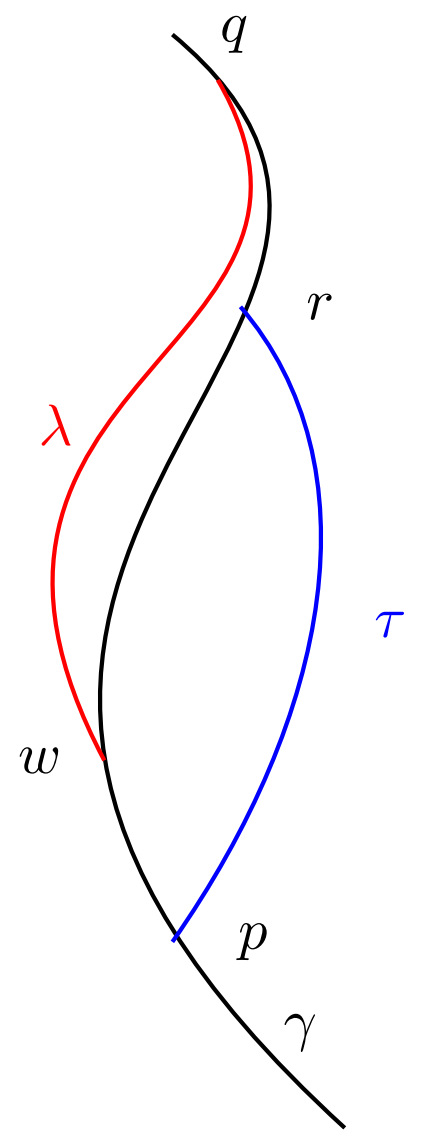

\begin{remark}
The above theorem \emph{does not} prove the existence of a geodesic  with maximal proper time that ends at the points $p$ and $q$. However, it states that if there is such a maximal geodesic $\gamma$, it cannot have a conjugate point between $p$ and $q$. Moreover, if $\gamma$ has a conjugate point between $p$ and $q$, it cannot be maximal.  The question of existence of such maximal geodesic $\gamma$  is a separate problem and requires different assumptions and techniques to prove it.\hfill{$\Diamond$}
\end{remark}

There is an analogous theorem where one of the points $p$ or $q$ of the statement above is replaced by a smooth spacelike hypersurface $\S$ of $\M$. In particular, the following can be shown:

\begin{theorem}\label{th:focusgeodesic1}
Let $\gamma$ be a smooth time-like curve from a space-like hypersurface $\S$ to a point $p\in M$. A necessary and sufficient condition for $\gamma$ to locally maximise the proper time  between $\S$ and $p$ over smooth variations of $\gamma$ is that $\gamma$ is a geodesic orthogonal to $\S$ that does not have a focus point between $\S$ and $p$.
\end{theorem}
\begin{proof}
Consider a family of curves $\gamma(t, s)$ with $t\in (a, b)$ the parameter of the curves and $s$ the parameter of the family. We impose the conditions that
\be
\gamma(t, 0)=\gamma(t)~,~~~\gamma(a, s)\in \S~,~~~\gamma(b, s)=p~,
\label{hgeodcon}
\ee
i.e. the point $\gamma(a, s)\in \S$ is allowed to vary on $\S$ with $s$.

To prove the first part of the theorem, let us vary the proper time functional as in (\ref{fvpt}) to find
\begin{align}
\frac{d}{ds}L_g(\gamma(s))&=-\int^b_a\, dt \partial_t g(\partial_s \gamma, f^{-1} \vec{\dot \gamma})+\int^b_a\, dt  g(\partial_s \gamma, \nabla_t( f^{-1}\vec{\dot \gamma}))
\cr
&=-g(\partial_s \gamma, f^{-1} \vec{\dot \gamma})|^b_a+\int^b_a\, dt  g(\partial_s \gamma, \nabla_t( f^{-1}\vec{\dot \gamma}))~.
\end{align}
From the boundary condition (\ref{hgeodcon}), it is reasonable to conclude that $\partial_s\gamma(t, s)\vert_{t=b}=0$ and $\partial_s\gamma(t, s)\vert_{t=a}$ can be an arbitrary vector field as $T_{\gamma(a, s)}\S$.  This gives that
\begin{align}
\frac{d}{ds}L_g(\gamma(s))\vert_{s=0}=g(\vz, \vec{\dot \gamma})\vert_{t=a'}+\int^{b'}_{a'}\, dt  g(\vz, \nabla_\tau\vec{\dot \gamma})~,
\end{align}
where we have also set $f=1$, i.e. we have parameterised $\gamma$ with an affine parameter $\tau\in [a', b']$.
Therefore, the first variation vanishes provided that $\gamma$ is a geodesic and it is orthogonal to $\vz$ for every tangent vector in $T_{\gamma(a)}\S$, i.e. the tangent vector of $\gamma$ at $\gamma(a)$ must be perpendicular to $\S$. This proves the first part of the statement.

To continue, one has to compute the second variation of the proper time functional.  This computation is similar to that done previously in
(\ref{svpt}) accounting for the new boundary conditions. In particular, one finds that
\begin{align}
\frac{d^2}{ds^2}L_g(\gamma(s))\Big|_{s=0}&=\big[-g(\vz, \nabla_\tau\vz)+ K(\vz, \vz)\big]|_{\tau=a'}+\int^{b'}_{a'}\, d\tau   g(\vz, \nabla^2_\tau \vz+R(\vz, \vec{\dot \gamma}) \vec{\dot \gamma} )
\end{align}
where $K$ is the second fundamental form of the hypersurface (\ref{sffh}).
For Jacobi fields $\vz$ the boundary term vanishes because they satisfy the boundary condition
\be
\big(\nabla_\tau \vz- K(\vz)^\flat\big)|_{a'}=0~,
\label{jfbc}
\ee
on the hypersurface $\S$.

The rest of the proof is similar to that explained in theorem \ref{th:conjugategeodesic1} and appendix \ref{ap:bb2}. In particular, one defines a variation of the Jacobi field $\vz$ which is determined by the velocity of $\vz$ and the conjugate point $\tau_0$ of the geodesic between $\S$ and $p$.
\end{proof}

\begin{remark}
To justify (\ref{jfbc}), we have that
\be
g(\partial_\tau \gamma(\tau, s), \vv)|_{\tau=a'}=0~,
\ee
 as all geodesics emanate orthogonally from $\S$
for every $\vv\in T_{\gamma(a',s)}\S$.  Taking the derivative of the above equation with respect to deformation parameter $s$ of the family and then setting $s=0$, we find that
\be
[g(\nabla_\tau \vz, \vv)+g(\partial_\tau \gamma(\tau, s), \nabla_s\vv)]_{\tau=a', s=0}=0~.
\ee
Notice that $\vu\equiv \vz|_{\tau=a', s=0}$ is tangent to $\S$ at $\gamma(a', 0)$ and $\nabla_s\vv=\nabla_{\vu}\vv$ with both $\vu$ and $\vv$ tangent vectors to $\S$.  Furthermore, $\vn\equiv \partial_\tau \gamma(\tau, s)|_{\tau=a', s=0}$ is normal to $\S$ and as $\tau$ is the affine parameter,  it can always by chosen to be normalised to $-1$.  But the normal component  of $\nabla_{\vu}\vv$ is given by the second fundamental form $K$ of $\S$, i.e.  $g(\vn, \nabla_{\vu}\vv)=-K(\vu, \vv)$. As the above equation must hold for every $\vv$, one derives (\ref{jfbc}).
\hfill{$\Diamond$}
\end{remark}

\begin{remark}
As it was in the case of theorem \ref{th:conjugategeodesic1}, the above theorem does not prove the existence of a maximum geodesic emanating for $\S$ and ending at $p$. It says that if such a  geodesic with maximum proper time were to exist, it cannot have a focus point between $\Sigma$ and $p$.  Moreover, if it has such a focus point, it cannot be maximal.\hfill{$\Diamond$}
\end{remark}

%\subsubsection{Null geodesics}

\subsection{Existence of maximal proper time geodesics}

The proof of existence of geodesics with maximal proper time between two causal related points or a hypersurface and a point in a spacetime proceeds in several steps. One of main difficulties is that the proper time functional $L_g$ is not a continuous function on either $\mC(p,q)$ or $\mC(\S, q)$. The typical way to show this is to observe that any time-like curve $\gamma$ can be approximated with a null curve possibly with corners. So if $L_g(\gamma)=\tau>0$, within a small neighbourhood of $\gamma$, there is a piecewise smooth null curve $\lambda$ with $L_g(\lambda)=0$.  Such a behaviour is not that of a continuous function.  However, $L_g$ is  upper semi-continuous which sufficient for the proof of the main statement.

\subsubsection{Semi-continuity of proper time functional} %$L_g$ is upper semi-continuous}

\begin{prop}\label{prop:usemi}
Let $\M$ be a strongly causal spacetime and $p,q$ be points in $\M$ with $q\in I^+(p)$. Then, the proper time functional  $L_g$ is upper semi-continuous in $\widetilde {\mC(p,q)}$.
\end{prop}
\begin{proof}

To prove the upper semi-continuity\footnote{There is a definition of upper semi-continuity using   sequences, see definition \ref{def:semicon}. In the present case given a sequence $(\gamma_n)$ that converges to $\gamma$,  $L_g$ is upper semi-continuous if $\lim_{n\rightarrow \infty}\sup L_g(\gamma_n)\equiv \lim_{n\rightarrow \infty} \sup\{ L_g(\gamma_k), k\geq n\}\leq L_g(\lambda)$.  } of $L_g$, we have to demonstrate that for every $\lambda\in \widetilde {\mC(p,q)}$, there is an open neighbourhood $U_\epsilon$ of $\lambda$, $\lambda\subset U_\epsilon$, $\epsilon>0$, such that for every $\gamma\subset U_\epsilon$
\be
L_g(\gamma)\leq L_g(\lambda)+\epsilon~.
\ee
First, we parameterise $\lambda$ with its proper time $\tau$. Then, as the image of $\lambda$ is compact, we cover it with normal convex neighbourhoods and extract a finite subcover.
   Next, consider a point $r\in \lambda$ and a normal neighbourhood $N_r$ of $r$. We foliate $N_r$  with spacetime hypersurfaces and  label them  by $\tau$. These hypersurfaces are constructed by considering space-like geodesics emanating orthogonally from each point of $\lambda$ contained in $N_r$.  Doing this for all the normal convex neighbourhoods of the finite cover of $\lambda$, we define an open set $U$, contained in the union of all normal convex neighbourhoods of $\Lambda$, and
a function $F$ on $U$ by setting
\begin{align}
F\,:\,\, &U\rightarrow \bR
\cr
&p\rightarrow F(p)=\tau~,
\end{align}
 where $\tau$ labels the hypersurface that passes through the point $p\in U$.

Clearly, $F(\lambda(\tau))=\tau$ and so $dF(\vec{\dot\lambda})=1$. As $dF^\flat\vert_\lambda$ and $\vec{\dot\lambda}$ are parallel choose $\vec{\dot\lambda}=-\partial^\mu F\partial_\mu\vert_\lambda$ which implies that $g^{\mu\nu}\partial_\mu F\partial_\nu F\vert_{\lambda}=-1$ provided the tangent vector field of $\lambda$ has been normalised to $-1$; clearly in the proper time parameterisation of $\lambda$, the speed of the curve is constant.

Then, parameterise nearby future directed curves $\gamma$ contained in $U$ with $F$.  In such a case,
\be
\vec{\dot\gamma}= f g^{\mu\nu}\partial_\mu F \partial_\nu|_{\gamma} + \vw~,
\ee
where $dF(\vw)=0$ and $\vw$ is spacelike. The function $f$ can be determined by taking
\be
dF(\vec{\dot\gamma})=1=f (g^{\mu\nu}\partial_\mu F \partial_\nu F)|_\gamma= f g^{-1}(dF, dF)|_{\gamma}~,
\ee
where the first equality follows because $\gamma$ is parameterised by $F$.  This gives
\be
\vec{\dot\gamma}=\frac{1}{g^{-1}(dF, dF)|_{\gamma}} (g^{\mu\nu}\partial_\mu F \partial_\nu)|_{\gamma}+\vw~,
\ee
and in turn we have
\be
g(\vec{\dot\gamma}, \vec{\dot\gamma})=\frac{1}{g^{-1}(dF, dF)|_{\gamma}}+g(\vw,\vw)\Longrightarrow \sqrt{-g(\vec{\dot\gamma}, \vec{\dot\gamma})}\leq \frac{1}{\sqrt{-g^{-1}(dF, dF)\vert_\gamma}}~.
\label{est11}
\ee
The 1-form $dF$ is continuous. Thus $\sqrt{-g^{-1}(dF, dF)}$ is also continuous and $\sqrt{-g^{-1}(dF, dF)}\vert_\lambda=1$. This implies that for every $\delta>0$, there is a $U_\delta\subset U$, such that
\be
|\big(-g^{-1}(dF, dF)\big)^{-1/2}-\big(-g^{-1}(dF, dF)\vert_{\lambda}\big)^{-1/2}|=|\big(-g^{-1}(dF, dF)\big)^{-1/2}-1|\leq \delta
\ee
i.e.
\be
-\delta+1\leq \big(-g^{-1}(dF, dF)\big)^{-1/2}\leq \delta+1~.
\ee
Substituting this in (\ref{est11}), we find that
\be
\sqrt{-g(\vec{\dot\gamma}, \vec{\dot\gamma})}\leq 1+\delta~.
\ee
Choosing $\delta=\epsilon/L_g(\lambda)$, $\epsilon>0$,  and integrating the above equation  over $\tau$, we find that
\be
L_g(\gamma)\leq L_g(\lambda)+\epsilon~,
\label{semisemisemi}
\ee
for $\gamma\in U_\epsilon$, which proves the upper semi-continuity of the proper time functional $L_g$.

\end{proof}

\begin{remark}
It turns out that $L_g$ is upper semi-continuous not only on $\widetilde {\mC(p,q)}$ but also on the space of all continuous causal curves $\mC(p,q)$ between the points $p$ and $q$ of a strongly causal spacetime $\M$.  Moreover, $L_g$ is also upper semi-continuous on $\mC(\S, q)$ provided that $\S$ is an achronal set in  a strongly causal spacetime $\M$. \hfill{$\Diamond$}
\end{remark}

\subsubsection{Geodesics}

Before we demonstrate that under certain conditions there are curves with maximal proper time between two causally related points $p$ and $q$ in a spacetime $\M$, we shall first produce an argument  to show that if  such curves exist  must be geodesics. In theorem \ref{th:conjugategeodesic1}, we have demonstrated  that amongst smooth curves the geodesics without conjugate points  maximise  the proper time between two points $p$ and $q$ in a  spacetime $\M$. The question that remains to be addressed is whether there are other continuous curves, not necessarily smooth,  that maximise the proper time between the points $p$ and $q$. First, let us answer this questions within a normal neighbourhood, see \cite{HE}.

\begin{prop}\label{prop:geodesicnormal}
Let $p$ and $q$ be points in a convex normal neighbourhood $U$ of the spacetime $\M$. If there is a causal curve $\gamma$ passing through $p$ and $q$ with $\gamma\in U$, then the curve with maximum proper time between $p$ and $q$ is the unique geodesic in $U$ that joins $p$ and $q$.
\end{prop}
\begin{proof}
Consider the map
\be
\varphi: D\subset \bR^2\rightarrow U
\ee
such that $\varphi(s, t)=\exp_p(s \vu(t))$, where $\exp_p$ is the exponential map, $\exp_p{\vec 0}=p$  and $g(\vu(t), \vu(t))=-1$, i.e. the vector field $\vu$ is time-like. Clearly for $t=t_0$, $\varphi$ is a geodesic.  Moreover the vector fields\footnote{The notation $\frac{\partial}{\partial t}|_\varphi$ means the push-forward of $\frac{\partial}{\partial t}$ under the map $\varphi$ and similarly for $\frac{\partial}{\partial s}|_\varphi$.} $\frac{\partial}{\partial t}|_\varphi$ for $s=s_0 $ and $\frac{\partial}{\partial s}|_\varphi$ for $t=t_0$ are orthogonal.  Indeed,
\be
\nabla_s g(\frac{\partial}{\partial s}|_\varphi, \frac{\partial}{\partial t}|_\varphi)=g(\frac{\partial}{\partial s}|_\varphi, \nabla_s\frac{\partial}{\partial t}|_\varphi)=\nabla_t g(\frac{\partial}{\partial s}|_\varphi, \frac{\partial}{\partial s}|_\varphi)=0~,
\ee
as the length of the tangent vector of geodesics is constant. Thus,  $g(\frac{\partial}{\partial s}|_\varphi, \frac{\partial}{\partial t}|_\varphi)$ depends only on $t$. But the vector field $\frac{\partial}{\partial t}|_\varphi$ at $s=0$ vanishes for every $t$. This in turn implies that $g(\frac{\partial}{\partial s}|_\varphi, \frac{\partial}{\partial t}|_\varphi)=0$.

The map $\varphi$ can be seen as a 1-parameter deformation of the geodesic $\lambda$ at $t=t_0$. The other causal curves in the deformation can be expressed as $\gamma(t)=\varphi(f(t), t)$. In such a case
\be
\vec{\dot\gamma}=\dot f  \frac{\partial}{\partial s}+\frac{\partial}{\partial t}~.
\ee
Thus
\be
g(\vec{\dot\gamma}, \vec{\dot\gamma})=- (\dot f)^2+ g(\frac{\partial}{\partial t}, \frac{\partial}{\partial t})\geq - (\dot f)^2~.
\ee
Therefore
\be
L_g(\gamma; p,q)\leq \int_p^q \dot f=L_g(\lambda; p,q)~,
\ee
with equality to hold iff $\frac{\partial}{\partial t}|_\varphi=0$ in which case $\gamma$ is a geodesic.  The proof applies to all piecewise  smooth curves $\gamma$. It only requires that $\dot f$ is defined almost everywhere.
\end{proof}

\begin{remark}
The above proof can be extended to all continuous causal curves contained in a convex normal neighbourhood $U$ of $\lambda$ with end points $p,q\in U$.  By the upper semi-continuity  $L_g$, one has that
\be
L_g(\gamma)\leq L_g(\lambda)~,
\ee
for every other continuous causal curve $\gamma$ in $U$, where we have neglected $\epsilon$ that appears in (\ref{semisemisemi}). This proves the statement with the only remaining issue   to be uniqueness. To settle this within $U$ suppose that $\gamma\not= \lambda$ with $L_g(\gamma)= L_g(\lambda)$. If $r\in \lambda$ and $r\notin \gamma$, consider the geodesics $\lambda_1$ and $\lambda_2$ from $p$ to $r$ and from $r$ to $q$. From the above semi-continuity argument $\lambda_1$ maximises the length of $\gamma$ from $p$ and $r$ and similarly $\lambda_2$ from $r$ to $p$. Thus
\be
L_g(\lambda_1;p,r)+L_g(\lambda_1;r,q)\leq L_g(\gamma;p,q)=L_g(\lambda;p,q)~.
\ee
But this  contradicts proposition \ref{prop:geodesicnormal} as the geodesic $\lambda$ maximises the proper time amongst all piece-wise smooth causal curves from $p$ to $q$. Thus $\lambda$ within the convex neighbourhood $U$ is unique.\hfill{$\Diamond$}
\end{remark}

\begin{remark}
More generally, one can show that geodesics without conjugate points maximise the proper time between two points $p$ and $q$ in a strongly causal spacetime.  A heuristic argument to demonstrate this  is that if at some point $r$ the curve $\lambda$ that maximises the proper time is not a geodesic within the normal neighbourhood of $r$,  one can construct a geodesic which will have greater proper time and use it to construct another curve $\tilde\lambda$ between $p$ and $q$ that has proper time greater than $\lambda$. This contradicts the assumption that $\lambda$
is the curve with the greatest proper time.  Thus $\lambda$ has to be a geodesic.

Even though we focused on the geodesics between two points $p$ and $q$ of a spacetime, the same applies for the geodesics between a set $\S$ and a point $q$. The curve that maximises the proper time between a spacelike hypersurface  $\S$ and a point $q$ is a spacetime is a geodesic orthogonal to $\S$ without a focus point. \hfill{$\Diamond$}
\end{remark}

\subsubsection{Existence of maximal proper time curves}

\begin{theorem}\label{th:maxgeodesic1}
Let $\M$ be a globally hyperbolic spacetime  and  points $p, q\in \M$. Then, there exists a curve $\lambda\in \mC(p,q)$ such that $L_g(\lambda;p,q)=L_g(p,q)$, i.e. $\lambda$ attains the maximum proper time between $p$ and $q$.
\end{theorem}
\begin{proof}
To outline a proof, recall that $\mC(p,q)$ is compact. From the definition \ref{def:properdist} of $L_g(p,q)$, there is a sequence of curves $(\gamma_n)$ such that their proper time
$L_g(\gamma_n;p,q)$ converges to $L_g(p,q)$.  As $\mC(p,q)$ is compact, $(\gamma_n)$ has an accumulation point $\lambda$ and there is a subsequence $(\gamma_{n_k})$ that converges to $\lambda$.  As $\lambda\in \mC(p,q)$, i.e. $\lambda$ is a causal curve between $p$ and $q$, it has proper time between $p$ and $q$ that is at most $L_g(p,q)$.  Therefore,
\be
L_g(\lambda; p,q)\leq L_g(p,q)~.
\ee
However, $L_g$ is upper semi-continuous.  This implies, after possibly choosing a further subsequence, that
\be
L_g(\gamma_{n_k}; p,q)\leq L_g(\lambda; p,q)~,
\ee
which upon taking the limit
\be
 L_g(p,q)\leq L_g(\lambda; p,q)~.
 \ee
 Therefore, one concludes that $L_g(p,q)= L_g(\lambda; p,q)$ and $\lambda$ attains the maximum proper time between $p$ and $q$.
\end{proof}

\begin{theorem}\label{th:maxgeodesic2}
Let $\M$ be a globally hyperbolic spacetime with a Cauchy surface $\Sigma$ and a point $q\in \M$. Then, there exists a curve $\lambda\in \mC(\Sigma, q)$ such that $L_g(\lambda;\Sigma, p)$ attains the value of maximum proper time between $\Sigma$ and $q$.
\end{theorem}
\begin{proof}
The proof is similar to that of the previous theorem concerning existence curves with maximal proper time between two points of a spacetime.
\end{proof}

Let us summarise the application of results we have obtained in this section with those in section \ref{sec:timeconjfocus} on globally hyperbolic manifolds. These are summarised without proof in the following theorem -- the proof is straightforward as such a manifold is strongly causal and both $\mC(p,q)$ and $\mC(\S, q)$ are compact spaces of causal curves.

\begin{theorem}\label{th:sumhyperbolic}
(i)~ Let $\M$ be a globally hyperbolic spacetime and two points $p,q\in \M$ with $q\in I^+(p)$. Then, there is a time-like curve $\lambda$ that attains the maximum proper time between $p$ and $q$.  Moreover, $\lambda$ is a  geodesic with no conjugate points between $p$ and $q$.

(ii)~Similarly, given a Cauchy surface $\Sigma$ in a globally hyperbolic spacetime $\M$ and a point $q\in \M$, there is a curve $\lambda$ that attains the maximal proper time between $\Sigma$ and $q$. Moreover, $\lambda$ is a time-like geodesic, orthogonal to $\Sigma$ without a focus point between $\Sigma$ and $q$. \hfill{$\square$}
\end{theorem}

\section{Singularity Theorems}

We shall demonstrate the proof of the Hawking cosmological singularity theorem and then the Penrose black hole singularity theorem. Both require to assume that the spacetime is globally hyperbolic. This is thought as a rather strong restriction. However, there are several generalisations that remove this assumption -- one such theorem is demonstrated below.  Nevertheless, these two theorems are  key results in the subject and the point of entry in the investigation of singularities in general relativity.

\subsection{Singularity theorems and spacelike hypersurfaces}

\begin{figure}[htbp]
\begin{center}
\begin{eqnarray*}
\begin{tikzpicture}[scale=0.6]
  % Coordinates
  \coordinate (A) at (0.00, 0.00);
  \coordinate (B) at (7.00, 0.00);
  \coordinate (C) at (2.75, -1.25);
  \coordinate (D) at (2.00, 2.25);
  \coordinate (E) at (6.00, 0.50);
  \coordinate (F) at (3.98, -5.01);
  \coordinate (G) at (5.00, -1.50);
  \coordinate (H) at (6.00, -3.50);
  \coordinate (I) at (2.00, 0.00);
  \coordinate (J) at (4.96, -3.79);
  \coordinate (K) at (2.75, -1.50);
  \coordinate (L) at (5.60, -1.47);

  % Drawing
  \draw[very thick] (A) .. controls (2.75, -1.25) and (2.00, 2.25) .. (B);
  \draw[red, thick] (E) .. controls (5.00, -1.50) and (6.00, -3.50) .. (F);
  \draw[blue, thick] (I) .. controls (2.75, -1.50) and (5.60, -1.47) .. (J);
  \fill[black] (F) circle (2pt);
  \fill[blue] (J) circle (2pt);
   \node[below] at (F) {$p$};
    \node[above] at (1.00,0.50) {$\Sigma$};
     \node[below, blue] at (3.00,-2.00) {$\tau$};
     \node[right, red] at (L) {$\lambda$};
     \node[below right] at (J) {$r$};
\end{tikzpicture}
\end{eqnarray*}
\end{center}
%\begin{figure}[htbp]
\caption{\small{The figure illustrates the proof of theorem \ref{th:singularity1}.  Both past directed time-like geodesics $\lambda$ and $\tau$ emanate orthogonally from the Cauchy hyper-surface $\Sigma$ and it is assumed that $\lambda$ is the maximal geodesic from $\Sigma$ to $p$. The point $p$ lies on $\lambda$ at proper time greater that $3|C|^{-1}$. But the condition on the extrinsic curvature of $\Sigma$ implies that  there must be a focus point $r$ between $\Sigma$ and $p$,  this leads to a contradiction because the geodesic $\lambda$ cannot be maximal.  }}
\centering
\label{fig:sin1}
\end{figure}
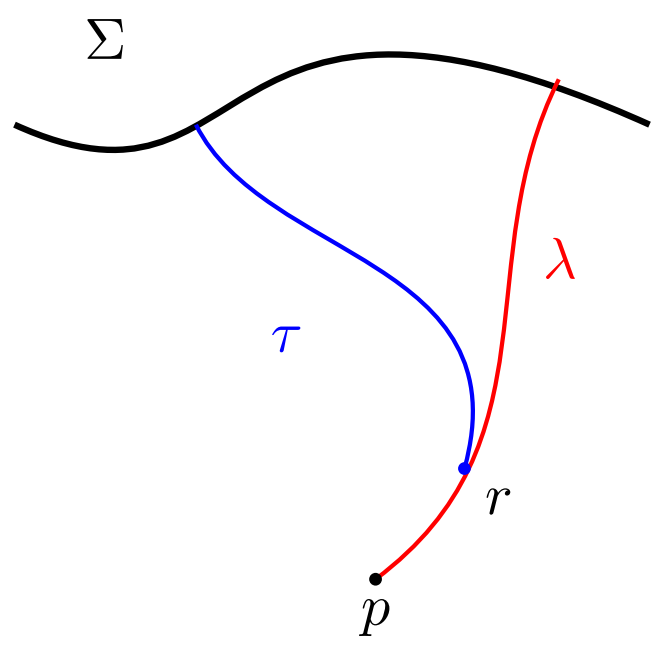

\begin{theorem}\label{th:singularity1}
Let $\M$ be a globally hyperbolic spacetime with $R_{\mu\nu} \vu^\mu\vu^\nu\geq 0$ for all time-like vector fields $\vu$. Assuming that there is an at least $C^2$ spacelike Cauchy surface $\Sigma$ with the trace of extrinsic curvature negative $K\leq C<0$. Then no past directed time-like curve from $\Sigma$ can have proper time greater  than $3|C|^{-1}$. In particular, all past directed time-like geodesics emanating from $\Sigma$ are incomplete.
\end{theorem}
\begin{proof}
The expansion $\theta$ of the past directed geodesic congruence emanating from $\Sigma$ evaluated at $\Sigma$ coincides with the trace of the trace of the extrinsic curvature of $\Sigma$, $\theta\vert_{\Sigma}=K\leq 0$. Therefore, all the geodesics of a congruence develop a focus point at most at affine distance $3|C|^{-1}$ from the Cauchy surface.

 Suppose that there is a past directed time-like curve $\lambda$ emanating from $\Sigma$ to a point $p\in \M$ with proper time more than  $3|C|^{-1}$. From theorem \ref{th:maxgeodesic2} or \ref{th:sumhyperbolic}, there must be  \emph{ maximal} geodesic $\gamma$ emanating orthogonally from $\Sigma$ and ending at $p$.  However, no such maximal geodesic can exist.  This is because if such a $\gamma$ were to exist, it would have proper time greater than  $3|C|^{-1}$.  As a result, $\gamma$ has to have a focus point between $\Sigma$ and $p$ and therefore it \emph{cannot be a maximal } geodesic as a consequence of theorem \ref{th:focusgeodesic1}.  This is a contradiction. Thus, the past directed time-like  geodesics emanating  from $\Sigma$ cannot extend to $p$ and the spacetime is geodesically incomplete.
It follows that all past directed time-like curves of $\M$ emanating from $\Sigma$  have to have proper time less than $3|C|^{-1}$ as they must be shorter than the maximal geodesics.
\end{proof}

\begin{figure}[htbp]
\begin{center}
\begin{eqnarray*}
\begin{tikzpicture}[scale=0.9]
  % Coordinates
  \coordinate (A) at (0.00, 0.00);
  \coordinate (B) at (10.00, 0.00);
  \coordinate (C) at (2.75, 1.50);
  \coordinate (D) at (6.50, -1.50);
  \coordinate (E) at (-1.00, -1.00);
  \coordinate (H) at (5.00, -7.00);
  \coordinate (F) at (11.00, -1.00);
  \coordinate (G) at (3.00, -5.00);
  \coordinate (I) at (9.00, -0.25);
  \coordinate (J) at (6.15, -4.75);
  \coordinate (K) at (9.00, -2.50);
  \coordinate (L) at (8.02, -0.45);
  \coordinate (M) at (5.00, -4.25);
  \coordinate (N) at (7.50, -2.25);
  \coordinate (O) at (6.75, -0.34);
  \coordinate (P) at (6.00, -2.50);
  \coordinate (Q) at (7.00, -1.25);
  \coordinate (R) at (9.75, -0.10);
  \coordinate (S) at (5.25, -5.75);
  \coordinate (T) at (8.75, -4.25);
  \coordinate (U) at (5.20, -4.50);
  \coordinate (V) at (4.49, -4.25);
  \coordinate (W) at (3.86, -4.25);
  \coordinate (X) at (7.25, -0.45);
  \coordinate (Y) at (5.50, -3.25);
  \coordinate (Z) at (7.25, -2.00);
  \coordinate (AA) at (6.00, -0.25);
  \coordinate (AB) at (5.00, -2.75);
  \coordinate (AC) at (5.75, -2.00);
  \coordinate (AD) at (5.50, -0.15);
  \coordinate (AE) at (4.25, -2.75);
  \coordinate (AF) at (5.25, -1.75);
  \coordinate (AG) at (5.25, -0.09);
  \coordinate (AH) at (3.50, -2.25);
  \coordinate (AI) at (4.75, -1.50);
  \coordinate (AJ) at (4.75, 0.00);
  \coordinate (AK) at (2.75, -3.00);
  \coordinate (AL) at (4.25, -1.75);

  % Drawing
  \draw[very thick] (A) .. controls (2.75, 1.50) and (6.50, -1.50) .. (B);
  \draw[very thick, green] (E) -- (H);
  \draw[very thick, green] (H) -- (F);
  \draw[thick] (G) .. controls (6.15, -4.75) and (9.00, -2.50) .. (I);
  \draw[thick] (G) .. controls (5.00, -4.25) and (7.50, -2.25) .. (L);
  \draw[thick] (G) .. controls (6.00, -2.50) and (7.00, -1.25) .. (O);
  \draw[thick] (G) .. controls (5.25, -5.75) and (8.75, -4.25) .. (R);
  \draw[blue, thick] (J) .. controls (5.50, -3.25) and (7.25, -2.00) .. (X);
  \draw[blue, thick] (U) .. controls (5.00, -2.75) and (5.75, -2.00) .. (AA);
  \draw[blue, thick] (V) .. controls (4.25, -2.75) and (5.25, -1.75) .. (AD);
  \draw[blue, thick] (W) .. controls (3.50, -2.25) and (4.75, -1.50) .. (AG);
  \draw[red, thick] (G) .. controls (2.75, -3.00) and (4.25, -1.75) .. (AJ);
  \fill[black] (G) circle (2.5pt);
  \fill[black] (J) circle (2pt);
  \fill[black] (U) circle (2pt);
  \fill[black] (V) circle (2pt);
  \fill[black] (W) circle (2pt);
  \fill[blue] (AG) circle (2pt);
  \fill[blue] (AD) circle (2pt);
  \fill[blue] (AA) circle (2pt);
  \fill[blue] (X) circle (2pt);
  \fill[red] (AJ) circle (2pt);
   \node[above, blue] at (AJ) {$r$};
   \node[above, blue] at (5.70, -0.09) {$(r_n)$};
   \node[above] at (7.90, -2.30) {$(\lambda_n)$};
    \node[above, blue] at (6.35, -1.20) {$(\gamma_n)$};
     \node[below left] at (3.00, -5.00) {$p$};
      \node[below right, green] at (8.00, -4.50) {$H^-(\S)$};
      \node[above] at (2.00, 0.80) {$\S$};
      \node[right, red] at (3.00, -2.00) {$\lambda$};
       %\node[right] at (5.20, -4.65) {$(q_n)$};
       \node[right] at (2.92, -4.00) {$(q_n)$};
        \node[below right] at (1.00, -1.00) {$D^-(\S)$};

\end{tikzpicture}
\end{eqnarray*}
\end{center}
%\begin{figure}[htbp]
\caption{\small{The figure depicts the construction of the maximal geodesic  $\gamma$ of theorem \ref{th:singularity2} that begins from the point $p$ of $H^-(\S)$ and ends at the point $r$ of $\S$. }}
\centering
\label{fig:sing2}
\end{figure}
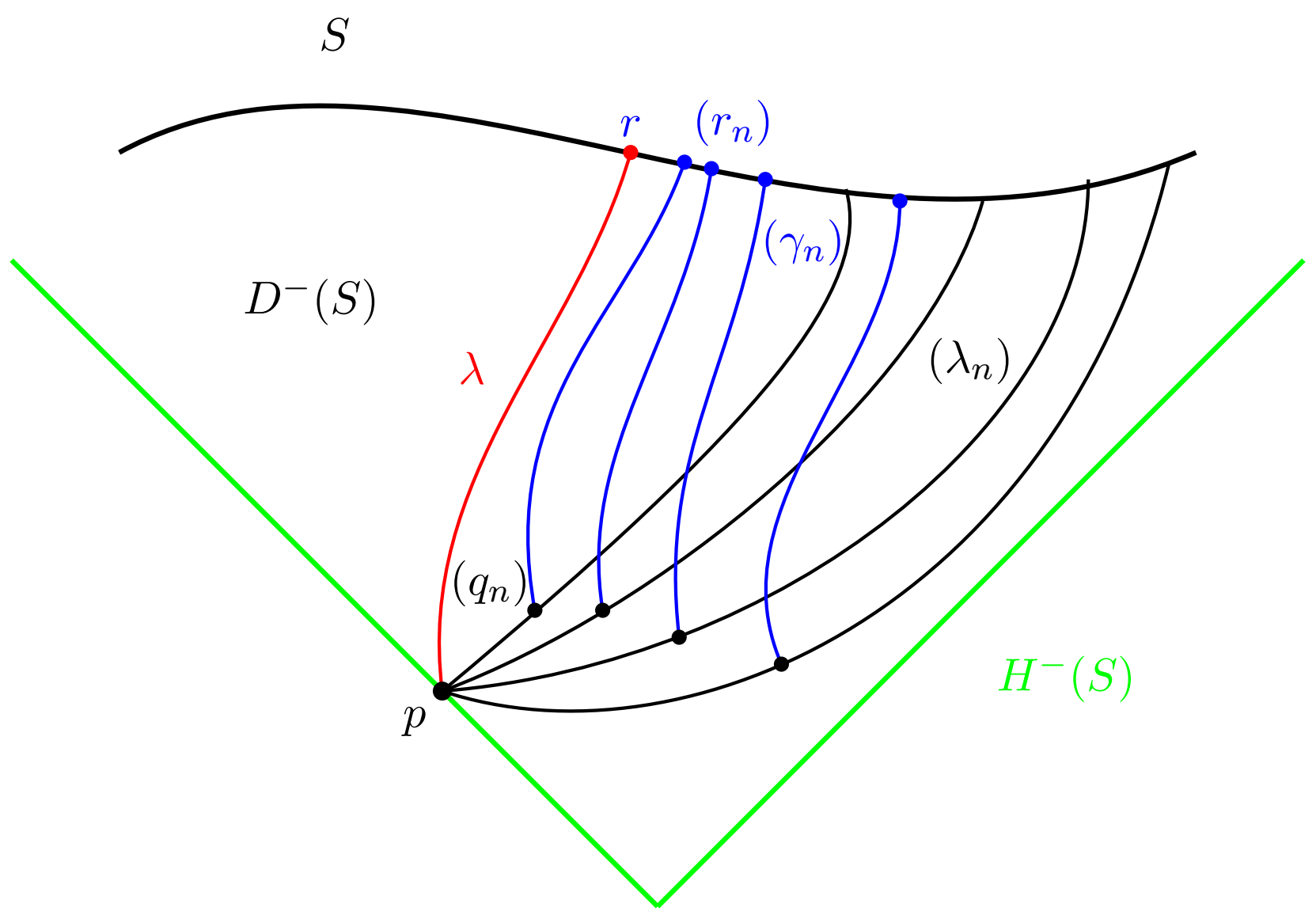

There is a adaptation of this theorem to spacetimes that are not globally hyperbolic.  It is also possible to remove the assumption that $\M$ is strongly causal mentioned below in the assumptions of the theorem  but we shall not explore this here.

\begin{theorem}\label{th:singularity2}
Let $\M$ be a strongly causal spacetime with an edgeless,  achronal, compact, smooth, spacelike hypersurface $\S$ and $R_{\mu\nu} \vu^\mu\vu^\nu\geq 0$ for all time-like vector fields $\vu$. If the trace $K$ of the second fundamental form of $\S$ is negative everywhere on $\S$ and $C$ is the maximum value of $K$, $K\leq C<0$.  Then at least one past directed time-like geodesic emanating from $\S$ has proper time no greater that $3|C|^{-1}$, i.e. it is incomplete.
\end{theorem}
\begin{proof}
Assuming that all past directed inextendible time-like geodesics emanating from $\S$ have proper time greater than $3|C|^{-1}$,  the main task to establish this theorem is to demonstrate that the past Cauchy horizon of $\S$, $H^{-}(\S)$ is compact. Then, this  gives a contradiction as theorem \ref{th:cauchyhorizons} implies that  every point $p\in H^-(\S)$ lies in a future inextendible null geodesic because $\S$ is edgeless  while theorem \ref{th:compactendpoints} implies that such geodesics must have ends points in $H^-(\S)$. The resolution of the contradiction is that the original assumption that all past directed inextendible time-like geodesics emanating from $\S$ have proper time greater than $3|C|^{-1}$ is false, i.e. one of them at least must be incomplete.

First, let us establish that $H^-(\S)$ is non-empty.  As $\S$ is compact, the trace of the second fundamental form of $\S$ is bounded and it has a maximum $C<0$.  For this consider $\N=\mathrm{int} D(\S)$ as a spacetime, i.e.  without viewing it as a subspace of $\M$.  In such a case, $\S$ is a Cauchy of surface of $\N$ and $\N$, as a spacetime,  satisfies all the requirements of the previous singularity theorem \ref{th:singularity1}.  As a result all past directed inextendible time-like curves emanating from $\S$ cannot have proper time greater than $3|C|^{-1}$, i.e. those points beyond $3|C|^{-1}$ are inaccessible from past inextendible causal curves from  $\S$ contained in $\N$. Thus, $\N$ has a non-trivial Cauchy horizon\footnote{The Cauchy horizon of $\N$, as globally hyperbolic spacetime, is empty. But $\S$ is not assumed to be a Cauchy surface in $\M$ and as such it can have an non-empty past Cauchy horizon $H^-(\S)$.} $H(\S)$ and all the past inextendible time-like geodesics must intersect $H^-(\S)$ before their length becomes greater than $3|C|^{-1}$.

Next consider $p\in H^-(\S)$.  As all future directed inextendible time-like curves $\lambda$ from $p$ intersect $\S$ and their proper time is bounded by  $3|C|^{-1}$, define
\be
\tau_0=\sup_{\lambda\in\widetilde {C(\S, p)}} L(\lambda)<\infty ~,~~~
\ee
where $\widetilde {C(\S, p)}$ are the smooth future directed time-like curves from $p$ to $\Sigma$. First prepare a sequence $(\lambda_n)$ in $\widetilde {C(\S, p)}$ such that
\be
\lim_{n\rightarrow\infty} L(\lambda_n) = \tau_0~.
\ee
Such a sequence exists from the definition of the least upper bound but this does not imply that an actual curve exist that has  proper time $\tau_0$. To establish the latter consider a sequence of points $(q_n)$ with $q_n\in \lambda_n$ whose limit is $p$. As each $q_n$ is a point in the globally hyperbolic spacetime $N$, there is a maximal geodesic $\gamma_n$ with emanates orthogonally for $\S$ and ends at $q_n$. Each $\gamma_n$ intersect $\S$ at a point $r_n$. As $(r_n)$ is a sequence of points in $\S$ and $\S$ is compact, the sequence $(r_n)$ has an accumulation point $r$ in $\S$. The orthogonal geodesic $\gamma$ from $r\in \S$ is maximal and ends at $p$. The latter is a consequence of the continuity of geodesics from their initial point and tangent vector.  This in particular implies that
\be
L(\gamma)=\tau_0~.
\ee
Having established the existence of a maximal geodesic from every point $p\in H^-(\S)$ to $\S$, we consider next a sequence of points $(p_n)$ in $H^-(\S)$. The maximal geodesic $\beta_n$ from $p_n$ to $\S$ intersects (orthogonally) $\S$ at $w_n$. Then, one repeats the previous argument to claim that the sequence $(w_n)$ has an accumulation point $w$ in $\S$, as $\S$ is compact. In turn the sequence $(p_n)$ has an accumulation point $p\in H^-(\S)$ which is the intersection of  the maximal geodesic $\beta$ emanating orthogonally from $w\in \S$ with $H^-(\S)$.  Thus, every sequence of points  $(p_n)$ in $H^-(\S)$ has a converging subsequence and so $H^-(\S)$ is compact.
\end{proof}

\subsection{Singularity theorems and trapped surfaces}

The two singularity theorems stated above are thought to apply to a cosmological setting.  In particular, the second establishes  the existence  of  singularities in a closed universe. The theorem described  below due to Penrose is associated  with the investigation of  singularities in the context of black holes and gravitational collapse. It relies on the notion of trapped surfaces.

\subsubsection{Trapped surfaces}

To explain what trapped surfaces are consider geodesic congruences generated by null geodesics emanating from a 2-dimensional spacelike submanifold $\T$ of the spacetime $\M$. Assuming that their normal 2-dimensional bundle of $\T$ in $\M$ is trivial, one can span it with two null future directed vector fields $\vk^{\pm}=\vt\pm \vw$, where $\vt$ is a future directed normalised time-like vector field and $\vw$ is a normalised space-like vector field. Usually, $\vk^+$ is referred to as the outgoing normal vector while $\vk^-$ is referred to as the ingoing normal vector -- in the example below we shall illustrate this terminology.
We shall refer to the null hypersurfaces generated by the null geodesics emanating from $\T$ with tangent vector either $\vk^+$ or $\vk^-$   as null geodesic congruences despite the fact that they span subspaces of $\M$ of co-dimension one.  However, their union is a geodesic congruence. One can define the second fundamental forms of these congruences with respect to both $\vk^\pm$ and the associated expansions $\theta^\pm$ as for null geodesic congruences in section \ref{ssec:nullcongruences}.

\begin{definition}
A trapped surface $\T$ in the spacetime $\M$ is a co-dimension 2 space-like compact submanifold such that both ingoing and outgoing null geodesic congruences have everywhere negative expansion $\theta^\pm<0$.\hfill{$\triangle$}
\end{definition}

  \begin{example}Take $\T$ to be a two sphere $S^2$ centered at the origin of a spacelike hypersurface $t=\mathrm{const.}$ of Minkowski spacetime, which is identified with $\bR^3$.  Adopting the usual angular coordinates in $\bR^3$ with radial coordinate $r$, the two future directed, null, normals to $\T=S^2$ are $\vk^\pm=\partial_t\pm \partial_r$. From the perspective of the 2-sphere, $\vk^+$ is pointing outwards while $\vk^-$ is pointing inwards. This is a justification of the terms outgoing and ingoing null geodesic congruences.  Clearly, the expansion of the outgoing null geodesic congruence is $\theta^+>0$ and that of the ingoing is $\theta^-<0$. As $\theta^+>0$, the 2-sphere $S^2$ expands along the outgoing null geodesic congruence as the affine parameter of the geodesics increases while $S^2$ shrinks along the ingoing null geodesic congruence as the affine parameter of the geodesics increases. Clearly, this $S^2$ submanifold of Minkowski space is not a trapped surface. For an extensive investigation of trapped surfaces see \cite{Sentrap}.\hfill{$\triangledown$}
  \end{example}

\subsubsection{Conjugate points of null geodesics}

We have seen that if a time-like geodesic $\gamma$ from a point $p$ to a point $q$ has a conjugate point $r$ between $p$ and $q$, then it is not the curve with maximal proper time between $p$ and $q$. In particular, there are curves $\lambda$ near $\gamma$ from $p$ to $q$  with proper time that exceeds that of $\gamma$.  This can be seen as a deformation of $\gamma$ by  ``rounding up the corner'' to produce the new curves $\lambda$.

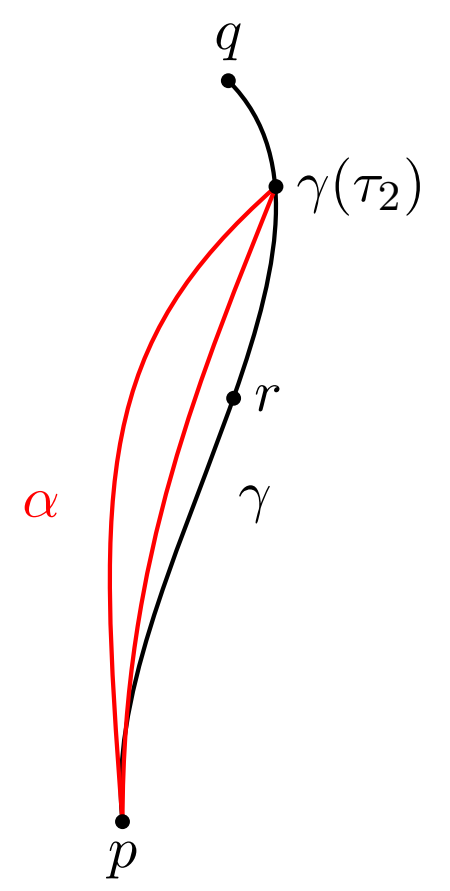
\begin{figure}[htbp]
\begin{center}
\begin{eqnarray*}
\begin{tikzpicture}[scale=0.7]
  % Coordinates
  \coordinate (A) at (0.00, 0.00);
  \coordinate (B) at (1.00, 7.00);
  \coordinate (C) at (-0.25, 2.25);
  \coordinate (D) at (2.50, 5.50);
  \coordinate (E) at (1.05, 4.00);
  \coordinate (F) at (1.45, 6.00);
  \coordinate (G) at (-0.25, 3.25);
  \coordinate (H) at (-0.25, 4.50);
  \coordinate (I) at (0.00, 2.50);
  \coordinate (J) at (0.75, 4.25);

  % Drawing
  \draw[thick] (A) .. controls (-0.25, 2.25) and (2.50, 5.50) .. (B);
  \draw[red, thick] (A) .. controls (-0.25, 3.25) and (-0.25, 4.50) .. (F);
  \draw[red, thick] (A) .. controls (0.00, 2.50) and (0.75, 4.25) .. (F);
  \fill[black] (A) circle (2pt);
  \fill[black] (B) circle (2pt);
  \fill[black] (E) circle (2pt);
  \fill[black] (F) circle (2pt);
   \node[right] at (F) {$\gamma(\tau_2)$};
    \node[right] at (E) {$r$};
     \node[above] at (B) {$q$};
      \node[below] at (A) {$p$};
        \node[right] at (0.90, 3.00) {$\gamma$};
    \node[left, red] at (-0.40, 3.00) {$\alpha$};
\end{tikzpicture}
\end{eqnarray*}
\end{center}
%\begin{figure}[htbp]
\caption{\small{The  figure depicts some general features of the proof of theorem \ref{th:nulldeformation}.  The null geodesic $\gamma$ from $p$ to $q$ with a conjugate point   $r$ to $p$  gets deformed with time-like curves $\alpha$. The null geodesics $\gamma$ is not achronal. }}
\centering
\label{fig:nulldef}
\end{figure}

A similar statement can be made for null geodesics. As ``rounding up the corner'' of a conjugate point of a null geodesic will produce a family of time-like curves, this can be expressed as follows:

\begin{theorem}\label{th:nulldeformation}
Let $\gamma$ be a smooth null geodesic from $p$ to $q$. Then, there does not exist a 1-parameter family deformation $\alpha$ of $\gamma$ with $\alpha(\tau, 0)=\gamma(\tau)$ and $\lambda_{s>0}=\alpha(\tau, s)$ time-like curves that pass through $p$ and $q$,  iff $\gamma$ is a null geodesic with no conjugate point between $p$ and $q$.
\end{theorem}
\begin{proof}
The proof of this statement has been given in \cite{HE}, proposition 4.5.12, with more proofs of the statement given in \cite{beem} and \cite{ONeill1983}.  The proof of this has a subtlety that one has to establish that the deformation is comprised by time-like curves.

 Here, we shall expand the proof of \cite{HE} and explicitly describe all the steps. Consider a null geodesic $\gamma$ from $p$ to $q$ and $r$ a conjugate point of $p$ between $p$ and $q$.   Set $\gamma(0)=p$, $\gamma(1)=q$ and $\gamma(\tau_0)=r$  -- it is assumed that $\gamma$ is parameterised with its affine parameter $\tau$.

By the stated assumptions, there is a Jacobi field $\vj$ such that
\be
\widehat{\vj}(0)=\widehat{\vj}(\tau_0)=0~,
\ee
and
\be
\nabla_\tau^2\widehat{\vj}+\widehat{R(\vj, \vk)\vk}=0
\ee
where $\vk$ is the tangent vector field of $\gamma$ and $\widehat{\vj}$ denotes the class of $\vj$ as section of $\widehat{ T\M}$, see section \ref{ssec:nullcongruences}. Consequently, $\vj$ has a space-like representative and $g(\vj, \vk)=0$.   Furthermore,
\be
\nabla_\tau \widehat{\vj}(\tau_0)\not =0~,
\ee
  as otherwise $\widehat{\vj}=0$ along the null geodesic $\gamma$. Similarly, $\nabla_\tau \widehat{\vj}(0)\not =0$.

  As $\widehat{\vj}$ does not vanish between $(0, \tau_0)$, we write
  \be
  \widehat{\vj}= f(\tau) \widehat{\vj}_n
  \ee
  where $g(\vj_n, \vj_n)=1$, i.e. $\vj_n$ is a normalised vector field. The function $f$ does not vanish in the interval $(0, \tau_0)$ and so it can be chosen to be positive
  \be
  f(\tau)>0~,~~~\tau\in (0,\tau_0)~.
  \ee
  However, for possibly a small interval  $\tau\in (\tau_0, \tau_1)$, $f$ will change sign, i.e.
  \be
  f(\tau)<0~,~~~\tau\in (\tau_0, \tau_1)~,~~~\tau_1<1.
  \ee
  Moreover, the geodesic deviation equation for $\vj$ and $g(\vj_n, \vj_n)=1$ give
  \be
  g(\vj, \nabla_\tau^2\widehat{\vj}+\widehat{R(\vj, \vk)\vk})=0\Longrightarrow \nabla_\tau^2 f+h f=0~,
  \ee
  where
  \be
  h=g(\widehat{\vj}_n, \nabla_\tau^2\widehat{\vj}_n+\widehat{R(\vj_n, \vk)\vk})~.
  \ee
  Let
  \be
  h_0=\mathrm{min} h(\tau)~,~~~\tau\in [0, \tau_1]
  \ee
  and choose $a$ such that $a^2+h_0>0$ for $\tau\in [0, \tau_1]$.

  Next, define
  \be
  \widehat{\vv} =r \widehat {\vj}_n
  \ee
  where
  \be
  r=b (e^{a\tau}-1)+f(\tau)~,~~~b=-\frac{f(\tau_1)}{e^{a\tau_1}-1}~.
  \ee
  Observe that $r(0)=0$ and $r(\tau_1)=0$.  Choose $0<\tau_2\leq \tau_1$, where $\tau_2$ is the first point after $0$ that $r$ vanishes. Observe that
  \be
  r(\tau)>0~,~~~\tau\in (0, \tau_2)
  \ee
  as $b>0$ because $f(\tau_1)<0$.

  It follows that
  \be
  Q(\tau)\equiv g(\widehat{\vv}, \nabla_\tau^2\widehat{\vj}+\widehat{R(\vj, \vk)\vk})= r (\frac{d^2}{d\tau^2} r+r h)=b r (e^{a\tau} (a^2+h)-h)~.
  \ee
  As $b>0$, $r(\tau)>0$ for $\tau\in (0,\tau_2)$ and
  \be
  e^{a\tau} (a^2+h)-h\geq a^2+h-h=a^2>0
  \ee
  one concludes that
  \be
  Q(\tau)>0~,~~~\tau\in (0, \tau_2)~.
  \ee
  One can always choose a lift $\tilde{\vv}$ of $\widehat{\vv}$ such that it vanishes at both $0$ and $\tau_2$. Indeed if
  \be
  \vv(0)= \lambda \vk(0)~,~~~\vv(\tau_2)=\mu \vk(\tau_2)~,
  \ee
 one can choose
  \be
  \tilde {\vv}(\tau)=\vv(\tau)-\mu \vk(\tau)+\frac{\mu-\lambda}{\tau_2} \tau \vk(\tau)~,
  \ee
  which vanishes at both $0$ and $\tau_2$.  From now on to simplify the notation, set $\vv=\tilde\vv$.

  To continue, choose a null vector field $\vz$ on $\gamma$ such that
  \be
  g(\vz, \vk)=-1~,~~~\nabla_\tau \vz=0~.
  \label{normconz}
  \ee
  Moreover, as $Q(\tau)$ is strictly positive in the compact interval $[\frac{1}{4} \tau_2, \frac{3}{4}\tau_2]$, consider
  \be
  \eta<\mathrm{min}\, Q(\tau)~,~~~\tau\in [\frac{1}{4} \tau_2, \frac{3}{4}\tau_2]~.
  \ee
    Then, define the continuous function
  \begin{align}
  \rho(\tau)&= -\eta \tau~,~~~\tau\in [0, \frac{1}{4}\tau_2]~;~~~
  \cr
  \rho(\tau)&= -\eta (\tau-\frac{1}{2}\tau_2)~,~~~\tau\in [\frac{1}{4} \tau_2, \frac{3}{4}\tau_2]~;~~~
  \cr
  \rho(\tau)&= \eta (\tau_2-\tau)~,~~~\tau\in [\frac{3}{4} \tau_2, \tau_2]~.
  \end{align}

  Now, we have prepared the ground to describe the family of curves that deform a null geodesic in the presence of a conjugate point.  For this consider that family $\alpha(\tau, s)$ of curves, where $s$ is the parameter. Set
  \be
  \vt(\tau, s)\equiv \partial_\tau \alpha~,~~~\vw(\tau, s)\equiv \partial_s \alpha(\tau, s)~,
  \ee
  where $\vt$ is the tangent vector field of the family and $\vw$ is the displacement vector field.  This is not an arbitrary family. Instead, it is required to have the following properties. First, one imposes that
  \be
  \vt(\tau, 0)=\vk~,~~~\vw(\tau, 0)=\vv~.
  \label{vvtau}
  \ee
  Furthermore, one prescribes that
  \be
  \nabla_s \vw(\tau, 0)\equiv \nabla_s \vw(\tau, s)|_{s=0}=\Big(g(\vv, \nabla_\tau\vv)-\rho(\tau)\Big) \vz~.
  \label{rhovw}
  \ee
  Clearly from (\ref{normconz}), we have
  \be
  g(\nabla_s \vw(\tau, 0), \vk)=-g(\vv, \nabla_\tau\vv)+\rho(\tau)~.
  \ee

  Let us prove now that for this family $F(\tau, s)\equiv g(\vt, \vt)<0$ for small enough $s$.  First,
  \be
  \frac{d}{ds} F(\tau, s)\vert_{s=0}=2 g(\vt, \nabla_s \vt)\vert_{s=0}=2 g(\vt, \nabla_\tau \vw)\vert_{s=0}= 2 g(\vk, \nabla_\tau \vv)=2\nabla_\tau g(\vk, \vv)=0~.
  \ee
  Thus the first derivative of $F$ along the variation parameter at the origin vanishes. Next consider the second variantion
  \begin{align}
  \frac{1}{2} \frac{\partial^2}{\partial s^2} F(\tau, s)\vert_{s=0}&= \Big(g(\nabla_s \vt, \nabla_s \vt)+ g(\vt, \nabla_s^2 \vt)\Big)\vert_{s=0}
  \cr
  &
  =\Big( g(\nabla_\tau\vw, \nabla_\tau\vw)+ g(\vt, \nabla_\tau\nabla_s\vw)+ g(\vt, R(\vw,\vt)\vw)\Big)\vert_{s=0}
  \cr
  &
  =\frac{d}{d\tau} \Big[g(\vt, \nabla_s \vw)+ g(\vw, \partial_\tau \vw)\Big]\vert_{s=0}- g(\nabla_\tau \vt, \nabla_s\vw)\vert_{s=0}- g(\vw, \nabla^2_\tau\vw+ R(\vw, \vt) \vt)\vert_{s=0}
  \cr
  &=\frac{d}{d\tau}\Big[ \frac{\partial}{\partial s} g(\vw, \vt)\Big]\vert_{s=0}-Q(\tau)=\frac{d}{d\tau}\rho(\tau)- Q(\tau)~,
  \label{Fss}
  \end{align}
  where we have used that $\nabla_\tau \vt\vert_{s=0}=0$ because $\gamma$ is a null geodesic, equation (\ref{vvtau}) and the prescription (\ref{rhovw}) to establish the last equality.

  The equation (\ref{Fss}) establishes that the second derivative of $F$ along the deformation parameter $s$ is universally bounded on the whole interval $[0, \tau_2]$ to be negative, i.e. there is a $c>0$ such that
  \be
  \frac{\partial^2}{\partial s^2} F(\tau, s)\vert_{s=0}< -c~,~~~\tau\in [0, \tau_2]~.
  \ee
  Using Taylor's theoren
  \be
  F(\tau, s)=g(\vk, \vk)+ \frac{d}{ds} F(\tau, s)\vert_{s=0} s+\frac{1}{2}\int^s_0 (s-u) \frac{\partial^2}{\partial u^2} F(\tau, u) du\leq -\frac{c}{2} \int^s_0 (s-u) du=-\frac{c}{4}<0~.
  \ee
  Thus,  the deformed curves for small $s$ are time-like. Considering one of these curves from $p$ to $\gamma(\tau_2)$ and continuing with $\gamma$ from $\gamma(\tau_2)$ to $q$, one constructs a causal curve, which is not null geodesics. This can be further deformed to a time-like curve from $p$ to $q$ that has greater proper length than a null geodesic.
  \end{proof}

\begin{remark}
One consequence of the above statement is that null geodesics with conjugate points are not achronal sets of a spacetime. Indeed, if a null geodesic $\gamma$ that passes through the points $p$ and $q$ with   $r$ a point between $p$ and $q$ that it is conjugate to $p$, then there is a point $x$ between $r$ and $q$ and a time-like curve $\lambda$ from $p$ to $x$.  Thus $\gamma$ is not achronal. \hfill{$\Diamond$}
\end{remark}

Using this, one can generalise the proposition \ref{prop:focuspoint} and  theorem \ref{th:focusgeodesic1}  as follows:
\begin{prop}\label{pro:nconjugate}
Let $\M$ be a spacetime with a smooth 2-dimensional spacelike submanifold $\S$ such that the expansion $\theta$ of, for example, outgoing null congruence is negative at a point $p\in \S$, $\theta(p)<0$. If in addition $R_{\mu\nu}\vk^\mu\vk^\nu\geq0$ for all null vector fields of $\M$, then the congruence from $\S$ has a focus point $q$ conjugate to $\S$ along the null geodesic $\gamma$ from $p$.
\end{prop}

\begin{theorem}\label{th:focusachr}
Let $\S$ be a smooth submanifold of the spacetime $\M$ as in the previous proposition and a point $p\in \M$, $p\notin \S$, such that there is a causal curve $\gamma$ from $\S$ to $p$. Then $\gamma$ cannot be deformed to a time-like curve from $\S$ to $p$, iff $\gamma$ is a null geodesic orthogonal to $\S$ without a focus point between $\S$ and $p$.
\end{theorem}

Furthermore, the above results lead to the statement below that will be used in the singularity theorem below.

\begin{theorem}\label{th:nboundary}
Let $\M$ be a globally hyperbolic spacetime and $\S$ be a compact, 2-dimensional spacelike submanifold of $\M$ with trivial normal bundle. Then every $p\in\partial I^+(\S)$ lies on a future directed null geodesic that it is orthogonal to $\S$ and  does not have a focus point between $\S$ and the point $p$.
\end{theorem}
\begin{proof}
It has been demonstrated in theorem \ref{prop:chronos} that $I^+(\S)\subset J^+(\S)\subseteq \overline{ I^+(\S)}$. Moreover, proposition \ref{prop:Iboundary} states that all the points  of  $\partial I^+(\S)$ lie on null geodesics which are entirely contained in $\partial I^+(\S)$ and either have an end point in $\S$ or are past inextendible. The latter phenomenon does not occur for globally hyperbolic spacetimes as for a compact space-like subset $\S$, $J^+(\S)$ is a closed subset of $\M$ and so $\partial I^+(\S)\subset J^+(\S)$. Thus, all points of $\partial I^+(\S)$ lie on null geodesics that end on $\S$.

Furthermore, suppose there is a focus point $r$ in the null geodesic $\gamma$ between $\S$ and $p\in \partial I^+(\S)$. From the discussion above $\gamma$ will not be achronal.  As $\gamma$ lies entirely in $\partial I^+(\S)$, $\partial I^+(\S)$ will not be achronal as well contradicting corollary \ref{cor:achronalbI}. Therefore, no such focus point can exist on $\gamma$.
\end{proof}

\subsubsection{Singularities of spacetimes with trapped surfaces}

\begin{figure}[htbp]
\begin{center}
\begin{eqnarray*}
\begin{tikzpicture}[scale=1]
  % Coordinates
  \coordinate (A) at (-6.00, 0.00);
  \coordinate (B) at (0.00, 0.00);
  \coordinate (C) at (-4.00, 1.00);
  \coordinate (D) at (-2.00, -0.50);
  \coordinate (E) at (6.00, 0.00);
  \coordinate (F) at (3.25, -1.50);
  \coordinate (G) at (1.83, 0.90);
  \coordinate (H) at (-2.00, 0.00);
  \coordinate (I) at (2.00, 0.00);
  \coordinate (J) at (0.00, 2.00);
  \coordinate (L) at (-5.00, 3.00);
  \coordinate (K) at (5.00, 3.00);
  \coordinate (M) at (-6.00, 3.00);
  \coordinate (N) at (6.00, 3.00);
  \coordinate (O) at (-4.50, 2.50);
  \coordinate (P) at (-4.50, -0.50);
  \coordinate (Q) at (-4.50, 1.50);
  \coordinate (R) at (-5.00, 0.75);
  \coordinate (S) at (-3.50, 1.50);
  \coordinate (T) at (-3.50, -0.50);
  \coordinate (U) at (-4.00, 0.75);
  \coordinate (V) at (-3.00, 0.00);
  \coordinate (W) at (-1.00, 1.00);
  \coordinate (X) at (-1.00, -0.50);
  \coordinate (Y) at (-1.25, 0.50);
  \coordinate (Z) at (-0.50, 0.25);
  \coordinate (AA) at (1.00, 1.00);
  \coordinate (AB) at (1.00, -0.50);
  \coordinate (AC) at (0.75, 0.75);
  \coordinate (AD) at (1.25, 0.00);
  \coordinate (AE) at (3.50, 1.50);
  \coordinate (AF) at (3.50, -0.50);
  \coordinate (AG) at (3.25, 0.75);
  \coordinate (AH) at (4.00, 0.25);
  \coordinate (AI) at (4.50, 2.50);
  \coordinate (AJ) at (4.50, -0.50);
  \coordinate (AK) at (4.25, 1.25);
  \coordinate (AL) at (5.00, 0.50);
  \coordinate (AM) at (0.00, -0.50);
  \coordinate (AN) at (-0.25, 1.00);
  \coordinate (AO) at (0.50, 0.25);

  % Drawing
  \draw[very thick] (A) .. controls (-4.00, 1.00) and (-2.00, -0.50) .. (B);
  \draw[very thick] (E) .. controls (3.25, -1.50) and (1.83, 0.90) .. (B);
  \draw[red, thick] (H) -- (J);
  \draw[red, thick] (I) -- (J);
  \draw[red, thick] (H) -- (L);
  \draw[red, thick] (I) -- (K);
  \draw[white] (M) -- (A);
  \draw[white] (M) -- (N);
  \draw[white] (N) -- (E);
  \draw[green, thick] (O) .. controls (-4.50, 1.50) and (-5.00, 0.75) .. (P);
  \draw[green, thick] (S) .. controls (-4.00, 0.75) and (-3.00, 0.00) .. (T);
  \draw[green, thick] (W) .. controls (-1.25, 0.50) and (-0.50, 0.25) .. (X);
  \draw[green, thick] (AA) .. controls (0.75, 0.75) and (1.25, 0.00) .. (AB);
  \draw[green, thick] (AE) .. controls (3.25, 0.75) and (4.00, 0.25) .. (AF);
  \draw[green, thick] (AI) .. controls (4.25, 1.25) and (5.00, 0.50) .. (AJ);
  \draw[green, thick] (J) .. controls (-0.25, 1.00) and (0.50, 0.25) .. (AM);
  \fill[blue] (H) circle (6pt);
  \fill[blue] (I) circle (6pt);
   \node[below, blue] at (-2.00, -0.50) {$\T$};
    \node[below, blue] at (2.00, -0.50) {$\T$};
     \node[above] at (0.00, 3.00) {$I^+(\T)$};
      \node[right, red] at (K) {$\partial I^+(\T)$};
        \node[below] at (6.00, -0.30) {$\Sigma$};
\end{tikzpicture}
\end{eqnarray*}
\end{center}
%\begin{figure}[htbp]
\caption{\small{The  figure depicts some general features of the proof of theorem \ref{th:singularity3}.  In particular the figure is a 2-dimensional cross section of the spacetime with the trapped surface $\T$ represented by the two blue dots, the boundary $\partial I^+(\T)$ by the red line, the Cauchy surface $\Sigma$ by the horizontal black curve.  The green lines at the integral curves of the vector field $\vt$. Each starts  from $\partial I^+(\T)$ and intersect the Cauchy surface $\Sigma$ only once. Thus establishing the map $\phi: \partial I^+(\T)\rightarrow \Sigma$ of the theorem. }}
\centering
\label{fig:nulldefx}
\end{figure}
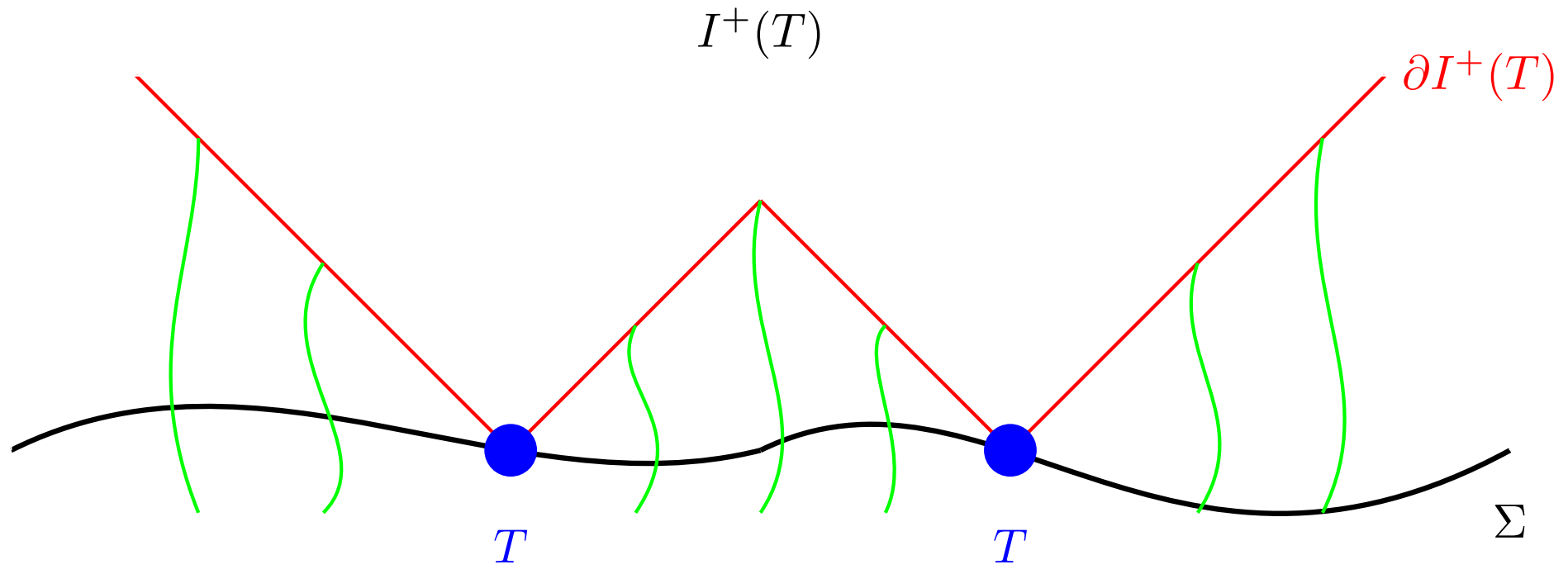

\begin{theorem}\label{th:singularity3}
Let $\M$ be a globally hyperbolic (connected) spacetime with a non-compact Cauchy surface $\Sigma$ that contains a trapped surface $\T$ such that $R_{\mu\nu} \vk^\mu\vk^\nu\geq 0$ for all null vector fields $\vk$. If $\theta_0$ is the maximum of the expansions of both ingoing and outgoing null geodesic congruences from $\T$, $\theta_0=\max\{ \theta^+, \theta^-\}$, then at least one inextendible future directed null geodesic orthogonal to $\T$ has affine length no greater that $2/\theta_0$, i.e. one or more null geodesics are incomplete.
\end{theorem}
\begin{proof}
Let us assume that all future directed  emanating from $\T$ null geodesics have affine length greater than $2/\theta_0$. This defines a map
\be
f_+ ~:~ T\times [0, 2/\theta_0]\rightarrow M~,
\ee
such that $f_+(q, a)=\gamma(a)$, where $\gamma$ is a null geodesic of the outgoing null congruence emanating from $q\in \T$ and $\gamma(a)\in \M$ is the point in $\M$ reached by $\gamma$ at affine parameter $\lambda=a$.  Similarly, we define $f_-$ using the ingoing null geodesic congruence. Both $f_\pm$ are continuous and since $T\times [0, 2/\theta_0]$ is compact, it follows that the set
\be
A\equiv f_+(T\times [0, 2/\theta_0])\cup f_-(T\times [0, 2/\theta_0])~,
\ee
is compact.

It has been demonstrated in proposition \ref{pro:nconjugate} and theorem \ref{th:nboundary} that $\partial I^+(\T)$ is contained in $A$ and since $\partial I^+(\T)$ is a closed subset of $\M$, it must be compact (Theorem \ref{th:clocom}).  As $\M$ is time oriented, there is a nowhere vanishing time-like vector field $\vt$ on $\M$. We have shown that  $\partial I^+(\T)$ is achronal, see corollary \ref{cor:achronal}.  So the  integral curves of $\vt$ that start from the points of $\partial I^+(\T)$ do not pass through  another point of $\partial I^+(\T)$ and intersect $\Sigma$ precisely once.  This gives a continuous map $\phi: \partial I^+(\T)\rightarrow \Sigma$.  As a result $\phi(\partial I^+(\T))$ is compact in  the topology of $\Sigma$ because it is the image of the compact set $\partial I^+(\T)$ (Theorem \ref{th:compactimage}) -- the compactness of $\partial I^+(\T)$  has already been established above. This, in particular, implies that $\phi(\partial I^+(\T))$ is closed (Theorem \ref{th:comclo}). However, we have shown in proposition \ref{prop:boundary} that $\partial I^+(\T)$ is a $C^{1-}$ manifold and therefore a union of charts that are open sets. As $\phi$ is a homeomorphism from $\partial I^+(\T)$ to the image $\phi(\partial I^+(\T))$, the same must apply for $\phi(\partial I^+(\T))$.  Thus $\phi(\partial I^+(\T))$ must also be open in $\Sigma$. Since $\phi(\partial I^+(\T))$ is both open and closed subset of $\Sigma$ and $\Sigma$ is connected, because $\M$ is, we conclude that $\partial I^+(\T)=\Sigma$.  However, this is a contradiction because $\partial I^+(\T)$ is compact while by the hypothesis of the theorem $\Sigma$ is not compact.  Therefore, there must exist at least one null geodesic in either the outgoing or  ingoing null congruence that cannot have affine length more than $2/\theta_0$.  The spacetime $\M$ is geodesically incomplete.
\end{proof}

\section{Concluding remarks } %\ref{th:singularity1}

The global techniques developed for the proof of the singularity theorems can be used in a variety of ways to prove other key results in general relativity. One such result is the Hawking \emph{black hole area theorem}. The future event horizon $\mathcal{H}^+$ of an asymptotically flat black hole is defined as  the boundary $\partial J^-(\mathscr{I}^+)$ of the causal past  of the future null infinity $\mathscr{I}^+$. We shall not define $\mathscr{I}^+$ here  -- roughly is the boundary that all null curves of a spacetime end in the future. It is a consequence of  corollary \ref{cor:achronalbI} that $\mathcal{H}^+$ is an achronal set. Moreover from proposition \ref{pro:nullgeo}, every point on $\mathcal{H}^+$ lies on a null geodesic contained in $\mathcal{H}^+$. In addition   from proposition \ref{prop:boundary}, $\mathcal{H}^+$ is a null hypersurface in $\M$.
Let $\vk$ be the tangent vector of the null geodesics that span $\mathcal{H}^+$.  As $\vk$ is null, it is also normal to $\mathcal{H}^+$.
Next suppose that the spacetime satisfies the null energy condition $R(\vk, \vk)\geq 0$ for all null vector fields $\vk$ on $\M$.  The equation (\ref{nullexpansion}) for the expansion  yields
\be
\frac{d}{d\tau}\theta+\frac{1}{2} \theta^2\leq 0~,
\ee
as the rotation $\omega$ vanishes and the rest of the terms in the right hand side are negative. This can be solved as
\be
\theta^{-1}(\tau)-\theta^{-1}(\tau_0)\geq \frac{1}{2} (\tau-\tau_0)~.
\ee
Notice that if $\theta(\tau_0)\geq 0$, the expansion remains positive as the affine parameter $\tau$ increases.  But if $\theta(\tau)\geq 0$, the infinitesimal volume element $\Delta A$ of the $\tau=\text{const}$ co-dimension two surfaces, the spatial horizon sections, are related to the expansion as
\be
\frac{d}{d\tau}\Delta A=\theta(\tau) \Delta A\geq 0~,
\ee
see (\ref{Atheta}). Therefore, the area of spatial horizon sections increases with the affine parameter $\tau$. This establishes the area theorem provided that $\theta(\tau)\geq 0$. This is indeed the case. We have seen that if $\theta$ becomes negative, the null geodesic congruence develops a focus point at some finite proper time $\tau$. In turn, this would imply from theorem \ref{th:focusachr} that $\mathcal{H}^+$ is not achronal, which is a contradiction. Thus $\theta(\tau)\geq 0$ and the area of event horizon spatial sections of a black hole increase with affine time.

We have described the proof of the Hawking cosmological singularity theorem  and the Penrose black hole singularity theorem. These are not the most general singularity theorems available. However, they are the stepping stones to understand the more general results.  An effort has been made to present the material from a pedagogical point of view without neglecting the technical part of the proof of the various statements.
To achieve this, we tried to close the gap between the logical steps required to prove the key theorems by providing the necessary information. Many examples have been given to illustrate the key mathematical  concepts from geometry, topology and analysis needed to establish the results -- some emphasis has been put on examples from analysis as they are not as frequently encountered.
We have also included several figures that aid the proof of some statements, illustrate the final conclusion of theorems or explain the reason behind some of the definitions.

The results have been described mostly following the traditional method as this has been explained in the book by Hawking and Ellis \cite{HE} and the book by Wald \cite{Wald}. However, some effort has been made to explain some of the results using the Arzel\'a-Ascoli theorem. This, especially, includes the proof of the limit curve theorem and the proof of existence of causal geodesics with maximum proper time between two causally related points in a spacetime or a point a hypersurface. This is a more modern approach to this area advocated in more recent publications like for example the book \cite{beem} and the review article  \cite{Minguzzi}.

The Hopf-Renow theorem is presented, which amongst other statements,  addresses the question of  geodesic completeness in Riemannian geometry.  The reason for this is twofold. One is to compare the properties of Lorentzian and Riemannian manifolds regarding the issue of geodesic completeness. The other is that some of the techniques in Riemannian geometry have a direct application in Lorentzian geometry, like for example in the proof of statements regarding the properties of geodesics with conjugate points.

The notes conclude with a number of appendices.  These collect the results from topology needed in the text. There is also a description of the Arzel\'a-Ascoli theorem and a detailed discussion on the properties of semi-continuous functions. Some of the statements of Hopf-Renow theorem are also proven as well as an alternative proof is described  regarding the existence of geodesics with maximal proper time in globally hyperbolic manifolds using the Arzel\'a-Ascoli theorem.

\vskip1cm
 \noindent {\it {Acknowledgements}}
 \vskip0.1cm
I thank Yuchen Zhou  for helping me  draw the diagrams  included in these notes and Eleni Alexandra Kontou for suggesting some improvements. This work has been partially supported by the STFC grants ST/X000753/1 and UKRI4231.

\newpage

\setcounter{section}{0}
\setcounter{subsection}{0}

\begin{appendices}
%\appendix

\section{Topological spaces}\label{ap:topspaces}

The topological properties of spacetime and those  of set of curves in a spacetime  have been proven essential for the arguments that lead to the singularity theorems. So it is useful to summarise some of the results of topology  that have a direct application in the proof of these theorems.

\begin{definition}
Topological space is a pair $(\X, \mathcal{O})$, where $\X$ is a set and $\mathcal{O}$ is a collection of subsets of $\X$ such that (i) the empty set $\emptyset$ and $\X$ are in the collection, $\emptyset, \X\in \mathcal{O}$, (ii) the union of \emph{any number} of sets in $\mathcal{O}$ is  in collection $\mathcal{O}$ and (iii)  the intersection of any \emph{finite number} of sets in $\mathcal{O}$ belongs to the collection $\mathcal{O}$. \hfill{$\triangle$}
\end{definition}

\begin{remark}
Some consequences and implications of the definition are the following:
\begin{itemize}
\item Let $(\X, \mathcal{O})$ be a topological space. A subset $\O\subset \X$ that belongs to the collection $\mathcal{O}$ is called \emph{open}. A subset $A$ of $\X$ is called \emph{closed}, iff its complement $\X-A$ is open.  It follows that the intersection of any number of closed sets is closed set and the union of finite number of closed sets is  a closed set.

    \item If $(\X, \mathcal{O})$ is a topological space, then the sets $\X$ and $\emptyset$ are both open and closed.  Moreover, $\X$ is \emph{connected}, iff the only open and closed sets in $\X$ are just $\X$ and $\emptyset$.

        \item A \emph{neighbourhood} $\W$ of a point $p$ in the topological space $(\X, \mathcal{O})$ is a subset of $\X$, not necessarily open, such that there is an open set $O\in \mathcal{O}$ with the property that $p\in O\subseteq W$.

    \item Let $\A$ be a subset of a topological space $(\X, \mathcal{O})$.  The \emph{closure} $\bar{\A}$ of $\A$ is the intersection of all closed set of $\X$ that contain $\A$. Similarly,
    the \emph{interior} $\mathrm{int} \A$ is the union of all open sets of  $\X$ contained in $\A$. It follows that $\bar{\A}$ is a closed set, $A\subseteq \bar{\A}$,  while $\mathrm{int} \A$ is an open set, $\mathrm{int} \A\subseteq A$.  The \emph{boundary}, $\partial \A$, of $\A$ is $\partial\A\equiv \bar{\A}-\mathrm{int} \A$.

    \item A basis $\mathcal{B}$ of the topological space $(\X, \mathcal{O})$ is  collection  of open subsets in $\X$ such that all the open sets of $\X$ can be constructed from those of $\mathcal{B}$  upon taking the union of any number and the intersection of finite number of elements in  $\mathcal{B}$. Let
        $(\X_1, \mathcal{O}_1)$ and $(\X_2, \mathcal{O}_2)$ be topological spaces, then $\X_1\times \X_2$ is also a topological space with basis of open sets  given by $\mathcal{B}=\mathcal{O}_1\times\mathcal{O}_2$.

        \item An \emph{open cover} of a topological space  $(\X, \mathcal{O})$ is a collection of open sets $(\O_\alpha)$ whose union is $\X=\cup_\alpha \O_\alpha$.

\item Let $\Y$ be a subset of the topological space  $(\X, \mathcal{O})$. Then, there is an  \emph{induced topology} on $\Y$ from $\X$ and an associated  collection of open sets given by
$\mathcal{O}_Y=\{O\cap Y\vert O\in \mathcal{O}\}$.
\end{itemize}
\hfill{$\Diamond$}
\end{remark}

\begin{remark}
There is a subtlety in the definition of the boundary of a subset of a topological space. To explain, apply the definition on the topological space $\X$ itself. As $\X$ is both closed and open set, the definition will give $\partial\X=\emptyset$. This is the conclusion for every topological space. However, this has to be distinguished from the definition of the boundary of a manifold which involves the use of charts modeled over upper half Euclidean spaces.  There is no straightforward relation between these two notions of boundary and both are used in the proof of the singularity theorems. \hfill{$\Diamond$}
\end{remark}

  \begin{example}Perhaps the most important example of a topological space is the Euclidean space $\bR^n$, where the basis of the topology $\mathcal{B}$ is given by the open balls
\be
B(x, r)=\{y\in \bR^n|\,\,\, |y-x|^2<r\}~,
\ee
with centre $x$ and radius $r$, where $|\cdot|$ is the usual Euclidean norm,  i.e.
\be
\mathcal{B}=\{ B(x, r)| x\in \bR^n, r\in \bR_{>0}\}~.
\ee
For example, the basis of open sets of the real line $\bR$ are the open intervals $(a,b)$ with $a<b$ and $a,b\in \bR$. This is the \emph{standard} or \emph{Euclidean} topology on $\bR^n$ often denoted with $\mathcal {O}_B$.  However, there are many other different topologies on $\bR^n$, e.g. one can take as a collection of open sets all the subsets of $\bR^n$ or the collection of open sets can be merely $\{\emptyset, \bR^n\}$. \hfill{$\triangledown$ }
\end{example}

Useful topologies are those that open sets can separate points. This naturally leads to $T_1$ and $T_2$ (equivalently Hausdorff) topological spaces.
\begin{definition}
Let $(\X, \mathcal{O})$ be a topological space. $\X$ is $T_1$, iff for every pair of distinct points $x, y\in \X$, there is a $O\in \mathcal{O}$, such that $x\in O$ and $y\notin O$. Moreover, $\X$ is Hausdorff iff for every pair of distinct points $x, y\in \X$, there are $O_1, O_2\in \mathcal{O}$ such that $x\in O_1$ and $y\in O_2$ with
$O_1\cap O_2=\emptyset$. \hfill{$\triangle$}
\end{definition}

\begin{definition}
 A topological space $(\X, \mathcal{O})$ is compact, iff for \emph{every open cover} of $\X$, there is a finite subcover, i.e a cover that contains a finite number of open sets.  \hfill{$\triangle$}
\end{definition}
\begin{remark}
The emphasis in the definition of a compact topological space is that \emph{every} cover has a finite subcover.  There are certainly many examples of topological spaces which are not compact and admit a finite cover. But not compact topological spaces, they admit covers  that do not have a finite subcover. \hfill{$\Diamond$}
\end{remark}

\begin{theorem}(Heine-Borel)
The compact subsets of $(\bR^n, \mathcal{O}_B)$, i.e. of $\bR^n$ equipped with the standard topology, are closed and bounded.
\end{theorem}

\begin{definition}
Let $(\X_1, \mathcal{O}_1)$ and $(\X_2, \mathcal{O}_2)$ be topological spaces. The map $\Phi: X_1\rightarrow X_2$ is \emph{continuous}, iff for every $\O\in \mathcal{O}_2$, the inverse image of $\O$, $\Phi^{-1}(\O)$, is in $\mathcal{O}_1$, i.e. the inverse images of open sets under $\Phi$ are open.\hfill{$\triangle$}
\end{definition}

\begin{theorem}\label{th:compactimage}
Let $(\K, \mathcal{O}_K)$ be a compact topological space and $\Phi: K\rightarrow X$ be a continuous map, where  $(\X, \mathcal{O})$ is a topological space. Then the image $\Phi(K)\subset X$ of $\Phi$ is a compact subset of $\X$.
\end{theorem}
\begin{remark}
The above theorem has important consequences. Suppose that $\X=\bR$, i.e. $\Phi$ is a real valued function. As $\Phi(K)$ is a compact subset of $\bR$, the the Heine-Borel theorem above implies  that it is closed and bounded. In other words, $\Phi$ attains a global maximum and a global minimum. This means that there exist points $x, y\in \K$ such that $\Phi(x)$ takes the maximum value of $\Phi$ and $\Phi(y)$ takes the minimum value of $\Phi$. Clearly, such a result can be applied to extremisation problems. This is nearly the case for the existence of minimal length (Riemannian geometry) or maximal proper time (Lorentzian geometry) geodesics between two points in a manifold, where $\K$ is an appropriate space of curves and $\Phi$ is the length functional.  The technical difficulty is that in these cases neither $K$ is compact nor $\Phi$ is continuous. One of the difficulties of proving the singularity theorems is to overcome these issues. \hfill{$\Diamond$}
\end{remark}

Compact topological spaces have many other properties. A summary of key theorems  is as follows:
 \begin {theorem}\label{th:compro}(Tychonoff)~
Let $(\K_1, \mathcal{O}_1)$ and $(\K_2, \mathcal{O}_2)$ be compact topological spaces. Then, the product $K_1\times K_2$ with a basis of open sets given by $\mathcal{O}_1\times \mathcal{O}_2$ is also compact. \end{theorem}

\begin{theorem}~\label{th:clocom}If $(\X, \mathcal{O})$ is Hausdorff and $\K$ is a compact subset of $\X$. Then, $\K$ is closed.\end{theorem}

\begin{theorem}\label{th:comclo}~ If $(\K, \mathcal{O})$ is compact and $\Y$ is a closed subset of $\K$. Then, $\Y$ is compact.\end{theorem}

Note also that a topological space $(\X, \mathcal{O})$ is \emph{locally compact}, iff every $p\in X$ has an open neighbourhood $\W$ with compact closure. All $C^k$, $k\geq0$, finite dimensional   manifolds are locally compact because locally are homeomorphic to $\bR^n$, which is a locally compact space.

Topological manifolds can admit collections of open sets which are very large. In particular, they may not even admit a countable basis of open sets. So it is natural to put some restrictions to such a complexity.

\begin{definition}
A topological space  $(\X, \mathcal{O})$  is \emph{first countable} iff for every $p\in \X$, there is a countable collection of open sets $(\O_n)$ such that every neighbourhood of $p$ contains at least one open set in the collection $(\O_n)$ -- the collection depends on $p$. Similarly,  $(\X, \mathcal{O})$ is \emph{second countable}, iff it admits a \emph{countable basis} $\mathcal{B}$ of open sets. \hfill{$\triangle$}
\end{definition}

 \begin{example}$\bR^n$ with the standard topology is second countable. As a basis of open sets, one can take the open balls with centres vectors in $\bR^n$ that have components rational numbers   and with radii that are  positive rational numbers.\hfill{$\triangledown$}
 \end{example}

 \begin{remark}
It is known that locally compact, second countable, Hausdorff topological spaces $(\X, \mathcal{O})$  are \emph{paracompact}.  This means that every open cover $(U_\alpha)$ of $\X$ has a  refinement, i.e. there is another open cover $(V_i)$ such that for every $U_\alpha$ there is a $V_i$ with $V_i\subset U_\alpha$,  and moreover every point $p\in \X$ has a neighbourhood $\W$ such that $\W$ intersects only a finite number of open sets of the cover $(V_i)$.

As manifolds are locally compact, every second countable Hausdorff manifold is paracompact. Paracompact manifolds admit partitions of unity which allow to patch together local information on the patches,  e.g. metrics, forms and tensor in general, to global objects on the manifold. In this way, one can prove the existence of a metric and other tensors on paracompact manifolds.  For more results on paracompact manifolds see \cite{BottTu}. As the existence of such tensors in the investigation of smooth manifolds is essential, it is assumed from the very beginning that smooth manifolds are paracompact.\hfill{$\Diamond$}
\end{remark}

\begin{remark}
It has already been mentioned that a point $p\in \X$, where  $(\X, \mathcal{O})$  is a topological space, is the \emph{limit}  of a sequence $(p_n)$, $p_n\in\X$, iff for every $\O\in \mathcal{O}$ with $p\in \O$, there is $N\in \bN$ such that $p_n\in O$ for all $n>N$. In turn, $p$ is the convergence point of the sequence $(p_n)$.  On the other hand an \emph{accumulation} point of the sequence $(p_n)$ is a point $p\in \X$ such that  every $\O\in \mathcal{O}$, with $p\in \O$, contains infinite number of points of the sequence.  For general topological spaces, If $p$ is the accumulation point of a sequence $(p_n)$, it is not apparent that one can extract a subsequence $(p_{n_k})$ of $(p_n)$ that has as a limit the accumulation point $p$. However, this is always possible if $\X$ is first countable. \hfill{$\Diamond$}
\end{remark}

Another useful result from topology is the set of criteria for a subset of a topological space to be compact. This is essential for the singularity theorems, especially for the proof of compactness of certain sets of curves.

\begin{theorem}\label{th:bw}(Bolzano-Weierstrass)
Let $(\X, \mathcal{O})$ be a topological space.  If $\K$ is a compact subset of $\X$, then every sequence of points $(p_n)$ in $\K$, $p_n\in \K$, has an accumulation point $p\in \K$. Moreover, if $(\X, \mathcal{O})$ is second countable and $\A$ a subset of $\X$, $\A\subset \X$, with the property that every sequence of points $(p_n)$ in $\A$, $p_n\in \A$, has an accumulation point $p\in \A$, then $\A$ is a compact subset of $\X$.
\end{theorem}

\begin{remark}
Consider the space of continuous maps $C(K, X)$ from the compact space $K$ to metric space $X$ with distance function $d$. $C(K, X)$ is a metric space with distance function
\be
d_\infty(f, g)\equiv \sup_{x\in K} d(f(x), g(x))~,~~~f, g\in C(K, X)~.
\label{dinfty}
\ee
As a result $C(K, X)$ is both Hausdorff and first countable. The convergence defined by this distance function on $C(K, X)$ is uniform convergence. If $K$ is not compact, as it is the case for some curves, then typically uniform convergence only applies to compact subspaces of the domain. \hfill{$\Diamond$}
\end{remark}

%\appendix
\section{Curves}\label{ap:aa}

\subsection{Topology on the space of curves and the Arzel\'a-Ascoli theorem}\label{ap:ab}

\subsubsection{The Arzel\'a-Ascoli theorem}
Existence proofs, like for example the existence  geodesics with length the minimal distance between two points in a manifold, rely on the existence of limits of sequences of functions. This in turn is related on the compactness properties of the relevant spaces. A central role in the proof of such statements has the Arzel\'a-Ascoli theorem that gives sufficient conditions for the existence of such limits. In preparation for stating the theorem, let $X$ be a compact space and $Y$ be a metric space with distance function $d$. Next, consider the space of continuous functions $C(X,Y)$ from $X$ to $Y$.  There are two  natural topologies that can be put on  $C(X,Y)$.  One is the \emph{uniform or equivalently the sup topology},  already described in (\ref{dinfty}), given by the distance function
\be
d_\infty(f, g)\equiv \sup_{x\in X} d(f(x), g(x))~.
\label{uni}
\ee
Existence of such a topology requires that $X$ to be compact so the continuous function $d(f(x), g(x))$, $f, g\in C(X,Y)$,  attains a maximum and so $d_\infty$ can exist.  This topology on  $C(X,Y)$ is called \emph{uniform} topology because a sequence of function $(f_n)$ in $C(X,Y)$ converges to a function $g$, iff for every $\epsilon>0$, there is a $N=N(\epsilon)$ such that
\be
d(f_n(x), g(x))<\epsilon
\ee
for $n\geq N(\epsilon)$ and all $x\in X$.  The important point is that $N(\epsilon)$ does not depend on $x\in X$ and so it can be chosen to be the same for all $x$.  If a sequence of continuous functions $(f_n)$ converges uniformly to a function $g$, $g$ is also continuous. This follows from the triangular inequality of the distance function $d_\infty$.

An alternative topology on $C(X,Y)$ is the \emph{compact-open topology}.  In this topology, the basis in the topology of $C(X,Y)$ is given by the open sets
\be
O_{K,U}\equiv \{f\in C(X,Y)\vert f(K)\subset U\}~,
\label{comopen}
\ee
where $K$ is a compact subset of $X$, $K\subseteq X$, and $U$ is an open subset of $Y$, $U\subseteq Y$. This topology on $C(X,Y)$ can be defined for $X$ non-compact. But if $X$ is compact, then the uniform topology and the compact-open topology on $C(X,Y)$ coincide.  For $X$ non-compact, convergence in the compact-open topology can be seen as uniform convergence on compact subsets of $X$. Typically, some control is lost over the convergence of sequences of functions outside the compact subsets of $X$, i.e. the convergence does not exhibit all the properties of uniform convergence.

\begin{theorem} (Arzel\'a-Ascoli)\label{th:aath}
Let $C(X,Y)$ be the set of all continuous function from $X$ to $Y$ with $X$ compact and $Y$ a metric space. Consider a subset $\mathcal{F}\subset C(X,Y)$ such that
\begin{enumerate}
\item $\mathcal{F}$ is \emph{ equicontinuous}. This means that for every $x\in X$ and $\epsilon>0$, there is a neigbourhood $U_\epsilon$ of $x$ such that $d(f(x), f(y))<\epsilon$ for all $f\in \mathcal{F}$. The key point is that $U_\epsilon$ does not depend on $f\in \mathcal{F}$,

\item The sets $\mathcal{F}(x)\equiv \{f(x)\in Y\vert f\in \mathcal{F}\}$ for every $x\in X$ have compact closure in $Y$,
\end{enumerate}
then $\mathcal{F}$ has a compact closure in $C(X,Y)$ with the uniform topology.
\end{theorem}

\begin{remark}
An alternative way to state the theorem is to say that if $(f_n)$ is a sequence of equicontinuous functions that also satisfy the second condition of the theorem, then there is a subsequence $(f_{n_k})$ of $(f_n)$ that converges to a continuous function.  The second condition of the theorem is not necessary if, in addition, $Y$ is compact as well.  This is because closed subsets of a compact set are compact, see Theorem \ref{th:clocom}. \hfill{$\Diamond$}
\end{remark}

\begin{remark}
Supposing  $X$ is also a metric space with distance function $d_X$ and denoting the distance function of $Y$ with $d_Y$, the equicontinuity condition on a sequence of continuous functions $(f_n)$ can be expressed as follows: For every function $f_n$ in the sequence and $\epsilon>0$, there is a $\delta=\delta(\epsilon)$ such that for $d_X(x,y)\leq \delta$, $d_Y(f_n(x), f_n(y))<\epsilon$. Again, $\delta$ does not depend  either on the individual functions $f_n$ of the sequence or on  the points $x,y\in X$ ($X$ compact).
\hfill{$\Diamond$}
\end{remark}

\subsubsection{Curves in a Riemannian manifold}\label{ap:b12}

The main aim in this appendix is to summarise some of the analytic tools that are used to investigate the properties of the space of curves in a manifold. These tools are utilised  in both the proof of  the singularity theorems and of the Hopf-Rinow theorem in appendix \ref{ap:bb}.  The new aspect here,  which goes beyond the discussion of the previous section, is that apart from the curves,  their velocities have to be controlled as well to establish the results.

Let $(\M, g)$ be a Riemannian manifold, $\M$ connected, with metric $g$ -- throughout we shall assume that $(\M,g)$ is Hausdorff and at least $C^2$.

\begin{definition}
Let $\gamma$ be a $C^1$ curve in $(\M, g)$, which is a map $\gamma: (a,b)\subset \bR \rightarrow \M$.  The \emph{length}, $L_g(\gamma)$, of $\gamma$ with respect to the metric $g$ is defined as
\be
L_g(\gamma; [a,b])\equiv \int^b_a\, dt\, |\vec{\dot \gamma}|_g =\int^b_a\, dt\, \sqrt {g(\vec{\dot \gamma}, \vec{\dot\gamma})}=\int^b_a \,dt\, \sqrt {g_{\mu\nu}(\gamma(t)) \,\vec{\dot\gamma}^\mu(t) \vec{\dot \gamma}^\nu(t)}~,
\ee
where $\vec{\dot\gamma}$ denotes the velocity of the curve, i.e.  derivative of the curve $\gamma$ with respect to the parameter $t$, and the last equality gives the coordinate or local expression of the the length in $\M$.  Similarly,
\be
E_g(\gamma; [a,b])\equiv \frac{1}{2} \int^b_a\,dt\, |\vec{\dot \gamma}|^2_g= \frac{1}{2}\int^b_a \,dt\, g_{\mu\nu} (\gamma(t)) \vec{\dot \gamma}^\mu(t) \vec{\dot \gamma}^\nu(t)~,
\label{energy}
\ee
is the \emph{energy} of a curve $\gamma$ in $\M$.
\hfill{$\triangle$}
\end{definition}

The length of curves can be used to define the distance $d_g$ between two point $p,q$ of a manifold $(\M,g)$. Let us denote with $\mC^1(p,q)$  the set of $C^1$ curves that begin from $p$ and end at $q$ in the manifold $\M$.

\begin{definition}
Let $(\M,g)$ be a Riemannian manifold and $p,q\in \M$.  Then, the metric $d_g$ on $\M$ or equivalently the distance function $d_g$ of $\M$, is given by
\be
d_g(p,q)=\inf_\gamma \{L_g(\gamma; p,q)\vert \gamma\in \mC^1(p,q)\}~,
\ee
where $L_g(\gamma; p,q)$ is the length of the curve $\gamma$ between the points $p$ and $q$.
$d_g(p,q)$ is identified with the distance between the points $p$ and $q$ given by the metric $g$.
\hfill{$\triangle$}
\end{definition}

\begin{remark}
The length $L_g(\gamma)$ of the curve $\gamma$ does not depend on the parameterisation of $\gamma$, e.g. it is invariant under at  least $C^1$ diffeomorphisms of $(a, b)$ that preserve the orientation of the interval.

It should be noted that although $C^1$ curves have been used to define the distance $d_g$ between two points in $\M$. The same distance would have been defined,  if either piecewise $C^1$ curves or  smooth curves were used instead, see also generalisations below.

It should also be emphasised that it is \emph{not apparent} that there is always a curve $\lambda$ that passes through $p$ and $q$ such that $d(p,q)=L_g(\lambda; p,q)$. The proof of existence of such a curve has to be demonstrated and it is one of questions that will be dealt with in these appendices. \hfill{$\Diamond$  }
\end{remark}

It is required  to introduce a variety of  different structures  on the space of curves with aim  to prove our various results. The difficulty stems from the fact that the functionals, like that of the length and energy of  curves,   depend non-linearly on the curves  and also on  their velocities.  The space of curves does not admit a natural linear structure for general Riemannian manifolds.  Here, we shall provide a description, mostly without detailed proofs, of the various spaces of curves that appear throughout these notes.

Having defined, the distance function $d_g$ on $\M$, one can define the length of a \emph{continuous curve}  as follows:
\begin{definition}\label{def:c0length}
Let $\gamma\in  C^0([a,b], \M)$ be a continuous curve from the interval $[a, b]$ to the Riemannian manifold $(\M, g)$. The length of $\gamma$ is defined as
\be
L_g(\gamma;[a,b])\equiv \sup_P \sum^N_{i=1} d_g(\gamma(t_{i-1}), \gamma(t_{i}))~,
\ee
where $P$ is the set of all finite ordered partitions  of $[a, b]$, $t_{i-1}<t_i$.\hfill{$\triangle$}
\end{definition}

\begin{remark}
The length of continuous curves can be infinite even if $[a,b]$ is compact. For example, the continuous curve $\gamma\in  C^0([0,1], \bR^2)$, where $\gamma(t)=(t, t^2 \sin(t^{-2}))$ for $t\in (0,1]$ and $\gamma(0)=0$, has infinite length, $L_g(\gamma;[0,1])=+\infty$.\hfill{$\Diamond$}
\end{remark}

To obtain curves with finite length over compact intervals additional conditions have to be imposed.
To begin, let us first define  the space $W^{1,1}([0,1], \M)$ of absolutely continuous curves; the interval $[0,1]$ can be replaced with any compact interval $[a,b]$. In particular, one has the following:
\begin{definition}
A curve $\gamma:  [0,1]\rightarrow \M$  is \emph{ absolutely continuous}, $\gamma\in W^{1,1}([0,1], \M)$,  iff there is a function $f\in L^1[0,1]$ such that
\be
d_g(\gamma(t), \gamma(s))\leq \int^t_s f(r) dr<+\infty
\ee
for every $s\leq t$ with $s,t\in [0,1]$, where $L^1[0,1]$ is the space of integrable functions on the interval $[0,1]$.\hfil{$\triangle$}
\end{definition}

\begin{remark}
All these curves are continuous, $\gamma\in C^0([0,1], \M)$.  It can also be shown that the \emph{metric speed} of these curves,
\be
|\gamma'(t)|\equiv \lim_{h\rightarrow 0}\frac{d_g(\gamma(t+h), \gamma(t))}{h}~,
 \ee
 can be defined almost everywhere on $[0,1]$ and its integral over $[0,1]$ is finite. Moreover, almost for every $t\in [0,1]$, there is $\vec{\dot\gamma}\in T_{\gamma(t)} \M$ such that $|\vec{\dot \gamma}(t)|_g=|\gamma'(t)|$,  i.e.  the definition of the speed above using the distance function $d$ coincides with the usual definition of the speed of the curve almost everywhere on $[0,1]$.\hfill{$\Diamond$}
 \end{remark}

Clearly $W^{1,1}$ curves have well defined length but not necessarily  a well-defined energy. This leads to following definition:
\begin{definition}
The $W^{1,2}$ space of curves is
\be
W^{1,2}([0,1], \M)\equiv \{\gamma\in W^{1,1}([0,1], \M)\, \text{and} \int_0^1\,dt\, |\gamma'|^2 <+\infty\}~,
\ee
where $|\gamma'|$ is the metric speed of the curve $\gamma$. \hfill{$\triangle$}
\end{definition}

\begin{remark}
It turns out that if $\gamma\in W^{1,2}([0,1], \M)$, almost for every $t\in [0,1]$, there is a $\vec{\dot \gamma}(t)\in T_{\gamma(t)}\M$ such that $|\vec{\dot \gamma}(t)|^2_g=|\gamma'(t)|^2$. So by definition all these curves have finite energy $E_g(\gamma)$  -- note that the energy of $\gamma$ depends of the velocity $\vec{\dot \gamma}$ of the curve $\gamma$.\hfill{$\Diamond$}
\end{remark}

\begin{lemma}\label{le:a1}
Let $\gamma$ be a curve such that its metric speed  $|\gamma'|$ is defined and  is in  $L^2$, or equivalently its energy is bounded, then $\gamma$ is absolutely continuous.
\end{lemma}
\begin{proof}
From the Cauchy-Schwarz inequality, one has
 \be
 \Big(\int^1_0 dr |\dot \gamma|_g\Big)^2\leq \int^1_0 dr 1^2 \int^1_0 dr |\dot \gamma|_g^2\leq 2 E_g(\gamma)~.
 \ee
 Thus the speed of $\gamma$ is in $L^1$ and so $\gamma$ is absolutely continuous.
 Therefore,  to show that a curve is absolutely continuous, it suffices to demonstrate that its energy is bounded provided that its metric speed can be defined.
 \end{proof}

 Another space that is useful in the investigation that follows is the space of $\alpha$-H\"older curves $H^\alpha([0,1], \M)$.
\begin{definition}
A curve $\gamma:  [0,1]\rightarrow \M$ is \emph{$\alpha$-H\"older  continuous},  iff there is a constant $C>0$ such that
\be
d(\gamma(t), \gamma(s))\leq C |t-s|^\alpha~,~~~0<\alpha\leq 1~,
\ee
for $t,s\in [0,1]$.
For $\alpha=1$ such a curve is called \emph{Lipschitz}. \hfill{$\triangle$}
\end{definition}

The difference between absolutely continuous curves and Lipschitz curves is that in the former the metric speed is \emph{integrable} over the interval $[0,1]$ while in the latter it is \emph{bounded}. The bound on the metric speed  is a stronger condition than the integrability. In particular, one can demonstrate that
\be
C^\infty\hookrightarrow C^1\hookrightarrow \text{Lipschitz}\hookrightarrow \alpha-\text{H\"older}
\ee
and
\be
C^\infty\hookrightarrow C^1\hookrightarrow \text{Lipschitz}\hookrightarrow \text{absolutely continuous}\hookrightarrow \text{continuous}
\ee
In the latter line of inclusions, the main restriction is on the speed of the curves and the condition becomes increasingly weaker from left to right. These restrictions make sure that the curves do not fluctuate too drastically and remain contained within well controlled subspaces of the manifold $(\M,g)$.

\begin{remark}
One can extend all the above definitions of the curves given on compact domains, like $[0,1]$ or more generally $[a,b]$ with $-\infty<a<b<+\infty$,  to non-compact ones, like $(a,b)$ with $a, b\in [-\infty, +\infty]$ and $a<b$. For this, we use the notion of \emph{local} absolutely continuous, Lipschitz and so on properties.  Such curves $\gamma$ are required to be absolutely continuous, Lipschitz and so on only on \emph{compact subspaces} of their domain. This means that if the domain is an open interval $(a,b)$ and $\gamma$ is say locally absolutely continuous, then it is absolutely continues only on compact subspaces of $(a,b)$. This results  to less control over the regularity properties of the curves $\gamma$ near either $a$ or $b$. \hfill{$\Diamond$}
\end{remark}

\subsubsection{ Re-parameterisation of curves and the Lipschitz property}\label{ap:b123}

A related question is under which conditions a curve in a Riemannian manifold $(\M, g)$ can be re-parameterised with its length. Keeping in mind applications in Lorentzian geometry, where the tangent vector of causal curves is non-vanishing almost everywhere, we shall consider the case that
\be
|\vec{\dot\gamma}(t)|_g>0~,~~~t\in [0,1]~,~~~\text{almost~everywhere}~,
\ee
with $\gamma$  an \emph{absolutely continuous} curve.  In such a case,
\be
\ell(t)\equiv L_g(\gamma; [t,0])=\int_0^t du |\vec{\dot\gamma}(u)|_g~,
\ee
is an increasing function of $t$,
\be
\ell(t_2)-\ell(t_1)=\int^{t_2}_{t_1} du |\vec{\dot\gamma}(u)|_g>0~,~~~t_2>t_1~,
\ee
and it has a continuous inverse, i.e. $\ell^{-1}: [0, L]\rightarrow [0,1]$, where $L=\ell(1)$ is the (total) length of $\gamma$.  The re-parameterised curve $\tilde \gamma\equiv \gamma\circ \ell^{-1}: [0, L]\rightarrow \M$ has unit speed with respect to the new parameter $s=\ell(t)$ and so it is Lipschitz in $[0,L]$. Indeed,
\be
d_g(\tilde\gamma(s_1), \tilde\gamma(s_2))\leq \int^{s_2}_{s_1} dv |\vec{\dot{\tilde\gamma}}(v)|_g=(s_2-s_1)~,~~~s_2>s_1~.
\ee
This is used often as the Lipschitz property of curves in crucial in many arguments involving limits of curves.
Note that for almost continuous curves on $[0,1]$, $L<+\infty$.

\begin{remark}
Re-parameterisations can also be considered for continuous curves. However, there are several difficulties that arise. One is that in their original parameterisation, say $t\in [0,1]$,  they can be constant over some measurable  subsets of their domain. Thus, the analogous map $\ell$ may not be invertible.  Another difficulty is that the total length $L$ of continuous curves even defined on compact intervals, like $[0,1]$,  can be infinite. In such a case the analogous $[0, L]$ domain of the length parameter may not be compact.  In any case, we shall not consider just continuous curves and these issues do not arise in our analysis.\hfill{$\Diamond$}
\end{remark}

\subsubsection{Re-parameterisation of causal curves}

 Let $(\M, g)$ be a Lorentzian manifold. It has already been mentioned below definition \ref{def:ccausal} that we may take the causal curves to be (locally) absolutely continuous with respect to the distance defined by some auxiliary Riemannian metric $g_e$ of $\M$. The proper time or equivalently proper distance  functional
\be
L_g(\gamma; [a,b])=\int_a^b dt\, \sqrt{-g(\vec{\dot\gamma}, \vec{\dot \gamma})}~,
\ee
is well defined.

The main topic is the derivation of a bound for the proper time functional of a causal curve. This is described in the following proposition.

\begin{prop}\label{prop:gepropertime}
Let $(\M, g)$ be a spacetime and $g_e$ an auxiliary Riemannian metric on $\M$. If $\gamma: [a, b]\rightarrow \M$ is a Lipschitz causal curve with $[a,b]$ a compact interval and moreover $\gamma([a,b])$ lies in a compact subset $\K$ of $\M$, then there is a constant that depends on $\K$ such that
\be
L_g(\gamma; [a,b])\leq \sqrt{C} L_{g_e}(\gamma; [a,b])<+\infty~.
\ee
Of course for a single Lipschitz causal curve, $\K$ can be chosen to be $\gamma([a,b])$.
\end{prop}
\begin{proof}
First, observe that there is a constant $C>0$ such that
\be
|g\vert_p(\vv, \vv)|\leq C\, g_{e}\vert_p(\vv, \vv)~,~~~p\in\K~,~~~\vv\in T_p\M
\label{boundptd}
\ee
where  $C$ does \emph{not} depend on $p\in \K$.

To see this, observe that $g$ is a continuous function from $S(T\M|_{K})=\{\vv\in T_p\M| g_e(\vv, \vv)=1, p\in \K\}$, which is a compact set for $\K$ compact, to $\bR$.  So it has a maximum and a minimum. As a result there is a constant $C$ such $|g\vert_p(\vv, \vv)|\leq C$ on $S(T\M|_{K})$, which in turn implies (\ref{boundptd}). Applying these to Lipschitz causal curves, one concludes that
\be
L_g(\gamma; [a,b])\leq \sqrt{C} L_{g_e}(\gamma; [a,b])~.
\label{rlbound}
\ee
This proves the proposition.
\end{proof}

\begin{remark}
A re-parameterisation of causal curves of interest is that in terms of their Riemannian length with respect to the auxiliary Riemannian metric $g_e$. As it has already been pointed out in section \ref{ap:b123}, if $\gamma$ is absolutely continuous on compact intervals then (1) such a re-parameterisation exists and (ii) the re-parameterised curve is Lipschitz. However, it should be pointed out that the bound (\ref{rlbound}), for curves over non-compact domains, may lead to causal curves of infinite Riemannian length that still have finite (total) proper time.\hfill{$\Diamond$}
\end{remark}

\begin{remark}
Re-parameterisations of causal curves with respect to proper time $\tau$ can also be considered. But as the proper time of null curves vanishes, such a parameterisation of causal curves will be problematic as the map of parameters cannot be one to one on the null sections of the causal curves. However, if $\gamma$ is a time-like causal curve, an argument similar to the one presented in the Riemannian case above allows $\gamma$ to be re-parameterised with the proper time $\tau$.  The re-parameterised curve will be  defined  on a interval $[0, T]$, where $T$ is the (total) proper time of the curve $\gamma$ defined on the compact interval $[a,b]$, i.e $T=L_g(\gamma; [a,b])$.
\hfill{$\Diamond$}
\end{remark}

\subsection{Semi-continuity, weak convergence and extremisation problems}

 \subsection{Semi-continuity } \label{ap:ac}

A key concept that it is used in the proof of both singularity theorems and the Hopf-Rinow theorem below is the semi-continuity  of energy, length and proper time functionals of curves. To begin, the definition of semi-continuity of a function is as follows:

\begin{definition}\label{def:semicon}
Let $X$ be a topological space and $(x_n)$ a sequence of points in $X$ that converge to $x$, $x_n\rightarrow x$. A function $f$ on $X$ is lower semi-continuous at $x$, iff for every sequence $(x_n)$ that converges to $x$
\be
f(x)\leq \lim_{n\rightarrow \infty}\inf f(x_n)\equiv \lim_{n\rightarrow \infty} \inf\{ f(x_k), k\geq n\}~.
\ee
If $f$ is semi-continuous at every point of $X$, then it is semi-continuous at $X$.

Similarly,  if
\be
f(x)\geq \lim_{n\rightarrow \infty}\sup f(x_n)\equiv \lim_{n\rightarrow \infty} \sup\{ f(x_k), k\geq n\}~,
\ee
then $f$ is upper-semi-continuous $x$.
\hfill{$\triangle$}
\end{definition}

 \begin {example} The function $f: [0,1]\rightarrow \bR$ with $f(x)=0$ for $x\not=0$ and $f(0)=1$ is not lower semi-continuous but it is upper semi-continuous at $x=0$.   \hfill{$\triangledown$}
\end {example}

  \begin {example}
  Consider $u_n=\sin (nt)/n$. Clearly as $n\rightarrow \infty$, $u_n\rightarrow 0$.  Thus $u=0$. On the other hand, the energy (\ref{energy})  of $u_n$ is
\be
E(u_n)=\frac{1}{2} \int^1_0 \cos^2(nt)=\frac{1}{2}~.
\ee
Thus $E(n_n)\rightarrow \frac{1}{2}$.  Clearly $E$ is not continuous because $\lim_{n\rightarrow +\infty} E(u_n)=1/2\not= E(\lim_{n\rightarrow +\infty} u_n)=E(u)=0$ but it is lower semi-continuous because $\lim_{n\rightarrow +\infty} E(u_n)>E(\lim_{n\rightarrow +\infty} u_n)$. The curve at the limit has lost all its oscillations but the energy of the oscillations of the curves survive the limit.\hfill{$\triangledown$}
\end {example}

\subsubsection{Weak convergence}

   To investigate the continuity  properties of the energy functional of curves, it is instructive to mention the concept of \emph{weak convergence} in Hilbert spaces and explore some of its aspects.   As an example, we shall apply the results in the next section  on the space of curves in $\bR^m$.

   Let $(\mathcal{H}, \langle\cdot, \cdot\rangle)$ be a separable Hilbert space, i.e. a Hilbert space with a countable basis. There two natural ways to define convergence of a sequence $(u_n)$ in the Hilbert space. One is \emph{ strong convergence}.
    \begin{definition}
     A sequence $(u_n)$ in $(\mathcal{H}, \langle\cdot, \cdot\rangle)$  converges \emph{strongly} to $u$, iff for every $\epsilon>0$, there is an $N$ such that $|| u_n-u||<\epsilon$ for all $n>N$, where $||\cdot||$ is the norm of  $(\mathcal{H}, \langle\cdot, \cdot\rangle)$. \hfill{$\triangle$}
     \end{definition}
    It turns out that in this definition of convergence \emph{ bounded} sequences, i.e. this means that there is a $C>0$ such that $||u_n||\leq C$ for all $n$,  may not have a convergent subsequence. In other words, bounded sequences in Hilbert spaces do not necessarily accumulate. This is not the case for finite dimensional vector spaces with the inner product topology.

Another notion of convergence in Hilbert spaces is \emph{ weak convergence}. This is defined as follows:

\begin{definition}
A the sequence $(u_n)$ in $(\mathcal{H}, \langle\cdot, \cdot\rangle)$ converges \emph{weakly} to $u$, iff for every $\epsilon>0$, there is an $N$ such that $|\langle u_n-u, v\rangle|<\epsilon$ for all $n>N$ and for every $v\in \mathcal{H}$.
\hfill{$\triangle$}
     \end{definition}

The main property of weak convergence, which  will be exploited later, is that \emph{ every bounded} sequence $(u_n)$ in the Hilbert space $(\mathcal{H}, \langle\cdot, \cdot\rangle)$, has a subsequence $(u_{n_k})$ that weakly  converges. One way to illustrate this is with an example.

\begin{example}  Consider the sequence $(u_n)$, where all the entries of the vector $u_n$ are zero apart from the n-th entry that it is $1$.  Therefore,
\be
u_n=(0, \dots, 0, 1, 0, \dots)~.
\ee
 Such a sequence is bounded in the usual Euclidean norm as
\be
| u_n|=1<+\infty~.
\ee
But, this sequence \emph{does not strongly converge} as it is not even Cauchy, i.e. for $n\not=m$
\be
|u_n-u_m|=\sqrt{1^2+1^2}=\sqrt2
\ee
which does not decrease for large enough $n$ and $m$.  However, $(u_n)$ \emph{weakly converges} to the $0$ vector. Indeed
\be
(u_n, v)=v_n~,
\ee
where $v_n$ is the n-th component of $v$. As $n$ becomes large the condition $|v|<+\infty$ forces the components of $v$ for large $n$ to become smaller and smaller tending to zero at the limit. Another way to see this is to observe that as we take $n$ large the non-zero component of $u_n$ moves to the right and disappears to infinity at the limit leaving the zero vector behind. \hfill{$\triangledown$}
\end{example}

\subsubsection{An instructive extremisation problem}\label{ap:flatenergy}

The purpose of this section is to prove that given two points $p$ and $q$ in the Euclidean space $\bR^m$, there is a curve $v$ that minimises the energy of all curves $u$ between $p$ and $q$.  This may appear as a rather too straightforward example. However, it exhibits all the complexity of the analogous question in Riemannian geometry without having to deal in addition with the non-linearity of the problem in the latter case. Because of this, it is very instructive to go systematically through over all the steps.

Let $\mC(p,q)$ be the space of  continuous curves between the points $p,q\in \bR^m$ and consider the space
\be
W^{1,2}([0,1], \bR^m; p,q)\equiv \{ \gamma\in  W^{1,2}([0,1], \bR^m)\cap \mC(p,q)| \gamma(0)=p, \gamma(1)=q\}~.
\ee
This is well defined as the uniform limit of sequences of curves preserves fixed points.

The energy functional of a curve $\gamma$
\be
E(\gamma)=\frac{1}{2} \int^1_0 dt\, |\vec{\dot \gamma}|^2~,
\ee
is well defined on  $W^{1,2}([0,1], \bR^m)$. From the definition, $W^{1,2}([0,1], \bR^m)$ is a Hilbert space, where the sequences of curves $(\gamma_n)$ are taken to converge uniformly while their velocities are taken to converge weakly.  Next define
\be
E_0=\inf_\gamma E(\gamma)~,~~~\gamma\in W^{1,2}([0,1], \bR^m; p,q)~.
\ee
Therefore, $E_0$ is the minimal value of the energy amongst all curves with fixed end points as indicated.

\begin{prop}\label{prop:flatenergy}
There is a curve $\lambda\in W^{1,2}([0,1], \bR^m; p,q)$ such that $E(\lambda)=E_0$.
\end{prop}
\begin{proof}
{\it Step 1:} To demonstrate of  such curve $\lambda$, let us pick the family $\mathcal{F}$ required by the Arzel\'a-Ascoli theorem \ref{th:aath}. For this consider the family of curves $\gamma$ in $W^{1,2}([0,1], \bR^m; p,q)$ that have  energy bounded by $C>E_0>0$.   Indeed given a curve $\gamma$ with bounded energy, one has that
 \be
 |\gamma(s)-\gamma(t)|^2\leq \Big(\int^s_t dr |\vec{\dot \gamma}(r)|\Big)^2\leq \int^s_t dr 1^2 \int^s_t  dr |\vec{\dot \gamma}(r)|^2\leq 2C  |s-t|~,
 \label{in1}
 \ee
 where the first inequality is the statement that the length of any curve between two points in $\bR^m$ is bigger or equal to the distance of the two points and the second inequality is an application of the Cauchy-Schwarz inequality.
 Thus,
\be
|\gamma(s)-\gamma(t)|\leq \sqrt{2C} |s-t|^{\frac{1}{2}}~,
\ee
and so all finite energy curves are $1/2$-H\"older.  This in turn implies that they are \emph{equicontinuous}. Indeed given $\epsilon>0$, set $\delta=\epsilon^2/2C$. Then
from the above inequality
\be
|\gamma(s)-\gamma(t)|\leq \sqrt{2C} |s-t|^{\frac{1}{2}}\leq \sqrt{2C} \delta^{\frac{1}{2}}=\epsilon~.
\ee
Furthermore, the second requirement of the Arzel\'a-Ascoli theorem holds as the length $L(\gamma)$ of all finite energy curves is bounded. Indeed from (\ref{in1}),
\be
L(\gamma)\equiv \int^1_0\, dr\, |\vec{\dot \gamma}(r)|\leq \sqrt{2 C}~.
\label{lebound}
\ee
Thus, all such curves lie in a ball of radius $\sqrt{2 C}$ with centre the point $p$. Moreover,  closed  balls are compact subsets of  $\bR^m$.  This implies that  the family of curves with finite energy satisfy all the requirements of Arzel\'a-Ascoli theorem.  This is turn implies that every sequence of such curves has a subsequence that converges.

Next, prepare a sequence of curves $(\gamma_n)$ with energy less that $C$ whose energies are gradually decreasing to $E_0$. Such a sequence exist as a virtue of the definition of the greatest lower bound that defines $E_0$. As such sequence satisfies the requirements of Arzel\'a-Ascoli theorem, there is a subsequence $(u_{n_k})$ that converges  to a curve $\lambda$.

{\it Step 2:} There are two problems with $\lambda$ to complete the proof of the statement.  First, the energy of $\lambda$ may not be defined because the sequence of velocities $(\vec{\dot \gamma}_{n_k})$ may not converge and second it is possible that, even if the sequence of velocities converge,  $E(\lambda)>E_0$. Let us deal with the first question. In order to define the energy of $\lambda$, we have to show that the sequence of  velocities $(\vec{\dot \gamma}_{n_k})$ converges as well. As the energy of the curves in the subsequence is bounded by $C$,
$(\vec{\dot \gamma}_{n_k})$ is a bounded sequence in a Hilbert space and so, upon choosing a further subsequence, it converges weakly to $\vec{\dot \lambda}$.  Thus,  $E(\lambda)$ is well defined.

{\it Step 3:} It remains to show that $E(\lambda)$ attains the minimum value $E_0$.
As $\lambda$ is another curve in $W^{1,2}([0,1], \bR^m; p,q)$, one has that
\be
E(\lambda)\geq E_0~,
\ee
because $E_0$ is the minimum energy of all curves from $p$ to $q$ including $\lambda$.

However,  below we shall show in lemma \ref{le:semienergy} that $E$ is semi-continuous. This gives the bound
\be
E(\lambda)\leq \lim_{n_k\rightarrow \infty} \inf E(\gamma_{n_k})=E_0~.
\ee
Therefore, $E(\lambda)=E_0$, which proves that there is a curve with minimal energy.  Of course this result for $\bR^m$ has been known for a very long time. It is also know that $\lambda$ is a straight line and so smooth. The novelty is that the same argument applies for general manifolds with hardly any substantial modification.
\end{proof}

Finally to complete the proof, let us demonstrate that the energy functional is lower semi-continuous. This results is proven in   the lemma below.
\begin{lemma}\label{le:semienergy}
The energy functional
\be
E(\gamma)=\frac{1}{2} \int dr\, |\vec{\dot \gamma}|^2~,
\ee
for $\gamma\in W^{1,2}([0,1], \bR^m)$ is lower semi-continuous.
\end{lemma}
\begin{proof}
  Suppose that there is a sequence $(\gamma_n)$ of curves that uniformly converges to a curve $\lambda$  such that $(\vec{\dot \gamma}_n)$ weakly converges to a velocity $\vec{\dot \lambda}$  -- we view the space of velocities as a Hilbert space.  This in particular implies  $\langle \vec{\dot \gamma}_n, \vec{\dot \lambda}\rangle\rightarrow ||\vec{\dot \lambda}||^2$.  Then,
\begin{align}
&0\leq ||\vec{\dot \lambda}-\vec{\dot \gamma}_n||^2= ||\vec{\dot \lambda}||^2+ ||\vec{\dot \gamma}_n||^2- 2\langle \vec{\dot \lambda},  \vec{\dot \gamma}_n\rangle \Longrightarrow \lim_{n\rightarrow \infty} \inf (||\vec{\dot \lambda}||^2+ ||\vec{\dot \gamma}_n||^2- 2\langle \vec{\dot \lambda}, \vec{\dot \gamma}_n\rangle)\geq 0
\cr
&
\Longrightarrow ||\vec{\dot \lambda}||^2\leq  \lim_{n\rightarrow \infty} \inf ||\vec{\dot \gamma}_n||^2 \Longrightarrow E(\lambda)\leq \lim_{n\rightarrow \infty} \inf E(\gamma_n)~,
\end{align}
where we have used that $(\vec{\dot \gamma}_n)$ weakly converges to $\vec{\dot \lambda}$ and $\lim_{n\rightarrow \infty} \inf \langle \vec{\dot \lambda}, \vec{\dot \gamma}_n\rangle=||\vec{\dot \lambda}||^2$ -- for the latter if all $\langle\dot \vec{\dot \gamma}_n, \vec{\dot \lambda}\rangle$ converge  to $||\vec{\dot \lambda}||^2$, after some $N$, so are the $\inf \langle\vec{\dot \gamma}_n, \vec{\dot \lambda}\rangle$.
\end{proof}

\begin{remark}
It is possible two curves to be very closed in the uniform topology but still have widely varying lengths  -- let us give an example. Indeed consider the curve $\lambda$ in $\bR^2$ given by the equation
\be
\lambda(t)=(x, y)=(t, 0)~,~~~ \text{where}~~~ t\in [0,1]~.
\ee
Then consider a subdivision of $[0,1]$ in $N$ equal intervals $I_k=[k/N, k+1/N]$, where $k=0, \dots N-1$.  In the first interval $I_0$,  join the point $(0,0)$ with $(1/2N, \epsilon)$ with straight line and similarly  the point $(1/2N, \epsilon)$ with the point $(1/N, 0)$, where $\epsilon>0$. Then a curve $\lambda_N$ can be constructed by periodically extending the curve constructed on the first interval $I_0$ to the rest of the intervals $I_k$.
For small $\epsilon$, the curves $\lambda$ and $\lambda_N$ are very close to each other in the $||\cdot ||_{\infty}$ norm because
\be
||\lambda_N-\lambda||_{\infty}=\sup_{t\in [0,1]} |\lambda_N(t)-\lambda(t)|=\epsilon~.
\ee
However, their lengths are widely different as $N$ becomes very large. Clearly, the length of $\lambda$ is $L(\lambda; [0,1])=1$, while an application of the Pythagorean theorem yields
\be
L(\lambda_N; [0,1])=2N \sqrt{\epsilon^2+\frac{1}{4 N^2}}=\sqrt{1+4 N^2 \epsilon^2}~,
\ee
for the length of $\lambda_N$.
If $N$ is much larger that $\epsilon$, the length of $\lambda_N$ is much larger than that of $\lambda$.  This is the case even though the two curves are close in the $||\cdot ||_{\infty}$ norm.
Thus, if we try to choose a  sequence of curves to approximate the length of $\lambda$ using the $||\cdot ||_{\infty}$ norm, we cannot exclude the presence of curves like $\lambda_N$ with $N$ large. This will make the lengths of the curves of the sequence to widely fluctuate from that of limit $\lambda$.  This is why in the construction above  the curves considered have bounded length as in (\ref{lebound}) and then a subsequence is chosen such that it eliminates from the calculation the curves like $\lambda_N$ with $N$ large. \hfill{$\Diamond$}
\end{remark}

\section{Minimal geodesics in Riemannian manifolds}\label{ap:minimal}

\subsection{The Hopf-Rinow Theorem}\label{ap:bb}

It is instructive in the context of the general relativity singularity theorems  to describe the Hopf-Rinow theorem. This,  in addition to other results,  provides a criterion for geodesic completeness for Riemannian manifolds. Also, some of the techniques used in the proof of the theorem are similar to  those employed  in the proof of singularity theorems but the setting is much more under control.  The detailed examination  reveals that there is a vast difference between the properties of geodesics  in Lorentzian and in Euclidean signature manifolds.

\begin{theorem}
Let $(\M,g)$ be a connected, without boundary, at least $C^2$-manifold. Then the following statements are equivalent

\begin{enumerate}

\item $(\M,g)$ is Cauchy complete or equivalently complete as a metric space with distance function $d_g$, i.e. every Cauchy sequence has a limit which is a point in $\M$.

\item $(\M,g)$ is geodesically complete, i.e. the geodesics can be extended to all affine  time.

\item The exponential map $\exp_p:  T_p\M\rightarrow \M$ is surjective.

\item Consider  balls of radius $R<\infty $ and with center at a point $p\in \M$, $B(p, R)\equiv \{x\in M\vert d_g(x, p)<R\}$.  The closure $\overline{B(p, R)}$ of $B(p, R)$  is a compact subset of $\M$.

\item Given two points $p,q\in \M$, there is, a non-necessarily unique,  curve $\gamma$ that starts from $p$ and ends at $q$ with length $L_g(\gamma; p,q)$ given by the distance $d_g(p,q)$ between the two points, $L_g(\gamma; p, q)= d(p,q)$.  Such a curve is a geodesic.
    \end{enumerate}
\end{theorem}

Before, we outline the proof of some of the statements made in the theorem, it is apparent that the issue of geodesic completeness in Riemannian manifolds has been resolved in the most elegant of ways. Cauchy completeness is a very natural assumption  -- this says that if  a sequence of points $(x_n)_{n\in \bN}$ in $\M$  has the property that for every $\epsilon>0$, there is a $N\in \bN$ such that for all $n,m\geq N$, $d_g(x_n, x_m)<\epsilon$, then $(x_n)_{n\in \bN}$   has a limit $x$ in $\M$, i.e. $d(x, x_n)\rightarrow 0$. It simply requires that all points should be included in the manifold that are limits of its points.  Therefore, the only reason that a geodesic cannot be extended beyond a certain affine time is whenever one or more points are removed from the manifold in such a way that the space becomes incomplete. Of course geodesic incompleteness can still arise provided one assumes that the manifold is not sufficiently smooth but this is not the case here or indeed in the proof of general relativity  singularity theorems.

\begin{proof}
The purpose here is to demonstrate some of the statements of this theorem that are most useful either in the investigation of the singularity theorems or in the illustration the difference in the behaviour of geodesics in Euclidean and Lorentzian signature manifolds.

$1\Longrightarrow 2$ Suppose that $\M$ is Cauchy complete and there is a geodesic $\gamma: (a,b)\rightarrow \M$ that cannot be extended beyond $b$.  First, $\gamma$ can be extended to $b$. For this, consider a sequence $(t_n)_{n\in \bN}$ of points in $(a,b)$ that converse to $b$.  As $\gamma$ is continuous, the sequence $(\gamma(t_n))_{n\in \bN}$ converges and it should converge to a point $p\in \M$ as $\M$ is (Cauchy) complete.  Moreover as $\gamma$ is at least $C^2$, the speed of $\gamma$, $|\vec{\dot \gamma}|_g$ is bounded and the acceleration of geodesics vanishes, the sequence of velocities $(\vec{\dot\gamma}(t_n))_{n\in\bN}$ also converges to a vector $\vec{v}\in T_p\M$. Given $(p, \vec{v})$ as initial conditions, there is a solution $\tilde \gamma$  of the geodesic equation on a small enough interval $(-\epsilon, \epsilon)$ with $\tilde \gamma(0)=p$ and $\vec{\dot{\tilde \gamma}}(0)=\vec{v}$.  Clearly the curve $\tilde \gamma\cup \gamma$ is an extension of $\gamma$, which is a contradiction. Thus $\gamma$ can be extended to the whole interval $(-\infty, \infty)$ as the extension beyond the point $a$ in the interval $(a,b)$ can be established in the same way.

There is a subtle point regarding the convergence of velocities mentioned above. For general $C^2$ curves $\gamma$ with bounded speeds $|\dot\gamma|<\infty$ defined on an interval $(a,b)$, the velocities do not converge because they can change too rapidly. Such behaviour is controlled by the acceleration -- for example a common condition is to assume that the velocities are absolutely continuous. However, geodesics by definition have zero acceleration and so their velocity does not vary rapidly along the curve-- in fact in normal coordinates the velocity of geodesics is a constant vector.

$4\Longrightarrow 5$  There are several ways to demonstrate that given two points $p,q\in \M$, there is  geodesic $\gamma$ that begins from $p$ and ends at $q$ such that has length $L_g(\gamma; p,q)$ equal to the distance $d_g(p,q)$ between the two points, $L_g(\gamma; p,q)=d_g(p,q)$. Here, we shall use the minimisation of the energy functional
 \be
 E_g(\gamma)\equiv \frac{1}{2}\int^1_0\, dt\, |\vec{\dot\gamma}|_g^2= \frac{1}{2}\int^1_0\, dt\, g_{\mu\nu}(x(t)) \dot x^\mu \dot x^\nu ~,
 \ee
 of curves with $\gamma(0)=p$ and $\gamma(1)=q$ to prove the result, where $x^\mu(t) \equiv x^\mu\circ \gamma(t)$ in local coordinates.  There are several reasons for this. One is the apparent observation that the critical points of this functional are geodesics. Apart from this, focusing on the curves that have finite energy puts a restriction on the curves and their velocities and leads to a compactness result based on the Arzel\'a-Ascoli theorem.   The space of continuous  curves $\mC(p,q)$ that start from $p$ and end at $q$ is too large of a set. The curves can vary wildly for the purpose of this problem and go far way from the region of $\M$ that contains the points $p$ and $q$. In addition, their velocities can  vary wildly too. So some a restriction is necessary to control their behaviour and also define the functional $E_g$ as well.
 The proof will follow the steps of proposition \ref{prop:flatenergy}.

 The energy functional is well-defined on the set $W^{1,2}([0,1], \M; p,q)$ of curves, see also appendix \ref{ap:flatenergy}. As it has been explained, these are curves that their velocities, which are defined almost everywhere,  are square integrable. Moreover, the topology of $W^{1,2}$ is such that convergence of curves is uniform on compact intervals and the velocities converge weakly.  The latter will be investigated in more detail below.

  As in proposition \ref{prop:flatenergy}, consider the family of curves with bounded energy $E_g(\gamma)\leq C$, where $C> d_g(p,q)$ is constant. Then, the Cauchy-Schwarz inequality implies that
 \be
 \Big(\int^t_s dr |\vec{\dot \gamma}|_g\Big)^2\leq \int^t_s dr 1^2 \int^t_s dr |\vec{\dot \gamma}|_g^2\leq 2 C |t-s|~.
 \label{lengthinex}
 \ee
  From this, one can conclude that
 \be
 d_g(\gamma(t), \gamma(s))\leq \vert\int^t_s dr |\vec{\dot \gamma}|_g\vert \leq \sqrt{2C} |t-s|^{\frac{1}{2}}~.
 \ee
Thus,  the family of finite energy curves bounded by $C$ are $1/2$-H\"older continuous. This in turn implies that they are equicontinuous. Indeed, given $\epsilon$ define $\delta= \frac{ \epsilon^2}{{2C}}$.  Then, for $|t-s|\leq \delta$,
\be
d(\gamma(t), \gamma(s))\leq \sqrt{2C} |t-s|^{\frac{1}{2}}\leq \sqrt{2C} \delta^{\frac{1}{2}}=\epsilon~.
\ee
Clearly $\delta$ does not depend on the individual curves of the finite energy family and so the family  is equicontinuous meeting the first condition of the Arzel\'a-Ascoli theorem \ref{th:aath}.

To apply the Arzel\'a-Ascoli theorem, one also has to show that all these curves lie in a compact subset of $\M$.  But we have already demonstrated that the length $L_g(\gamma; p.q)$ of curves in this family  is bounded
\be
L_g(\gamma;p,q)\leq \sqrt{2C}~,
\ee
see equation (\ref{lengthinex}).
Therefore, all of them lie in a ball $B(p, R)$ of radius  $R>\sqrt{2C}$ with $\gamma(0)=p$.  But from the Hopf-Rinow theorem $\overline{B(p, R)}\subset \M$ is compact and the second condition of the Arzel\'a-Ascoli theorem \ref{th:aath}  is met as well. Therefore, for every sequence $(\gamma_n)$ of finite energy curves, there is a subsequence $(\gamma_{n_k})$ that converges uniformly to a continuous curve $\lambda$.  This establishes the compactness properties of the family of finite energy curves.

As the closure of the family of finite energy curves is compact, if  the energy of the curves were continuous, it would exist a curve such that the energy functional attained a minimal value which would have been the distance $d_g(p,q)$. But, the energy functional $E_g$ is not continuous. However as we shall demonstrate, it is lower semi-continuous that suffices to prove the statement.

Before, we proceed to prove semi-continuity of the energy functional $E_g$ for curves on a manifold Riemannian $(\M,g)$, it is instructive to recall that we have already proven the statement for the energy functional for curves in $W^{1,2}([0,1],\bR^m)$.  In the topology given on $W^{1,2}([0,1], \M)$, a  sequence  of curves $(\gamma_n)$  converges  uniformly to curve $\lambda$ (in the distance topology)  while that of  velocities $(\vec{\dot \gamma}_n)$ weakly converges to $\vec{\dot \lambda}$. The latter property means that for  $L^2$ test functions $\chi$
\be
\int_0^1\, dt\, \chi(\vec{\dot \gamma}_n-\vec{\dot \lambda})=\int_0^1\, dt\, \chi_\mu(\vec{\dot \gamma}_n-\vec{\dot \lambda})^\mu \rightarrow 0~,
\ee
for large enough $n$.

%for the velocities requires some explanation. This is because although there is an inner product on the space of velocities, this depends on the curves, unlike for the case that $\M=\bR^m$, and therefore there is an underlying non-linear structure.

 To outline a proof that $E_g$ is semi-continuous, we shall work on a patch $(U, x)$ of $(\M,g)$ and at the end use the result of proposition \ref{prop:flatenergy}.  For this consider a sequence of curves $(\gamma_n)$  in $W^{1,2}([0,1], \M)$ that converges to a curve $\lambda$, there is a corresponding sequence of the components of the curves $(u_n(t)=x\circ \gamma_n(t))$ that converges uniformly in $t$  to $v=x\circ \lambda$ and a corresponding sequence of velocities $(\vec{\dot u}_n)$ that weakly converge to $\vec{\dot v}$.
  This implies that the components of the metric $(g_{\mu\nu})$ will uniformly  converge as
\be
(g_{\mu\nu}(u_n))\rightarrow (g_{\mu\nu}(v))~,
\ee
as $g$ is at least $C^2$.
Introduce a co-frame $e$ for $(g_{\mu\nu})=(\delta_{ab} e^a_\mu e^b_\nu)$, i.e. a ``square root'' for $(g_{\mu\nu})$, then again
\be
(e(u_n))\rightarrow (e(v))
\ee
uniformly in $t$.  Next define
\be
\vw^a_n= e^a_\mu(u_n) \vec{\dot u}_n^\mu~.
\ee
The task is to demonstrate that $\vec w_n$ converges weakly to $\vz^a= e^a_\mu (v) \vec{\dot v}^\mu$.  For this consider a test function $\phi$ in $L^2$, then
\be
\int \vec w_n \phi =\int \vec{\dot u}_n e(u_n) \phi=\int \vec{\dot u}^\mu_n e^a_\mu(u_n) \phi_a~.
\ee
Next,
\be
|| e(u_n)\phi-e(v)\phi||_{L^2}\leq ||e(u_n)-e(v)||_{L^\infty} ||\phi||_{L^2}\rightarrow  0~,
\ee
as $e(u_n)$ converges uniformly to $e(v)$ and $||\phi||_{L^2}$ is bounded.
Thus, $ (e(u_n)\phi)$ converges strongly in $L^2$ to $(e(v)\phi)$.
Moreover,
\be
\int \vec{\dot u}_n e(u_n) \phi=\int \vec{\dot u}_n e(u) \phi+\int \vec{\dot u}_n (e(u_n)- e(v)) \phi~.
\ee
As $e(v)\phi$ is another test function, the first term converges to $\int\vec{\dot v} e(v) \phi$ while the second term
\be
\Big|\int \vec{\dot u}_n (e(u_n)- e(v)) \phi\Big|\leq ||\dot u_n||_{L^2} ||(e(u_n)- e(v)) \phi||_{L^2}~.
\ee
The first factor in the right hand side is uniformly bounded and the second tends to zero. Thus $(\vec w_n)$  weakly converges to $\vec z$.  Note that the energy written in terms of $\vec w_n$ and $\vec z$ is proportional to the standard $L^2$ inner product on $\bR^m$. We have already shown that this is semi-continuous.

The statement has been shown for one patch. Working at a compact subset of the manifold and since the curve at the limit is compact, we can cover it with finite number of patches. Then we subdivide the interval $[0,1]$ as $\{[t_i, t_{i+1}]\}$ such that for each $[t_i, t_{i+1}]$ the curves take value at a single patch in $\M$. Then, we collect the local results for each patch  and prove the semi-continuity of the energy functional globally.  There is also another elegant proof of this using the Nash isometric embedding theorem and proposition \ref{prop:flatenergy}, see also appendix \ref{ap:dx} for a slightly different approach.

It remains to prove regularity. We have shown that there is a curve $\lambda$ that minimises the energy functional $E_g$ for all curves with fixed end points. We shall now demonstrate that all these curves are smooth. The minimisation has been achieved in the $W^{1,2}([0,1], \M; p,q)$ space and $\lambda$ must be a critical point of the functions $E_g$. Taking the first variation of $E_g$ by setting
\be
\gamma_\epsilon=\lambda+\epsilon \eta
\ee
where we require that the end points remain fixed and so take  $\eta(0)=\eta(1)=0$. The weak E\"uler-Lagrange equation requires that
\be
\int_0^1\, dt\,\big( g_{\mu\nu}(v(t))\, \vec{\dot v}^\mu {\dot \eta}^\nu+\frac{1}{2} \partial_\rho g_{\mu\nu}(v(t)) \vec{\dot v}^\mu \vec{\dot v}^\nu \eta^\rho\big)=0~,
\ee
where $v=x\circ \lambda$.
Define the momentum
\be
p_\rho= g_{\rho\nu}(v(t)) \vec{\dot v}^\nu~.
\ee
The weak derivative of $p$ is
\be
\int\, dt\, \dot p_\rho \eta^\rho\equiv -\int \, dt\, p_\rho {\dot \eta}^\rho=\frac{1}{2} \int\, dt\,  \partial_\rho g_{\mu\nu}(v(t)) \vec{\dot v}^\mu \vec{\dot v}^\nu \eta^\rho~.
\ee
The metric is at least $C^2$ and  the curve is in $\lambda\in W^{1,2}$ and so continuous. Therefore, the coefficients $\partial_\rho g_{\mu\nu}$ are bounded along $\lambda$. As the velocities $\vec{\dot \lambda}$ are in $L^2$, their product $\vec{\dot v}^\mu \vec{\dot v}^\nu $ is in $L^1$ as a result of the H\"older inequality $|| \vec{\dot v}^\mu \vec{\dot v}^\nu ||_{L^1}\leq ||\vec{\dot v}^\mu||_{L^2} ||\vec{\dot v}^\nu ||_{L^2}$.  Thus $\partial_\rho g_{\mu\nu}(v(t)) \vec{\dot v}^\mu \vec{\dot v}^\nu$ is in $L^1$. Thus,  $\dot p$ is represented by an $L^1$ function, i.e. the weak derivative of $p$ is in $L^1$. As a result $p$ is in $W^{1,1}$,  i.e. $p$ is a collection of absolute continuous functions on  $[0,1]$.  As $p$ is continuous, then $ \gamma$ is $C^1$. This also implies from the geodesic equation that $\ddot u$ is also continuous and so $\gamma$ must be in $C^2$. Continuing in the same way, one concludes that $\gamma$ is $C^\infty$ provided that the metric is smooth as well.
\end{proof}

 \begin{example}To give an example of a curve that cannot be extended beyond a point utilising the non-vanishing tangent vector at that point, consider the curve $\gamma: (0,a)\rightarrow \bR$ with $\gamma(t)= \exp(-1/t)$. Clearly $\gamma(0)=0$ for $t_+\rightarrow 0$. But also $\vec{\dot \gamma}(0)=0$. In fact,the acceleration and all other derivatives vanish as well in the limit $t_+\rightarrow 0$, i.e. as $0$ is approached from positive values of $t$.  So although the curve can be extended to $[0,a)$, it cannot be extended to $t<0$ as for $t_-\rightarrow 0$ the function blows up. \hfill{$\triangledown$}
 \end{example}

\begin{remark}
One may wonder why the argument used to prove the statement 2 in the Hopf-Renow theorem regarding the extension of geodesics cannot be used in the context of Lorentzian geometry. The issue is with the extension of the velocity of such geodesics. To explain this consider such a geodesics in the Lorentzian manifold $\M$ with metric $g$, $\gamma: (a,b)\rightarrow \M$ and attempt to  extend it beyond $b$.  First, $\gamma$ can be extended to $b$. For this, consider a sequence $(t_n)_{n\in \bN}$ of points in $(a,b)$ that converse to $b$,  As $\gamma$ is continuous, the sequence $(\gamma(t_n))_{n\in \bN}$ converges and it should converge to a point $p\in \M$, assuming that $\M$ is (Cauchy) complete.  For the velocities, first suppose  that $\gamma$ is a null geodesics. In such a case,  the velocities of the sequence satisfy the null condition  $g(\vec{\dot\gamma}(t_n), \vec{\dot\gamma}(t_n))=0$. This condition does not constrain the velocities sufficiently and it perfectly possible that in the limit $\lim_{n\rightarrow \infty}\vec{\dot\gamma}(t_n)=0$ while the null condition is maintained. If the velocity at the limit vanishes, the geodesic equation cannot be used to find a solution in a neighbourhood of $b$ and the extension argument breaks down. The similar argument can apply to time-like geodesics. In this case the components of the  velocities of the sequence $(\gamma(t_n))_{n\in \bN}$ can diverge at the limit, i.e. the length of these velocities with respect to a Riemannian metric $g_e$ on $\M$ diverges at the limit, while $g(\vec{\dot\gamma}(t_n), \vec{\dot\gamma}(t_n))$ remains constant. As there can be no well defined velocity as the limit at $b$ is reached, again the geodesic equation cannot be used to extend the curve in the neighbourhood of $b$.

One may argue to consider a Riemannian metric $g_e$ on $\M$ and explore a possible extension of the geodesics of the Lorentzian signature metric $g$ from the perspective of $g_e$. In such a case, the Lorentzian geodesic equation on $\M$ is a second order differential equation without any particular special properties. Such an equation admit solutions for a small interval $[a, b)$ of their parameter given some initial conditions for the position and velocity  at the origin $a$. However, there is not necessarily an extension beyond $b$. What goes wrong is that the acceleration, $\nabla^{g_e}\vec{\dot \gamma}_n$, of the Lorentzian geodesic $\gamma$ as $t_n\rightarrow b$ measured by the Riemannian metric $g_e$, which does not vanish, can vary widely prohibiting a well defined limit for the velocity at $b$. At the end, the root of the difficulty is that the space of velocities at each point on the geodesic is compact for Riemannian geometries but not compact for causal Lorentzian ones.\hfill{$\Diamond$}
\end{remark}

\subsection{Conjugate points and minimal geodesics}\label{ap:bb2}

The implications of the presence of conjugate points in a geodesic are key in the proof of singularity theorems. There is an analogue of these results in Riemannian geometry described in the theorem below.

\begin{theorem}\label{th:lengthmin}
Let $(\M, g)$ be a Riemannian manifold and $\gamma: [0, a]\rightarrow \M$ be a geodesic parameterised by its length $t$. If $\gamma$ has a conjugate point between $[0,a]$ at $t=t_0$, then $\gamma$ is not length minimising between $\gamma(0)$ and $\gamma(a)$.
\end{theorem}
\begin{proof}
The strategy of the proof is to demonstrate that $\gamma$ between $\gamma(0)$ and $\gamma(a)$ is not a minimum of the energy functional.  It is just a saddle point.
If $\gamma$ were a minimum, then the second variation of the energy functional, $\delta^2 E$ must be positive definite in every direction. However, as we shall demonstrate, if the geodesic $\gamma$ has a conjugate point between $\gamma(0)$ and $\gamma(a)$, the $\delta^2 E$  exhibits a negative direction.  We shall present two proofs for this one of which will be distributional and the other smooth.

\emph{A distributional proof:} As we have assumed that there is a conjugate point at $\gamma(t_0)$, there is a Jacobi field $\vz$ such that
\be
\vz_{\gamma(0)}=\vz_{\gamma_(t_0)}=0~.
\ee
However, $\nabla_s\vz_{\gamma_(t_0)}\not=0$ as otherwise $\vz=0$ which is a contradiction as we have assumed that $\gamma$ exhibits  conjugate points.

Next we extend $\vz$ to the whole geodesic $\gamma$ as follows
\be
\vy(t)=\vz(t)~,~~~0\leq t\leq t_0~;~~~\vy(t)=0~,~~~t_0<t\leq a~.
\ee
The extension $\vy$ is continuous because $\vz$ vanishes as $t=t_0$.  Moreover, $\vy$ satisfies the same equation as the Jacobi vector field $\vz$, i.e.
\be
\nabla^2_{\vec{\dot\gamma}}\vy+ R(\vy, \vec{\dot\gamma}) \vec{\dot\gamma}=0~.
\ee

The second variation of the energy functional evaluated on $\gamma$ is given by
\begin{align}
I(\vv,\vw)&=\int^a_0 dt \Big(g(\nabla_{\vec{\dot \gamma}}\vv, \nabla_{\vec{\dot \gamma}}\vw)-g(R(\vv, \vec{\dot\gamma}) \vec{\dot\gamma}, \vw)\Big)
\cr
&= - \int^a_0 dt \Big(g(\vv, \nabla^2_{\vec{\dot \gamma}}\vw)+g(R(\vv, \vec{\dot\gamma}) \vec{\dot\gamma}, \vw)\Big)+ (g(\vv, \nabla_{\vec{\dot \gamma}}\vw)\Big|^a_0~,
\label{a:svf}
\end{align}
where the second line follows after integrating by parts.

To find a negative direction set
\be
\vv_\epsilon=\vy+\epsilon \vl~,
\ee
where $\epsilon$ is a parameter and $\vl$ is a vector field on $\gamma$.  Then
\be
I(\vv_\epsilon, \vv_\epsilon)=I(\vy, \vy)+2 \epsilon I(\vy, \vl)+\epsilon^2 I(\vl,\vl)~.
\ee
Let us evaluate each term separately. As $\vy$ vanishes in the interval $[t_0, a]$,
\be
I(\vy,\vy)=I(\vz,\vz)=0~.
\ee
This follows from the second line in (\ref{a:svf}) upon using the equation for the Jacobi field $\vz$. The surface term vanishes as well because $\vz$ vanishes both at $t=0$ and $t=t_0$.

Next to evaluate $I(\vy, \vl)$, we use the first line in (\ref{a:svf}) and split the integral between $[0, t_0]$ and $[t_0, a]$. The  latter integral must vanish as $\vy=0$. To evaluate the rest, we use the second line of (\ref{a:svf}) and as $\vy$ satisfies the equation of a Jacobi vector field, the only contribution comes from the boundary term that reads
\be
I(\vy, \vl)=g(\nabla_{\vec{\dot\gamma}}\vz, \vl)|_{t=t_0}~.
\ee
In fact, one can choose $\vl$ such that $\vl|_{t=t_0}=\nabla_{\vec{\dot\gamma}}\vz|_{t=t_0}$ in which case
\be
I(\vy, \vl)=g(\nabla_{\vec{\dot\gamma}}\vz, \nabla_{\vec{\dot\gamma}}\vz)|_{t=t_0}>0~.
\ee
As $I(\vz, \vz)$ is bounded, there is always an $\epsilon$ such that $I(\vv_\epsilon, \vv_\epsilon)<0$ and so $\gamma$ is a saddle point.

\emph{A smooth proof:}
The proof given above has a discontinuity   in the derivative of $\vy$ at $t=t_0$.  However, this can be smoothen out as follows:
First, define a smooth vector field $\vy_\epsilon$ as
\be
\vy_\epsilon(t)=\vz(t)~,~~~t\in [0, t_0-\epsilon]~; \vy_\epsilon(t)= A \epsilon \phi(\frac{t-t_0}{\epsilon})~,~~~t\in [t_0-\epsilon, t_0+\epsilon]~; \vy_\epsilon(t)=0~,~~t\in [t+\epsilon, a]~,
\ee
where $A=\nabla_{\vec{\dot\gamma}}\vz|_{t=t_0}$ and the smoothing function $\phi$ has been chosen such that
\be
\phi(-1)=-1~,~~~\phi'(-1)=1~,~~~\phi(1)=0~,~~~\phi'(1)=0~.
\ee
The choice of conditions on $\phi$ is justified from the way that $\vz$ approaches $t_0$. Observe that $\phi$ is of order $\epsilon$ while $\phi'$ is of order $1$ and
\be
\nabla_{\vec{\dot\gamma}}\vy_\epsilon=A \phi'(\frac{t-t_0}{\epsilon})
\ee
for $t\in [t_0-\epsilon, t_0+\epsilon]$.

Next let as evaluate $I(\vy_\epsilon, \vy_\epsilon)$ by splitting the integration in the three intervals $[0, t_0-\epsilon]$, $[t_0-\epsilon, t_0+\epsilon]$ and $[t+\epsilon, a]$. Indeed using the equation for the Jacobi field and the second line in (\ref{a:svf}), we find that
\be
I_{[0, t_0-\epsilon]}(\vy_\epsilon, \vy_\epsilon)=I_{[0, t_0-\epsilon]}(\vz, \vz)=g(\nabla_{\vec{\dot\gamma}}\vz, \vz)|_{t=t_0-\epsilon}=-\epsilon g(A,A)+\mathcal{O}(\epsilon^2)~.
\ee
Next
\be
I_{[t_0-\epsilon, t_0+\epsilon]}(\vy_\epsilon, \vy_\epsilon)=\epsilon g(A,A)\int^1_{-1} ds (\phi')^2 +\mathcal{O}(\epsilon^3)~,
\ee
where notice that the curvature term in the first line of (\ref{a:svf}) is of order $\epsilon^3$  and we have used that the derivative of $\vy_\epsilon$ is of order $1$.
Finally,
\be
I_{[t_0+\epsilon, a]}(\vy_\epsilon, \vy_\epsilon)=0~,
\ee
as $\vy_\epsilon=0$ in this interval.

An optimal choice of smoothing function is $\phi(s)=-(1-s)/2$ -- it satisfies the stated criteria. With this choice, one finds that
\begin{align}
I(\vy_\epsilon, \vy_\epsilon)&=I_{[0, t_0-\epsilon]}(\vy_\epsilon, \vy_\epsilon)+I_{[t_0-\epsilon, t_0+\epsilon]}(\vy_\epsilon, \vy_\epsilon)=-\epsilon g(A,A)+\frac{1}{2} \epsilon g(A,A)+\mathcal{O}(\epsilon^2)
\cr
&= -\frac{1}{2} \epsilon g(A,A)+\mathcal{O}(\epsilon^2)~.
\end{align}
Thus, for small enough $\epsilon$, $I(\vy_\epsilon, \vy_\epsilon)<0$ and so the geodesic $\gamma$ is a saddle point instead of a minimum.
\end{proof}

\begin{remark}
The calculation described above for Riemannian manifolds also applies for Lorentzian manifolds and for time-like geodesics with conjugate points. The only difference is that now the geodesics are those curves that extremise the  proper time between two spacetime points instead of minimising the distance. The existence of a saddle point in the Lorentzian case requires the existence of positive directions in the Hessian of the proper time  functional instead of negative directions. This is taken care by an overall sign that appears in the calculation described in theorem \ref{th:conjugategeodesic1}. Otherwise, the calculation is identical. For the null geodesics with conjugate points, a similar calculation  applies. However, in addition one  has to demonstrate that the deformed curve is causal, see theorem \ref{th:nulldeformation}. \hfill{$\Diamond$}
\end{remark}

\section{Existence of causal  curves with maximal proper time} \label{ap:dx}

Here, we shall sketch an argument using the Arzel\'a-Ascoli theorem to demonstrate the existence of  curves whose proper time is maximal between two points in a  Lorentzian manifold. Let $(\M, g)$ be a spacetime, $p,q\in \M$ with $q\in I^+(p)$  and $g_e$ be  an auxiliary Riemannian metric on $\M$. Define
\be
T=\sup_{\gamma} L_g(\gamma; p.q)
\ee
where $\gamma$ are the absolutely continuous future directed causal curves from $p$ to $q$, i.e. $\gamma\in W^1(\M;[0,1], p,q)$.

Next let us make the following assumptions: (1) There is a family $\mathcal{F}$ of curves in $W^1(\M;[0,1], p,q)$ that contains at least a  maximizing sequence. (2) There is a parameterisation, e.g. in terms of $g_e$ length, that the speeds of the causal curves in $\mathcal{F}$ are uniformly bounded with respect to the $g_e$ metric. (3) The images of curves of the family are contained in a compact subset $\K$ of the spacetime.  Under these conditions, we shall demonstrate using the Arzel\'a-Ascoli theorem that there is a curve $\lambda$ such that $T=L_g(\lambda; p,q)$.

 First using proposition \ref{prop:gepropertime}, there is a constant $C_K>0$ such that
 \be
 L_g(\gamma;[0,1])\leq C_K  L_{g_e}(\gamma;[0,1])<+\infty~,
 \ee
 as all the speeds of $\gamma\in  L_g(\gamma;[0,1])$ are uniformly bounded and $\mathcal{F}$ has a compact containment. As $\mathcal{F}$ contains a maximizing subsequence $T<+\infty$.

 The family $\mathcal{F}$ is equicontinuous as the speeds of the curves are uniformly bounded, i.e.
 \be
 d_{g_e}(\gamma(t), \gamma(s))\leq \int_t^s |\vec{\dot\gamma}|_{g_e}\leq C_s |s-t|~,
 \ee
 where $C_s$ is the uniform bound of the speeds of the family.

 Since the images of curves in $\mathcal{F}$ are contained in a compact set $\K$, all the condition of the Arzel\'a-Ascoli theorem are met and every sequence $(\gamma_n)$ in $\mathcal{F}$ has a subsequence $(\gamma_{n_k})$ that converges uniformly to a curve $\lambda$.  In particular, if $(\gamma_n)$ is a maximising sequence, i.e. a sequence such that $L_g(\gamma_n; p,q)$ converges  to $T$, it will admit a subsequence that uniformly converges to a curve $\lambda$.  (Because of this, $\lambda$ will have end points $p$ and $q$.)
There is a subtle point here to prove that $\lambda$ is causal.  This is indeed the case but we shall not explore this further.

Although, we have demonstrated that every sequence $(\gamma_n)$ of curves has a convergent subsequence, the associated sequence of velocities $(\vec{\dot\gamma}_{n_k})$ may not be under control. However, the sequence $(\vec{\dot\gamma}_n)$ is bounded in $L^1$, i.e. the integral of their speeds is bounded. (In fact the sequence is bounded in $L^\infty$, as their speeds are bounded. But it is better to work in $L^1$ as the dual of $L^\infty$ is rather complicated while the dual of $L^1$ is $L^\infty$ and moreover $L^\infty[0,1]\subset L^1[0,1]$.)  Thus, there is a subsequence of $(\vec{\dot\gamma}_n)$ that converges weakly in $L^1$.

It remains to show the the Lorentzian length functional is upper semi-continuous.  We have already demonstrated that  $
L_g(\gamma, \vec{\dot\gamma}; [0,1])<+\infty$.
 To continue consider a sequence $(\gamma_n)$ of curves that uniformly converge to $\lambda$ with the sequence of velocities $(\vec{\dot\gamma}_n)$ to weakly converge in $L^1$. Next, we shall calculate
 \be
 L_g(\gamma_n, \vec{\dot\gamma}_n)- L_g(\lambda, \vec{\dot\lambda})=P_n+Q_n~,~~~P_n=L_g(\gamma_n, \vec{\dot\gamma}_n)-L_g(\lambda, \vec{\dot\gamma}_n)~,~~~Q_n=L_g(\lambda, \vec{\dot\gamma}_n)-L_g(\lambda, \vec{\dot\lambda})~.
 \ee
 In particular, we shall demonstrate first that $P_n\rightarrow 0$ as $\gamma_n\rightarrow \lambda$. For this set
 \be
 F(x, \vu)=\sqrt{-g_x(\vu,\vu)}~,
 \ee
 and observe that
 \be
 |g_x(\vu, \vu)-g_y(\vu, \vu)|\leq L^2_{\K} |x-y|_e |\vu|_e^2~,
 \ee
 where we have used that $g_x(\vu,\vu)$ is a smooth function of $x$, so Lipschitz,  and $L_{\K}>0$ is a uniform bound over the compact set $K$. (Instead of the Euclidean distance $|x-y|_e $, we could have used $d_{g_e}(x,y)$ but locally these definitions of the distance are equivalent for convergence considerations.)  Taking the square root of this expression and using the inequality $|{a}-{b}|\leq \sqrt{|a^2-b^2|}$, $a,b\geq 0$, we find that
 \be
 |F(x, \vu)-F(y,\vu)|\leq L_K |x-y|_e^{\frac{1}{2}} |\vu|_e
 \ee
 Thus
 \be
 |F(\gamma_n, \vec{\dot\gamma}_n)-F(\lambda, \vec{\dot\gamma}_n)|\leq L_K |\gamma_n-\lambda|_e^{\frac{1}{2}} |\vec{\dot\gamma}_n|_e
 \ee
 which vanishes as $(\gamma_n)$ converges to $\lambda$ and the speeds of the curves are bounded.

 It remains to calculate $Q_n$. For this consider the inequality
 \be
 F(x, \vu)\leq F(x, \vw)-\frac{g_x(\vw, \vu-\vw)}{F(x,\vw)}
 \ee
where both $\vw$ and $\vu$ are future directed and $\vw$ is time-like. This is a consequence of the Lorentzian version of the Cauchy-Schwarz inequality for future directed causal vectors. Setting $\vw=\vec{\dot\lambda}$,  $\vu=\vec{\dot\gamma}_n$ and $x=\lambda$, observe that the second term in the above inequality, after integration, converges to zero because $-\frac{g_x(\vw,\cdot
)}{F(x,\vw)}$ is in $L^\infty$, which is the dual of $L^1$, and the  velocities weakly converge in $L^1$. Thus we find that
\be
\lim_{n\rightarrow \infty} \sup L_g(\gamma_n, \vec{\dot\gamma}_n; [0,1]) \leq L_g(\lambda, \vec{\dot\lambda}; [0,1])~,
\ee
which proves the upper semi-continuity of the proper time functional.  Then it follows  using the standard argument that $T=L_g(\lambda;p,q)$.

\begin{remark}
Let us apply the above theorem to golbally hyperbolic spacetimes. Let $(\M, g)$ be a globally hyperbolic spacetime and $p,q\in \M$ with
$q\in I^+(p)$. The casual diamonds are compact in $\M$ and so we can take $\K=J^-(q)\cap J^+(p)$.  Therefore all the causal curves from $p$ to $q$ lie in a compact subset of the spacetime.

Next, let us consider the parameterisation of these causal curves. The globally hyperbolic spacetimes admit a temporal function
\be
t:\, \M\rightarrow \bR~,
\ee
 such that $dt(\vv)>0$ for all future directed time-like vector fields $\vv$.
Therefore, the function $t$ is strictly increasing along forward directed time-like curves $\gamma_n$. Suppose that $\gamma_n: [a_n, b_n]\rightarrow \M$ with $\gamma_n(a_n)=p$ and $\gamma_n(b_n)=q$.  Clearly, $t\circ \gamma_n (a_n)=t(p)$ and
$t\circ \gamma_n (b_n)=t(q)$. Moreover, $\mathrm{Im} t=[t(p), t(q)]$ with $t(p)< t(q)$.  Define now a new parameter $u$ as
\be
u=\frac{t(\gamma_n(s))-t(p)}{t(q)-t(p)}~,~~t(\gamma_n(s))>t(p)~,~~s>a_n~,
\ee
where $s$ is the original parameter of the curves and $u\in [0,1]$. Let us denote the curves $\gamma_n$ with the new parameter $u$ as $\tilde \gamma_n$. Then
\be
t(\tilde \gamma_n(u))=t(p)+ u (t(q)-t(p))~,
\ee
and so
\be
dt(\dot{\tilde \gamma}_n)=t(q)-t(p)~.
\ee
Next introduce Riemannian metric $g_e$ and consider $dt(\vv)$ as a function on $S(TK)=\{(x, \vv)| x\in \K ~\text{and}~|\vv|_{g_e}=1\}$, where $\K=J^+(p)\cap J^-(q)$. As $\M$ is globally hyperbolic $S(TK)$ is compact. Moreover, $dt(\vv)$ is strictly positive defined over a compact set. Therefore, it has a minimum $c>0$, $dt(\vv)>c$. This gives
\be
c |\dot{\tilde \gamma}_n|_{g_e} \leq dt (\dot{\tilde \gamma}_n)= t(q)-t(p)~.
\ee
Thus, the speeds of $\gamma_n$ are uniformly bounded as
\be
|\dot{\tilde \gamma}_n|_{g_e}\leq c^{-1} (t(q)-t(p))=C_s~,
\ee
over the same interval $u\in [0,1]$, i.e.  the constant $C_s$ does not depend on $n$.
%Thus
%\be
%d_{g_e}(\tilde \gamma_n(u), \tilde \gamma_n(r)\leq \int^s_r |\dot{\tilde \gamma}_n|_{g_e}\leq  C_s |u-r|
%\ee
%and the sequence is equicontinuous as required by the Arzel\'a-Ascoli theorem.
It follows from the discussion above in this appendix that there is a curve $\lambda$ that maximises the proper time between $p$ and $q$.  This is an alternative proof that  globally hyperbolic  spacetime always admit geodesics that maximise the proper time between any causally related points.\hfill{$\Diamond$}
\end{remark}

\end{appendices}

 \vskip0.5cm
%\noindent{\it Data availability statement}

%All data generated or analyzed during this study are included in this published article.

\vskip0.5cm
%\noindent{\it Conflict of interest statement}

%The author has no relevant financial or non-financial interests to disclose.

\vskip0.5cm

 \bibliographystyle{unsrt}

\end{document}